\documentclass[11pt,a4paper]{article}
\pdfoutput=1
\usepackage{amssymb,amsmath,amsfonts, mathtools, mathrsfs}
\usepackage[utf8]{inputenc} 
\usepackage[dvipsnames]{xcolor}
\usepackage{stmaryrd}
\SetSymbolFont{stmry}{bold}{U}{stmry}{m}{n}

\newif\ifnatbibsort\natbibsorttrue

\relax

\ifnatbibsort\RequirePackage[numbers,sort&compress]{natbib}\else\RequirePackage[numbers,compress]{natbib}\fi
\RequirePackage[colorlinks=true
,urlcolor=blue
,anchorcolor=blue
,citecolor=blue
,filecolor=blue
,linkcolor=blue
,menucolor=blue
,pagecolor=blue
,linktocpage=true
,pdfproducer=medialab
,pdfa=true
]{hyperref}

\usepackage{graphicx}
\usepackage{caption}
\usepackage{tensor}
\usepackage{subfigure}
\usepackage{enumerate}
\usepackage{dsfont}
\usepackage{verbatim}
\usepackage{amsfonts,euscript,amssymb,stmaryrd,braket}
\usepackage{tikz}
\usetikzlibrary{arrows,decorations.markings,patterns}
\usepackage{slashed}

\def\clock{{\count0=\time
		\divide\count0 60
		\ifnum\count0<10 0\fi\the\count0
		\multiply\count0 -60 \advance\count0 \time
		:\ifnum\count0<10 0\fi \the\count0
}}
\newcommand{\timestamp}{{\small\vbox{\hbox{\tt\jobname.tex}
			\hbox{\the\day/\the\month/\the\year, \clock}}}}

\newcommand{\bea}{\begin{eqnarray}}
\newcommand{\eea}{\end{eqnarray}}

\newcommand{\be}{\begin{equation}}
\newcommand{\ee}{\end{equation}}

\makeatletter
\let\old@startsection=\@startsection
\let\oldl@section=\l@section
\renewcommand{\@startsection}[6]{\old@startsection{#1}{#2}{#3}{#4}{#5}{#6\mathversion{bold}}}
\renewcommand{\l@section}[2]{\oldl@section{\mathversion{bold}#1}{#2}}
\makeatother

\makeatother

\numberwithin{equation}{section}

\usepackage{color}

\def \RR {{\mathbb R}}

\def\ri {{\rm i}}
\def\rd {{\rm d}}
\def\e {{\rm e}}

\begin{document}
	\renewcommand{\thefootnote}{\arabic{footnote}}

	\overfullrule=0pt
	\parskip=2pt
	\parindent=12pt
	\headheight=0in \headsep=0in \topmargin=0in \oddsidemargin=0in

	\vspace{ -3cm} \thispagestyle{empty} \vspace{-1cm}
	\begin{flushright} 
		\footnotesize
		\textcolor{red}{\phantom{print-report}}
	\end{flushright}

\begin{center}
	\vspace{.0cm}

	{\Large\bf \mathversion{bold}
	Entanglement Hamiltonian of two disjoint blocks
	}
	\\
	\vspace{.25cm}
	\noindent
	{\Large\bf \mathversion{bold}
	in the harmonic chain}

	\vskip  0.8cm
	{
		Francesco Gentile,
		Andrei Rotaru
		and Erik Tonni
	}
	\vskip  1.cm
	
	\small
	{\em
		SISSA and INFN Sezione di Trieste, via Bonomea 265, 34136, Trieste, Italy 
	}
	\normalsize

\end{center}

\graphicspath{{Figures/}}
\vspace{0.3cm}
\begin{abstract} 

We study the entanglement Hamiltonian of two disjoint blocks in the harmonic chain on the line and in its ground state. 
In the regime of large mass, 
the non vanishing terms are only the on-site and the nearest-neighbour ones.
Analytic expressions are obtained for their profiles, which are written in terms of piecewise linear functions 
that can be discontinuous and display sharp transitions as the separation between the blocks changes. 
In the regime of vanishing mass,
where the matrices characterising the entanglement Hamiltonian contain couplings at all distances, 
we explore the location of the subdominant terms 
and some combinations of matrix elements that are useful for the continuum limit,
comparing the results with the corresponding ones for the free chiral current.
The single-particle entanglement spectra of these entanglement Hamiltonians are also investigated. 

\end{abstract}

\newpage
\tableofcontents

\newpage
\section{Introduction}


The entanglement Hamiltonian is an important operator to investigate in order to understand the bipartite entanglement in quantum systems
\cite{Haag:1992hx, EislerPeschel:2009review, Casini:2009sr}.
Consider a quantum system in a state characterised by the density matrix $\rho$
and whose space is bipartite into two regions $A$ and $B$.
Assuming that its Hilbert space can be factorised accordingly, i.e. $\mathcal{H} = \mathcal{H} _A \otimes \mathcal{H}_B$,
the reduced density matrix of the subsystem $A$ is defined as $\rho_A \equiv \textrm{Tr}_{\mathcal{H}_B} \rho$,
which is a positive, non-singular and self adjoint operator, 
hence it can be written as $\rho_A \equiv \e^{- K_A}$, 
where $K_A$ is, by definition, the entanglement Hamiltonian for the subsystem $A$.
The von Neumann entropy of $\rho_A$ gives the entanglement entropy $S_A \equiv - \,\textrm{Tr}_{\mathcal{H}_A} \big(\rho_A \log \rho_A \big)$.
When the system is in a pure state (e.g. its ground state), 
$\rho$ is the projector on this state and the Schmidt decomposition tells us that $S_A = S_B$.
It is insightful to study the operator $K_A$ both in lattice models and in quantum field theories, 
exploring also the continuum limit procedure that relates them.


The first crucial result for the entanglement Hamiltonians has been obtained by Bisognano and Wichmann
\cite{Bisognano:1975ih, Bisognano:1976za} within the context of algebraic quantum field theory \cite{Haag:1992hx}.
Considering a relativistic quantum field theory in the ground state, 
whose  space is bipartite into two equal parts by a hyperplane,
they found that $K_A$ is  the generator of the boosts along the direction identifying the bipartition. 
As for the lattice models, 
in a one-dimensional infinite harmonic chain in its ground state, 
the entanglement Hamiltonian of the half chain has been studied through the corner transfer matrix in \cite{Peschel:1991xeo,Peschel:1999xeo},
and an operator given by the energy density multiplied by a linear weight function has been found, 
in agreement with the result of Bisognano and Wichmann in the continuum limit.


Entanglement Hamiltonians in quantum field theories have been mainly explored for conformal field theories (CFT)
because the symmetry allows to study various cases by combining the result of Bisognano and Wichmann and the proper conformal mapping. 
For instance, 
$K_A$ of the sphere in Minkowski spacetime when the entire CFT is in its ground state has been studied \cite{Hislop:1981uh, Casini:2011kv},
for a generic number of spacetime dimensions. 
Instead, since in two-dimensional CFT the conformal symmetry is infinite dimensional,
more cases can be explored has been studied in this $1+1$ dimensional spacetimes, 
like e.g. the case of the interval on the circle when the system is in the ground state 
and the case of the interval on the line at finite temperature \cite{Wong:2013gua, Cardy:2016fqc}.
In all these two-dimensional CFT setups, the entanglement Hamiltonian takes the form 
$K_A = \int_a^b \beta(x)\, T_{tt}(x) \,\rd x$;
i.e. it is a local operator given by the integral over the interval $A=[a,b]$ of the energy density $T_{tt}$ multiplied by the proper weight function $\beta(x)$.
Some of these CFT results have been obtained also through lattice computations, mainly in free models,
where the properties of Gaussian states can be employed
\cite{Arias:2016nip, Eisler:2017cqi, Eisler:2018ugn, DiGiulio:2019cxv, Javerzat:2021hxt, Eisler:2019rnr}.
The continuum limit procedure is highly non trivial because 
the quadratic operator to consider is inhomogeneous and contains long-range hopping terms. 
In particular, the continuum limit procedure involves all the diagonals of the matrices characterising the quadratic operator.
%


In order to gain insights into the structure of entanglement Hamiltonians, 
it is important  to investigate free massive quantum systems in one spatial dimension bipartite by an interval.
The analytic expression of this operator in the continuum limit is not available in the literature,
but some studies indicate that it is fully non-local 
\cite{Arias:2016nip, Longo:2020amm, Bostelmann:2022yvj, Cadamuro:2023eli}.
In one-dimensional infinite lattices, the entanglement Hamiltonians of a single block 
in the massive regime have been explored when the mass parameter takes very large values, 
both in the harmonic chain and in the free fermionic model,
finding that the corresponding matrices significantly simplify in this regime.
In particular, they become tridiagonal
and the inhomogeneity is described by a simple triangular profile \cite{Eisler:2020lyn},
whose analytic expression can be constructed from the above mentioned results 
for the half chain \cite{Peschel:1991xeo,Peschel:1999xeo}.


Another important class of entanglement Hamiltonians
corresponds to a subsystem $A$ made by many disjoint regions. 
In one spatial dimension, the simplest case is $A = A_1 \cup A_2$,
where  $A_1$ and $A_2$ are two disjoint blocks of contiguous sites in the infinite line. 
In the continuum, very few entanglement Hamiltonians in this class are available in the literature
and all of them involve non-local terms. 
For the massless Dirac fermion on the line and in the ground state,
when $A$ is the union of two disjoint intervals on the line,
$K_A$ is a quadratic operator that includes also a bilocal term \cite{Casini:2009vk}.
This result has been recovered also from a free fermionic chain 
through a non trivial continuum limit procedure \cite{Eisler:2022rnp},
along the lines of the method developed for the single block \cite{Arias:2016nip, Eisler:2019rnr}.
For a massless Dirac field in the ground state, 
bilocal terms in the entanglement Hamiltonians have been found also 
in some cases where translation invariance does not occur,
like e.g. when $A$ is either 
an interval in the half line \cite{Mintchev:2020uom} (see \cite{Eisler:2022rnp} for the corresponding lattice computations)
or the union of two disjoint equal intervals at the same distance from a point-like defect on the line \cite{Mintchev:2020jhc}.
bilocal terms occur also in the entanglement Hamiltonian of an interval on the circle 
for the massless Dirac field at finite temperature
\cite{Blanco:2019xwi, Fries:2019ozf}.
As for the entanglement Hamiltonian of two disjoint intervals on the line for a free bosonic model, 
the case of the chiral current of the massless scalar field in the ground state has been explored in \cite{Arias:2018tmw}, 
finding that it contains also fully non-local terms. 
This model  \cite{Arias:2018tmw, Sonnenschein:1988ug, Berenstein:2023tru} is different from the massless scalar field
(indeed, e.g. it does not satisfy the Haag duality \cite{Arias:2018tmw})
and a numerical check of $K_A$ through a lattice computation is not available in the literature. 

Further lattice results on entanglement Hamiltonians have been reviewed in \cite{Dalmonte:2022rlo}.


The bipartite entanglement of a one-dimensional quantum system in its ground state 
when the subsystem $A$ is the union of disjoint intervals 
has been largely explored in the literature  through the entanglement entropies.
These scalar quantities are simpler and easier to evaluate than the corresponding entanglement Hamiltonians;
hence they are sometimes accessible also for more complicated models,
like e.g. some non trivial CFT models
\cite{Calabrese:2004eu, Caraglio:2008pk, Furukawa:2008uk, Calabrese:2009ez,Calabrese:2010he,Coser:2013qda, DeNobili:2015dla, Grava:2021yjp}.
The entanglement entropy of disjoint regions has been studied also in the 
context of the gauge/gravity correspondence 
through the formula proposed in \cite{Ryu:2006bv, Ryu:2006ef} 
for the gravitational side of the duality. 
Interestingly, this quantity displays sharp transitions 
as the separation distances between the disjoint regions change
\cite{Hubeny:2007re, Headrick:2010zt, Tonni:2010pv, Fonda:2014cca}.
These transitions provide a characteristic feature of the 
CFT models with holographic duals at strong coupling.


In this manuscript we perform some numerical analyses of the 
entanglement Hamiltonian of two disjoint blocks in the harmonic chain on the line and in its ground state.
We mainly focus on the two regimes where the mass parameter is either very large or vanishing.


The outline of this manuscript is as follows. 
In Sec.\,\ref{sec-correlators-EH} we describe the setting 
by defining the harmonic chain in the infinite line, 
its ground state correlators  and the quantities of interest.
In Sec.\,\ref{sec-crossover} we briefly provide some general features 
of the entanglement Hamiltonian of two disjoint blocks in the massive regimes. 
The remaining part of the manuscript can be divided in two parts, 
corresponding to the limiting regimes of large and vanishing mass.

In the first part the large mass regime is explored
by first reviewing the existing results about the entanglement Hamiltonian of the half chain and of the single block (Sec.\,\ref{sec-single-interval-massive})
and then discussing our findings about the entanglement Hamiltonian of two disjoint blocks
having either equal or unequal lengths 
(Sec.\,\ref{sec-2int-massive-equal} and Sec.\,\ref{sec-2int-generic-EH} respectively).
In this regime of large mass, we extend the results for the single block reported in \cite{Eisler:2020lyn},
finding that the matrices characterising the entanglement Hamiltonian become three-diagonal
with inhomogeneity profiles that display sharp transitions as the separation distance changes. 
In the second part we report some results 
for the entanglement Hamiltonians of two generic blocks in the line 
when the mass parameter is vanishing.
In this regime, both the harmonic chain in the massless limit (Sec.\,\ref{sec-2int-massless})
and the free chiral current model \cite{Arias:2018tmw} (Sec.\,\ref{sec-chiral-current})
are investigated.  
%
Some conclusions are drawn in Sec.\,\ref{sec-conclusions}.
The Appendices\;\ref{app-disk-picture}, \ref{app-A1-fixed},  \ref{app-fermions}, \ref{app-massless}, \ref{app-massive:1block},
\ref{app:chiral} and  \ref{app:mutual-info}
contain technical details and supplementary results supporting the discussions reported in the main text.

\section{Entanglement Hamiltonians in the infinite harmonic chain}
\label{sec-correlators-EH}

In this section we first introduce the model 
and then report the expressions providing the entanglement Hamiltonian,
that will be employed throughout our numerical analyses.


The Hamiltonian of the harmonic chain with nearest neighbour spring-like interaction on the infinite line reads
\be
\label{ham-HC}
\widehat{H}  = \sum_{i \,\in\, \mathbb{Z}}
\left(
\frac{1}{2m}\, \hat{p}_i^2 + \frac{m \omega^2}{2}\, \hat{q}_i^2 +\frac{K}{2} \big(\hat{q}_{i+1} - \hat{q}_i \big)^2
\right) 
\ee
where $\hat{q}_i$ and $\hat{p}_i$ are hermitean operators satisfying the 
canonical commutation relations given by 
$[\hat{q}_i , \hat{p}_j ] = \ri \delta_{i,j}$ and $[\hat{q}_i , \hat{q}_j ] = [\hat{p}_i , \hat{p}_j ] = 0$,
 in units where  $\hbar =1$.
The parameters $m$, $K$ and $\omega$ in (\ref{ham-HC}) 
correspond respectively to the mass of the oscillators,
the nearest neighbour coupling and the frequency characterising the confining potential at each site. 

A canonical transformation 
implementing the replacement of
$\hat{p}_i $ and $\hat{q}_i $ with 
$( K m)^{1/4} \,\hat{p}_i $ and  $( K m)^{-1/4} \,\hat{q}_i $ respectively
leads to write (\ref{ham-HC}) as follows
\be
\label{eq:HamiHarmChain}
\widehat{H} = 
\sqrt{K/m} \,
\sum_{i \,\in\, \mathbb{Z}}
\frac{1}{2} \left(
\hat{p}_i^2 + \frac{\omega^2}{K/m}\, \hat{q}_i^2 + \big(\hat{q}_{i+1} - \hat{q}_i \big)^2
\right) 
\ee
where $\sqrt{K/m}$ has the dimension of the energy, 
in the units where $\hbar =1$ that we are considering.
In (\ref{eq:HamiHarmChain}), the positions and momenta are dimensionless
and the frequency is measured in units of $\sqrt{K/m}$.
For simplicity, we set $K=m=1$ throughout  our numerical analyses;
hence $\omega$ is the only parameter characterising the Hamiltonian of the harmonic chain on the line. 
The continuum limit of (\ref{ham-HC}) gives  the Hamiltonian 
the massive scalar field, whose massless limit is a prototypical example of 
a conformal field theory (CFT) with central charge $c=1$.


In our analyses we focus  on the ground state of the harmonic chain on the line,
which is fully characterised by the corresponding correlation functions
for the positions and momenta, that are given respectively by 
\bea
\label{qq-corr-int}
\langle \, \hat{q}_i \, \hat{q}_j  \rangle
&=&
\int_{-\pi}^{+\pi} \! \frac{\cos[q(i-j)] }{ \sqrt{\omega^2 +4[\sin(q/2)]^2} }\; \frac{\rd q}{4\pi}
\\
\rule{0pt}{.8cm}
\label{pp-corr-int}
\langle \, \hat{p}_i \, \hat{p}_j  \rangle
&=&
\int_{-\pi}^{+\pi} \!  \sqrt{\omega^2 +4[\sin(q/2)]^2} \; \cos[q(i-j)] \; \frac{\rd q}{4\pi} \,.
\eea
These expressions can be equivalently written as integrals over $q\in (0,2\pi)$,
by first splitting the integration domain as $(-\pi, 0\,]\cup[\, 0, \pi)$
and then redefining the integration variable as $\tilde{q} = q+2\pi$ when $q\in (-\pi, 0\,]$.
Calculating the integrals in (\ref{qq-corr-int}) and (\ref{pp-corr-int}), 
one finds respectively \cite{Botero:2004vpl}
\bea
\label{qq-corr-hyper}
\langle \, \hat{q}_i \, \hat{q}_{i+r}  \rangle
&=&
\frac{\kappa^{r+1/2} }{ 2 } \; \frac{ \Gamma(r+1/2) }{\Gamma(1/2) \; \Gamma(r+1)}\;
_2F_1(1/2 , r+1/2 ; r+1 ; \kappa^2)
\\
\rule{0pt}{.8cm}
\label{pp-corr-hyper}
\langle \, \hat{p}_i \, \hat{p}_{i+r}  \rangle
&=&
\frac{\kappa^{r- 1/2} }{ 2 } \; \frac{ \Gamma(r-1/2) }{\Gamma(-1/2) \; \Gamma(r+1)}\;
_2F_1( -1/2 , r-1/2 ; r+1 ; \kappa^2)
\eea
where $_2F_1$ is the hypergeometric function and the parameter $\kappa $ is defined as 
\be
\label{kappa-from-omega}
\kappa \equiv \frac{\big( \sqrt{\omega^2 + 4} - \omega \big)^2}{4}
\ee
that satisfies $0 < \kappa < 1$.
The correlators (\ref{qq-corr-hyper}) and (\ref{pp-corr-hyper})
provide the generic elements of the correlations matrices $Q$ and $P$ respectively.
These matrices allow to construct the covariance matrix $\gamma \equiv Q \oplus P$,
which fully characterises the ground state of this system. 
We remark that the translation invariance of the model induces the occurrence of a zero mode,
which prevents us from setting $\omega=0$;
indeed, the correlator (\ref{qq-corr-int}) is divergent in this case. 


Consider a bipartition of the chain given by the subsystem $A$ and its complement $B$,
where $A$ is the union of disjoint blocks of contiguous sites. 
This leads to the natural assumption that, correspondingly, the Hilbert space can be written as 
$\mathcal{H} = \mathcal{H}_A  \otimes \mathcal{H}_B$.
Since we are considering the ground state of the harmonic chain (\ref{ham-HC}), 
the reduced density matrix $\rho_A \equiv \textrm{Tr}_{ \mathcal{H}_B} ( |0 \rangle \langle 0 | ) $ 
of the subsystem $A$ remains Gaussian.

The bipartite entanglement associated to this bipartition of the harmonic chain in its ground state
\cite{Casini:2009sr, Eisler:2009vye,Eisert:2008ur,Weedbrook:2012cjy,Peschel:2002yqj,Botero:2004vpl,Audenaert:2002xfl,Plenio:2004he,Cramer:2005mx,Schuch:2006he} 
can be studied by constructing the  reduced covariance matrix $\gamma_A \equiv Q_A \oplus P_A$,
where $Q_A$ and $P_A$ are the correlation matrices reduced to the subsystem $A$,
whose generic element is given by $(Q_A)_{i,j} \equiv Q_{i,j}$ and $(P_A)_{i,j} \equiv P_{i,j}$ respectively,
where $i, j \in A$;
hence $\gamma_A$ is a $(2L_A) \times (2L_A)$ matrix, where $L_A$ is the number of sites in  $A$.
Since the reduced covariance matrix $\gamma_A$ is real, symmetric and positive definite, 
its Williamson decomposition provides the symplectic spectrum $\{ \sigma_k \, ; 1\leqslant k \leqslant L_A \}$ of $\gamma_A$,
which can be obtained from the eigenvalues of $Q_A P_A$ in a standard way for the cases explored in our analyses.

The R\'enyi entropies $S_A^{(n)}$ and the entanglement entropy $S_A $
are scalar quantities defined respectively as  
$S_A^{(n)}  \equiv \tfrac{1}{1-n}  \log\! \big( \textrm{Tr} \rho_A^n\big)$ 
and $S_A \equiv  - \textrm{Tr} ( \rho_A \log \rho_A ) = \lim_{n \,\to\, 1} S_A^{(n)} $,
in terms of $\rho_A$.
For the case we are considering, these entanglement quantifiers  
can be obtained from the symplectic spectrum of $\gamma_A$ as follows
\bea
\label{renyi-from-ss}
S_A^{(n)} 
&=&
\frac{1}{n-1} \,\sum_{k=1}^{L_A} 
\log\!\big[(\sigma_k+1/2)^n - (\sigma_k-1/2)^n \big]
\\
\label{EE-from-ss}
S_A 
&=&
\sum_{k=1}^{L_A} 
\Big[ (\sigma_k+1/2)\, \log(\sigma_k+1/2) - (\sigma_k-1/2)\, \log(\sigma_k-1/2) \Big] \,.
\eea


Since the reduced density matrix $\rho_A = \e^{- K_A}$ is Gaussian, 
the corresponding entanglement Hamiltonian $K_A$ is a quadratic operator
\cite{Casini:2009sr}
\be
\label{KA-T-V-matrices}
\widehat{K}_A
\,=\, 
\frac{1}{2} 
\sum_{i,j \,\in\, A} \! \Big( T_{i,j}\, \hat{p}_i \, \hat{p}_j + V_{i,j}\, \hat{q}_i \, \hat{q}_j  \Big)
\,=\,
\frac{1}{2} \, \hat{\boldsymbol{r}}^{\textrm{t}} H_A \, \hat{\boldsymbol{r}}
\;\;\;\qquad\;\;\;
\hat{\boldsymbol{r}} \equiv 
\bigg(\begin{array}{c}
\hat{\boldsymbol{q}} \\ \hat{\boldsymbol{p}}
\end{array}
\bigg)
\ee
where the $(2L_A)\times (2L_A)$ symmetric matrix $H_A = V \oplus T$ is block diagonal
and the $L_A \times L_A $ symmetric matrices $T$ and $V$
correspond to the kinetic and potential energy terms,
whose explicit expression in terms of the reduced correlation matrices $Q_A$ and $P_A$
read respectively
\bea
\label{V-mat-QP}
V &=&
h\big(\sqrt{P_A Q_A}\, \big) \, P_A 
\,=\,
P_A\, h\big(\sqrt{Q_A P_A}\, \big) 
\\
\label{T-mat-QP}
\rule{0pt}{.5cm}
T &=&
h\big(\sqrt{Q_A P_A}\, \big) \, Q_A 
\,=\,
Q_A\, h\big(\sqrt{P_A Q_A}\, \big) 
\eea
being $h(y)$ the function defined as follows
\be
\label{eq:hfunction}
h(y) \equiv \frac{1}{y} \, \log \! \left( \frac{y + 1/2}{ y-1/2} \right) .
\ee
The expressions (\ref{KA-T-V-matrices})-(\ref{eq:hfunction}) have been employed in various analyses
(see e.g. \cite{Banchi:2015rmr, Arias:2016nip, Arias:2017dda, DiGiulio:2019lpb, Eisler:2020lyn}).


The entanglement Hamiltonian (\ref{KA-T-V-matrices}) can be written as
$ \widehat{K}_A = \frac{1}{2} \sum_{k=1}^{L_A} \varepsilon_k \hat{f}^{\dagger}_k \hat{f}_k  $,
where $f_k$ are the proper bosonic operators
and $\varepsilon_k $ are the single-particle entanglement energies,
which can be found from the symplectic eigenvalues of $\gamma_A$ as \cite{Peschel:1999xeo}
\be
\label{epsilon-symplectic-spectrum}
\varepsilon_k = 2\, \textrm{arccoth}(2 \sigma_k) = \log \left( \frac{\sigma_k+1/2}{\sigma_k-1/2} \right)
\ee
which can be inverted, finding $\sigma_k = \tfrac{1}{2} \coth(\varepsilon_k /2)$.
The single-particle entanglement spectrum,
whose elements are (\ref{epsilon-symplectic-spectrum}), 
has been largely discussed in the literature \cite{EislerPeschel:2009review}.



In the case mainly considered in this manuscript,
the bipartition of the line is given by the union of the blocks $A_1$ and $A_2$,
made by $L_1$ and $L_2$ sites respectively, separated by $D$ sites;
hence the subsystem  $A=A_1 \cup A_2$ contains $L_A = L_1 + L_2$ sites. 
We find it convenient to introduce the dimensionless parameter 
\be
\label{delta-chi-ratios}
\delta \equiv \frac{D}{L_A}
\ee
corresponding to the separation distance,
and the dimensionless parameter characterising the relative size of the two blocks
\be
\label{rho-chi-def}
\rho \,\equiv\, \frac{L_1 }{L_A} = \frac{1}{1+\chi}
\;\; \qquad\;\; \Longleftrightarrow \;\; \qquad\;\; 
\chi  \equiv \frac{L_2}{L_1} = \frac{1-\rho}{\rho} \,.
\ee
Since $L_1 \leqslant L_2$ can be assumed without loss of generality,
$\rho \in (0, 1/2\,] $ and $\chi \geqslant 1$ hereafter. 



Throughout this manuscript, 
the indices $i$ and $j$ label the physical sites in the subsystem $A=A_1 \cup A_2$ (see e.g. (\ref{KA-T-V-matrices})).
Considering the index $\tilde{i} \in \big[ 1, L_A \big]$ labelling either the rows or the columns of the 
$ L_A \times L_A $ matrices in (\ref{V-mat-QP}) and (\ref{T-mat-QP}), 
the relation between $i$ and $\tilde{i}$ depends on the choice of the origin for the index $i$ in the chain. 
For instance, taking the origin as the first endpoint of $A_1$, we have 
\be
\label{indices-choice-1}
\left\{ \begin{array}{lll}
i = \tilde{i}  &  i \in A_1  &   \tilde{i} \in \big[ 1, L_1 \big]
\\
\rule{0pt}{.5cm}
i = \tilde{i}   + D \hspace{1cm} &  i \in A_2  \hspace{1cm}  &   \tilde{i} \in \big[ L_1+1 , L_1+ L_2 \big] \,.
\end{array}\right.
\ee
Since both the sizes of the blocks and their separation distances are even in our analyses,
a natural choice for the origin is the middle of the finite block separating $A_1$ and $A_2$.
This gives
\be
\label{indices-choice-2}
\left\{ \begin{array}{lll}
i = \tilde{i} - L_1 - \lfloor D/2 \rfloor \hspace{1cm} &  i \in A_1  \hspace{1cm}  &   \tilde{i} \in \big[ 1, L_1 \big]
\\
\rule{0pt}{.5cm}
i = \tilde{i}  -L_1  +  \lfloor (D+1)/2 \rfloor \hspace{1cm} &  i \in A_2  \hspace{1cm}  &   \tilde{i} \in \big[ L_1+1 , L_1+ L_2 \big]
\end{array}\right.
\ee
where  $\lfloor \,\dots \rfloor$ denotes the floor function.
%

\section{Massive regime}
\label{sec-crossover}

In this section we briefly discuss some numerical results for the matrices $T$ and $V$ 
characterising the entanglement Hamiltonian (\ref{KA-T-V-matrices}) in the infinite harmonic chain in its ground state,
evaluated through the expressions (\ref{V-mat-QP})-(\ref{eq:hfunction}),
when  the subsystem $A= A_1 \cup A_2$ is made by the union of two disjoint blocks (see Sec.\,\ref{sec-correlators-EH}),
and the mass parameter $\omega$ is finite and does not take extreme values.

\begin{figure}[t!]
\vspace{-1cm}
\includegraphics[width=1.05\textwidth]{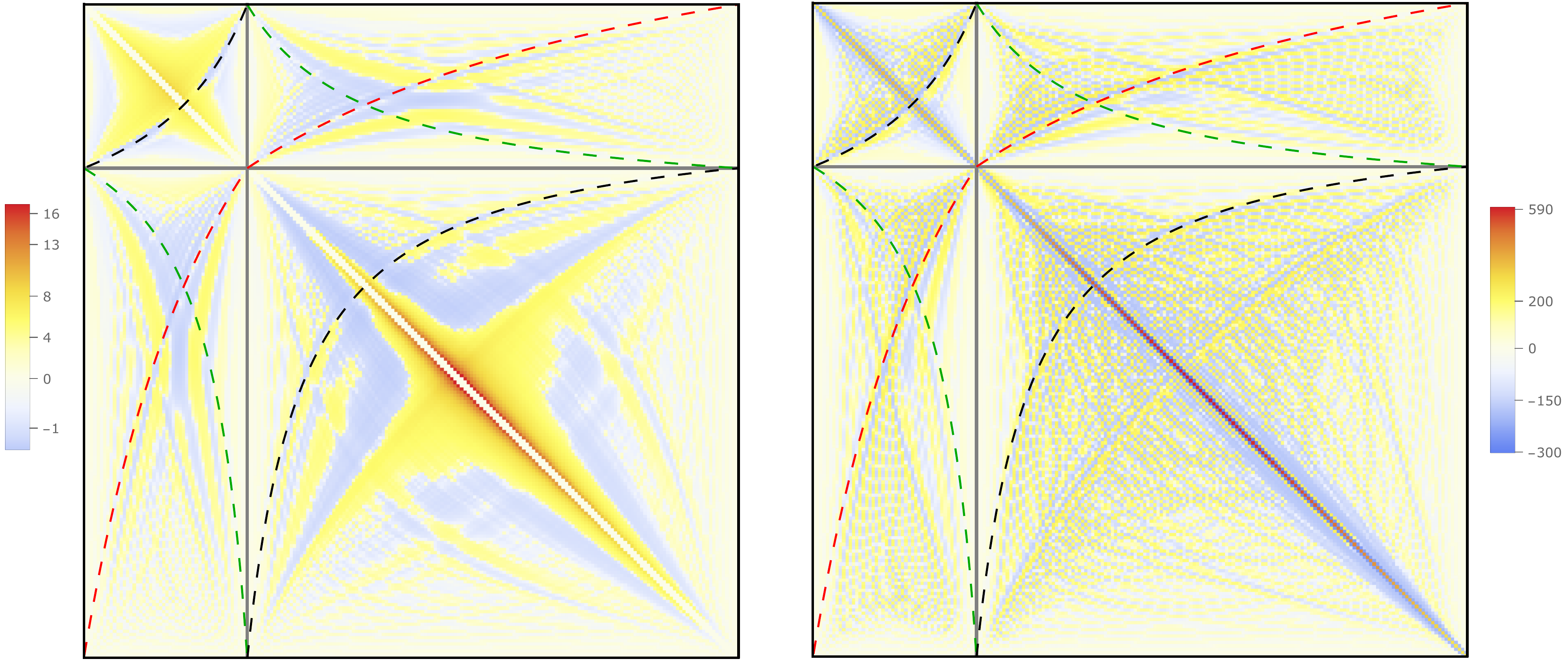}
\\
\rule{0pt}{7.7cm}
\hspace{-.16cm}
\includegraphics[width=1.05\textwidth]{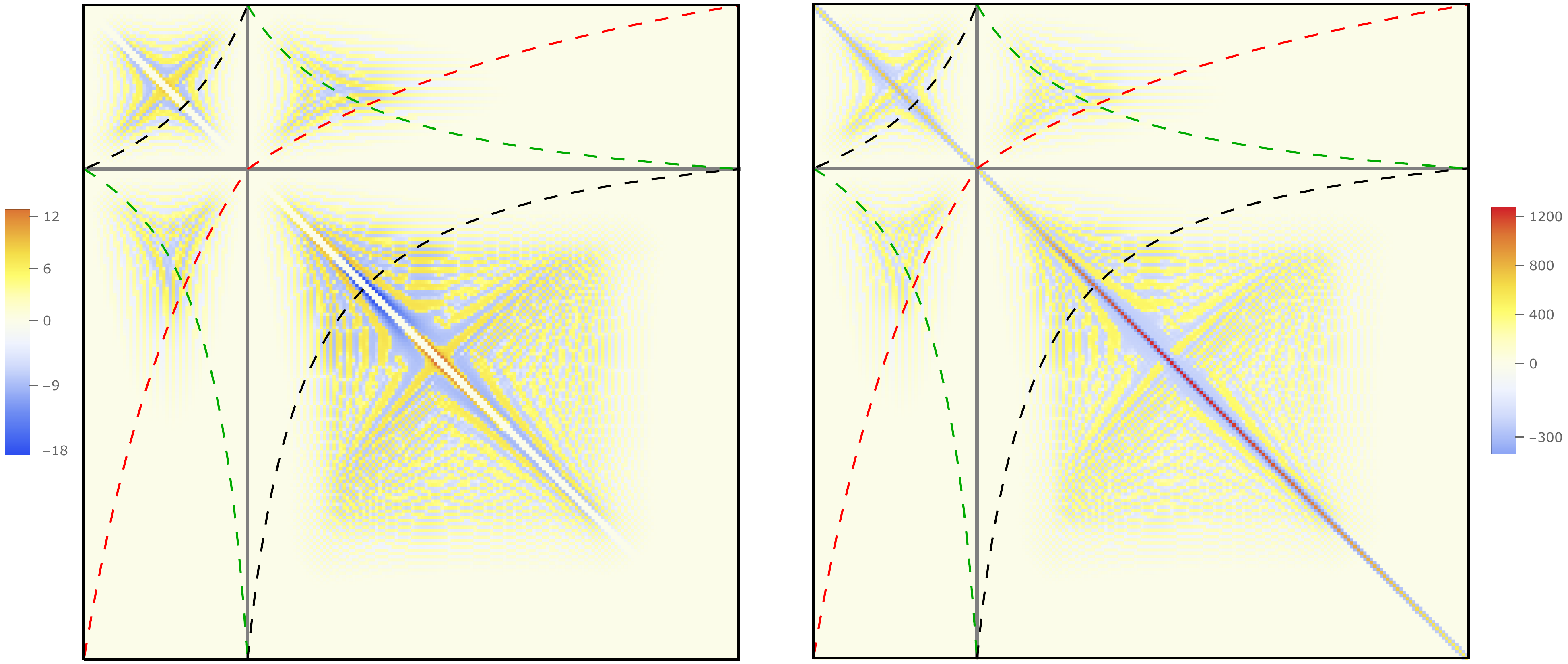}
\\
\rule{0pt}{7.7cm}
\hspace{-.16cm}
\includegraphics[width=1.05\textwidth]{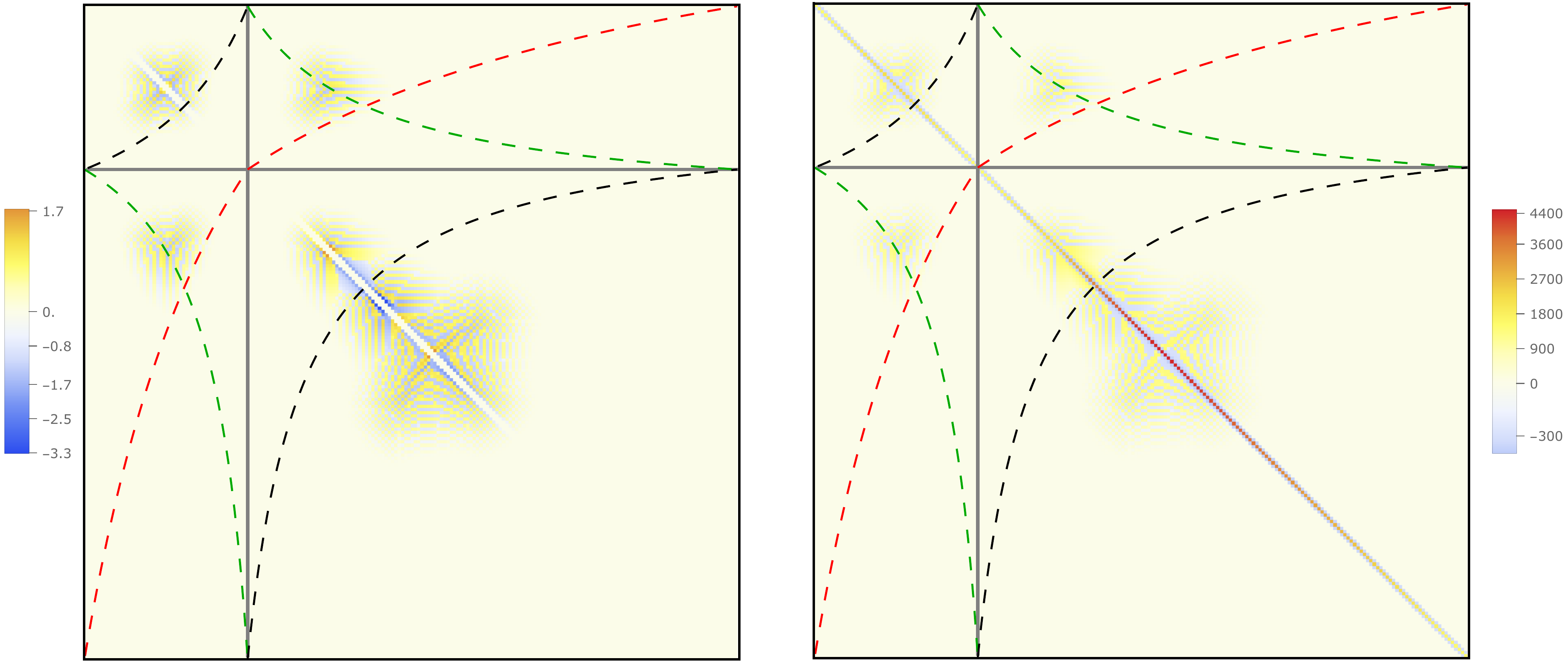}
\vspace{-.2cm}
\caption{Matrix $T$ (left) and $V$ (right) in the massive regime
with either $\omega = 0.1$ (top) or $\omega = 1$ (middle) or $\omega = 3$ (bottom),
for $L_2 = 3L_1$ with $L_1=50$ and $\delta =1/4$.
}
\label{fig:EH-density-massive}
\end{figure}

Since the subsystem $A$ is the union of the two blocks $A_1$ and $A_2$,
it is natural to introduce the block decomposition of $T$ and $V$
provided by such bipartition of $A$, namely
\begin{equation}
\label{eq:MatrixTandVdecomposition}
T=\left(
\begin{array}{c|c}
T^{\textrm{\tiny  $(1,\!1)$} } & \, T^{\textrm{\tiny  $(1,\!2)$} } 
\\
\hline 
T^{\textrm{\tiny  $(2,\!1)$} } & \,T^{\textrm{\tiny  $(2,\!2)$} }
\end{array}
\right)
\;\;\;\;\qquad\;\;\;
V=\left(
\begin{array}{c|c}
V^{\textrm{\tiny  $(1,\!1)$} } & \, V^{\textrm{\tiny  $(1,\!2)$} } 
\\
\hline 
V^{\textrm{\tiny  $(2,\!1)$} } & \,V^{\textrm{\tiny  $(2,\!2)$} }
\end{array}
\right)
\end{equation}
where the diagonal blocks correspond to the sites in $A_1$ and $A_2$ 
(hence they are $L_1 \times L_1 $ and $L_2 \times L_2 $ symmetric matrices respectively)
and the off-diagonal blocks of $T$ and $V$ are rectangular matrices 
whose elements are identified by one site in $A_1$ and the other one in $A_2$.
In Fig.\,\ref{fig:EH-density-massive}, the block decomposition (\ref{eq:MatrixTandVdecomposition}) 
is given by the grey straight lines.

%

In Fig.\,\ref{fig:EH-density-massive} we show the matrices $T$ (left panels) and $V$ (right panels),
for increasing values of the mass parameter $\omega \in \{0.1\,, 1, 3\}$, going from the top to the bottom panels. 
The blocks $A_1$ and $A_2$ contain $L_1 = 50$ and $L_2 =150 $ sites respectively 
and are separated by $D=50$ sites. 
The main diagonal and the two ones next to it of the matrix $T$ provide the largest amplitudes;
hence in the left panels of Fig.\,\ref{fig:EH-density-massive} these three diagonals 
have been removed, in order to make visible the remaining elements of the matrix. 

\clearpage

\begin{figure}[t!]
\vspace{-.5cm}
\hspace{0.cm}
\includegraphics[width=1\textwidth]{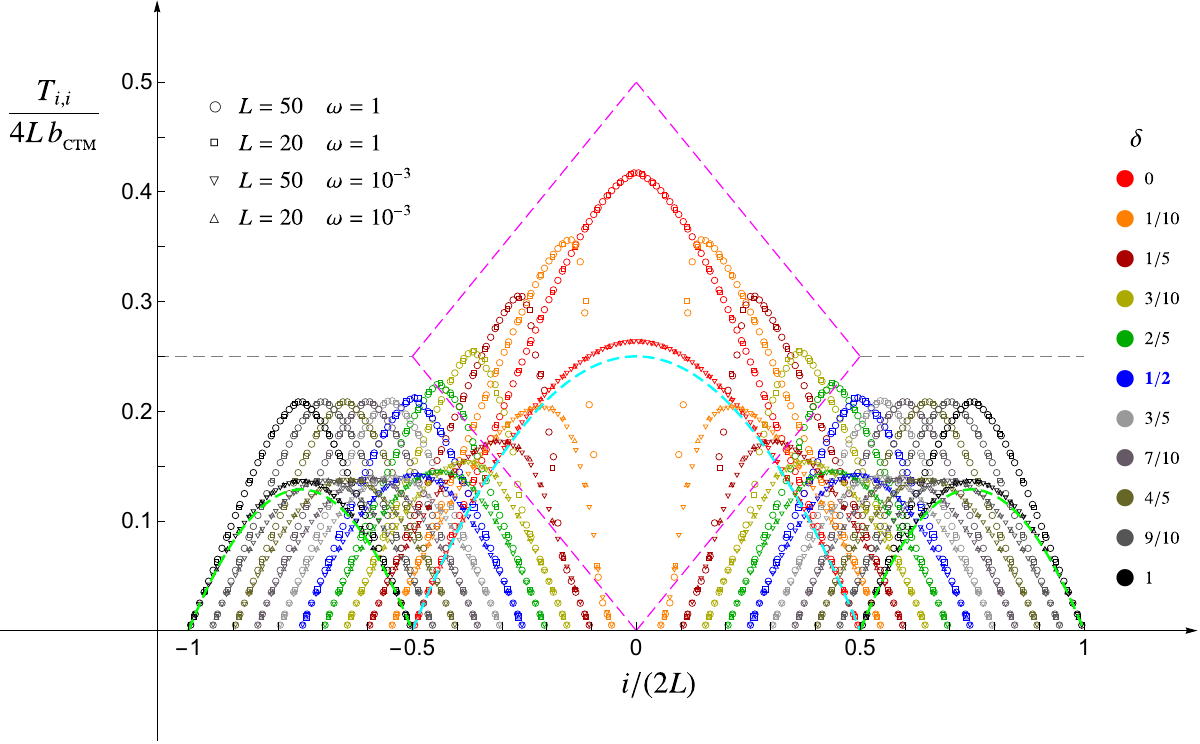}
\vspace{-.3cm}
\caption{Main diagonal of $T$ for two equal blocks made by $L$ sites when $\omega = 1$ and $\omega = 10^{-3}$,
for various values of $\delta$ (see (\ref{delta-chi-ratios})). 
The magenta and grey dashed segments refer to the large $\omega$ regime (see Fig.\,\ref{fig:2int-equal-massive-diag}),
while the green and cyan dashed lines correspond to the massless regime 
(see e.g. Fig.\,\ref{fig:DiagOffDiag_massless_lratio2_Tmat} and Fig.\,\ref{fig:EHmassless_lratio1}). 
}
\label{fig:2int-equal-massive-diag-crossover}
\end{figure}

Some plots of these matrices in the massive regime have been shown also in \cite{Arias:2016nip}.
The main feature to highlight about these matrices is their non-local and inhomogeneous nature. 
Since finding analytic results about the matrices $T$ and $V$ for this bipartition  in the massive regime is extremely difficult, 
in this manuscript we mainly focus on the limiting regimes of large $\omega$
(see Sec.\,\ref{sec-2int-massive-equal} and Sec.\,\ref{sec-2int-generic-EH})
and vanishing $\omega$ (see Sec.\,\ref{sec-2int-massless}).
The dashed curves in Fig.\,\ref{fig:EH-density-massive} 
correspond to analytic curves occurring in the massless regime 
and will be introduced in Sec.\,\ref{sec-2int-massless}.

It is worth remarking that, in Fig.\,\ref{fig:EH-density-massive},  
the matrix elements corresponding to non-local terms vanish as $\omega$ increases
and that a local operator is obtained in the extreme regime where $\omega \gg 1$.
In the limit of large $\omega$, the matrices $T$ and $V$ become three-diagonal;
hence they are fully described by simple profiles. 
These functions, 
which have been discussed in \cite{Eisler:2020lyn} for the case of the single block,
will be investigated in Sec.\,\ref{sec-2int-massive-equal} and Sec.\,\ref{sec-2int-generic-EH}
for the case of the union of two disjoint blocks.


In Fig.\,\ref{fig:2int-equal-massive-diag-crossover} we consider blocks of equal length 
and show the main diagonal of $T$ for either $\omega=1$ or $\omega=10^{-3}$
(the constant $b_{_\textrm{\tiny CTM}}$ depends on $\omega$ and is introduced below,  in (\ref{b-CTM-def})). 
The data points in Fig.\,\ref{fig:2int-equal-massive-diag-crossover} obtained for $\omega=1$ 
should be combined with the ones displayed in the middle panels of Fig.\,\ref{fig:EH-density-massive}.
Nice collapses are observed for the data sets corresponding to different values of $L$
and it would be very interesting to find analytic expressions for these curves. 
A similar result has been found for the main diagonal of $V$, 
but we have not reported it here.
In this figure,  the grey and magenta dashed straight lines 
provide the structure obtained in the large $\omega$ regime and described in Sec.\,\ref{sec-2int-massive-equal}
(these curves are displayed also in Fig.\,\ref{fig:2int-equal-massive-diag}),
while the cyan and green dashed curves correspond to the opposite limiting regime $\omega \to 0^+$,
for vanishing and large separation distance respectively
(e.g. in the top panel of Fig.\,\ref{fig:EHmassless_lratio1},
see the black dashed curve for $\delta=0$ and the solid green curves respectively).

\section{Large $\omega$ regime: Half chain and single block}
\label{sec-single-interval-massive}


Considering the ground state of the harmonic chain (\ref{ham-HC}),
in this section we review the analytic results 
for the entanglement Hamiltonian (\ref{KA-T-V-matrices})
of the half chain \cite{Peschel:1991xeo,Peschel:1999xeo}
and of a single block in the regime of large mass \cite{Eisler:2020lyn}.


The entanglement Hamiltonian (\ref{KA-T-V-matrices})
when the subsystem $A$ is the half chain with sites $i \in [\,1 , +\infty)$
has been investigated in \cite{Peschel:1999xeo},
through the corner transfer matrix (CTM) technique developed in \cite{Peschel:1991xeo}.
In the regime of large $\omega$, 
the matrices $T$ and $V$ become diagonal and  tridiagonal respectively (see also \cite{Eisler:2020lyn}).
Their  non vanishing elements are well described by 
\be
\label{T-diag-half-chain}
T_{i,i} = 2b_{_\textrm{\tiny CTM}} \big(i- 1/2\big)
\ee
and
\be
\label{V-diag-half-chain}
V_{i,i} = (\omega^2 + 2)\,2b_{_\textrm{\tiny CTM}}  \big(i- 1/2\big)
\;\;\;\;\;\qquad\;\;\;\;\;
V_{i,i+1} = -\,2b_{_\textrm{\tiny CTM}} \, i
\ee
where $i \geqslant 1 $ and the coefficient $b_{_\textrm{\tiny CTM}}$ depends on $\omega$ as follows
\be
\label{b-CTM-def}
b_{_\textrm{\tiny CTM}} \equiv 2\, \sqrt{\kappa} \; I \big(\kappa ' \big)
\;\;\;\qquad\;\;\;
\kappa ' \equiv \sqrt{1-\kappa^2}
\ee
being $\kappa$ defined as in (\ref{kappa-from-omega})
and $I(y)$ the complete elliptic integral of the first kind.
In this setup the single-particle entanglement spectrum is given by  \cite{Peschel:1999xeo} 
\be
\label{eps-half-line-ctm}
\varepsilon_k =  2\varepsilon_{_\textrm{\tiny CTM}} \,  \big(k- 1/2\big)
\;\;\;\qquad\;\;\;
 \varepsilon_{_\textrm{\tiny CTM}}\, \equiv \, \pi \,\frac{I(\kappa ')}{I(\kappa)}
\;\;\;\qquad\;\;\;
k \geqslant 1 \,.
\ee

We remark that in the massless limit $\omega\to 0$ we have 
\be
b_{_\textrm{\tiny CTM}} \rightarrow\, \pi
\;\;\;\;\;\;\qquad\;\;\;\;\;
 \varepsilon_{_\textrm{\tiny CTM}} \rightarrow\, 0
\ee
for the expressions introduced in (\ref{b-CTM-def}) and (\ref{eps-half-line-ctm}) respectively.


The entanglement Hamiltonian (\ref{KA-T-V-matrices})
when $A$ is a block made by $L$ sites in the infinite chain has been studied in \cite{Eisler:2020lyn},
finding that, in the regime of large $\omega$, 
the matrices $T$ and $V$ become diagonal and  tridiagonal respectively also in this case. 
The profiles characterising the non vanishing diagonals are (see Fig.\,3 of \cite{Eisler:2020lyn})
\be
\label{T-diag-1int}
\frac{T_{i,i}}{L} = 2 b_{_\textrm{\tiny CTM}}\, F_1 \! \left( \frac{i -1/2}{L} \right)
\ee
and
\be
\label{V-diag-1int}
\frac{V_{i,i}}{L} = (\omega^2 + 2)\, 2b_{_\textrm{\tiny CTM}}\, F_1 \! \left( \frac{i -1/2}{L} \right)
\;\;\;\qquad\;\;\;
\frac{V_{i,i+1}}{L} = -\,2b_{_\textrm{\tiny CTM}}\, F_1 \! \left( \frac{i }{L} \right)
\ee
where $1 \leqslant i \leqslant L$,
the coefficient $b_{_\textrm{\tiny CTM}}$ is given in (\ref{b-CTM-def})
and the function $F_1(x)$ can be  introduced
through $\Delta(a,b; x)$,
a continuous function having a triangular shape defined as follows for $x\in \RR$
\be
\label{Delta-function-ab-def}
\Delta(a,b; x)
\, \equiv \, 
\Theta_{[a,b]}(x) \left( \frac{b-a}{2} - \left| x -\frac{a+b}{2}\right|\, \right)
\,=\,
\left\{ \,
\begin{array}{ll}
x -a \hspace{1.cm} & x \in \big[a \, , \tfrac{a+b}{2} \big]
\\
\rule{0pt}{.45cm}
b-x &x \in \big[ \tfrac{a+b}{2}  , b \big]
\\
\rule{0pt}{.45cm}
0 & x \notin  \big[a,b\big]
\end{array}
\right.
\ee
in terms of the characteristic function $\Theta_{[a,b]}(x)$ for the interval $[a,b]$.
Setting the origin of the spatial coordinate  either in the middle point of $A$ or in its first endpoint, 
the function $F_1(x)$ to employ  in (\ref{T-diag-1int})-(\ref{V-diag-1int}) 
reads respectively 
%
\be
\label{F1-from-Delta}
F_1(x) \equiv \Delta(-1/2,1/2; x) 
\;\;\;\qquad\;\;\;
F_1(x) \equiv \Delta(0,1; x)  \,.
\ee
We remark that (\ref{T-diag-half-chain}) and (\ref{V-diag-half-chain}) 
can be obtained by replacing the function $F_1(x)$ with the function $x$
in (\ref{T-diag-1int}) and (\ref{V-diag-1int}) respectively.
Further progress about the entanglement Hamiltonian of a single block in the large $\omega$ regime 
has been recently made in \cite{Baranov:2024vru}.

In the same regime of large $\omega$,
the single-particle entanglement spectrum for this setup is nicely approximated by 
the following linear behaviour \cite{Eisler:2020lyn}
\be
\label{epsilon-diag-1int}
\frac{ \varepsilon_k }{L} =  2\varepsilon_{_\textrm{\tiny CTM}} \,  \frac{ k- 1/2 }{L}
\ee
where $1 \leqslant k \leqslant L$, 
whose slope is determined by (\ref{eps-half-line-ctm}),
as shown in Fig.\,4 of \cite{Eisler:2020lyn}.

By employing the single-particle entanglement spectrum for the half line given in (\ref{eps-half-line-ctm}),
it has been found that  the entanglement entropy at large $\omega$ reads 
\cite{Eisler:2020lyn} 
\be
\label{entropy-interval-massive}
S_A \,=\,
 - \frac{1}{12} \left[\,
\log\!\left(\frac{16\, \kappa'^4}{\kappa^2}\right) 
-\big(1+\kappa^2\big) \, \frac{4\, I(\kappa)\, I(\kappa')}{\pi}
\,\right]
\ee
which is independent of $L$, as expected from the area law. 
%

\section{Large $\omega$ regime: Two equal  blocks}
\label{sec-2int-massive-equal}

In this section we explore the entanglement Hamiltonian (\ref{KA-T-V-matrices})
for the harmonic chain in its ground state,
in the case where the subsystem $A = A_1 \cup A_2$ is the union of two disjoint blocks and $\omega$ is very large.
This analysis extends the result obtained in \cite{Eisler:2020lyn} in this regime for the single block, 
that has been reviewed in Sec.\,\ref{sec-single-interval-massive}.
In this section we focus on configurations having equal blocks, i.e. $L_1 = L_2 \equiv L$,
implying that (\ref{delta-chi-ratios}) becomes $\delta = D/(2L)$,
while the  case $L_1 \neq L_2$  is investigated  in Sec.\,\ref{sec-2int-generic-EH}.

\subsection{Entanglement Hamiltonian}
\label{sec-EH-2int-equal}


In the regime of large $\omega$ that we are considering, 
when $\delta =0$
the result for the single block 
given by (\ref{T-diag-1int})-(\ref{V-diag-1int}) with $L$ replaced by $2L$ 
must be obtained. 
Instead, in the opposite regime where $\delta$ is sufficiently large,
the two blocks become independent and therefore we expect to find 
(\ref{T-diag-1int})-(\ref{V-diag-1int}) for each block.
It is worth exploring the interpolation between these two regimes 
by studying the entanglement Hamiltonian (\ref{KA-T-V-matrices}) 
for $\delta$ that varies between $\delta=0$ and values 
such that $A_1$ and $A_2$ become independent.

When $\omega$ is large enough, 
one observes that the non vanishing elements of the matrices (\ref{V-mat-QP}) and (\ref{T-mat-QP})
occurring in the entanglement Hamiltonian (\ref{KA-T-V-matrices}) 
follow the same three-diagonals approximation found in \cite{Eisler:2020lyn} for the single block,
but the profiles along the diagonals now encode the spatial configuration of the bipartition
that we are considering. 
In the special case where the two disjoint blocks have the same size,
we find that these non vanishing diagonals are nicely described by 
\be
\label{T-diag-2int}
\frac{T_{i,i}}{2L} = 2 b_{_\textrm{\tiny CTM}}\, \widetilde{F}_2\! \left( \delta  \,; \frac{i -1/2}{2L} \right)
\ee
and
\be
\label{V-diag-2int}
\frac{V_{i,i}}{2L} = (\omega^2 + 2)\, 2b_{_\textrm{\tiny CTM}}\, \widetilde{F}_2\!  \left( \delta \,; \frac{i -1/2}{2L} \right)
\;\;\;\qquad\;\;\;
\frac{V_{i,i+1}}{2L} = -\,2b_{_\textrm{\tiny CTM}}\, \widetilde{F}_2\!  \left( \delta \,; \frac{i}{2L} \right)
\ee
in terms of the coefficient $b_{_\textrm{\tiny CTM}}$ defined in (\ref{b-CTM-def})
and of the function $\widetilde{F}_2(\delta ; x )$ with $x\in [0,1]$, parameterised by $\delta = D/(2L)$.
We remark that the index $i$ in (\ref{T-diag-2int})-(\ref{V-diag-2int}) labels the physical sites (see \eqref{indices-choice-2}). 
Notice that (\ref{T-diag-2int})-(\ref{V-diag-2int}) have
the same form of the corresponding quantities for the single block and for the half chain, 
reported in (\ref{T-diag-1int})-(\ref{V-diag-1int}) and (\ref{T-diag-half-chain})-(\ref{V-diag-half-chain}) respectively. 
The crucial difference is given by the function $\widetilde{F}_2(\delta ; x )$.
%

\begin{figure}[t!]
\vspace{-.5cm}
\hspace{-1.1cm}
\includegraphics[width=1.14\textwidth]{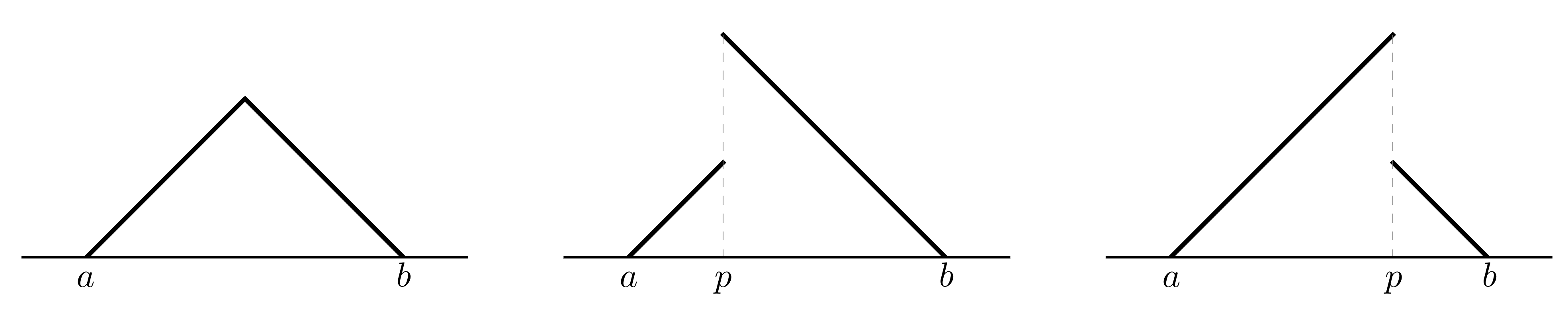}
\vspace{-.7cm}
\caption{
The function (\ref{lambda-function-ab-def}) 
for $p = \tfrac{a+b}{2}$ (left), $p < \tfrac{a+b}{2}$ (middle) and $p > \tfrac{a+b}{2}$ (right).}
\label{fig:2int-equal-aux-functions-shapes}
\end{figure}

In order to define the function $\widetilde{F}_2(\delta ; x )$ in (\ref{T-diag-2int})-(\ref{V-diag-2int}),
let us introduce the following auxiliary function 
\be
\label{lambda-function-ab-def}
\lambda(a,b; p; x)
\, \equiv \, 
\bigg\{ \,
\begin{array}{ll}
x -a \hspace{1.cm} & x \in \big[a , p\big]
\\
\rule{0pt}{.45cm}
b-x &x \in \big[ p   , b \big]
\end{array}
\;\;\;\;\;\qquad\;\;\;
p \in \big[a,b\big]
\ee
which is a real function of $x\in \RR$ with support in the interval $\big[a,b\big]$.
For a generic $p \in (a,b)$,
the function (\ref{lambda-function-ab-def}) is discontinuous at $x=p$ 
and its name has been chosen because 
its shape resembles the form of the greek letter $\lambda$ when $p < \tfrac{a+b}{2}$.
For $p =\tfrac{a+b}{2}$,
the function (\ref{lambda-function-ab-def}) becomes the continuous triangular function (\ref{Delta-function-ab-def}).
In Fig.\,\ref{fig:2int-equal-aux-functions-shapes},
the function (\ref{lambda-function-ab-def}) is shown 
for the three relevant regimes of its parameter $p$.

\begin{figure}[t!]
\vspace{-.5cm}
\hspace{0.cm}
\includegraphics[width=1\textwidth]{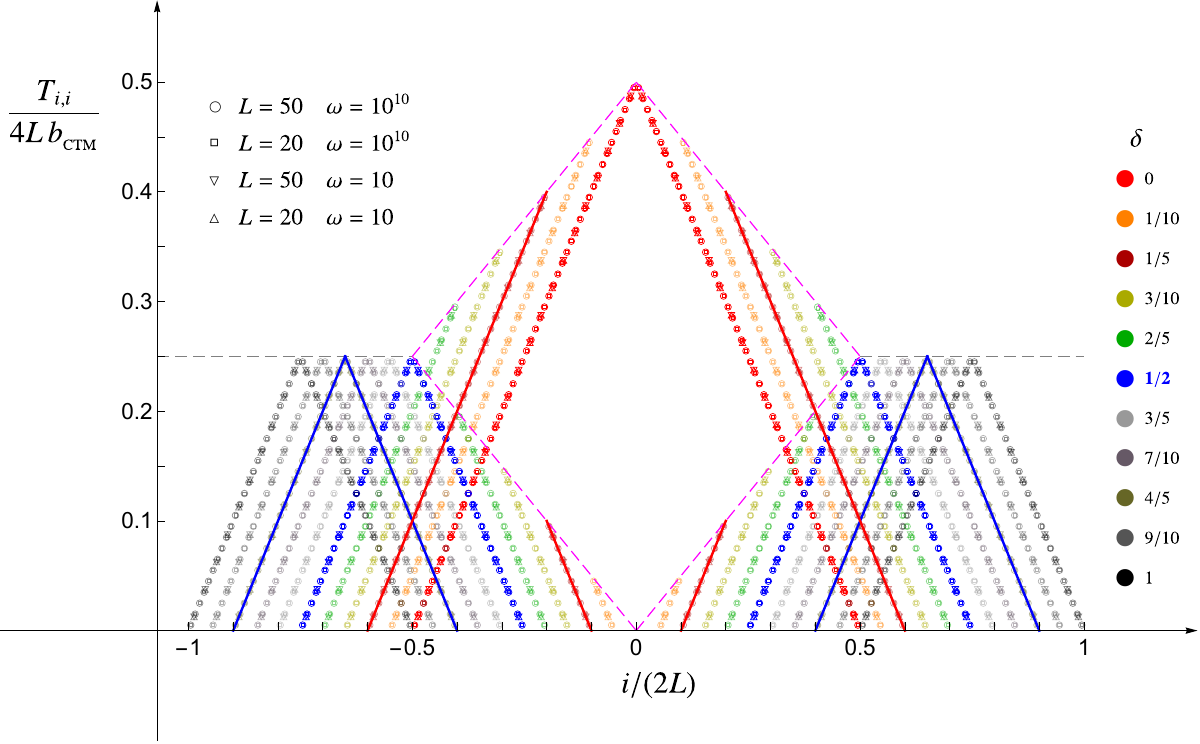}
\vspace{-.3cm}
\caption{Main diagonal of $T$ for two equal blocks made by $L$ sites in the regime of large $\omega$,
for various values of $\delta$. 
The solid curves correspond to the analytic expression in (\ref{T-diag-2int})
(see also (\ref{tilde-F2-def-equal-2int-A1A2-move-V0}) or (\ref{tilde-F2-def-equal-2int-compact})),
which has a sharp transition at the critical value (\ref{delta-critical-equal})
(blue data points).
}
\label{fig:2int-equal-massive-diag}
\end{figure}

The function $\widetilde{F}_2(\delta ; x ) $ occurring in (\ref{T-diag-2int})-(\ref{V-diag-2int}) 
can be written through  the functions defined in (\ref{Delta-function-ab-def}) and (\ref{lambda-function-ab-def}) as follows
\be
\label{tilde-F2-def-equal-2int-A1A2-move-V0}
\widetilde{F}_2(\delta ; x ) 
\,\equiv\,
\left\{
\begin{array}{ll}
\Delta_1( x) + \Delta_2(x)
 \hspace{1cm}
 & 
\delta \geqslant \delta_{\textrm{c}} 
\\
\rule{0pt}{.5cm}
\lambda_1\big( b_1 - s(\delta); x \big) 
+ 
\lambda_2\big(a_2 + s(\delta) ; x \big)
 \hspace{1cm}
 & 
\delta \leqslant \delta_{\textrm{c}} 
\end{array}
\right.
\ee
where $b_1 - a_1 = b_2 - a_2 = L $
and we have introduced the notation
$\Delta_j(x)  \equiv \Delta(a_j ,b_j ; x) $ and $\lambda_j (p ; x )  \equiv \lambda (a_j,b_j; p ; x ) $
to enlighten the formula
and highlight the fact that the first and the second term in the r.h.s. correspond to $A_1$ and $A_2$ respectively
(the dimensionless parameters $a_j < b_j$ refer to the endpoints of the interval $A_j$).
To match the above mentioned expectations  in the two limiting regimes 
of vanishing and large separation distance,
$s(\delta)$ in (\ref{tilde-F2-def-equal-2int-A1A2-move-V0})
must satisfy $s(0) = 0$ and $s(\delta) \to 1/4$ as $\delta \to +\infty$ respectively. 
The function $s(\delta)$ in (\ref{tilde-F2-def-equal-2int-A1A2-move-V0})
fulfilling these consistency conditions and that nicely fits our numerical data reads
\be
\label{q-func-equal-data}
s(\delta) \equiv 
\left\{ \begin{array}{cc}
\delta/2  \hspace{.8cm} & \delta \leqslant \delta_{\textrm{c}} 
\\
\rule{0pt}{.5cm}
1/4 \hspace{.8cm} & \delta \geqslant \delta_{\textrm{c}} 
\end{array}
\right.
\ee
where the critical value of $\delta$ corresponds to 
\be
\label{delta-critical-equal}
\delta_{\textrm{c}} = \frac{1}{2} \,.
\ee
In Fig.\,\ref{fig:2int-equal-massive-diag} 
we report some numerical data points for the main diagonal of $T$ for various values of the parameter $\delta$
(obtained through (\ref{T-mat-QP})),
showing that they are well described by the function (\ref{T-diag-2int}),
given by the solid curves. 
The same analysis has been carried out for $V_{i,i}$ and $V_{i, i+1}$, 
by employing (\ref{V-mat-QP}) for the numerical data points 
and (\ref{V-diag-2int}) for the analytic curve. 
The results are not reported here because they are 
qualitatively identical to the ones in Fig.\,\ref{fig:2int-equal-massive-diag}.

A heuristic picture explaining the critical value (\ref{delta-critical-equal}) is described in Appendix\;\ref{app-disk-picture}.

The expression (\ref{tilde-F2-def-equal-2int-A1A2-move-V0})
highlights the fact that a transition occurs at the critical value (\ref{delta-critical-equal}).
An equivalent form for (\ref{tilde-F2-def-equal-2int-A1A2-move-V0}), just in terms of $s(\delta) $ introduced in  (\ref{q-func-equal-data}),
reads
\be
\label{tilde-F2-def-equal-2int-compact}
\widetilde{F}_2(\delta ; x ) 
\,\equiv\,
\lambda \big( a_1 ,  b_1 ; b_1 - s(\delta) ; x \big)
+
\lambda \big( a_2 ,  b_2 ; a_2 + s(\delta) ; x \big) \,. 
\ee

Summarising, the function $\widetilde{F}_2(\delta ; x )$ introduced 
in (\ref{tilde-F2-def-equal-2int-A1A2-move-V0}) or  in (\ref{tilde-F2-def-equal-2int-compact}), 
displays two phases separated by the critical value (\ref{delta-critical-equal}) for the dimensionless ratio $\delta$.
Correspondingly, qualitative different shapes are observed for the diagonals in (\ref{T-diag-2int})-(\ref{V-diag-2int}),
which are described by two equal triangles when $\delta \geqslant \delta_{\textrm{c}} $  (see the left panel in Fig.\,\ref{fig:2int-equal-aux-functions-shapes}) 
and by two discontinuous piecewise linear functions when $\delta \leqslant \delta_{\textrm{c}} $
(see the middle and right panels in Fig.\,\ref{fig:2int-equal-aux-functions-shapes}).


We remark that the specific expression for $s(\delta)$ given by (\ref{q-func-equal-data})-(\ref{delta-critical-equal}) 
cannot be determined just by imposing the 
asymptotic behaviours at $\delta=0$ and $\delta \to +\infty$ (see the text above (\ref{q-func-equal-data})).
Indeed, for instance, let us consider the following  one-parameter family of functions 
\be
s\big(\tilde{\delta}_{\textrm{c}}  ; \delta \big) 
\equiv 
\left\{ \begin{array}{ll}
\delta/(4 \tilde{\delta}_{\textrm{c}} )  \hspace{.8cm} & \delta \leqslant \tilde{\delta}_{\textrm{c}} 
\\
\rule{0pt}{.5cm}
1/4 \hspace{.8cm} & \delta \geqslant \tilde{\delta}_{\textrm{c}} 
\end{array}
\right.
\ee
which becomes $s(\delta)$ in (\ref{q-func-equal-data})-(\ref{delta-critical-equal}) when $\tilde{\delta}_{\textrm{c}} = 1/2$.
By replacing $s(\delta)$ with $s\big(\tilde{\delta}_{\textrm{c}}  ; \delta \big) $ 
in (\ref{tilde-F2-def-equal-2int-A1A2-move-V0}) and (\ref{tilde-F2-def-equal-2int-compact}),
the resulting $\widetilde{F}_2(\delta ; x ) $ satisfy the same asymptotic behaviours at $\delta=0$ and $\delta \to +\infty$ 
for any finite value of the parameter $\tilde{\delta}_{\textrm{c}}$.
However, since it displays the transition at $\delta = \tilde{\delta}_{\textrm{c}} $,
it disagrees with the numerical data points for $\tilde{\delta}_{\textrm{c}} \neq 1/2$.


In order to compare the analytic expression in
(\ref{tilde-F2-def-equal-2int-A1A2-move-V0}) and (\ref{tilde-F2-def-equal-2int-compact})
against the corresponding numerical data,
a choice of the endpoints of the blocks $A_1$ and $A_2$ of equal length is needed.

A convenient choice for the parameters providing  the endpoints of the blocks 
in (\ref{tilde-F2-def-equal-2int-A1A2-move-V0}) and (\ref{tilde-F2-def-equal-2int-compact})
reads
\be
\label{2int-equal-move-symmetrically}
a_1 = - \frac{\delta}{2}  - \frac{1}{2}
\;\qquad\;
b_1 =  - \frac{\delta}{2} 
 \;\;\;\;\; \qquad \;\;\;\;\;
 a_2 =   \frac{\delta}{2} 
\;\qquad\;
b_2 =  \frac{\delta}{2} + \frac{1}{2}
\ee
which tells us that the two blocks $A_1$ and $A_2$ behave in the same way, 
symmetrically with respect to the origin, as $\delta$ decreases
and become adjacent at the origin when $\delta=0$.
Setting (\ref{2int-equal-move-symmetrically}) into (\ref{tilde-F2-def-equal-2int-compact}),
one finds that the  asymptotic behaviours of $\widetilde{F}_2(\delta ; x ) $
in the limiting regimes of vanishing and large separation between the two blocks 
can be written in terms of (\ref{Delta-function-ab-def}) respectively as 
\be
\label{adj-cond-equal2int}
\lim_{\delta \to 0} \widetilde{F}_2(\delta ; x ) = \Delta\! \left( \! -\frac{1}{2} \,, \frac{1}{2}\,  ; x \right)
\;\qquad
\lim_{\delta \to +\infty} \! \widetilde{F}_2(\delta ; x ) =
\Delta\! \left( \! -\frac{\delta}{2}-\frac{1}{2} \,, -\frac{\delta}{2}\,  \, ; x \right)
+
\Delta\! \left( \, \frac{\delta}{2} \,, \frac{\delta}{2}+\frac{1}{2}\,  ; x \right) .
\ee


A different choice for the parameters determining the endpoints of the blocks 
in (\ref{tilde-F2-def-equal-2int-A1A2-move-V0}) and (\ref{tilde-F2-def-equal-2int-compact}) 
corresponds to keep e.g. $A_1$ fixed, namely to choose its endpoints independent of $\delta$.
If the origin coincides with the first endpoint of $A_1$, this parameterisation is given by 
\be
\label{2int-equal-A1-fixed}
a_1 = 0
\;\qquad\;
b_1 =  \frac{1}{2}
 \;\;\;\;\; \qquad \;\;\;\;\;
 a_2 =   \frac{1}{2}+ \delta
\;\qquad\;
b_2 =  \delta + 1 \,.
\ee
When (\ref{2int-equal-A1-fixed}) is imposed in (\ref{tilde-F2-def-equal-2int-compact}),
the  asymptotic behaviours of $\widetilde{F}_2(\delta ; x ) $ for vanishing and large separation between the two blocks
become respectively
\be
\label{adj-cond-equal2int-v2}
\lim_{\delta \to 0} \widetilde{F}_2(\delta ; x ) = \Delta(0,1; x) 
\;\;\qquad\;\;
\lim_{\delta \to +\infty} \! \widetilde{F}_2(\delta ; x ) = \Delta(0 ,1/2\,; x) + \Delta(1/2+\delta , 1 + \delta\,; x) \,.
\ee

Our numerical analyses are based on the matrices (\ref{V-mat-QP}) and (\ref{T-mat-QP}).
The main technical difficulty at numerical level is that high precision is required. 
In particular, a symplectic eigenvalue very close to $(1/2)^+$ is approximated to $1/2$ by the software
(Mathematica, in our case),
but this is forbidden because a divergence occurs in \eqref{epsilon-symplectic-spectrum}.
In order to avoid this approximation, very high precision is needed.
This issue has already been highlighted in previous works on entanglement Hamiltonians in harmonic lattices
\cite{Arias:2016nip, DiGiulio:2019cxv, DiGiulio:2019lpb, Eisler:2020lyn, Javerzat:2021hxt}.
Our numerical results have been obtained by employing
2000 digits when $\omega = 10^2$ and 5000 digits when $\omega = 10^{10}$.


In Fig.\,\ref{fig:2int-equal-massive-diag} 
we report some numerical results for $T_{i,i}$ 
in the regime of large $\omega$,
for blocks made by $L \in \{20, 50 \}$ sites
and various values of $\delta$. 
Our analysis shows that $\omega = 10^2$ can be already considered a large value for $\omega$ in this setup.
However, also the data for $\omega = 10^{10}$ has been reported, 
for consistency with the corresponding analyses involving unequal blocks (see Sec.\,\ref{sec-2int-generic-EH}),
where larger values for $\omega$ are required  to get a stable picture 
(see the discussion of Fig.\,\ref{fig:2int-massive-EH-diag-chi2-chi4}).
The parameterisation of the endpoints given by (\ref{2int-equal-move-symmetrically})
has been chosen to obtain Fig.\,\ref{fig:2int-equal-massive-diag}.

    The numerical results in Fig.\,\ref{fig:2int-equal-massive-diag} strongly support the validity of (\ref{T-diag-2int});
    indeed the solid curves correspond to the analytic expression (\ref{tilde-F2-def-equal-2int-A1A2-move-V0}) or (\ref{tilde-F2-def-equal-2int-compact}),
    with the choice (\ref{2int-equal-move-symmetrically}).
    In Fig.\,\ref{fig:2int-equal-massive-diag} we have reported the analytic curve for a typical value $\delta < \delta_{\textrm{c}} $ and for another typical value $\delta > \delta_{\textrm{c}} $
    (see the solid red and blue curves respectively), 
    but it has been checked that also the data sets corresponding to the other values of $\delta$ in the figure are nicely reproduced 
    by the analytic expression mentioned above as well.
As already mentioned below (\ref{delta-critical-equal}),
in the same setup, 
we have checked that the corresponding numerical results for the matrix $V$ satisfy (\ref{V-diag-2int})
and nicely agree with the analytic expression of $\widetilde{F}_2(\delta ; x ) $ in the same way.
In Fig.\,\ref{fig:2int-equal-massive-diag}
we have highlighted the blue and the red markers 
because they correspond respectively
to the critical value of $\delta$ (see (\ref{delta-critical-equal})), 
where the transition between the two phases occurs,
and to $\delta = 0$, i.e. to adjacent intervals.
The grey and magenta dashed straight lines indicate the interpolation of the relevant vertices as $\delta$ varies. 
These straight lines have been shown also in Fig.\,\ref{fig:2int-equal-massive-diag-crossover} as a reference for the large $\omega$ regime. 
The slopes of the solid lines and of the magenta dashed lines in Fig.\,\ref{fig:2int-equal-massive-diag}, 
are $\pm 1$ and $\pm 1/2$ respectively.
Thus, this figure also shows that the expectations mentioned 
at the beginning of this subsection are fulfilled. 

It is insightful to compare Fig.\,\ref{fig:2int-equal-massive-diag} against Fig.\,\ref{fig:2int-equal-massive-diag-crossover},
where the same setups have been considered for different regimes of $\omega$.
From this comparison, one observes that the sharp transitions in $\delta$ and discontinuous profiles for $\delta < \delta_{\textrm{c}} $
are characteristic features of the large $\omega$ regime. 
Indeed, they do not occur in Fig.\,\ref{fig:2int-equal-massive-diag-crossover}, where smaller values of $\omega$ are considered. 
However, let us remark that the three-diagonalapproximation does not hold for the data points reported in Fig.\,\ref{fig:2int-equal-massive-diag-crossover}
(see e.g. the middle panels of Fig.\,\ref{fig:EH-density-massive} for $\omega =1$).

\subsection{Single-particle entanglement spectrum}
\label{sec-2int-equal-ES}


\begin{figure}[t!]
\vspace{-.5cm}
\hspace{0.cm}
\includegraphics[width=1\textwidth]{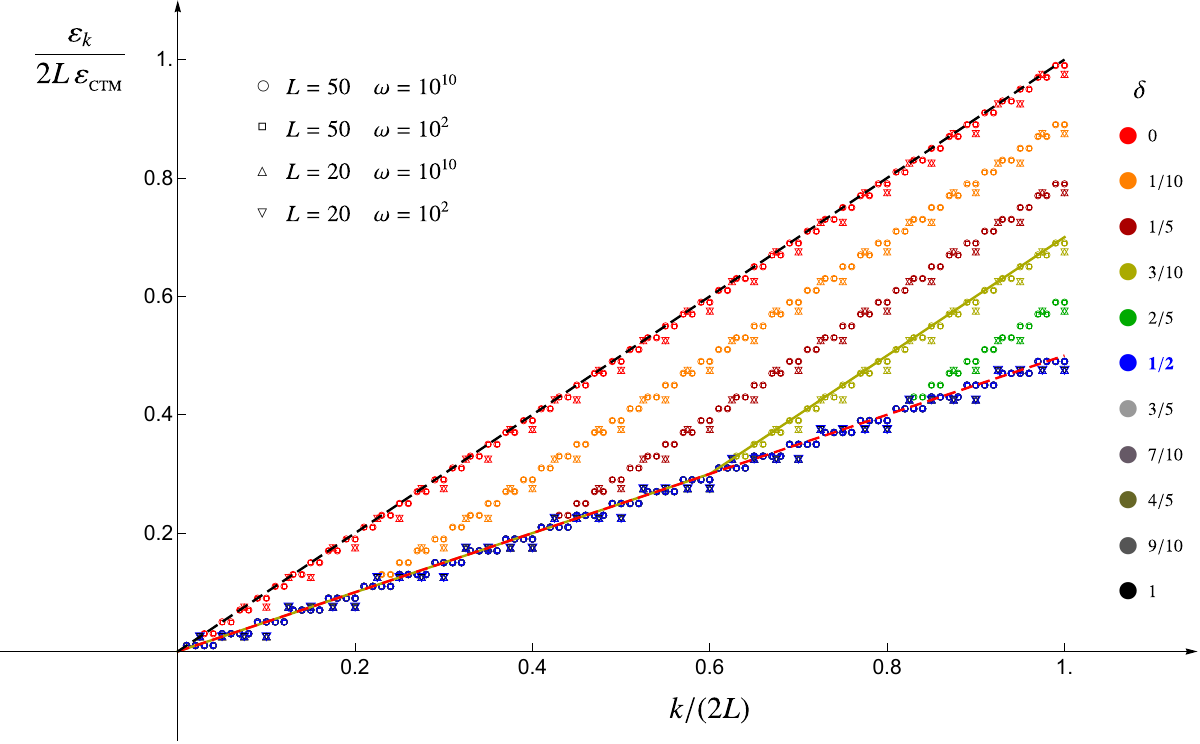}
\vspace{-.2cm}
\caption{
Single-particle entanglement spectrum 
for two equal blocks in the regime of large $\omega$,  for various values of $\delta$
and in the same setup of Fig.\,\ref{fig:2int-equal-massive-diag}.
The solid yellow curve corresponds to the analytic expression in (\ref{es-2int-equal})-(\ref{eta-critical-def}). 
The critical value of $\delta$ is (\ref{delta-critical-equal}), like in Fig.\,\ref{fig:2int-equal-massive-diag}.
}
\label{fig:2int-equal-massive-spectrum}
\end{figure}

We find it worth exploring the effect 
of the occurrence of the two phases described in Sec.\,\ref{sec-EH-2int-equal}
on the single-particle entanglement spectrum,
which is obtained from the symplectic spectrum of the reduced covariance matrix
through (\ref{epsilon-symplectic-spectrum}), with $1 \leqslant k \leqslant 2L$ in this case.

The numerical results for the single-particle entanglement spectrum
are displayed in Fig.\,\ref{fig:2int-equal-massive-spectrum},
for the same setup of  Fig.\,\ref{fig:2int-equal-massive-diag}.
These numerical data points are nicely described by the following analytic expression
\be
\label{es-2int-equal}
\frac{ \varepsilon_k }{2L} =  \varepsilon_{_\textrm{\tiny CTM}}\, \tilde{f}_2\! \left( \delta  \,; \frac{k}{2L} \right)
\ee
where $ \varepsilon_{_\textrm{\tiny CTM}}$ has been introduced (\ref{eps-half-line-ctm})
and $\tilde{f}_2(\delta; \eta) $ is a  continuous piecewise linear function defined as follows
\be
\label{2int-equal-function-spectrum-data}
\tilde{f}_2(\delta; \eta) 
\equiv
\left\{ \begin{array}{ll}
\eta / 2   & \eta \leqslant  \eta_{\textrm{c}}(\delta)
\;\;\qquad\;\;
\\
\rule{0pt}{.5cm}
\eta - \eta_{\textrm{c}}(\delta) / 2  \hspace{.8cm} & \eta >  \eta_{\textrm{c}} (\delta)
\end{array} \right.
\qquad
\eta \in \big[ 0,1\big] 
\ee
being $ \eta_{\textrm{c}}(\delta) $ defined through the critical value $\delta_{\textrm{c}}$ in (\ref{delta-critical-equal}) as 
\be
\label{eta-critical-def}
 \eta_{\textrm{c}}(\delta) 
 \equiv
\left\{ \begin{array}{ll}
\delta/ \delta_{\textrm{c}}  & \delta \leqslant  \delta_{\textrm{c}}
\\
\rule{0pt}{.5cm}
\; 1  \hspace{1.cm} & \delta >  \delta_{\textrm{c}} \,.
\end{array} \right.
\ee

The results reported in Fig.\,\ref{fig:2int-equal-massive-spectrum} show that, 
in the large $\omega$ regime that we are considering, 
the rescaled single-particle entanglement spectrum $\varepsilon_k / (2L\,\varepsilon_{_\textrm{\tiny CTM}}) $ as a function of $k/(2L)$
displays two qualitatively different behaviours for $\delta > \delta_{\textrm{c}}$ and $\delta < \delta_{\textrm{c}}$.
First, let us remark that, as a consistency check,  
when $\delta = 0$ the result for the single block reported in Fig.\,4 of  \cite{Eisler:2020lyn} (bottom panel) 
is recovered (see the dashed black line in Fig.\,\ref{fig:2int-equal-massive-spectrum}, whose slope is equal to $1$)
and in this case the degeneracy is equal to $2$ because two endpoints occur. 
When $\delta > \delta_{\textrm{c}}$, 
the rescaled single-particle entanglement spectrum
is described by a straight line with slope equal to $1/2$ (see the dashed red line in Fig.\,\ref{fig:2int-equal-massive-spectrum})
and the degeneracy of the data points along this line is equal to $4$.
This corresponds to the fact that, since the two blocks are independent at these separation distances,
four endpoints occur at effective level. 
%
Instead,  when $\delta < \delta_{\textrm{c}}$ the rescaled single-particle entanglement spectrum
is described by a continuous piecewise linear function made by two segments
having slope $1/2$ for $0< k/(2L) <   \eta_{\textrm{c}}(\delta) $ and slope $1$ for $ \eta_{\textrm{c}}(\delta) < k/(2L) \leqslant 1$,
where $ \eta_{\textrm{c}}(\delta)$ has been introduced in (\ref{eta-critical-def}).
In this phase the degeneracy of the data points is equal to $4$ for $0< k/(2L) <   \eta_{\textrm{c}}(\delta) $
and equal to $2$ for $ \eta_{\textrm{c}}(\delta) < k/(2L) \leqslant 1$.
This degeneracy pattern seems to tell us that the effective number of endpoints is reduced when $\delta < \delta_{\textrm{c}}$,
becoming equal to $2$ when $\delta = 0$, as expected. 


We remark that the mutual information corresponding to the 
single-particle entanglement spectra reported in Fig.\,\ref{fig:2int-equal-massive-spectrum}
takes the same value for all the different values of $\delta$,
as expected from the fact that the area law holds in this regime. 
In particular, we have checked that 
the entanglement entropy of the union of two equal disjoint blocks 
is twice the value given by (\ref{entropy-interval-massive}) for the single block,
for all values of $\delta$ and $L$ that we have considered. 
Thus, in the limit of large $\omega$, 
the entanglement entropies or the mutual information 
cannot capture the different regimes of $\delta$ discussed above.
This confirms that the entanglement Hamiltonian 
and its single-particle entanglement spectrum 
contain more information than the corresponding entanglement entropies.

\section{Large $\omega$ regime: Two generic blocks}
\label{sec-2int-generic-EH}

In this section, the analyses performed in Sec.\,\ref{sec-2int-massive-equal} for two equal blocks
are extended to the case where the blocks have different lengths
(we can assume $L_1 \leqslant L_2$ without loss of generality);
hence also the relative size of the two blocks (see (\ref{rho-chi-def})) must be considered.

\subsection{Entanglement Hamiltonian}
\label{sec-EH-2int-different}


In the regime of large $\omega$, 
also for $L_1 \leqslant L_2$.
we find that the matrices $T$ and $V$ characterising 
the entanglement Hamiltonian of $A= A_1 \cup A_2$ in the quadratic form (\ref{KA-T-V-matrices})
are well described by the three-diagonals approximation,
discussed in Sec.\,\ref{sec-2int-massive-equal} for equal blocks. 
Our numerical analyses also show that the profiles characterising the three non vanishing diagonals 
can be written conveniently as 
\be
\label{T-diag-2int-different}
\frac{T_{i,i}}{ L_A } 
= 
2 b_{_\textrm{\tiny CTM}}\, F_2\! \left( \delta   , \rho\,; \frac{i -1/2}{ L_A } \right)
\ee
and
\be
\label{V-diag-2int-different}
\frac{V_{i,i}}{ L_A } 
\,=\,
(\omega^2 + 2)\, 2b_{_\textrm{\tiny CTM}}\,F_2\!  \left( \delta   , \rho\,;\frac{i -1/2}{ L_A } \right)
\;\;\;\qquad\;\;\;
\frac{V_{i,i+1}}{ L_A } 
\,=\,
-\,2b_{_\textrm{\tiny CTM}}\, F_2\!  \left( \delta   , \rho\,; \frac{i}{L_A} \right)
\ee
in terms of coefficient $b_{_\textrm{\tiny CTM}}$ introduced in (\ref{b-CTM-def})
and of the function $F_2(\delta,\rho; x)$,
parameterised by the dimensionless ratios (\ref{delta-chi-ratios}) and (\ref{rho-chi-def}).
The special case of equal intervals, discussed in Sec.\,\ref{sec-2int-massive-equal}, corresponds to $\rho = 1/2$.
In order to write an explicit form for $F_2(\delta,\rho; x)$,
let us first discuss the phases encountered for the different values of the parameters 
$\rho$ and $\delta$.


For a given value of $\rho$, as $\delta$ decreases from large positive values to $\delta=0$, 
in our numerical analyses we have observed four different phases, that are labelled by I, II, III and IV in the following
(this increasing order is defined by starting from $\delta \gg 1$ and arriving to $\delta=0$).
These four phases are separated by three critical values of $\delta$, 
denoted by $ \delta_{\textrm{c}}^{\,\textrm{\tiny I/II}} > \delta_{\textrm{c}}^{\,\textrm{\tiny II/III}} > \delta_{\textrm{c}}^{\,\textrm{\tiny III/IV}} $.
We find that these critical values for $\delta$ can be written in terms of the parameter $\rho$ as follows
\be
\label{delta-critical-values-rho}
\delta_{\textrm{c}}^{\,\textrm{I/II}} = \frac{1}{2}
\;\;\qquad\;\;
\delta_{\textrm{c}}^{\,\textrm{\tiny  II/III}} \, , \,
\delta_{\textrm{c}}^{\,\textrm{\tiny  III/IV}} 
\in \bigg\{   \rho \; , \, \frac{1}{2}- \rho \, \bigg\}
\ee
and the corresponding critical values $D_{\textrm{c}}$ for the separation distance $D$ are
\be
\label{critical-distances-values}
D_{\textrm{c}}
\in 
 \bigg\{   \frac{L_1 + L_2}{2} \; , \, \frac{L_2 - L_1}{2} \; , \, L_1  \, \bigg\}  \,. 
\ee
From (\ref{delta-critical-values-rho}), 
one observes the occurrence of a critical ratio $\rho = \rho_{\textrm{c}} $ 
such that $\rho_{\textrm{c}} = 1/2 - \rho_{\textrm{c}}$,
meaning that
\be
\label{critical-distances-L12}
\rho_{\textrm{c}} = \frac{1}{4}
\ee
i.e. $L_2 = 3 L_1$.
When $\rho = \rho_{\textrm{c}} $,  only three phases occur because $\delta_{\textrm{c}}^{\,\textrm{\tiny  II/III}} = \delta_{\textrm{c}}^{\,\textrm{\tiny  III/IV}} $.

A heuristic picture explaining the critical values (\ref{delta-critical-values-rho}) has been described in Appendix\;\ref{app-disk-picture},
but it would be very interesting to find them through analytic calculations.

In order to define the function $F_2(\delta,\rho; x)$ in (\ref{T-diag-2int-different}) and (\ref{V-diag-2int-different}),
we have to introduce four piecewise linear functions with support in $x\in [a,b]$
made by segment with slope equal to either $\pm 1$ or $\pm s$, with $s>0$,
written in terms of the parameters $p$ and $q$ 
satisfying $a < p < q < \tfrac{a+b}{2}$ when  $L_1 \leqslant L_2$.
In the definition of this function, the parameter $s>0$ for the slope is kept generic for the sake of generality, 
but the numerical data are nicely reproduced for
\be
\label{s-value-data}
s=3
\ee
It would be insightful to obtain (\ref{s-value-data}) by employing analytic methods, 
e.g. through a derivation based on an extension of the corner transfer matrix techniques. 

\begin{figure}[t!]
\vspace{-.5cm}
\hspace{-.5cm}
\includegraphics[width=1.07\textwidth]{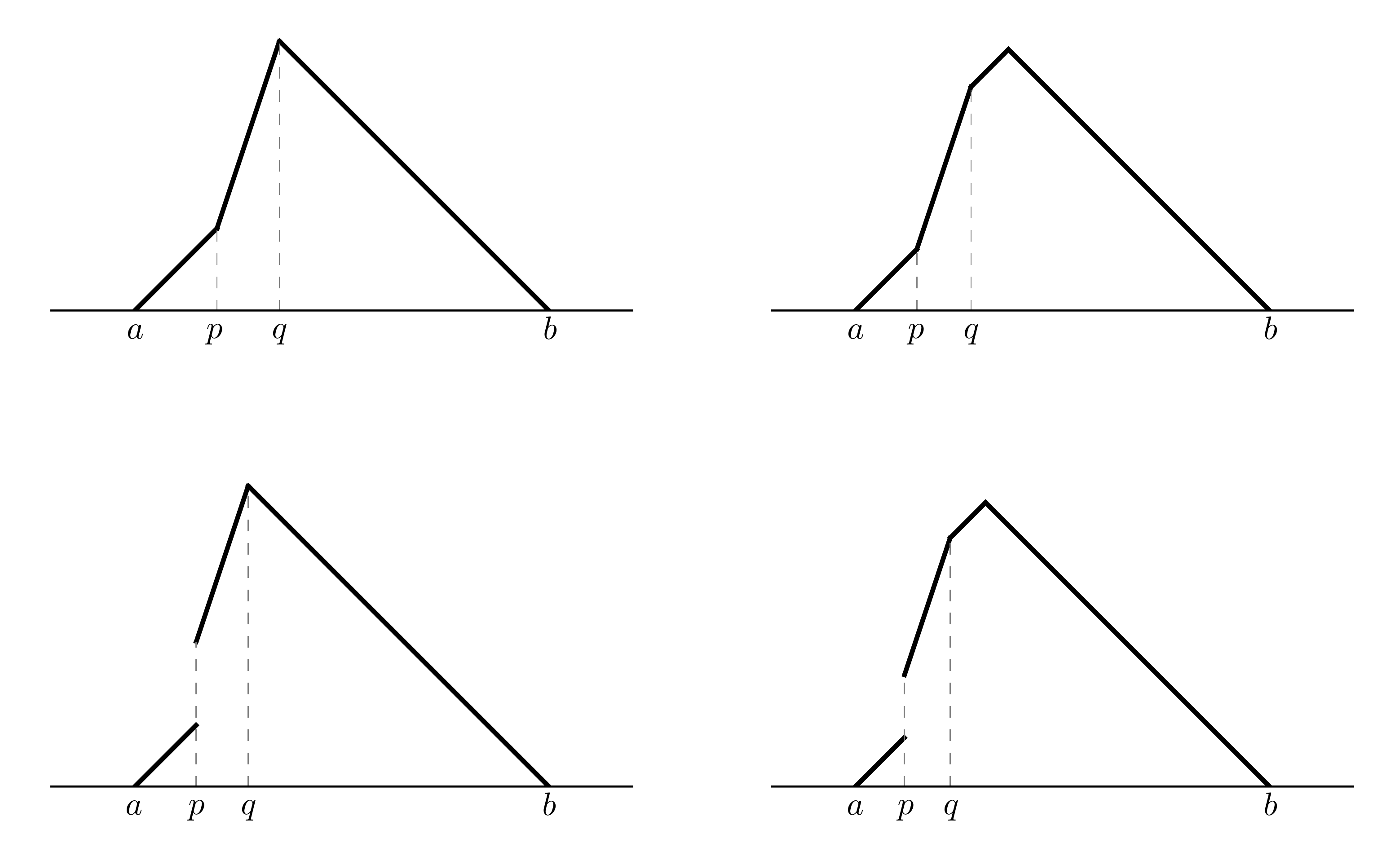}
\vspace{-.6cm}
\caption{
The auxiliary functions occurring in the definition of $F_2(\delta,\rho; x)$ 
(see (\ref{T-diag-2int-different}) and (\ref{V-diag-2int-different})) for two generic intervals,
defined in (\ref{s-value-data})-(\ref{tilde-p-q-lambda-def}).
The slopes of the segments in these piecewise linear functions are either $\pm 1$ or $3$.
}
\label{fig:2int-generic-aux-functions-shapes}
\end{figure}

The continuous piecewise linear functions needed to define $F_2(\delta,\rho; x)$ 
in (\ref{T-diag-2int-different}) and (\ref{V-diag-2int-different}) are
\be
\label{Delta-bar-def}
\bar{\Delta}(a,b; p; x)
\, \equiv \, 
\left\{ \,
\begin{array}{ll}
x -a  & x \in \big[a , p\big]
\\
\rule{0pt}{.45cm}
s\, x-(s-1) p -a \hspace{1cm} &x \in \big[ p   , \tilde{p} \big]
\\
\rule{0pt}{.45cm}
b-x &x \in \big[ \tilde{p}   , b \big]
\end{array}
\right.
\ee
and
\be
\label{Delta-tilde-def}
\tilde{\Delta}(a,b; p, q; x)
\, \equiv \, 
\left\{ \,
\begin{array}{ll}
x -a \hspace{1.cm} & x \in \big[a , p\big]
\\
\rule{0pt}{.45cm}
s\, x-(s-1) p -a \hspace{1cm} &x \in \big[ p   , q \big]
\\
\rule{0pt}{.45cm}
x-(s-1) (p-q) -a \hspace{.5cm}  & x \in \big[ q, \tilde{q}  \big]
\\
\rule{0pt}{.45cm}
b-x &x \in \big[ \tilde{q}   , b \big]
\end{array}
\right.
\ee
where the parameters $\tilde{p}$ and $\tilde{q}$ read
\be
\label{tilde-p-q-Delta-def}
\tilde{p} \equiv \frac{(s-1) p + a+ b}{s+1}
\;\;\;\qquad\;\;\;
\tilde{q} \equiv \frac{(s-1) (p-q) + a+ b}{2}  \,. 
\ee
The auxiliary functions (\ref{Delta-bar-def}) and (\ref{Delta-tilde-def})
are shown in the top left and top right panels 
of Fig.\,\ref{fig:2int-generic-aux-functions-shapes} respectively. 
Moreover, notice that $\bar{\Delta}$ is a deformation of $\Delta$, while $\tilde{\Delta}$ is  a deformation of $\bar{\Delta}$.
The other two piecewise linear functions defining $F_2(\delta,\rho; x)$ 
are discontinuous at $x=p$, where a finite jump of height $h$ occurs, and continuous at $x=q$.
They read
\be
\label{lambda-bar-def}
\bar{\lambda}(a,b; p, h ; x)
\, \equiv \, 
\left\{ \,
\begin{array}{ll}
x -a  & x \in \big[a , p\big]
\\
\rule{0pt}{.45cm}
s\, x-(s-1) p - a +h  \hspace{1cm} &x \in \big[ p   , \tilde{p} \big]
\\
\rule{0pt}{.45cm}
b-x &x \in \big[ \tilde{p}   , b \big]
\end{array}
\right.
\ee
 and
\be
\label{lambda-tilde-def}
\tilde{\lambda}(a,b; p, h, q; x)
\, \equiv \, 
\left\{ \,
\begin{array}{ll}
x -a \hspace{1.cm} & x \in \big[a , p\big]
\\
\rule{0pt}{.45cm}
s\, x-(s-1) p - a + h  &x \in \big[ p   , q \big]
\\
\rule{0pt}{.45cm}
x-(s-1) (p-q) -a + h\hspace{1cm}  & x \in \big[ q, \tilde{q}  \big]
\\
\rule{0pt}{.45cm}
b-x &x \in \big[ \tilde{q}   , b \big]
\end{array}
\right.
\ee
where the parameters $\tilde{p}$ and $\tilde{q}$ are given by 
\be
\label{tilde-p-q-lambda-def}
\tilde{p} \equiv \frac{(s-1) p + a+ b-h}{s+1}
\;\;\;\qquad\;\;\;
\tilde{q} \equiv \frac{(s-1) (p-q) + a+ b-h}{2}  \,. 
\ee
The discontinuous auxiliary functions (\ref{lambda-bar-def}) and (\ref{lambda-tilde-def})
are shown in the bottom left and bottom right panels 
of Fig.\,\ref{fig:2int-generic-aux-functions-shapes} respectively. 
Notice that  $\bar{\lambda}$ is a deformation of $\lambda$, while $\tilde{\lambda}$ is  a deformation of $\bar{\lambda}$.
Moreover, (\ref{lambda-bar-def}) becomes (\ref{Delta-bar-def}) when $h=0$.
In the following, we often enlighten the notation by adding the subindex $j$ to the functions introduced in 
(\ref{s-value-data})-(\ref{tilde-p-q-lambda-def}) 
whenever they are supported in $A_j$, suppressing the explicit dependence on  $a_j$ and $b_j$;
hence e.g. $\bar{\Delta}_1(p; x) \equiv \bar{\Delta}(a_1,b_1; p; x)$
and $\tilde{\lambda}_2(p, h, q; x) \equiv \tilde{\lambda}(a_2,b_2; p, h, q; x)$.

\begin{figure}[t!]
\vspace{-.5cm}
\hspace{0.cm}
\includegraphics[width=1.\textwidth]{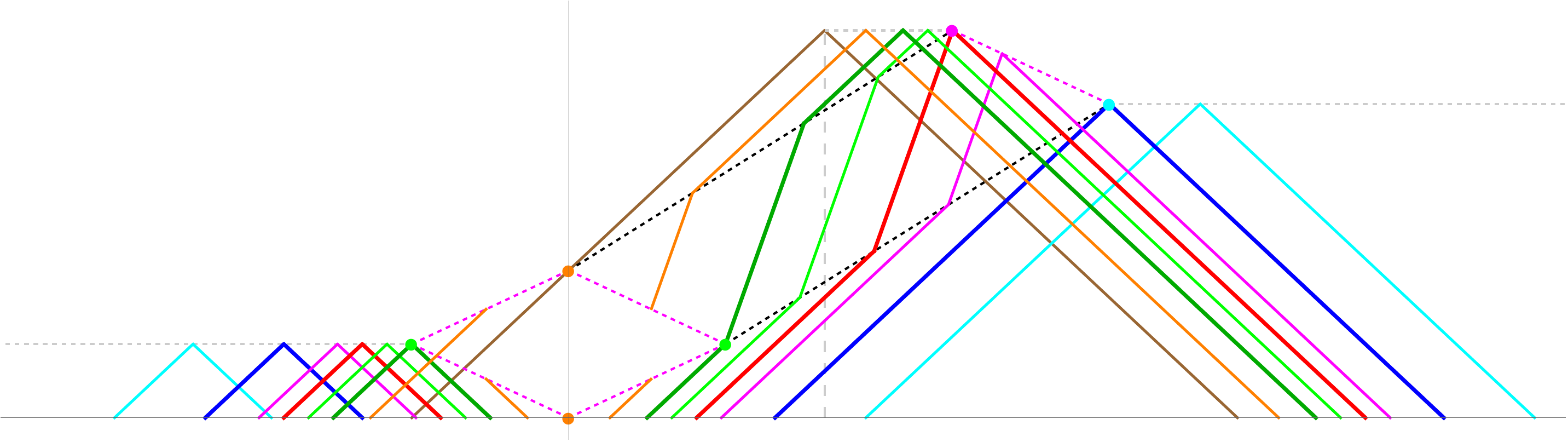}
\\
\rule{0pt}{5.5cm}
\includegraphics[width=1.\textwidth]{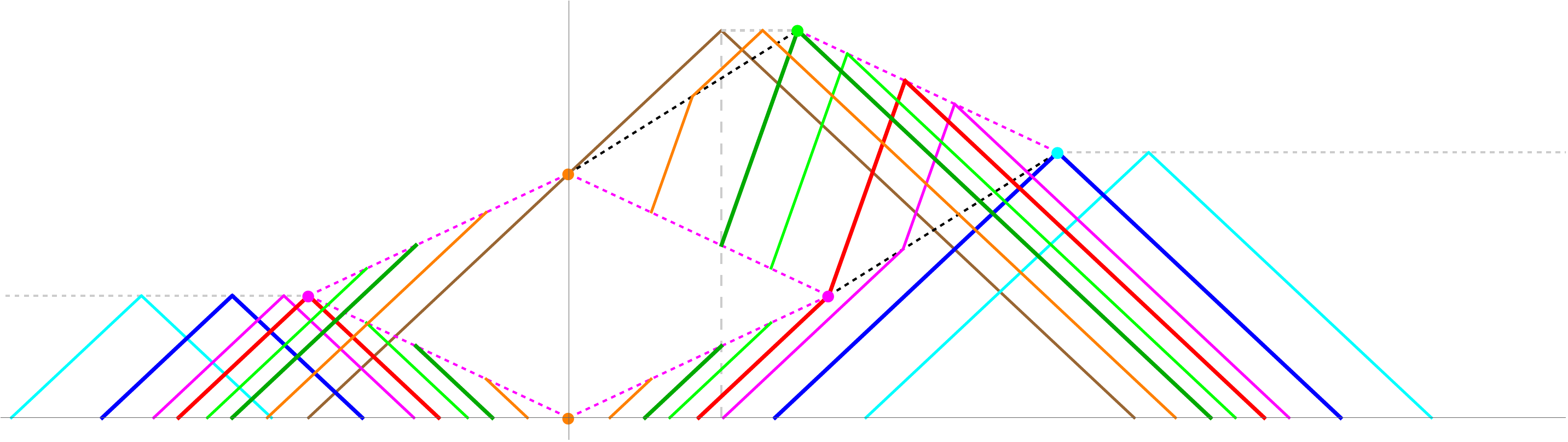}
\vspace{-.2cm}
\caption{The function $F_2(\delta,\rho; x)$ for $\rho  < \rho_{\textrm{c}}$ (top panel) and  $\rho  > \rho_{\textrm{c}}$ (bottom panel),
given in (\ref{tilde-F2-def-different-2int-below-rhoc}) and (\ref{tilde-F2-def-different-2int-above-rhoc}) respectively,
for various values of $\delta$ corresponding to different colours. 
Four phases occur and each of them is described by one of the auxiliary functions shown in Fig.\,\ref{fig:2int-generic-aux-functions-shapes}.
The dots denote the points in  (\ref{P-points-rhombi-main}) and (\ref{Q-points-rhombi-main}).
}
\label{fig:2int-skeleton}
\end{figure}


As already discussed in Sec.\,\ref{sec-2int-massive-equal},
the choice of the parameterisation of the endpoints of the blocks $A_1$ and $A_2$ 
is crucial to perform a comparison between 
the analytic expression in (\ref{T-diag-2int-different}) and (\ref{V-diag-2int-different})
and their numerical data points. 
The results discussed in this section correspond to the following parameterisation
\be
\label{endpoints-A-symm-move}
a_1 = - \frac{\delta}{2}  - \rho
\;\qquad\;
b_1 =  - \frac{\delta}{2} 
 \;\;\;\;\; \qquad \;\;\;\;\;
 a_2 =   \frac{\delta}{2} 
\;\qquad\;
b_2 =  \frac{\delta}{2} + 1-\rho
\ee
meaning that $A_1$ and $A_2$ behave in the same way, 
symmetrically with respect to the origin, as $\delta$ decreases until $\delta=0$,
where they become adjacent at the origin.
Indeed,  (\ref{endpoints-A-symm-move}) becomes (\ref{2int-equal-move-symmetrically}) in the special case of $\rho=1/2$, i.e. for $L_1 = L_2$.

Instead, in Appendix\;\ref{app-A1-fixed} a parameterisation where the endpoints of $A_1$ are independent of $\delta$ is considered.
In particular, one uses
\be
\label{endpoints-A1-fixed-A2-move}
a_1 =  - \,\rho
\;\qquad\;
b_1 =  0
 \;\;\;\;\; \qquad \;\;\;\;\;
 a_2 =   \delta
\;\qquad\;
b_2 =  \delta + 1-\rho
\ee
meaning that, when $\delta =0$, 
the origin corresponds to the shared endpoint of the two adjacent blocks.


We find it worth providing further details about the construction of the function 
$F_2(\delta,\rho; x)$ in (\ref{T-diag-2int-different}) and (\ref{V-diag-2int-different}) 
by introducing the following rhombi, 
which correspond to the dashed magenta and dashed black segments in Fig.\,\ref{fig:2int-skeleton},
whose slopes are $\pm 1/2$ and $2/3$ respectively. 
Without loss of generality, we consider the parameterisation (\ref{endpoints-A-symm-move}),
where the origin $O \equiv (0,0)$ is fixed in the point 
where the two blocks join when their separation distance vanishes.
Consider the rhombus $\mathcal{R}_0$ having vertices $O$, $P_0$ and $P_\pm$, where
\be
\label{P-points-rhombi-main}
P_0 \equiv (0 \,,  \rho )
\;\;\;\;\;\qquad \;\;\;\;
P_\pm \equiv ( \pm \rho \,,  \rho/2 )
\ee
which is made by the contiguous dashed magenta segments in Fig.\,\ref{fig:2int-skeleton}.
Since $L_1 < L_2$, the other rhombus $\widetilde{\mathcal{R}}$
have its vertices in $P_0$, $P_+$, $Q_{<}$ and $Q_{>}$, where
\be
\label{Q-points-rhombi-main}
Q_< \equiv \bigg( \, \frac{3}{2} \big(1/2 - \rho \big)\, , \, 1/2 \bigg)
\;\;\;\;\;\qquad \;\;\;\;
Q_> \equiv \bigg( \rho+ \frac{3}{2} \big(1/2 - \rho \big) \, , \, \frac{1-\rho}{2} \bigg)  \,. 
\ee
The segment whose endpoints are $P_0$ and $Q_> $
and the one having its endpoints in $P_+$ and $Q_< $ have slope equal to $2/3$
(see the dashed black segments in Fig.\,\ref{fig:2int-skeleton}),
while the slope of the other two segments of $\widetilde{\mathcal{R}}$ is $-1/2$. 
In the special case of  $L_1 = L_2$, i.e. for $\rho=1/2$, 
we have that $Q_<  = P_0$ and $Q_>  = P_+$;
hence only $\mathcal{R}_0$ occurs because $\widetilde{\mathcal{R}}$ shrinks to a segment (see Fig.\,\ref{fig:2int-equal-massive-diag}).

As for the parameterisation (\ref{endpoints-A1-fixed-A2-move}), 
the corresponding rhombi are described in Appendix\;\ref{app-A1-fixed}.

In Fig.\,\ref{fig:2int-skeleton} we show that, 
in order to define $F_2(\delta,\rho; x)$ in (\ref{T-diag-2int-different}) and (\ref{V-diag-2int-different}),
the vertices of the auxiliary functions in Fig.\,\ref{fig:2int-generic-aux-functions-shapes}
must belong to the edges of the rhombi $\mathcal{R}_0$ and  $\widetilde{\mathcal{R}}$ above introduced
and to the horizontal dashed grey half lines.
%
These constraints allow us to determine the parameters in the auxiliary functions (\ref{s-value-data})-(\ref{tilde-p-q-lambda-def})
and this provides the following analytic expressions for 
the function $F_2(\delta,\rho; x)$ occurring in (\ref{T-diag-2int-different}) and (\ref{V-diag-2int-different}),
which hold for the parameterisation of the endpoints of $A$ given by (\ref{endpoints-A-symm-move}).

For $\rho  \in \big(0 ,\rho_{\textrm{c}} \big]$,
the function $F_2(\delta,\rho; x)$ can be written as follows
\be
\label{tilde-F2-def-different-2int-below-rhoc}
F_2( \delta, \rho \, ; x ) 
\,\equiv\,
\left\{
\begin{array}{ll}
\Delta_1( x) + \Delta_2(x)
 & 
\delta \in \big[\,\delta_{\textrm{c}}^{\,\textrm{\tiny I/II}} , +\infty\,\big)
\\
\rule{0pt}{.6cm}
\Delta_1( x) +  \bar{\Delta}_2( 3a_2 -\rho/2 ; x)
 & 
\delta \in \big[\,\delta_{\textrm{c}}^{\,\textrm{\tiny II/III}} , \delta_{\textrm{c}}^{\,\textrm{\tiny I/II}} \,\big]
\\
\rule{0pt}{.6cm}
\Delta_1( x) + \tilde{\Delta}_2( 3a_2 -\rho/2  , 3a_2 ; x)
 & 
\delta \in \big[\,\delta_{\textrm{c}}^{\,\textrm{\tiny III/IV}} , \delta_{\textrm{c}}^{\,\textrm{\tiny II/III}} \,\big]
\\
\rule{0pt}{.6cm}
\lambda_1(2b_1;x) + \tilde{\lambda}_2( 2a_2, \rho - 2a_2, 3a_2 ; x )
 \hspace{.8cm}
 & 
\delta \in \big[\,0\,, \delta_{\textrm{c}}^{\,\textrm{\tiny III/IV}} \,\big]
\end{array}
\right.
\ee
while for $\rho  \in \big[\rho_{\textrm{c}} \, , 1/2\big]$ it is given by 
\be
\label{tilde-F2-def-different-2int-above-rhoc}
F_2( \delta, \rho \, ; x ) 
\,\equiv\,
\left\{
\begin{array}{ll}
\Delta_1( x) + \Delta_2(x)
 & 
\delta \in \big[\,\delta_{\textrm{c}}^{\,\textrm{\tiny I/II}} , +\infty\,\big)
\\
\rule{0pt}{.6cm}
\Delta_1( x) +  \bar{\Delta}_2( 3a_2 -\rho/2 ; x)
 & 
\delta \in \big[\,\delta_{\textrm{c}}^{\,\textrm{\tiny II/III}} , \delta_{\textrm{c}}^{\,\textrm{\tiny I/II}} \,\big]
\\
\rule{0pt}{.6cm}
\lambda_1(2b_1;x) + \bar{\lambda}_2(2a_2, \rho - 2a_2;x)
 & 
\delta \in \big[\,\delta_{\textrm{c}}^{\,\textrm{\tiny III/IV}} , \delta_{\textrm{c}}^{\,\textrm{\tiny II/III}} \,\big]
\\
\rule{0pt}{.6cm}
\lambda_1(2b_1;x) + \tilde{\lambda}_2( 2a_2, \rho - 2a_2, 3a_2 ; x )
 \hspace{.8cm}
 & 
\delta \in \big[\,0\,, \delta_{\textrm{c}}^{\,\textrm{\tiny III/IV}} \,\big]
\end{array}
\right.
\ee
in terms 
of the auxiliary functions introduced in (\ref{s-value-data})-(\ref{tilde-p-q-lambda-def})
and of the occurrence of the four phases I, II, III and IV,
which are separated by the critical values of $\delta$ defined in (\ref{delta-critical-values-rho}).

\begin{figure}[t!]
\vspace{-.5cm}
\hspace{-.27cm}
\includegraphics[width=1\textwidth]{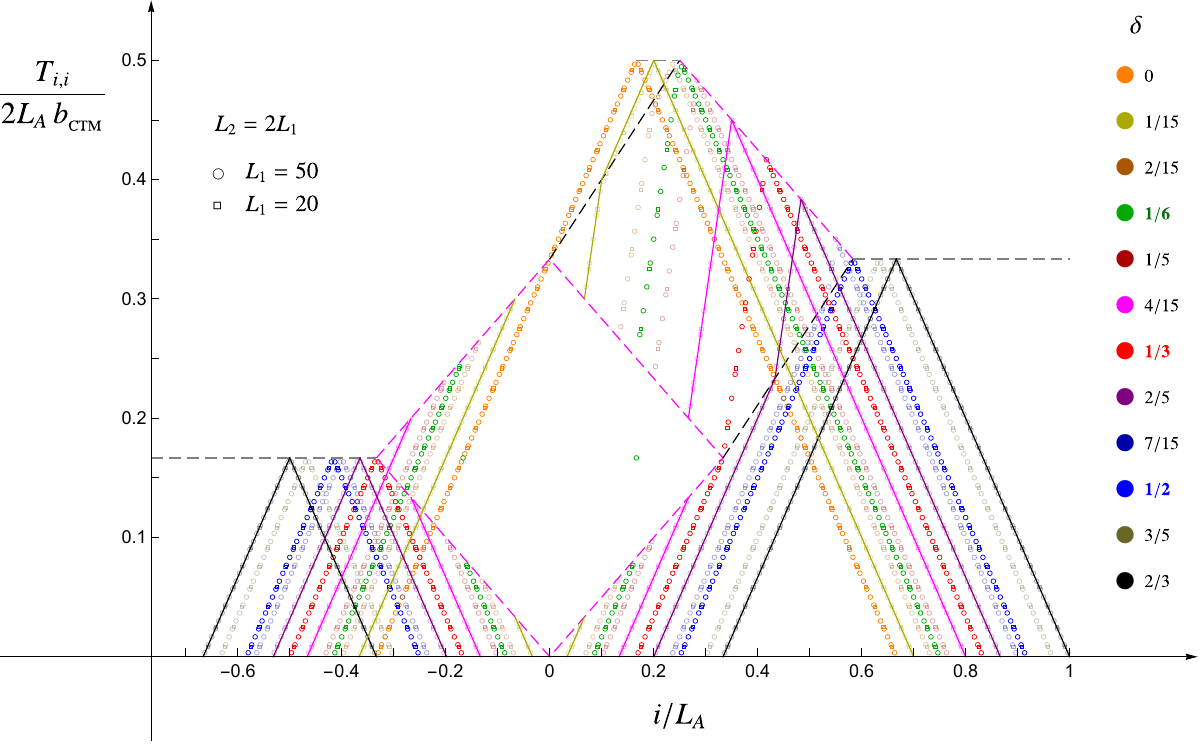}
\\
\rule{0pt}{10.2cm}
\hspace{-.4cm}
\includegraphics[width=1\textwidth]{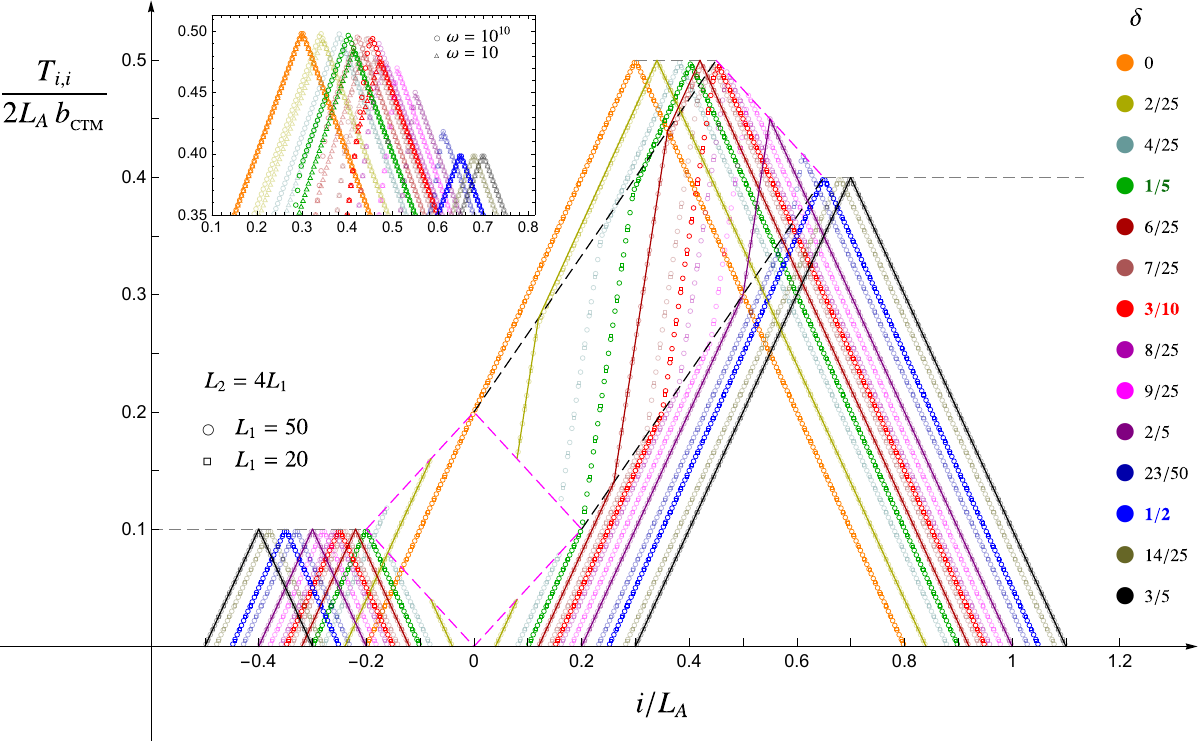}
\vspace{.4cm}
\caption{Main diagonal of $T$ for two generic blocks when $\omega = 10^{10}$, for various values of $\delta$. 
The solid curves correspond to (\ref{T-diag-2int-different}).
Here  $\rho = 1/3 $ (top) and  $\rho = 1/5 $ (bottom), 
hence (\ref{tilde-F2-def-different-2int-below-rhoc}) and (\ref{tilde-F2-def-different-2int-above-rhoc})
are used respectively for the analytic prediction (see also Fig.\,\ref{fig:2int-skeleton}). 
The critical values (\ref{delta-critical-values-rho}) are coloured in the legenda. 
}
\label{fig:2int-massive-EH-diag-chi2-chi4}
\end{figure}

\clearpage

\begin{figure}[t!]
\vspace{-.5cm}
\hspace{0.cm}
\includegraphics[width=1\textwidth]{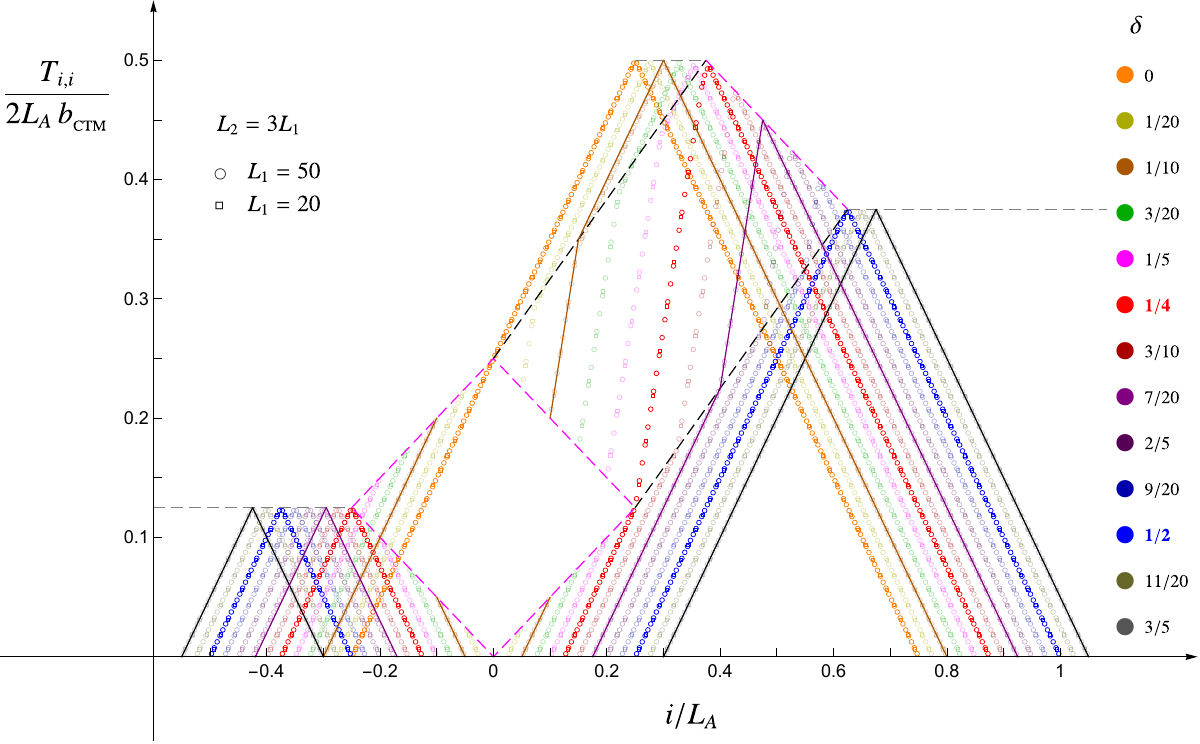}
\vspace{-.2cm}
\caption{
Main diagonal of $T$ for two generic blocks for $\omega = 10^{10}$ and various values of $\delta$. 
The solid curves correspond to (\ref{T-diag-2int-different}) and (\ref{tilde-F2-def-different-2int-below-rhoc})-(\ref{tilde-F2-def-different-2int-above-rhoc}).
Here  $\rho = \rho_{\textrm{c}}= 1/4 $ takes the critical value (\ref{critical-distances-L12}); hence only three phases occur.
}
\label{fig:2int-massive-EH-diag-chi3}
\end{figure}

When $\rho  < \rho_{\textrm{c}}$ the function  $F_2(\delta,\rho; x)$ is discontinuous only in phase IV, 
while for $\rho  > \rho_{\textrm{c}}$ it is discontinuous in both phases III and IV.
The expressions (\ref{tilde-F2-def-different-2int-below-rhoc}) and (\ref{tilde-F2-def-different-2int-above-rhoc})
provide the piecewise linear curves in the top and bottom panel of Fig.\,\ref{fig:2int-skeleton} respectively. 

In the limit of either vanishing or large separation distance between the blocks, 
the function $F_2(\delta,\rho; x)$ defined in (\ref{tilde-F2-def-different-2int-below-rhoc}) and (\ref{tilde-F2-def-different-2int-above-rhoc})
satisfies respectively the following consistency conditions 
\bea
\label{adj-cond-diverse2int}
\lim_{\delta \to 0} F_2(\delta, \rho \,; x )  
& = & \Delta(-\rho\,,1-\rho\,; x) 
\\
\label{distant-cond-diverse2int}
\rule{0pt}{.4cm}
\lim_{\delta \to +\infty} \! F_2(\delta, \rho \,; x )  
& = &
\Delta ( a_1, b_1 ; x )
+
\Delta ( a_2, b_2 ; x )
\eea
that become the ones in (\ref{adj-cond-equal2int}) in the special case of equal intervals, as expected. 

In Appendix\;\ref{app-A1-fixed}  we report the definition of the function $F_2(\delta, \rho \,; x )$
for the parameterisation of the endpoints of $A$ given by (\ref{endpoints-A1-fixed-A2-move}).
Notice that the consistency conditions (\ref{adj-cond-diverse2int}) and (\ref{distant-cond-diverse2int}) hold 
also for this choice.



In Fig.\,\ref{fig:2int-massive-EH-diag-chi2-chi4} and Fig.\,\ref{fig:2int-massive-EH-diag-chi3}
we report our numerical results for the main diagonal of the matrix $T$ in (\ref{KA-T-V-matrices}),
evaluated through (\ref{T-mat-QP}),
in the regime of large $\omega$ and for $L_1 < L_2$.
The data points are nicely described by the analytic expression given by (\ref{T-diag-2int-different}),
(\ref{tilde-F2-def-different-2int-below-rhoc}) and (\ref{tilde-F2-def-different-2int-above-rhoc}).
In particular, these numerical results have been obtained for $\omega = 10^{10}$,
by setting either $L_2 = 2L_1$ or $L_2 = 4L_1$
(top and bottom panel of Fig.\,\ref{fig:2int-massive-EH-diag-chi2-chi4} respectively)
or $L_2 = 3L_1$ 
(Fig.\,\ref{fig:2int-massive-EH-diag-chi3}, which corresponds to the critical value (\ref{critical-distances-L12}) for $\rho$).
The numerical values for the critical values for $\delta$ (see (\ref{delta-critical-values-rho})) 
have been highlighted through coloured and bold numbers in the legenda of Fig.\,\ref{fig:2int-massive-EH-diag-chi2-chi4} and Fig.\,\ref{fig:2int-massive-EH-diag-chi3}
and the corresponding markers in the plots have been made more visible. 
In the same setup, we have checked that the corresponding numerical results
for the matrix $V$ in (\ref{KA-T-V-matrices}), evaluated through (\ref{V-mat-QP}),
are nicely described by (\ref{V-diag-2int-different}), (\ref{tilde-F2-def-different-2int-below-rhoc}) and (\ref{tilde-F2-def-different-2int-above-rhoc}),
but the outcomes of these analyses have not been reported here because basically the same curves shown in 
Fig.\,\ref{fig:2int-massive-EH-diag-chi2-chi4} and Fig.\,\ref{fig:2int-massive-EH-diag-chi3} have been found.



The (large) numerical values of  $\omega$ have been chosen by requiring reliable collapses for the data points. 
In the inset in the bottom panel of Fig.\,\ref{fig:2int-massive-EH-diag-chi2-chi4},
we show that, while data collapses are observed for $\delta\geqslant 1/2$ already at $\omega =10$, 
for smaller separation distance this is not the case;
hence a larger value of $\omega$ is needed to find the profiles described above. 
Unfortunately, larger values of $\omega$ require higher numerical precisions.   
For instance, at $\omega =10^{10}$, about $10^4$ digits of precision are needed to obtain reliable data points.

\subsection{Single-particle entanglement spectrum}
\label{sec-2int-generic-ES}


The single-particle entanglement spectrum is obtained from the symplectic spectrum of $\gamma_A$ through (\ref{epsilon-symplectic-spectrum}),
hence it contains less information than the entanglement Hamiltonian. 
In the following this quantity is explored in the case where $L_1 \leqslant L_2$,
extending the corresponding analysis performed in Sec.\,\ref{sec-2int-massive-equal} for two equal blocks.

\begin{figure}[t!]
\vspace{-.5cm}
\hspace{0.cm}
\includegraphics[width=1\textwidth]{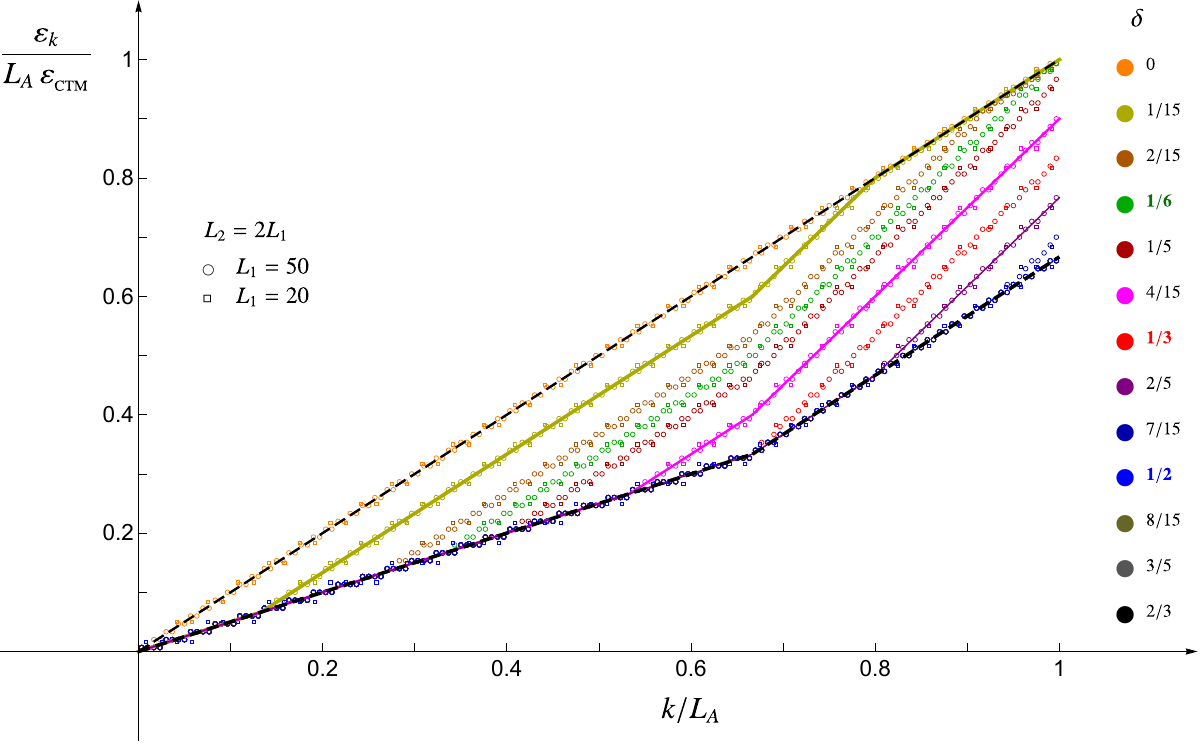}
\\
\rule{0pt}{10.2cm}
\hspace{-.1cm}
\includegraphics[width=1\textwidth]{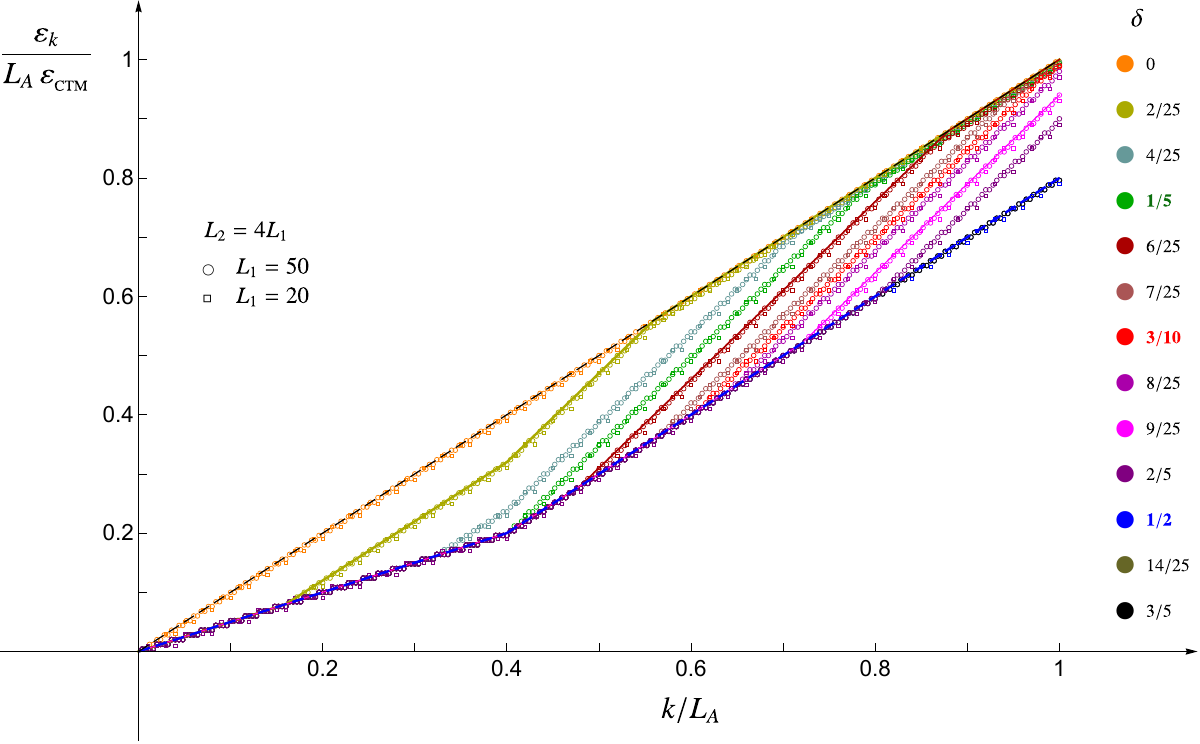}
\vspace{-.2cm}
\caption{
Single-particle entanglement spectrum for $\rho=1/3$ (top) and $\rho=1/5$ (bottom) 
in the regime of large $\omega$ and for various values of $\delta$,
in the same setups of the corresponding panels of Fig.\,\ref{fig:2int-massive-EH-diag-chi2-chi4} .
The solid curves are obtained from the analytic expression 
defined in (\ref{es-2int-different})-(\ref{g2-function-def}).  
The critical values (\ref{delta-critical-values-rho}) are coloured in the legenda. 
}
\label{fig:2int-equal-massive-spectrum-chi2-chi4}
\end{figure}
\begin{figure}[t!]
\vspace{-.5cm}
\hspace{0.cm}
\includegraphics[width=1\textwidth]{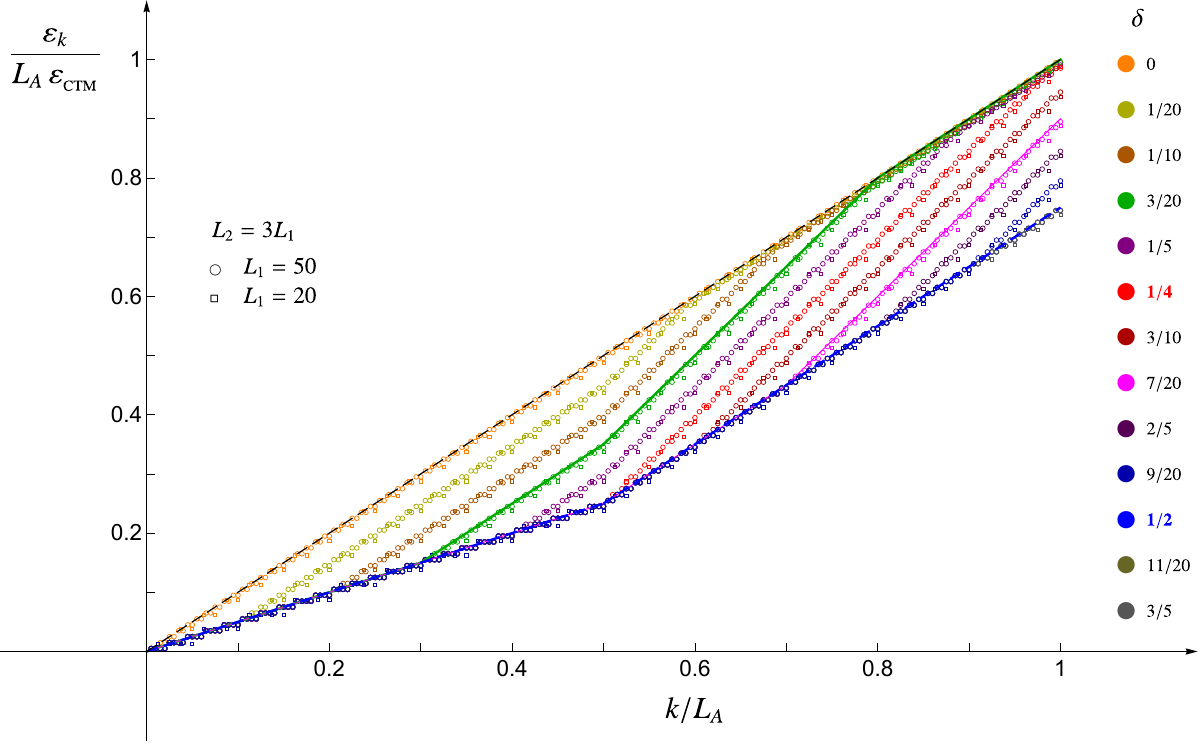}
\vspace{-.2cm}
\caption{
Single-particle entanglement spectrum for two blocks
at the critical value $\rho = \rho_{\textrm{c}}= 1/4 $ (see (\ref{critical-distances-L12}))
in the regime of large $\omega$ and for various values of $\delta$,
in the same setups of Fig.\,\ref{fig:2int-massive-EH-diag-chi3}.
The solid curves are obtained from the analytic expression 
defined in (\ref{es-2int-different})-(\ref{g2-function-def}). 
}
\label{fig:2int-equal-massive-spectrum-chi3}
\end{figure}

In the regime of large $\omega$, our numerical results are reported
in Fig.\,\ref{fig:2int-equal-massive-spectrum-chi2-chi4} and Fig.\,\ref{fig:2int-equal-massive-spectrum-chi3},
showing the single-particle entanglement spectra for the entanglement Hamiltonians employed in 
Fig.\,\ref{fig:2int-massive-EH-diag-chi2-chi4} and Fig.\,\ref{fig:2int-massive-EH-diag-chi3} respectively. 
These data points in the regime of large $\omega$ are nicely described by 
\be
\label{es-2int-different}
\frac{ \varepsilon_k }{L_A} =  \varepsilon_{_\textrm{\tiny CTM}}\, f_2\! \left( \rho , \delta  \,; \frac{k}{L_A} \right)
\ee
where $1 \leqslant k \leqslant L_A$, 
the coefficient $\varepsilon_{_\textrm{\tiny CTM}}$ is (\ref{eps-half-line-ctm})
and $f_2(\rho, \delta; \eta) $ is the continuous piecewise linear function defined as follows
\be
\label{2int-function-spectrum-data}
f_2(\rho, \delta; \eta) 
\equiv
\left\{ \begin{array}{ll}
g_2(\rho, \delta; \eta)  \hspace{1cm} & \delta \leqslant  \rho
\\
\rule{0pt}{.5cm}
g_2(\delta, \rho; \eta)  \hspace{.8cm} & \delta \geqslant  \rho
\end{array} \right.
\;\;\;\qquad\;\;\;\;
\eta \in \big[ 0,1\big] 
\ee
in terms of
\be
\label{g2-function-def}
g_2(\rho, \delta; \eta)
\,\equiv\,
\left\{
\begin{array}{ll}
\eta/2    &    \eta \in \big[\,0 \, , 2\delta \,\big]
\\
\rule{0pt}{.5cm}
\eta - \delta    &    \eta \in \big[\,2\delta \, , 2\rho \,\big]
\\
\rule{0pt}{.5cm}
3\eta/2 - \delta - \rho    \hspace{1cm} &    \eta \in \big[\,2\rho \, ,  \textrm{min} \big\{ 2(\rho + \delta), 1\big\}  \,\big]
\\
\rule{0pt}{.5cm}
\eta     &    \eta \in \big[\, 2(\rho + \delta) \, , 1 \,\big]
\end{array}
\right.
\ee
where the fourth segment, given by $\eta$, does not occur whenever $2(\rho+\delta) \geqslant 1$,
i.e. for $\delta \geqslant 1/2 - \rho$.
As a consistency check for (\ref{2int-function-spectrum-data}), 
we observe that (\ref{2int-equal-function-spectrum-data}) is recovered 
in the special case of equal intervals, namely
\be
f_2(1/2, \delta; \eta) = \tilde{f}_2(\delta; \eta)   \,. 
\ee

\clearpage

In Fig.\,\ref{fig:2int-equal-massive-spectrum-chi2-chi4}
we show the single-particle entanglement spectra 
(rescaled through $\varepsilon_{_\textrm{\tiny CTM}}$ in (\ref{eps-half-line-ctm}), as suggested by (\ref{es-2int-different})) 
corresponding to the entanglement Hamiltonians considered in Fig.\,\ref{fig:2int-massive-EH-diag-chi2-chi4}.
Since $\rho=1/3$ and $\rho=1/5$ in the top and bottom panel respectively, 
the three critical values for $\delta$ in (\ref{delta-critical-values-rho}) are distinct
and in the legenda their values are bold and coloured. 
The numerical data points are nicely described by 
the analytic expression defined in (\ref{es-2int-different})-(\ref{g2-function-def}),
as shown by the solid lines in Fig.\,\ref{fig:2int-equal-massive-spectrum-chi2-chi4}.
%
In Fig.\,\ref{fig:2int-equal-massive-spectrum-chi3}
the single-particle entanglement spectra of the entanglement Hamiltonians 
in Fig.\,\ref{fig:2int-massive-EH-diag-chi3} are reported
and,
since $\rho = \rho_{\textrm{c}}= 1/4 $ takes the critical value (\ref{critical-distances-L12}) in this case, 
only two distinct critical values of $\delta$ occur
(see the coloured and bold values of $\delta$ in the legenda).
Different degeneracies are observed in the domains of $k/L_A$ 
where the linear behaviour of the single-particle entanglement spectrum follows different slopes. 
In particular, the data points following the slope $1/2$ and $1$ have degeneracy equal to $4$ and $2$ respectively,
while along the segments with slope $3/2$ an unusual degeneracy pattern occurs.


The mutual information of two blocks for a given value of $\rho$
does not change with their separation distance because
the area law holds in this regime of large $\omega$,
similarly to the equal blocks case discussed in the last paragraph
of Sec.\,\ref{sec-2int-equal-ES}.
Moreover, also for disjoint blocks of unequal sizes
the entanglement entropy of their union
is twice the value given by (\ref{entropy-interval-massive}) for the single block,
and therefore it is independent of $\delta$, $L_1$ and $L_2$, 
as expected from the area law.  
This tells us that 
the various regimes for $\delta$ observed for large $\omega$
through the entanglement Hamiltonian and its single-particle entanglement spectrum
cannot be captured by employing the entanglement entropies,
as already remarked in Sec.\,\ref{sec-2int-equal-ES}.

\section{Massless regime: Two generic blocks}
\label{sec-2int-massless}


In this section we explore 
the entanglement Hamiltonian of the union of two disjoint blocks $A = A_1 \cup A_2$ 
in the massless regime of the infinite harmonic chain.

\begin{figure}[t!]
\vspace{-1.1cm}
\includegraphics[width=1.05\textwidth]{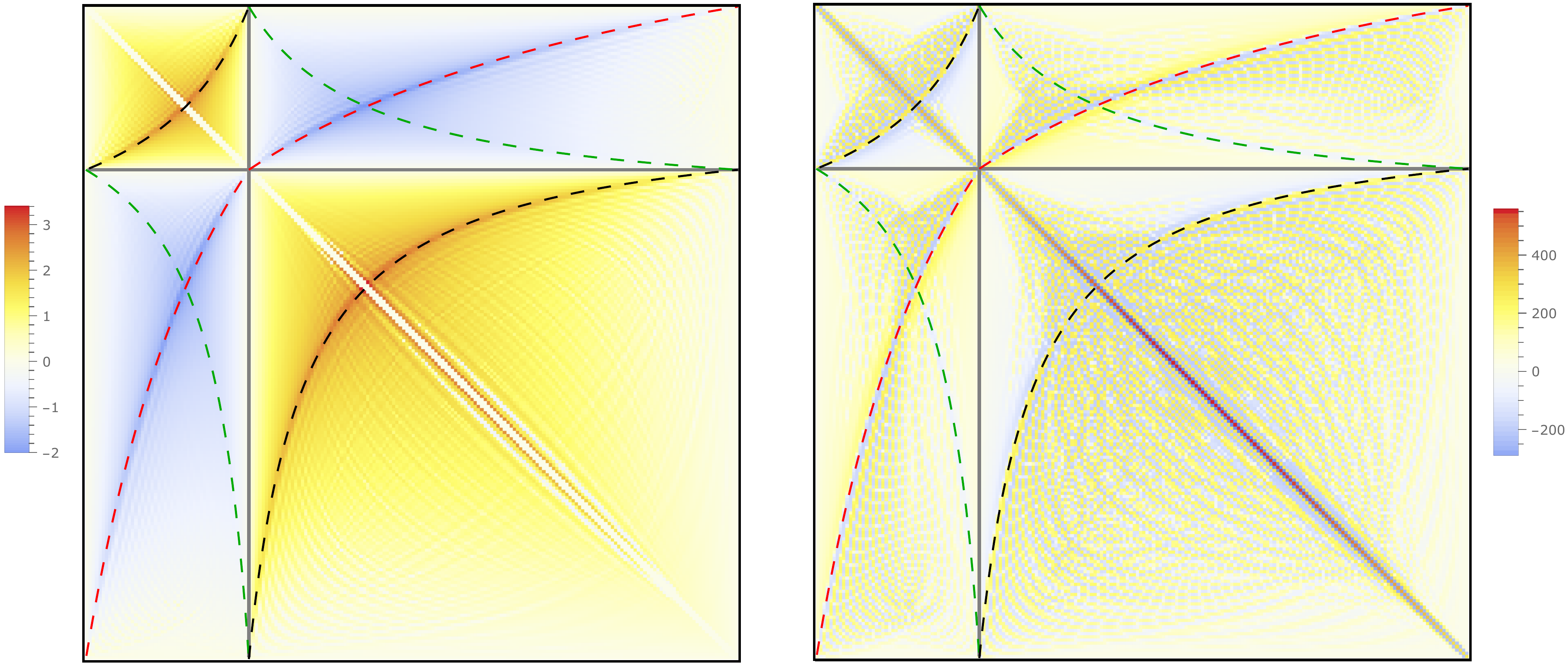}
\\
\rule{0pt}{7.7cm}
\hspace{-.16cm}
\includegraphics[width=1.05\textwidth]{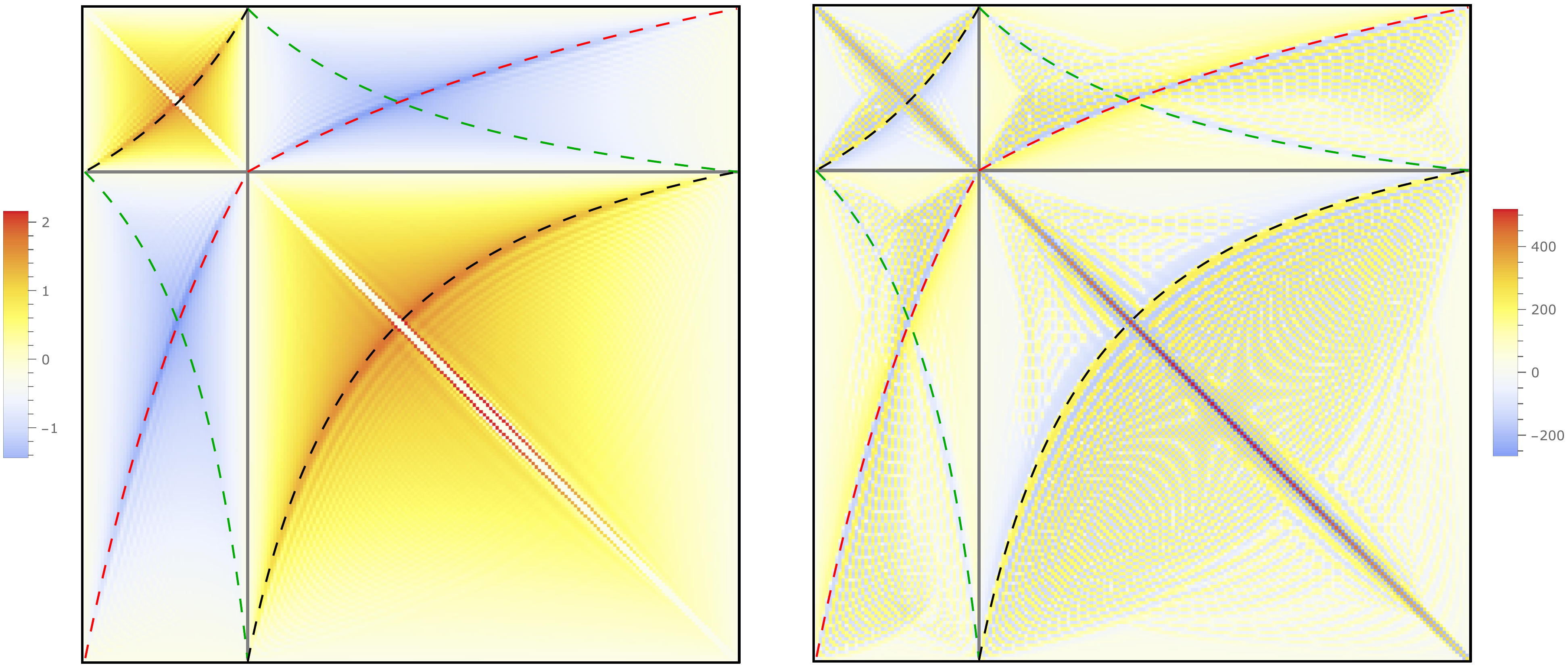}
\\
\rule{0pt}{7.7cm}
\hspace{-.16cm}
\includegraphics[width=1.05\textwidth]{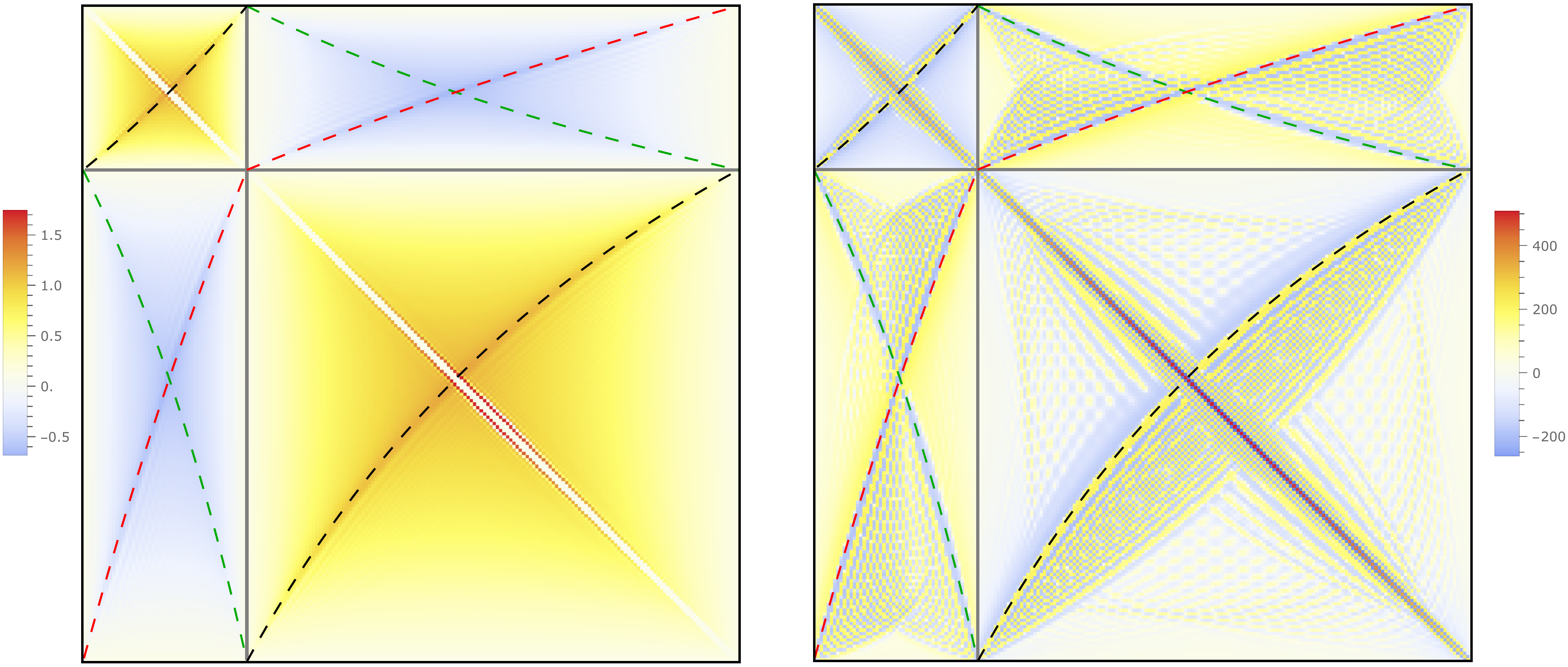}
\vspace{-.3cm}
\caption{Matrices $T$ (left) and $V$ (right) in the massless regime ($\omega=10^{-500 }$),
for $L_2 = 3L_1$ with $L_1=50$ and separation is given by
$\delta = 1/4$ (top), $\delta = 1/2$ (middle) and $\delta = 1$ (bottom).  
}
\label{fig:EH-density}
\end{figure}

While in the large $\omega$ regime the matrices $T$ and $V$ providing $\widehat{K}_A$
through  (\ref{KA-T-V-matrices})  become diagonal and tridiagonal respectively, 
as discussed in Sec.\,\ref{sec-2int-massive-equal} and Sec.\,\ref{sec-2int-generic-EH},
in the opposite limit of vanishing mass we observe that all the matrix elements of $T$ and $V$ are non vanishing;
hence in this limiting regime $\widehat{K}_A$ becomes a genuine long-range and inhomogeneous quadratic operator. 
We stress that the occurrence of the zero mode,
associated to the translation invariance of the model,
does not allow to set $\omega =0$, 
as already mentioned in Sec.\,\ref{sec-correlators-EH}.
In Fig.\,\ref{fig:EH-density} we show the matrices $T$ (left panels) and $V$ (right panels)
when $\omega=10^{-500}$ 
and the bipartition of the line is given by the union of the blocks $A_1$ and $A_2$ 
containing $L_1 = 50$ and $L_2 =150 $ sites respectively,
and separated by either $D=50$ (top panels) or $D=100$ (middle panels) or $D=200$ (bottom panels) 
contiguous sites.
In the left panels of Fig.\,\ref{fig:EH-density}, 
the main diagonal and the two ones next to it (i.e. three diagonals in total)
of the matrix $T$ have been removed,
in order to make visible the remaining elements of the matrix,
as done also in Fig.\,\ref{fig:EH-density-massive}.
It is instructive to compare the top panels of Fig.\,\ref{fig:EH-density} 
with all the panels in Fig.\,\ref{fig:EH-density-massive} 
because the same setups have been considered, except for the value of $\omega$.
Hence, this comparison allows us to gain insights into the massive regime
by visualising the elements of $T$ and $V$ that vanish as $\omega$ increases
until the regime where the tri-diagonal approximation starts to hold. 
%


The block decomposition in (\ref{eq:MatrixTandVdecomposition}) leads 
to write the entanglement Hamiltonian (\ref{KA-T-V-matrices}) as the following sum
of two quadratic operators
\be
\label{KA-T-V-matrices-2int-dec}
\widehat{K}_A
\,=\, 
\widehat{K}_{A, \textrm{\tiny \,diag}} + \widehat{K}_{A, \textrm{\tiny \,off}}
\ee
where $\widehat{K}_{A, \textrm{\tiny \,diag}} $ comes from the diagonal blocks of $T$ and $V$ and reads
\be
\label{KA-diag-hat}
\widehat{K}_{A, \textrm{\tiny \,diag}}
=
\sum_{r=1}^2 \widehat{K}_{A_r, \textrm{\tiny \,diag}} 
\;\;\;\qquad\;\;\;
\widehat{K}_{A_r, \textrm{\tiny \,diag}}
\,\equiv\,
\frac{1}{2} 
 \sum_{i,j \,\in\, A_r} \!\! \Big( T_{i,j}^{\textrm{\tiny $(r,\!r)$}} \hat{p}_i \, \hat{p}_j + V_{i,j}^{\textrm{\tiny $(r,\!r)$}} \hat{q}_i \, \hat{q}_j  \Big)
\ee
while $\widehat{K}_{A, \textrm{\tiny \,off}}$ is provided by the off-diagonal blocks of $T$ and $V$ 
and is given by 
\be
\label{KA-off-hat}
\widehat{K}_{A, \textrm{\tiny \,off}}
\,=\,
\widehat{K}_{A, \textrm{\tiny \,off}}^{\textrm{\tiny $(1,\!2)$}} + \widehat{K}_{A, \textrm{\tiny \,off}}^{\textrm{\tiny $(2,\!1)$}} 
\;\;\;\qquad\;\;\;
\widehat{K}_{A, \textrm{\tiny \,off}}^{\textrm{\tiny $(m,\!n)$}} 
\,\equiv\,
\frac{1}{2}  \sum_{\substack{  i \,\in\, A_m \\  j \,\in\, A_n }}
 \!\! \Big( T_{i,j}^{\textrm{\tiny $(m,\!n)$}}  \hat{p}_i \, \hat{p}_j + V_{i,j}^{\textrm{\tiny $(m,\!n)$}}  \hat{q}_i \, \hat{q}_j  \Big)  \,. 
\ee

\begin{figure}[t!]
	\vspace{-.5cm}
	\hspace{-.24cm}
	\includegraphics[width=1.05\textwidth]{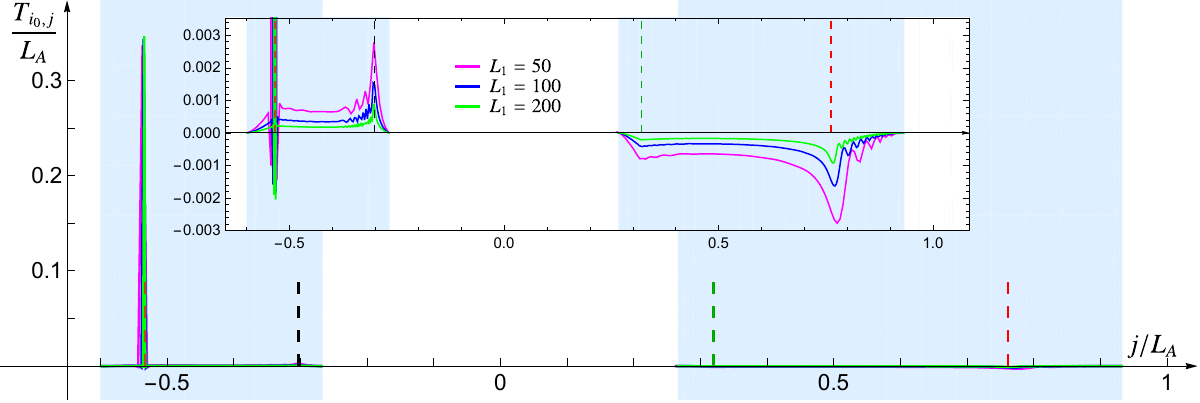}
	\\
	\rule{0pt}{6.6cm}
	\hspace{-.32cm}
	\includegraphics[width=1.05\textwidth]{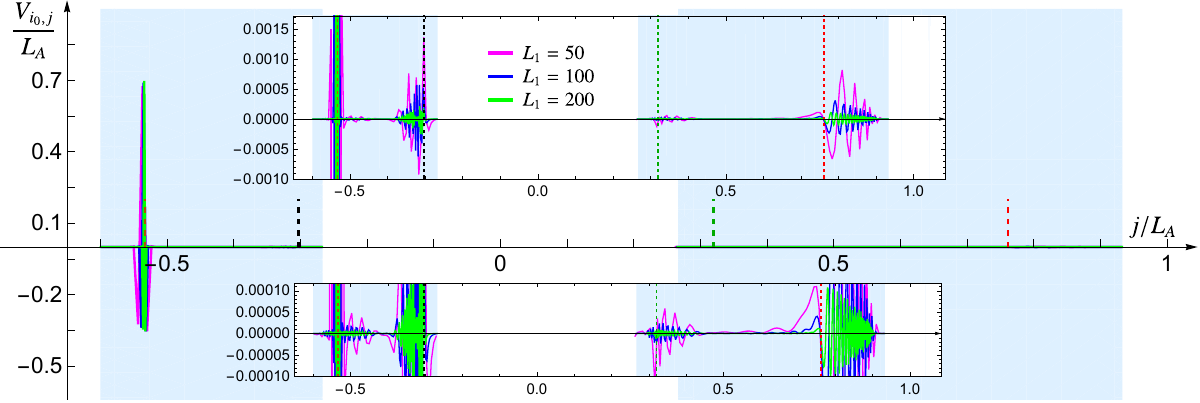}
	\vspace{.2cm}
	\caption{
	 Matrix elements of $T$ (top) and $V$ (bottom) along a row corresponding to the site labelled by $i=i_0$.
	 Here $\omega= 10^{-500}$, $L_2=2L_1$ and $i_0=L_1/5$, and $\delta=8/15$.
	 The green, red, black and brown vertical dashed segments correspond to the intersections of the row with the curves given by 
	 \eqref{x-conj-def}, \eqref{x-gamma-def}, \eqref{x-chi-def} and the main diagonal respectively
	 (see also the dashed lines in Fig.\,\ref{fig:EH-density}). 
	 The insets zoom in to highlight the oscillatory behaviour around these particular values and their relative amplitudes. 
	}
	\label{fig:all_elements}
\end{figure}

In Fig.\,\ref{fig:all_elements} we show the profiles 
along a given row (corresponding to the site labelled by $i=i_0$) 
of the matrices $T$ and $V$ (top and bottom panels respectively).
%
The largest contribution comes from the main diagonal in both these matrices,
which remains non vanishing also in the opposite regime of large $\omega$,
as discussed in  Sec.\,\ref{sec-2int-massive-equal} and Sec.\,\ref{sec-2int-generic-EH}.
By zooming in (see the insets in Fig.\,\ref{fig:all_elements}),
one realises that three other regions provide a significant contribution 
with respect to the other elements along the row.
These regions are located around the intersections of the row  with the curves introduced below
(see \eqref{x-conj-def}, \eqref{x-gamma-def} and \eqref{x-chi-def},
which correspond to the dashed lines in Fig.\,\ref{fig:EH-density}),
indicated by the vertical dashed segments in Fig.\,\ref{fig:all_elements},
with the same color code adopted for the corresponding dashed curves in Fig.\,\ref{fig:EH-density}.

\clearpage

It is very instructive to compare Fig.\,\ref{fig:EH-density} and Fig.\,\ref{fig:all_elements}
with the corresponding ones for the entanglement Hamiltonian of two disjoint blocks
in the infinite fermionic hopping chain 
(see Fig.\,\ref{fig:2int-fermion-density} and Fig.\,\ref{fig:2int-fermion-row} respectively,
where the same colour code has been adopted),
whose continuum limit \cite{Eisler:2022rnp}
provides the entanglement Hamiltonian for the massless Dirac 
free field found in \cite{Casini:2009vk}
(these results are reviewed in Appendix\;\ref{app-fermions} for completeness).
From this comparison, the main feature to highlight 
is that the subleading front in the diagonal blocks corresponding to \eqref{x-chi-def} 
(see the black dashed curves in Fig.\,\ref{fig:EH-density} and Fig.\,\ref{fig:2int-fermion-density})
does not occur in the fermionic case. 
Let us anticipate that in Sec.\,\ref{sec-chiral-current} 
(see Fig.\,\ref{density-plots-M-chiral-current} and Fig.\,\ref{3dplots-MatrixTVMabs})
we find that the absence of this front is observed also in the free chiral current model 
considered in \cite{Arias:2018tmw}.
This indicates that a non-local contribution to the entanglement Hamiltonian 
for the massless scalar field in the continuum 
comes also from the diagonal blocks, in contrast with both the fermionic model 
and the bosonic model given by the chiral current.

\begin{figure}[t!]
	\vspace{-1cm}
	\hspace{0.cm}
	\includegraphics[width=1\textwidth]{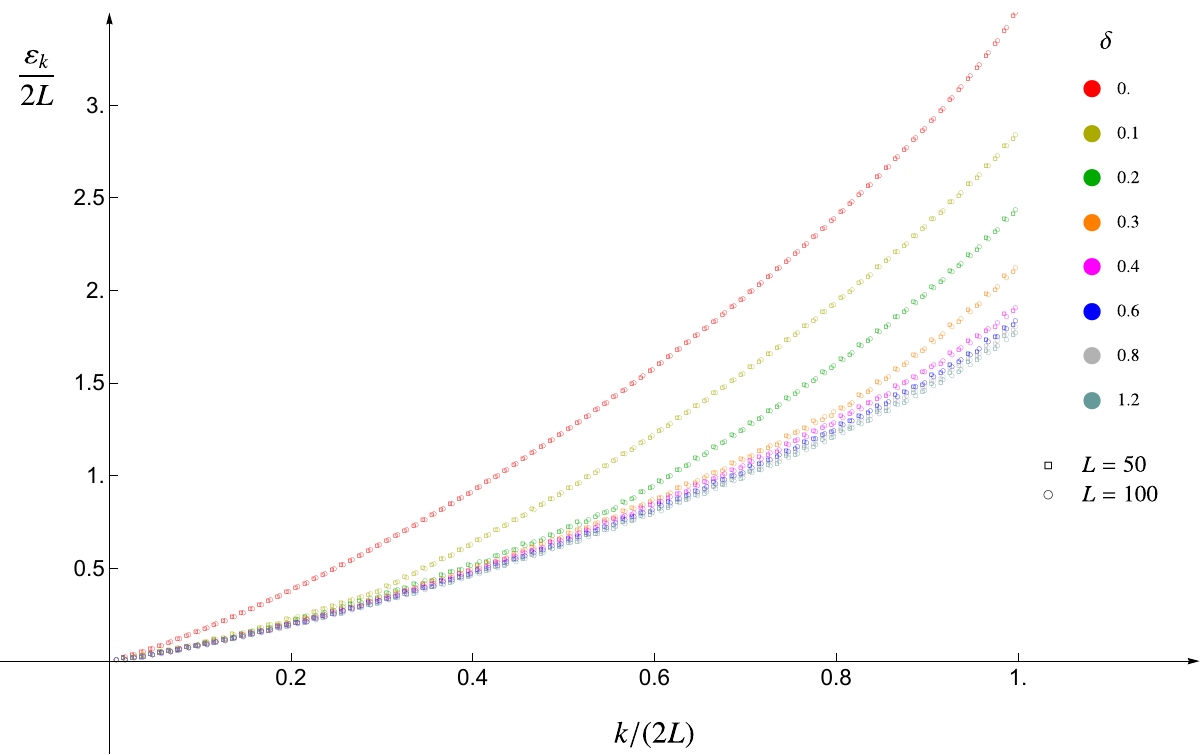}
	\\
	\rule{0pt}{10.7cm}
	\hspace{-.08cm}
	\includegraphics[width=1\textwidth]{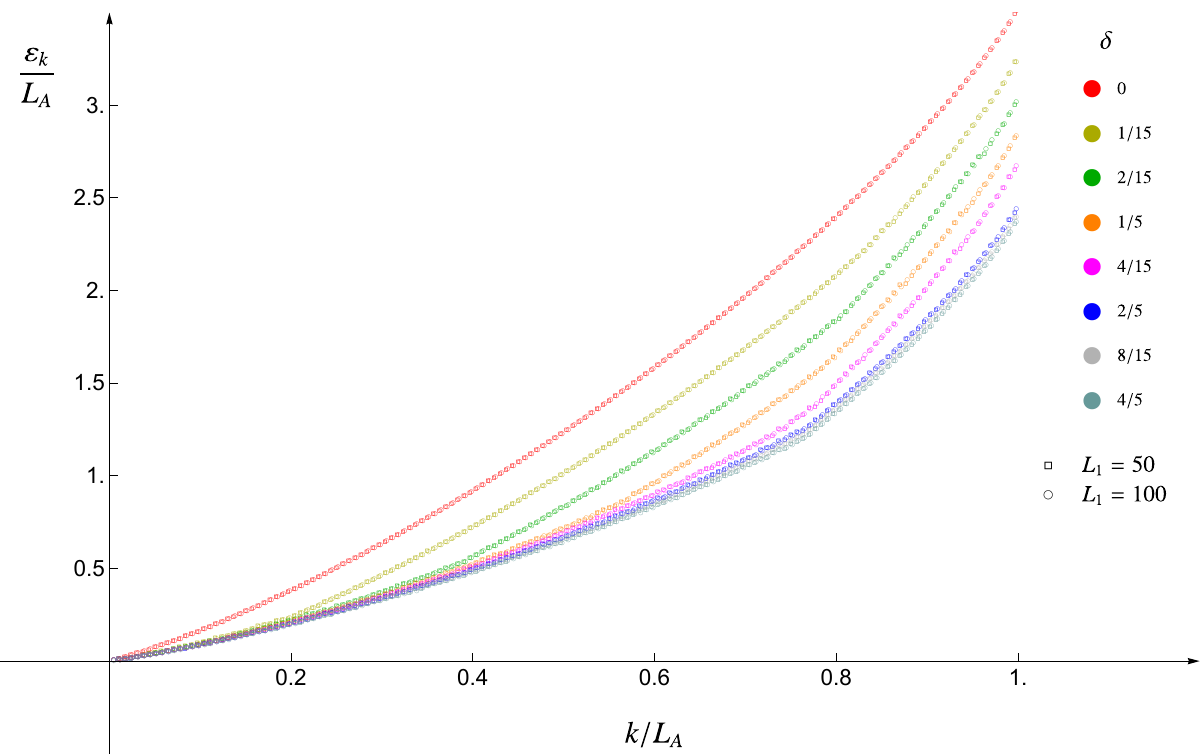}
	\vspace{-.2cm}
	\caption{
	Single-particle entanglement spectrum in the massless regime ($\omega L_A = 10^{-50}$)
	when $A$ is made by two blocks 
	with either $L_2 = L_1$ (top) or $L_2 = 2L_1$ (bottom).
	}
	\label{fig:EntSpec_massless_lratio1}
\end{figure}

\clearpage


It is worth considering the single-particle entanglement spectrum in the massless regime.
In Fig.\,\ref{fig:EntSpec_massless_lratio1} 
we report the numerical results obtained for this quantity
when $\omega L_A = 10^{-50} $
(we checked that the same values are obtained for $\omega L_A = 10^{-500} $),
with either $L_2 = L_1$ (top panel) or  $L_2 = 2L_1$ (bottom panel),
for various values of the separation distance. 
These results should be compared with the ones for the same quantities in the  large $\omega$ regime,
displayed  in Fig.\,\ref{fig:2int-equal-massive-spectrum}, Fig.\,\ref{fig:2int-equal-massive-spectrum-chi2-chi4} and Fig.\,\ref{fig:2int-equal-massive-spectrum-chi3}.
The perfect collapses of the numerical data points  in Fig.\,\ref{fig:EntSpec_massless_lratio1}  
for different values of $L_1$ show that in the  limit $L_A \to \infty$
the ratio $\varepsilon_k/L_A$ becomes a continuous function of $k/L_A$ 
parameterised by the ratios $L_2/L_1$ and $\delta$.
It would be interesting to obtain an analytic expression for this function.
Notice that a qualitative difference occurs between 
the curve for $L_2 = L_1$ and the one for $L_2 = 2L_1$,
namely the lack of smoothness in the latter case.
This could be  a footprint of $L_2 \neq L_1$.
%

\begin{figure}[t!]
\vspace{-.2cm}
	\hspace{-.87cm}
	\begin{minipage}{0.53\textwidth}
		\includegraphics[width=1.0\textwidth]{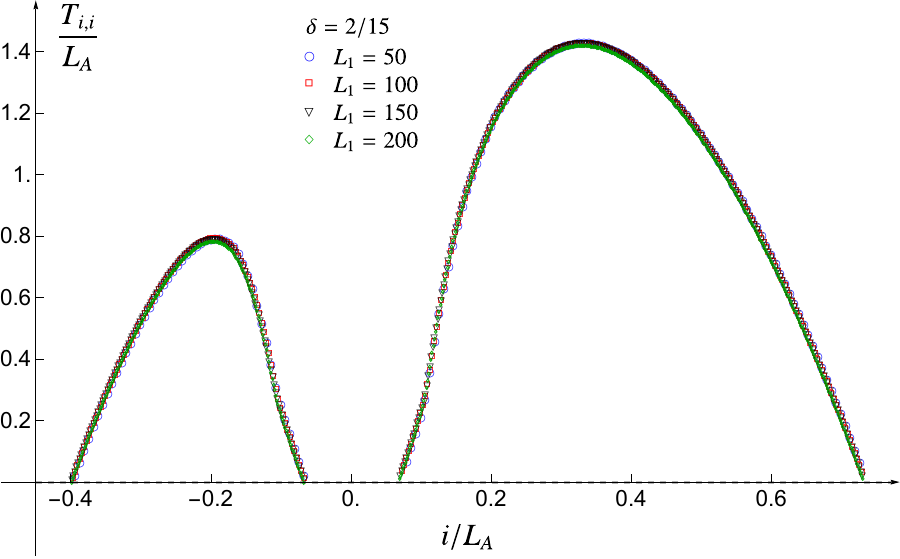}
	\end{minipage}
	\hspace{.38cm}
	\begin{minipage}{0.53\textwidth}
		\vspace{0.3cm}
		\includegraphics[width=1.0\textwidth]{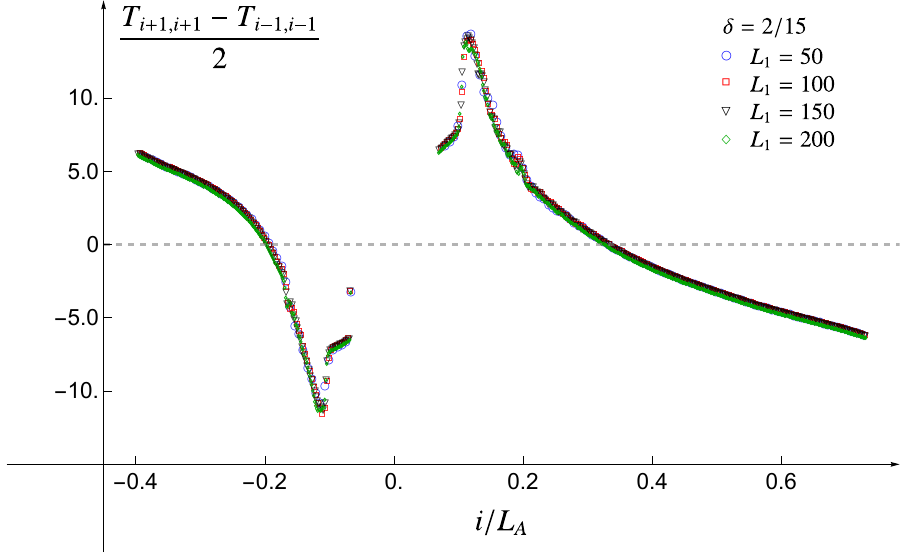}
	\end{minipage}
\vspace{.5cm}
	\\
	\begin{minipage}{0.53\textwidth}
		\hspace{-.92cm}
		\includegraphics[width=1.0\linewidth]{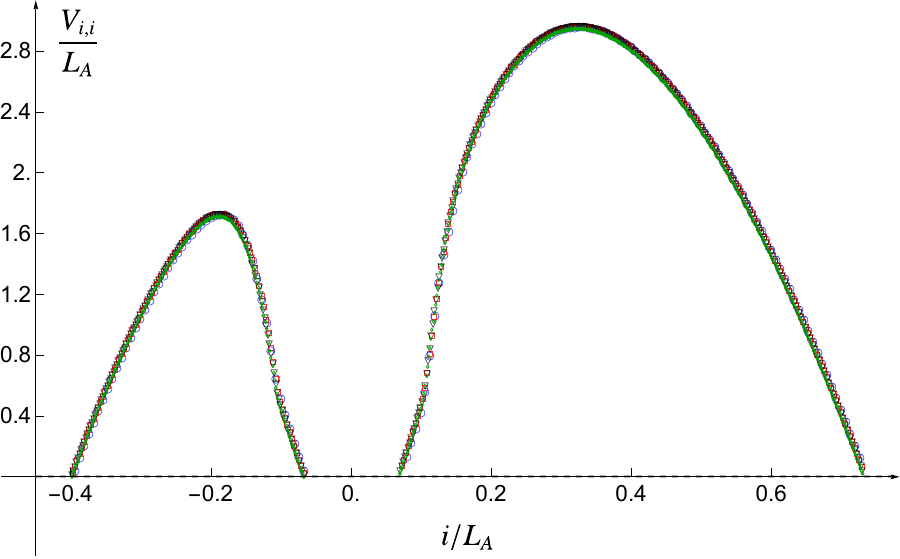}
	\end{minipage}
	\hspace{-1.81cm}
	\begin{minipage}{0.53\textwidth}
		\vspace{0.14cm}
		\includegraphics[width=1.2\linewidth]{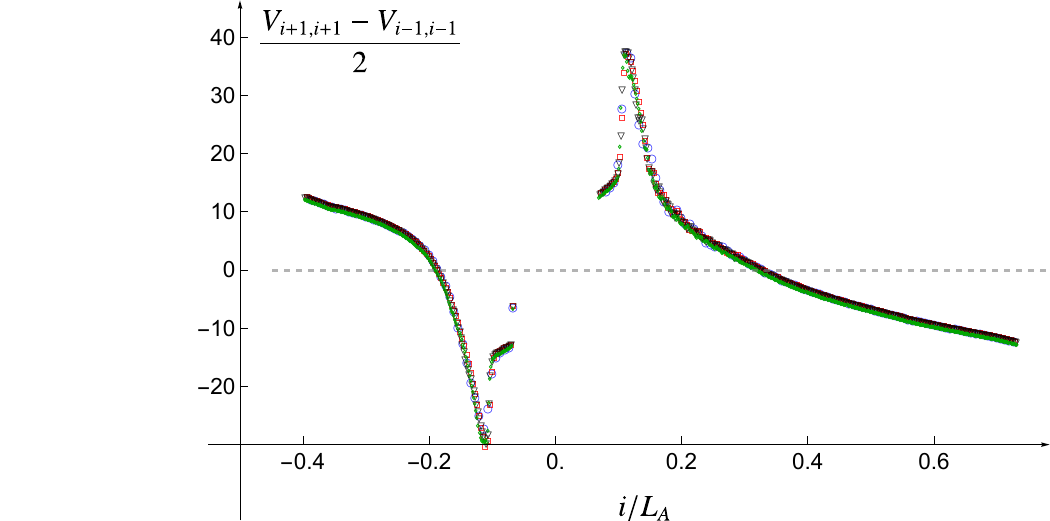}
	\end{minipage}
\vspace{.7cm}
	\\
	\begin{minipage}{0.53\textwidth}
		\hspace{-1.04cm}
		\includegraphics[width=1.0\textwidth]{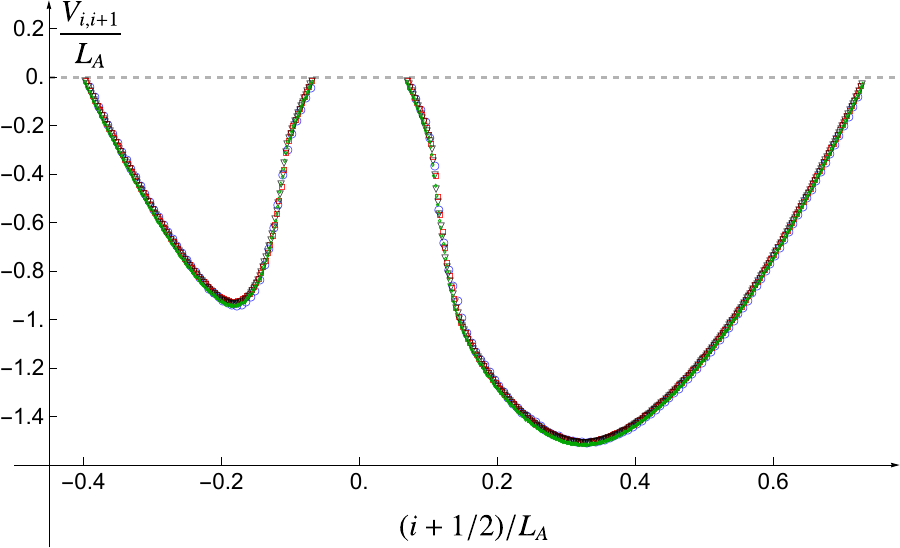}
	\end{minipage}
	\hspace{-.005cm}
	\begin{minipage}{0.53\textwidth}
		\hspace{-1.28cm}
		\vspace{-0.05cm}
		\includegraphics[width=1.2\linewidth]{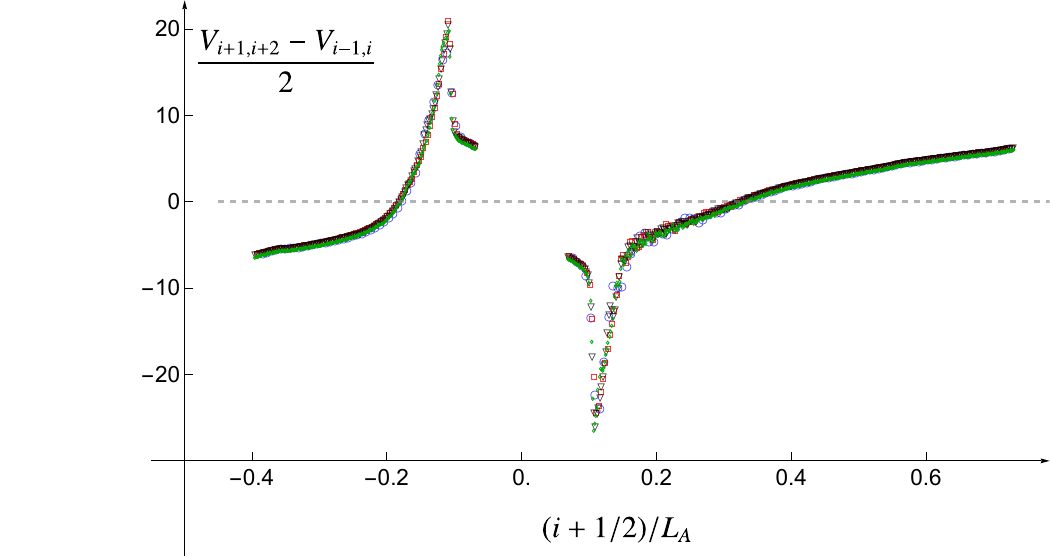}
	\end{minipage}
\vspace{.2cm}
	\caption{Diagonals $T_{i,i}$ (top left), $V_{i,i}$ (middle left) and $V_{i,i+1}$ (bottom left) for $L_2 = 2L_1$ and $\delta = 4/30$.
		Their discrete first derivative are displayed in the corresponding right panels.
	}
	\label{MainDiagonals-lratio2}
\end{figure}


It is important to investigate the operator coming from the continuum limit 
of the quadratic operators on the lattice given by (\ref{KA-diag-hat}) and (\ref{KA-off-hat}).
Previous analyses about the entanglement Hamiltonian of a single block
in the infinite fermionic and harmonic chains 
\cite{Eisler:2017cqi, Eisler:2019rnr, Arias:2016nip, DiGiulio:2019cxv, Eisler:2018ugn} 
showed that it is worth considering the diagonals of the matrices $T$ and $V$
in the thermodynamic limit. 
In Fig.\,\ref{MainDiagonals-lratio2} (left panels)
we show some numerical results for 
the three diagonals given by $T_{i,i}$, $V_{i,i}$ and $V_{i,i+1}$,
which provide the largest contributions in the massless regime
and the only non vanishing elements in the large $\omega$ regime,
as discussed in Sec.\,\ref{sec-2int-massive-equal} and Sec.\,\ref{sec-2int-generic-EH}.
These data points have been obtained for 
$L_2 = 2L_1$ and $\delta = 4/30$, 
with $\omega L_A = 10^{-50} $, 
but we checked that the same results are obtained with $\omega L_A = 10^{-500} $.
The nice collapses of the data sets corresponding to different values of $L_1$
show that the thermodynamic limit of the ratios 
$T_{i,i} / L_A$, $V_{i,i} / L_A$ and $V_{i,i+1} / L_A$ provide well defined continuous functions,
similarly to the single block case \cite{DiGiulio:2019cxv}.
In the right panels of Fig.\,\ref{MainDiagonals-lratio2}
we show the numerical results for the discrete first derivative 
of the curves in the corresponding panels on the left,
and the collapses of these data points tell us that
the first derivative of the continuous functions 
coming from the thermodynamic limit of 
$T_{i,i} / L_A$, $V_{i,i} / L_A$ and $V_{i,i+1} / L_A$ 
considered in the left panels have two spikes;
hence  their second derivative is not continuous.
Thus, the continuous functions 
in the left panels of Fig.\,\ref{MainDiagonals-lratio2} are not smooth and 
this is in contrast with the results obtained for these quantities in the case of the single block.
We remark that, 
in the case of the infinite harmonic chain in the ground state bipartite by a single block
considered in \cite{DiGiulio:2019cxv},
the smoothness of the functions obtained from 
the thermodynamic limit of the diagonals of $T$ and $V$
is a crucial ingredient to obtain the expected CFT result in the continuum limit 
(see also \cite{Javerzat:2021hxt} for the sphere in higher dimensions).
The lack of smoothness of the continuous functions 
obtained from the thermodynamic limit of the diagonals of $T/ L_A$ and $V / L_A$ highlighted above 
tells us that the procedure discussed in 
\cite{Eisler:2019rnr, Arias:2016nip, DiGiulio:2019cxv, Javerzat:2021hxt} 
cannot be applied  straightforwardly
to the case of the infinite harmonic chain in the ground state bipartite by 
the union of two disjoint blocks.

In Appendix\;\ref{app-massless} 
we report other numerical results about the diagonals of $T/ L_A$ and $V / L_A$, 
focussing on the special case of equal intervals for the sake of simplicity
(see Fig.\,\ref{Tdiagonals} and Fig.\,\ref{Vdiagonals}).
The occurrence of non-smooth functions in the thermodynamic limit
is observed also for the diagonals of higher order.


The continuum limit of (\ref{KA-diag-hat}) and (\ref{KA-off-hat})
where $A_1$ and $A_2$ contain $L_1$ and $L_2$ sites respectively and are separated by $D$ sites, 
can be studied in the standard way 
by first introducing the lattice spacing $s$ as the infinitesimal ultraviolet (UV) cutoff
and then taking the continuum limit, namely
$s\to 0^+$, $L_1 \to +\infty$, $L_2 \to +\infty$ and $D \to +\infty$
while $\ell_1 \equiv L_1 \, s$, $\ell_2 \equiv L_2 \, s$ and $d \equiv D \, s$ are kept fixed. 
In the continuum, these parameters provide respectively
the lengths of the finite segments in the real line
given by $A_1$, $A_2$ and the domain separating them. 
The position along the line supporting the model in the continuum 
is labelled by $x= i\, s$, being $i \in \mathbb{Z}$ the discrete index  in (\ref{ham-HC}) 
and the fields $\Phi(x)$ and $\Pi(x)$ are introduced through 
the operators $\hat{q}_i$ and $\hat{p}_i$  in (\ref{ham-HC})  respectively as 
\be 
\label{qp-field-replacement}
\hat{q}_i \, \longrightarrow \,  \Phi(x)
\;\;\;\; \qquad \;\;\;\;
\hat{p}_i \, \longrightarrow \, s \, \Pi(x)
\ee
where the UV cutoff 
guarantees the validity of the canonical commutation relations in the continuum limit,
which involve Dirac delta functions. 
It is convenient to rewrite the sums in (\ref{KA-diag-hat})-(\ref{KA-off-hat}) in the form 
$L_A \sum_{i \in A} \! \big(\dots\big) = \tfrac{(L_A s)}{s^2}  \sum_{i \in A} \!\big(\dots\big) s$,
that is suitable for the continuum limit
(with a slight abuse of notation, here we denote by $A \equiv A_1 \cup A_2$ also the union of two disjoint intervals $A_1$ and $A_2$ in the real line).
%
%
%
The continuum limit 
$\widehat{K}_{A, \textrm{\tiny \,diag}}  \to K_{A, \textrm{\tiny \,diag}} $ 
and $ \widehat{K}_{A, \textrm{\tiny \,off}} \to  K_{A, \textrm{\tiny \,off}}$
of the quadratic operators in (\ref{KA-diag-hat}) and (\ref{KA-off-hat}) respectively,
provides $K_{A, \textrm{\tiny \,diag}} $ and $K_{A, \textrm{\tiny \,off}}$, 
which are quadratic operators of the fields $\Phi(x)$ and $\Pi(x)$.
It is highly non trivial to establish whether $K_{A, \textrm{\tiny \,diag}} $ and $K_{A, \textrm{\tiny \,off}}$ 
are either local or bilocal or fully non-local operators. 
%


For the sake of completeness, 
in Appendix\;\ref{app-fermions} we have reviewed the known results 
about the entanglement Hamiltonian of two disjoint intervals in the line 
for the massless Dirac field (which is a free fermionic CFT model with central charge is $c=1$)
in its ground state \cite{Casini:2009vk}
and its derivation from the entanglement Hamiltonian of two disjoint blocks
in the infinite fermionic hopping chain at half-filling 
through a continuum limit procedure described in \cite{Eisler:2022rnp} (see Fig.\,\ref{fig:2int-fermion-beta}),
obtained by adapting the corresponding analysis for the single block \cite{Eisler:2019rnr}.
This fermionic example provides useful tools to understand some
important features of the bosonic case explored in this section. 
The analytic expression of this fermionic entanglement Hamiltonian in the continuum
is a quadratic operator in the Dirac field that can be written as the sum of a local term and a bilocal term \cite{Casini:2009vk}
(see (\ref{mod-ham-2int-KA})).
The local term  of this entanglement Hamiltonian (see (\ref{K_A-2int-terms-local}))
is given by  the energy density of the massless Dirac field 
weighted by the following function
\be
\label{beta-loc-2int}
\beta_{\textrm{\tiny loc}}(x) = \frac{2\pi}{w'(x)}
\ee
where 
\be
w(x) \equiv \log\left(\! - \frac{(x-a_1)(x-a_2) }{ (x-b_1)(x-b_2) }\right)  . 
\ee
Instead, the bilocal term (see (\ref{beta-bi-loc-2int}))
is a quadratic operator in the field (see (\ref{T-bilocal-fermions-app}))
where the two fields are evaluated in $x \in A$ and in its conjugate point $x_{\textrm{c}} \in A$
defined as 
(see also \cite{Mintchev:2022fcp})
\be
\label{x-conj-def}
x_{\textrm{c}} \equiv x_0 - \frac{r_0^2}{x - x_0}
\ee
where 
\be
\label{q0-r0-def}
x_0  \equiv 
\frac{b_1\, b_2 - a_1\, a_2}{ b_1 - a_1 + b_2 - a_2 } 
\;\;\;\qquad\;\;\;
r_0 \, \equiv \, 
\frac{ \sqrt{ (b_1 - a_1) (b_2 - a_2) (b_2 - a_1) (a_2 - b_1)} }{ b_1 - a_1 + b_2 - a_2  }   \,. 
\ee
We remark that  $x_{\textrm{c}} \in A_j$ when $x \in A_i$, with $i \neq j$.


As for the bosonic case that we are exploring in this section,
the expression $x_{\textrm{c}}$ in (\ref{x-conj-def}) as a function of $x \in A$
(obtained for the fermionic case)
provides the green dashed curves in the off-diagonal blocks of the matrices 
displayed in Fig.\,\ref{fig:EH-density}.

From (\ref{x-conj-def}), let us introduce also the functions of $x\in A$ given by 
\be
\label{x-gamma-def}
x_{_\Gamma}
\equiv \, x_{\textrm{c}} \big|_{a_2 \leftrightarrow b_2} 
= \, x_{\textrm{c}} \big|_{a_1 \leftrightarrow b_1} 
\ee
and 
\be
\label{x-chi-def}
x_{_\chi} 
\equiv \, x_{\textrm{c}} \big|_{a_2 \leftrightarrow b_1} 
= \, x_{\textrm{c}} \big|_{a_1 \leftrightarrow b_2}   \,. 
\ee
In all the matrices displayed in Fig.\,\ref{fig:EH-density},
the functions $x_{_\Gamma}$ and $x_{_\chi} $ provide respectively
the red and black dashed curves, in the off-diagonal and diagonal blocks respectively.
The function $x_{_\Gamma}$ in (\ref{x-gamma-def}) 
is obtained by applying the prescription proposed in \cite{Calabrese:2012ew,Calabrese:2012nk,Calabrese:2014yza}
to study the negativity 
(a quantifier of the bipartite entanglement in mixed states obtained from the partial transposition
of the reduced density matrix \cite{Peres:1996dw,Vidal:2002zz})
in quantum field theory
and, in the case of the massless Dirac field in the vacuum and on the line bipartite through the union of two disjoint intervals,
it occurs in the analyses of the corresponding operator \cite{Murciano:2022vhe}.


The two solutions of $x_{_\chi}( \tilde{x}) = \tilde{x}$ are
\be
\label{tilde-x-pm-def}
\tilde{x}_\pm  = \frac{a_2\, b_2 - a_1\, b_1}{ a_2 - a_1 + b_2 - b_1 }  
\pm 
\frac{ \sqrt{ (a_2 - a_1) (b_2 - b_1) (b_2 - a_1) (a_2 - b_1) } }{  a_2 - a_1 + b_2 - b_1   }   \,. 
\ee
By writing the endpoints as $b_1 = a_1 +\ell_1$, $a_2 = a_1 +\ell_1 + d$ and $b_2 = a_1 +\ell_1 + d+ \ell_2$, 
in terms of the positive parameters $\ell_1$, $d$ and $\ell_2$,
it is straightforward to prove that $\tilde{x}_{-} \in A_1 $ while $\tilde{x}_{+} \in A_2$.
Interestingly, the expressions in (\ref{tilde-x-pm-def}) provide also the two solutions of $x_{_\Gamma}( \tilde{x}) = x_{\textrm{c}}(\tilde{x})$.
This implies that, in Fig.\,\ref{fig:EH-density},  
the intersections between the two dashed curves in the off-diagonal blocks 
and between the dashed curve and the main diagonal in the diagonal blocks form a square.
Indeed, since one can show analytically $x_{_\Gamma}(\tilde{x}_{\pm})=\tilde{x}_{\mp}$,  
the four intersection points correspond to $(\tilde{x}_\pm,\tilde{x}_\pm)$ and  $(\tilde{x}_\mp,\tilde{x}_\pm)$. 
In the special case of the symmetric configuration $A = A_{\textrm{\tiny sym}} = (-b, -a) \cup (a,b)$ with $0<a<b$, 
the expressions in (\ref{tilde-x-pm-def}) drastically simplify to 
\be
\tilde{x}_\pm \big|_{A = A_{\textrm{\tiny sym}}}  =\,  \pm \, \sqrt{a\, b}  \;. 
\ee


In the continuum limit procedure 
for the entanglement Hamiltonian of two disjoint blocks in the infinite fermionic chain in its ground state,
discussed in \cite{Eisler:2022rnp} and reviewed in Appendix\;\ref{app-fermions},
the weight function of the bilocal term given in \eqref{velocity_fund-2int}
has been obtained by introducing a specific combination of all the elements along a line of the off-diagonal blocks 
(see \eqref{eq:fermion-EH-offdiag-lattice}), finding excellent agreement
(see also the right panel of Fig.\,\ref{fig:2int-fermion-beta}).
This result for the fermionic case suggests 
that also for the bosonic case that we are exploring
one should consider specific combinations of all the elements along a line.

For the diagonal blocks occurring  in (\ref{KA-diag-hat}), let us introduce 
\be 
\label{row-wise-sums-diag}
\mathsf{T}_{\! \textrm{\tiny diag}}(i) 
\equiv
\left\{\begin{array}{ll}
\displaystyle	\sum_{j\,\in\, A_1} T_{i,j}
\hspace{.7cm}& i \in A_1
	\\
	\rule{0pt}{.6cm}
\displaystyle	\sum_{j\,\in\, A_2} T_{i,j}
& i \in A_2
\end{array}
\right.
\hspace{1.5cm}
\mathsf{V}_{\! \textrm{\tiny diag}}(i) 
\equiv
\left\{\begin{array}{ll}
\displaystyle	\sum_{j\,\in\, A_1} V_{i,j}
\hspace{.7cm}& i \in A_1
	\\
	\rule{0pt}{.6cm}
\displaystyle	\sum_{j\,\in\, A_2} V_{i,j}
& i \in A_2
\end{array}
\right.
\ee
and, similarly, for the off-diagonal blocks in (\ref{KA-off-hat}) we define the following row-wise combinations
\be 
\label{row-wise-sums-off-diag}
\mathsf{T}_{\! \textrm{\tiny off}}(i) 
\equiv
\left\{\begin{array}{ll}
\displaystyle	\sum_{j\,\in\, A_2} T_{i,j}
\hspace{.7cm}& i \in A_1
	\\
	\rule{0pt}{.6cm}
\displaystyle	\sum_{j\,\in\, A_1} T_{i,j}
& i \in A_2
\end{array}
\right.
\hspace{1.5cm}
\mathsf{V}_{\! \textrm{\tiny off}}(i) 
\equiv
\left\{\begin{array}{ll}
\displaystyle	\sum_{j\,\in\, A_2} V_{i,j}
\hspace{.7cm}& i \in A_1
	\\
	\rule{0pt}{.6cm}
\displaystyle	\sum_{j\,\in\, A_1} V_{i,j}
& i \in A_2
\end{array}
\right.
\ee

\begin{figure}[t!]
	\vspace{-.3cm}
	\hspace{-.54cm}
	\includegraphics[width=1.05\textwidth]{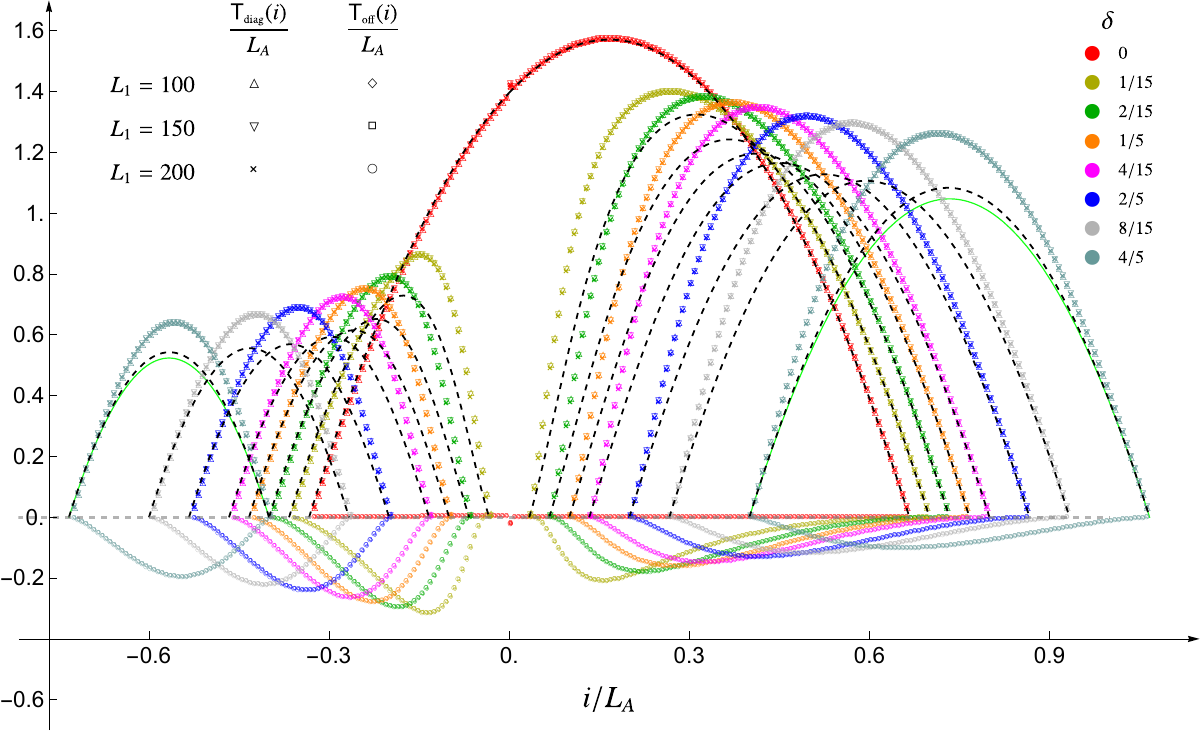}
	\\
	\rule{0pt}{11.2cm}
	\hspace{-.45cm}
	\includegraphics[width=1.05\textwidth]{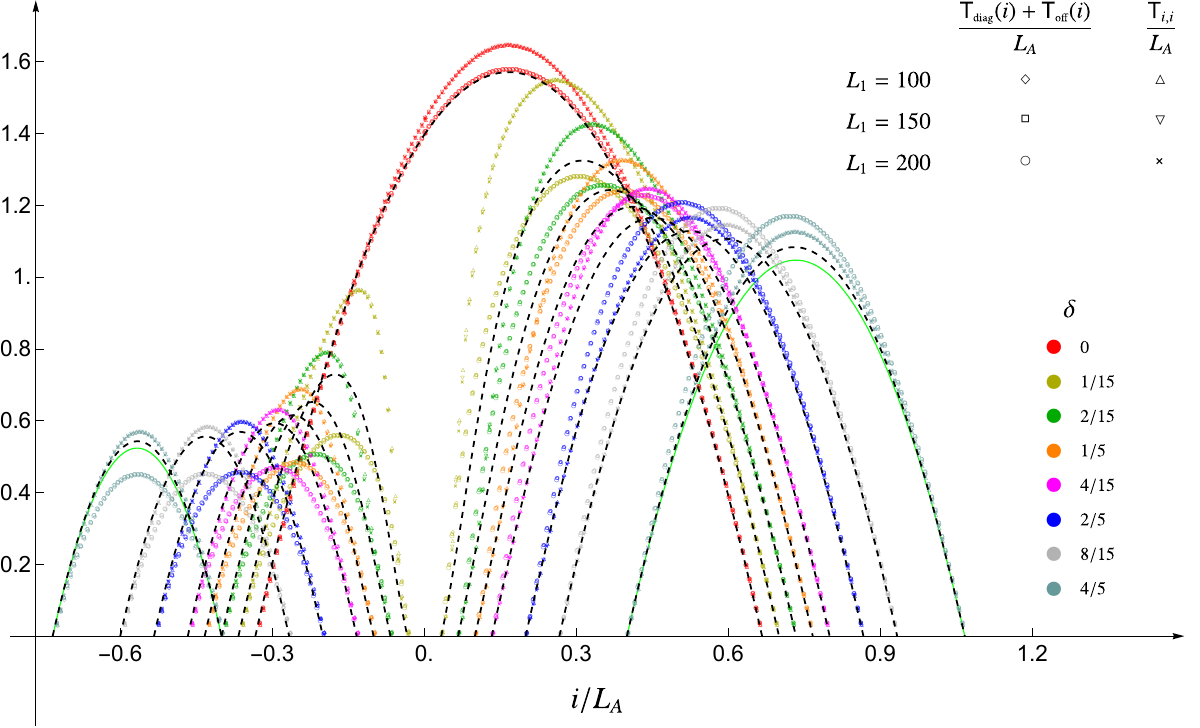}
	\vspace{.0cm}
	\caption{Combinations of elements in $T$ defined in (\ref{row-wise-sums-diag}) and (\ref{row-wise-sums-off-diag}), for $L_2 = 2L_1$. 
	}
	\label{fig:DiagOffDiag_massless_lratio2_Tmat}
\end{figure}
\begin{figure}[t!]
	\vspace{-.3cm}
	\hspace{-3.4cm}
	\includegraphics[width=1.23\textwidth]{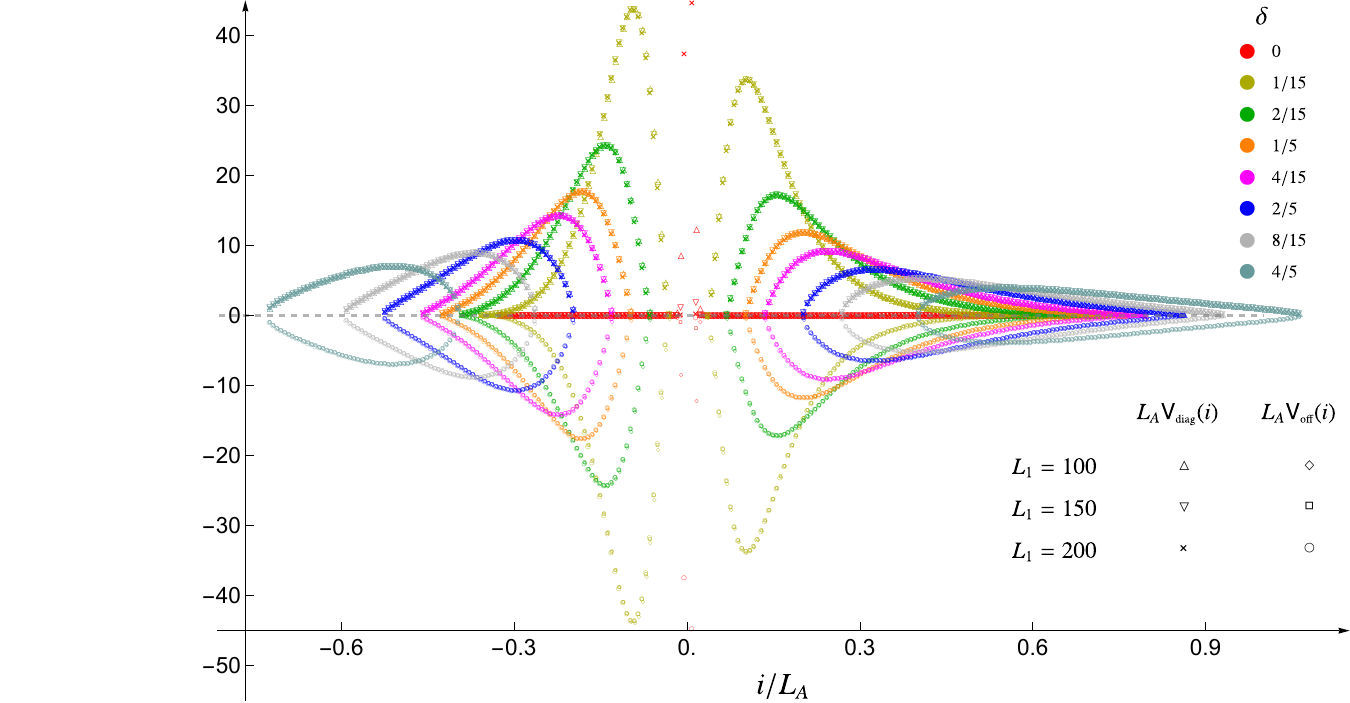}
	\\
	\rule{0pt}{11.2cm}
	\hspace{-.82cm}
	\includegraphics[width=1.055\textwidth]{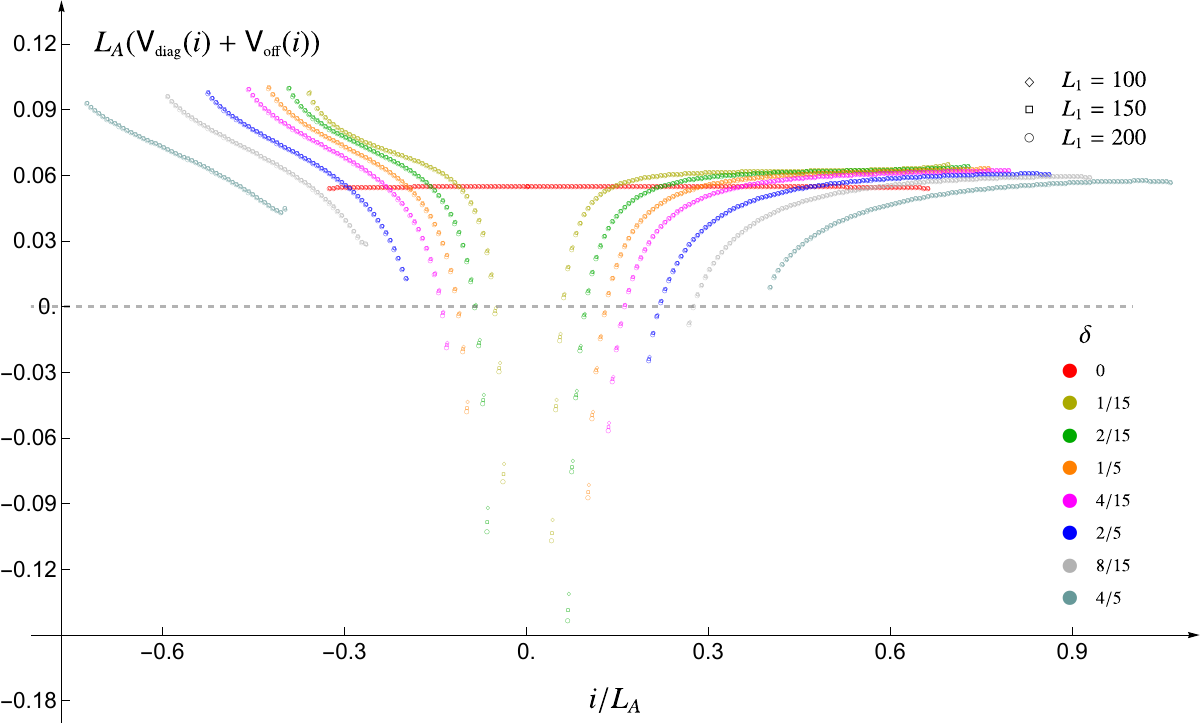}
	\vspace{.0cm}
	\caption{
		Combinations of elements in $V$ defined in (\ref{row-wise-sums-diag}) and (\ref{row-wise-sums-off-diag}), for $L_2 = 2L_1$. 
	}
	\label{fig:DiagOffDiag_massless_lratio2_Vmat}
\end{figure}

In Fig.\,\ref{fig:DiagOffDiag_massless_lratio2_Tmat} and Fig.\,\ref{fig:DiagOffDiag_massless_lratio2_Vmat} 
we show some numerical results for the combinations 
of the elements along the $i$-th row of $T$ and $V$ 
introduced in (\ref{row-wise-sums-diag}) and (\ref{row-wise-sums-off-diag}) respectively,
when $L_2 = 2L_1$ and for various values of $\delta$.
Notice  that these combinations scale in different ways for the two different matrices.
%
The black dashed curves in Fig.\,\ref{fig:DiagOffDiag_massless_lratio2_Tmat} correspond to (\ref{beta-loc-2int})
occurring in the fermionic case mentioned above (see also Appendix\;\ref{app-fermions}),
while the solid green curve is its limit for large separation distance $d \to +\infty$,
which is given by $2\pi (b_1 - x)(x- a_1)/(b_1 - a_1)$ when $x \in A_1$
and by $2\pi (b_2 - x)(x- a_2)/(b_2 - a_2)$ when $x \in A_2$.
%
In each figure, the top panel shows separately the contribution of the diagonal block 
and of the off-diagonal block for a given row, 
while in the bottom panel the sum of these two contributions is considered. 
In the bottom panel of Fig.\,\ref{fig:DiagOffDiag_massless_lratio2_Tmat} also the main diagonal of $T$ is shown.
The nice collapses of the data points observed 
in Fig.\,\ref{fig:DiagOffDiag_massless_lratio2_Tmat} and Fig.\,\ref{fig:DiagOffDiag_massless_lratio2_Vmat} 
indicate that these combinations, once scaled in the proper way, 
provide well defined functions in the thermodynamic limit.
Finding the analytic expressions of these functions is an interesting problem for future studies.

As for  Fig.\,\ref{fig:DiagOffDiag_massless_lratio2_Tmat} about $T$,
the data points collapse on the weight function (\ref{beta-loc-2int}) 
near the entangling points of the subsystem $A$, but a discrepancy with this analytic curve
is observed in the central part of $A_1$ and $A_2$.
When $\delta=0$, the red data points in Fig.\,\ref{fig:DiagOffDiag_massless_lratio2_Tmat} 
nicely reproduces the result of \cite{DiGiulio:2019cxv} for the single block, as expected,
and from the bottom panel we see that
the main diagonal of $T$ is not enough to capture the parabolic function occurring in the continuum limit,
hence the contribution of the elements outside the main diagonal is also needed.


In Fig.\,\ref{fig:DiagOffDiag_massless_lratio2_Vmat}, where the combinations of elements of $V$ defined in
(\ref{row-wise-sums-diag}) and (\ref{row-wise-sums-off-diag}) are considered, 
we observe that the diagonal blocks and the off-diagonal blocks provide very similar curves, except for an overall sign. 
However, their sum is not noisy, although two orders of magnitude smaller than each term, 
as shown in the bottom panel of the figure. 
Moreover, the curves in the top panel of Fig.\,\ref{fig:DiagOffDiag_massless_lratio2_Vmat}
display  the same qualitative  behaviour of $\big| \beta_{\textrm{\tiny biloc}}(x) \big|$, 
where $\beta_{\textrm{\tiny biloc}}(x) $ is the weight function (\ref{velocity_fund-2int})
occurring in the bilocal term (\ref{beta-bi-loc-2int}), found in \cite{Casini:2009vk} for the massless Dirac field.
The special case of adjacent intervals can be understood by employing e.g. the panel about the matrix $V$ 
corresponding to $\omega = 10^{-30}$ in Fig.\,$10$ of \cite{Eisler:2020lyn}.
%

Given the results of \cite{Eisler:2022rnp} for the fermionic case (see also Appendix\;\ref{app-fermions}),
we conclude that the collapses 
observed in Fig.\,\ref{fig:DiagOffDiag_massless_lratio2_Tmat} and Fig.\,\ref{fig:DiagOffDiag_massless_lratio2_Vmat} 
strongly indicate that a non-local term should occur in the entanglement Hamiltonian of the massless scalar field,
but we cannot establish whether this term is bilocal, like in the fermionic case of \cite{Casini:2009vk} 
(see (\ref{beta-bi-loc-2int})-(\ref{T-bilocal-fermions-app})), 
or fully non-local, like in the case of the free chiral current model considered in \cite{Arias:2018tmw} (see also Sec.\,\ref{sec-chiral-current}).


It is worth exploring the local term of the entanglement Hamiltonian in the continuum limit
by adapting to our case 
the method developed in \cite{Arias:2016nip, Eisler:2017cqi, Eisler:2019rnr, DiGiulio:2019cxv, Eisler:2018ugn} 
for the single block in the infinite free chains
(see \cite{Javerzat:2021hxt} for the higher dimensional case of the ball),
which has provided the corresponding CFT results \cite{Hislop:1981uh, Casini:2011kv}.
%

This local term originates from the diagonal blocks in $T$ and $V$.
As first step of this analysis, we consider the thermodynamic limit of the $k$-th diagonal of these diagonal blocks, namely
\be
\label{diagonal-collapse-assumption}
\lim_{L_A \to \infty} \! \frac{T_{i, i+k}}{L_A} \equiv  \tau_k(x_k)
\;\;\;\qquad\;\;\;
\lim_{L_A \to \infty} \! \frac{V_{i, i+k}}{L_A} \equiv \nu_k(x_k)
\;\;\;\qquad\;\;\;
x_k \equiv \frac{1}{L_A} \left( i + \frac{k}{2}\right)
\ee
where $ i + k/2$ corresponds to the midpoint between the $i$-th and the $(i+k)$-th site.


\clearpage

\clearpage


%
Notice that $x_k$ can be also written in the form $x_k = (i \,s +k\,s/2)/(L_A s)$, 
where now also the lattice spacing occurs, 
and this form tells us that $x_k$ becomes a function of the position $x\in A$ in the continuum limit. 
For  the single block, 
the thermodynamic limits (\ref{diagonal-collapse-assumption}) have been studied 
in \cite{Eisler:2017cqi, Eisler:2018ugn} for free fermionic chains,
finding also some analytic results, 
and in \cite{DiGiulio:2019cxv} for the harmonic chain,
but in the latter case the corresponding analytic expressions are not available in the literature. 
In the case where $A$ is the union of two disjoint blocks of the harmonic chain,
the numerical results shown in Fig.\,\ref{MainDiagonals-lratio2} for $L_2 = 2L_1$
and in Appendix\;\ref{app-massless} for $L_2 = L_1$  (see Fig.\,\ref{Tdiagonals} and Fig.\,\ref{Vdiagonals})
provide some evidence that the limits in (\ref{diagonal-collapse-assumption}) exist
and that the functions $\tau_k(x_k)$ and $\nu_k(x_k)$ are not smooth.
This feature prevents us from applying
the continuum limit procedure  described in \cite{DiGiulio:2019cxv} for the single block case in a straightforward way;
indeed, in that analysis the smoothness of the functions obtained from the thermodynamic limit of the diagonals
is a crucial assumption.
This lack of smoothness induces us to consider only 
the combinations of matrix elements providing the discrete approximations 
of the first derivatives of $\tau_k(x_k)$ and $\nu_k(x_k)$.
Instead, we do not consider the combination of matrix elements of $V$ 
corresponding to the discrete approximation of $\nu_k''(x)$,
which occurs in the single block analysis of \cite{DiGiulio:2019cxv}.

Following \cite{DiGiulio:2019cxv}, let us introduce the following combinations of diagonals
\be
\label{T-V-summation-0}
\mathsf{T}_{k_\textrm{\tiny max}}^{(0)}(i)
\equiv
\left\{\begin{array}{l}
\displaystyle	T_{i,i} + 2 \sum_{k=1}^{k_\textrm{\tiny max}} T_{i,i+k} 
	\\
	\rule{0pt}{.9cm}
\displaystyle		T_{i,i} + 2 \sum_{k=1}^{k_\textrm{\tiny max}} T_{i-k,i} 
\end{array}
\right.
\hspace{1.cm}
\mathsf{V}_{k_\textrm{\tiny max}}^{(0)}(i)
\equiv
\left\{\begin{array}{ll}
\displaystyle	V_{i,i} + 2 \sum_{k=1}^{k_\textrm{\tiny max}} V_{i,i+k}  \hspace{.8cm}& \tilde{i} \in A_<
	\\
	\rule{0pt}{.9cm}
\displaystyle		V_{i,i} + 2 \sum_{k=1}^{k_\textrm{\tiny max}} V_{i-k,i}  & \tilde{i} \in A_>
\end{array}
\right.
\ee
which include the main diagonals of $T$ and $V$ respectively, and
\be
\label{V-summation-2}
\mathsf{V}_{k_\textrm{\tiny max}}^{(2)}(i)
\equiv
\left\{\begin{array}{ll}
\displaystyle	\sum_{k=1}^{k_\textrm{\tiny max}} k^2\,V_{i,i+k} 
\hspace{.8cm}& \tilde{i} \in A_<
	\\
	\rule{0pt}{.9cm}
\displaystyle		\sum_{k=1}^{k_\textrm{\tiny max}} k^2 \, V_{i-k,i} 
& \tilde{i} \in A_>
\end{array}
\right.
\ee
where the relation (\ref{indices-choice-2}) 
between the matrix index $\tilde{i} \in [1, L_A]$ and the index $i$ labelling the sites of the chain  
has been used and
\be
\label{A-sets-matrix-indices-def}
\begin{array}{l}
A_< \equiv \big[1\,, L_1/2  \big] \cup \big[ D +  L_1 +1  \, , D + L_1 + L_2 /2 \big] 
\\
\rule{0pt}{.6cm}
A_>  \equiv  \big[ L_1 /2 + 1\, , L_1 \big] \cup \big[ D + L_1 + L_2/2 + 1\, , D + L_A \big] 
\end{array}
\ee
which are well defined for even values of $L_1$ and $L_2$.
The branches $A_<$ and $A_>$ in (\ref{A-sets-matrix-indices-def})
have been introduced by adapting the analysis for the single block \cite{DiGiulio:2019cxv} 
(where a symmetry with respect to the center of the subsystem occurs)
to this case. 
However, a different choice of $A_<$ and $A_>$ could provide better results.
In particular, a definition of $A_<$ and $A_>$  based on $x_{_\chi}$ seems natural,
but we leave this analysis for future studies.

In this limiting procedure, the limit $L_A \to + \infty$ should be taken before the limit $k_\textrm{\tiny max} \to +\infty$.
In the numerical analysis, where both $L_A$ and $k_\textrm{\tiny max} $ are finite,
this prescription corresponds to keeping $k_\textrm{\tiny max}  \ll L_A$.
Considering $L_A \to + \infty$ first,
from (\ref{T-V-summation-0}) and (\ref{V-summation-2}) 
one introduces 
\be
\label{T0-V0-large-L}
\mathcal{T}_{k_\textrm{\tiny max}}^{(0)}(x) 
 \, \equiv 
\lim_{L_A \to \infty} \! \frac{ \mathsf{T}_{k_\textrm{\tiny max}}^{(0)}(i) }{L_A} 
\;\qquad\;
\mathcal{V}_{k_\textrm{\tiny max}}^{(0)}(x) 
 \, \equiv 
\lim_{L_A \to \infty} \! \frac{ \mathsf{V}_{k_\textrm{\tiny max}}^{(0)}(i) }{L_A} 
\;\qquad\;
\mathcal{V}_{k_\textrm{\tiny max}}^{(2)}(x) 
 \, \equiv 
\lim_{L_A \to \infty} \! \frac{ \mathsf{V}_{k_\textrm{\tiny max}}^{(2)}(i) }{L_A} 
\ee
where the r.h.s.'s are functions of $x$; indeed
(\ref{diagonal-collapse-assumption}) can be employed 
to write them
through the functions $\tau_k(x_k)$ and $\nu_k(x_k)$,
which become functions of $x$ in the continuum limit,
as highlighted above (see the text below (\ref{diagonal-collapse-assumption})).
%
Then, taking $k_\textrm{\tiny max} \to +\infty$ in (\ref{T0-V0-large-L}) leads us to introduce respectively
\be
\label{T0-V0-large-L-large-k}
\mathcal{T}_{\infty}^{(0)}(x) 
 \, \equiv 
\lim_{k_\textrm{\tiny max} \to \infty} \!
\mathcal{T}_{k_\textrm{\tiny max}}^{(0)}(x) 
\;\qquad\;
\mathcal{V}_{\infty}^{(0)}(x) 
 \, \equiv 
\lim_{k_\textrm{\tiny max} \to \infty} \!
\mathcal{V}_{k_\textrm{\tiny max}}^{(0)}(x) 
\;\qquad\;
\mathcal{V}_{\infty}^{(2)}(x) 
 \, \equiv 
\lim_{k_\textrm{\tiny max} \to \infty} \!
\mathcal{V}_{k_\textrm{\tiny max}}^{(2)}(x)   \,. 
\ee
Assuming that these functions  are well defined 
and also neglecting all the terms coming from 
the second and higher derivatives of $\tau_k(x_k)$ and $\nu_k(x_k)$,
for the local term of the entanglement Hamiltonian 
in the continuum limit we find 
\be
\label{KA-diag-continuum-structure}
K_{A, \textrm{\tiny \,loc}} 
=\,
\frac{\ell_A}{s^2}
\int_A \frac{\mathcal{V}_{\infty}^{(0)}(x) }{2} \;
\Phi(x)^2 \, \rd x
+ \ell_A
\int_A \frac{1}{2} \bigg[ \mathcal{T}_{\infty}^{(0)}(x)\, \Pi(x)^2   -  \mathcal{V}_{\infty}^{(2)}(x)  \big(\Phi'(x)\big)^2  \bigg]  \rd x
+ O(s)  \,. 
\ee

\begin{figure}[t!]
	\vspace{-1.5cm}
	\hspace{0.cm}
	\includegraphics[width=1\textwidth]{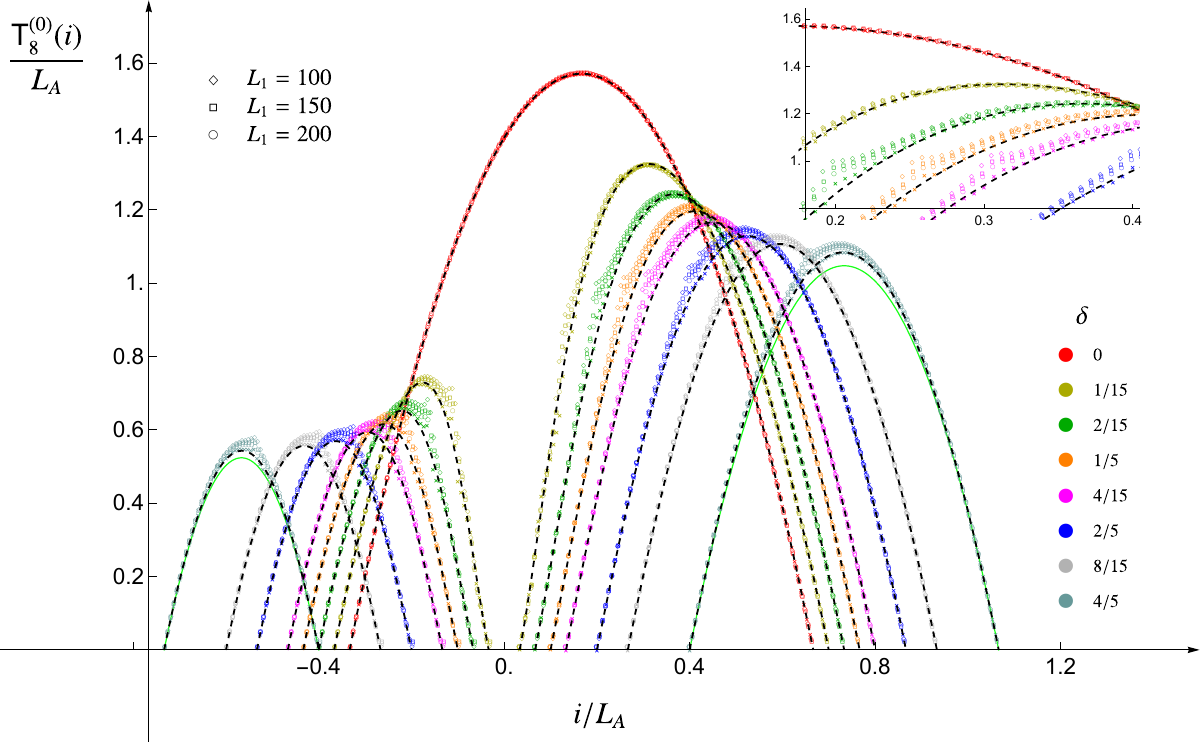}
	\\
	\rule{0pt}{10.2cm}
	\hspace{-.08cm}
	\includegraphics[width=1\textwidth]{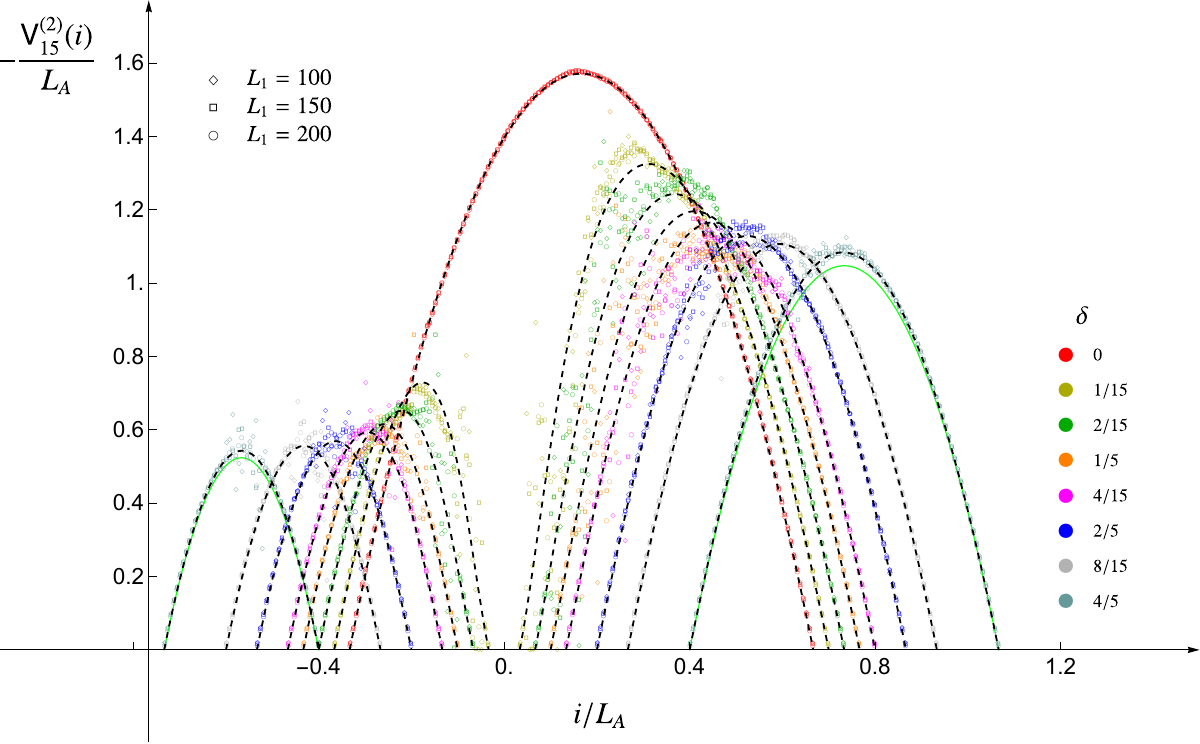}
	\vspace{-.2cm}
	\caption{The combinations of diagonals defined 
	in the first expression of (\ref{T-V-summation-0}) (top) and in (\ref{V-summation-2}) (bottom) 
	for $k_\textrm{\tiny max}  \ll L_A$ (see also (\ref{T0-V0-large-L})), when $L_2 = 2L_1$.
	The dashed black curve corresponds to (\ref{beta-loc-2int})
	and the solid green curve to its limit $d \to +\infty$ 
	(see also Fig.\,\ref{fig:DiagOffDiag_massless_lratio2_Tmat}).
	}
	\label{fig:EHmassless_lratio2}
\end{figure}

\begin{figure}[t!]
	\vspace{-.5cm}
	\hspace{0.cm}
	\includegraphics[width=1\textwidth]{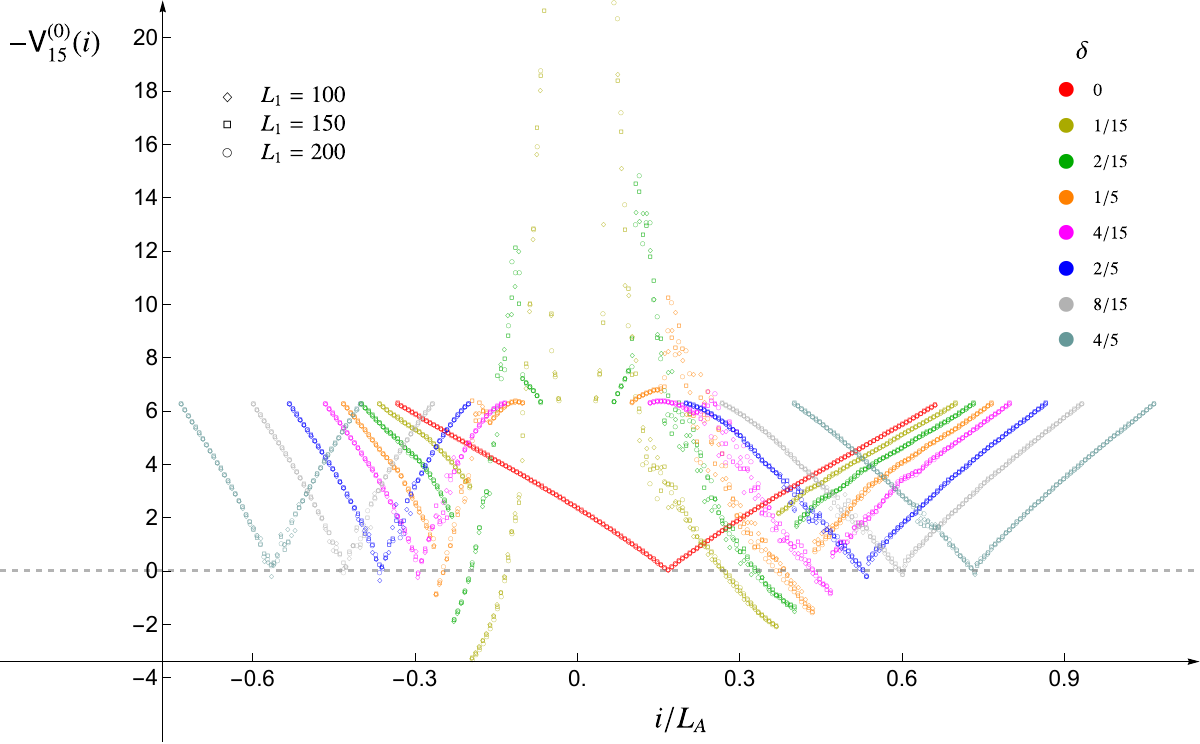}
	\vspace{-.2cm}
	\caption{
	The combination of diagonals defined in the second expression of (\ref{T-V-summation-0}) 
	for $k_\textrm{\tiny max}  \ll L_A$ (see also (\ref{T0-V0-large-L})), when $L_2 = 2L_1$.
	}
	\label{fig:EHmassless_V0term_lratio2}
\end{figure}

In Fig.\,\ref{fig:EHmassless_lratio2} and Fig.\,\ref{fig:EHmassless_V0term_lratio2}
we test numerically the functions 
$\mathcal{T}_{\infty}^{(0)}(x)$ , $\mathcal{V}_{\infty}^{(0)}(x)$ and $\mathcal{V}_{\infty}^{(2)}(x)$
introduced in (\ref{T0-V0-large-L-large-k}) when $L_2 = 2L_1$
by reporting the numerical data obtained for the ratios
$ \mathsf{T}_{k_\textrm{\tiny max}}^{(0)}(i) / L_A $, $ \mathsf{V}_{k_\textrm{\tiny max}}^{(2)}(i) / L_A $
and $ \mathsf{V}_{k_\textrm{\tiny max}}^{(0)}(i)  $ 
(see (\ref{T-V-summation-0}) and (\ref{V-summation-2}))
with  increasing values of $L_A$ (see (\ref{T0-V0-large-L}))
in the regime where $k_\textrm{\tiny max}  \ll L_A$,
for various values of $\delta$ and $\omega L_A=10^{-50}$.
In Appendix\;\ref{app-massless},
the same quantities have been considered for $L_2 = L_1$
(see Fig.\,\ref{fig:EHmassless_lratio1} and Fig.\,\ref{fig:EHmassless_V0term_lratio1}) for $\omega L_A=10^{-50}$.
%
The black dashed curves in Fig.\,\ref{fig:EHmassless_lratio2} and Fig.\,\ref{fig:EHmassless_lratio1}
denote the weight function (\ref{beta-loc-2int})
and  its large separation distance limit corresponds to the solid green curves.

The values of $k_\textrm{\tiny max}$ chosen 
in Fig.\,\ref{fig:EHmassless_lratio2}, Fig.\,\ref{fig:EHmassless_V0term_lratio2},
Fig.\,\ref{fig:EHmassless_lratio1} and Fig.\,\ref{fig:EHmassless_V0term_lratio1}
correspond to the best collapses of the numerical data points among the ones that we have explored. 
In Appendix\;\ref{app-massless}, considering the case of equal intervals,  
 the crucial role of the parameter $k_\textrm{\tiny max}$ has been investigated
by showing $ \mathsf{T}_{k_\textrm{\tiny max}}^{(0)}(i) / L_A $ and $ \mathsf{V}_{k_\textrm{\tiny max}}^{(2)}(i) / L_A $
for smaller and still non trivial values of $k_\textrm{\tiny max}$
(see Fig.\,\ref{fig:EHmassless_lratio1_kmax3} and Fig.\,\ref{fig:EHmassless_lratio1_kmax6}).

In our numerical analyses we have observed that
less precision is needed in the massless regime 
with respect to the massive cases considered in Sec.\,\ref{sec-2int-generic-EH}.
Indeed, about 4000 digits have been employed 
when $L_2 \neq L_1$ in the massless regime,
while about $5000$ have been used to explore the large $\omega$ regime. 

The  collapses of the numerical data points 
in Fig.\,\ref{fig:EHmassless_V0term_lratio2} and Fig.\,\ref{fig:EHmassless_V0term_lratio1}
suggest that the weight function for the $O(1/s^2)$ term in (\ref{KA-diag-continuum-structure}) 
vanishes identically, namely $\mathcal{V}_{\infty}^{(0)}(x) = 0$
(for the analogue of this result in the single interval case, see the left panel of Fig.\,4 in \cite{DiGiulio:2019cxv}).
Instead, the numerical results displayed in Fig.\,\ref{fig:EHmassless_lratio2} and Fig.\,\ref{fig:EHmassless_lratio1}
tell us that a reasonable conjecture for the weight functions occurring in the $O(1)$ terms in (\ref{KA-diag-continuum-structure}) 
can be written through  (\ref{beta-loc-2int}) as
\be
\label{T0-V2-beta-loc}
\mathcal{T}_{\infty}^{(0)}(x) \,=\, \beta_{\textrm{\tiny loc}}(x) 
\;\;\;\;\qquad\;\;\;\;
\mathcal{V}_{\infty}^{(2)}(x) \,=\, -\, \beta_{\textrm{\tiny loc}}(x)   \,. 
\ee

\clearpage

Following \cite{Arias:2016nip}, an extrapolation of the numerical data has been performed  in some cases.
In particular, assuming 
$\mathsf{T}_{k_\textrm{\tiny max}}^{(0)}(i)  \approx  \alpha_{\textrm{\tiny $\mathsf{T}$}}   L_A + \beta_{\textrm{\tiny $\mathsf{T}$}}$ 
and $\mathsf{V}_{k_\textrm{\tiny max}}^{(2)}(i) \approx  \alpha_{\textrm{\tiny $\mathsf{V}$}}   L_A + \beta_{\textrm{\tiny $\mathsf{V}$}}$ 
as $L_A \to \infty $, a fitting procedure has been employed to extract $\alpha_{\textrm{\tiny $\mathsf{T}$}} $, $\beta_{\textrm{\tiny $\mathsf{T}$}}$, 
$\alpha_{\textrm{\tiny $\mathsf{V}$}} $ and  $\beta_{\textrm{\tiny $\mathsf{V}$}}$
from the data sets corresponding to large enough values of $L_A$.
In the inset of the top panels in 
Fig.\,\ref{fig:EHmassless_lratio2}, Fig.\,\ref{fig:EHmassless_lratio1}, 
Fig.\,\ref{fig:EHmassless_lratio1_kmax3} and Fig.\,\ref{fig:EHmassless_lratio1_kmax6},
the crosses correspond to $\alpha_{\textrm{\tiny $\mathsf{T}$}} $ obtained through this extrapolation procedure
and a nice agreement is observed with (\ref{beta-loc-2int}).
A similar analysis for $\alpha_{\textrm{\tiny $\mathsf{V}$}} $ has been carried out 
only in the regions of $A$  where the data points have a clear trend 
(e.g., in the bottom panels of Fig.\,\ref{fig:EHmassless_lratio2} and Fig.\,\ref{fig:EHmassless_lratio1},
around the endpoints of the blocks),
finding agreement with (\ref{beta-loc-2int}) again.

Summarising, 
while the numerical data points in the top panels 
of Fig.\,\ref{fig:EHmassless_lratio2} and Fig.\,\ref{fig:EHmassless_lratio1}
support the conjecture for $\mathcal{T}_{\infty}^{(0)}(x)$ in (\ref{T0-V2-beta-loc}),
the numerical results displayed in the bottom panels of these figures are quite noisy 
and therefore they do not provide a convincing numerical evidence for the conjecture 
proposed in (\ref{T0-V2-beta-loc}) for $\mathcal{V}_{\infty}^{(2)}(x)$.


Since the data points in Fig.\,\ref{fig:EH-density} show the occurrence of the hyperbolic front \eqref{x-gamma-def},
it is worth exploring the scaling of the elements in the off-diagonal blocks along this front.
The results of this analysis have been reported in Appendix\;\ref{app-massless}.
Here we just mention that we find reasonable collapses of the data points when the elements of $T$ along this front are divided by $\log L_A$,
when both $L_1 = L_2$ and $L_1 \neq L_2$ 
(see the top panel in Fig.\,\ref{fig:AntiDiag_massless_lratio1}  and  Fig.\,\ref{fig:T_hyperbola_massless_lratio2} respectively).
In \cite{Arias:2018tmw}, a similar logarithmic scaling has been observed in the free chiral current model (see Sec.\,\ref{sec-chiral-current}).
Instead, for the matrix $V$, 
while a reasonable collapse of these elements is observed when  $L_1 = L_2$ 
(see the bottom panel of Fig.\,\ref{fig:AntiDiag_massless_lratio1}),
i.e. when the front \eqref{x-gamma-def} corresponds to the antidiagonal of $V$,
it becomes unclear when $L_1 \neq L_2$ 
(see the bottom panel of Fig.\,\ref{fig:T_hyperbola_massless_lratio2}).
%

Let us conclude our analysis of the entanglement Hamiltonians in the harmonic chain by remarking that, 
while the regimes of large and vanishing $\omega$ have been explored quite extensively in the previous sections, 
for the most important one where the mass parameter takes finite and non vanishing values has been poorly discussed,
except for the brief comments in Sec.\,\ref{sec-crossover}.
This is because exploring this regime is very difficult. 
In order to gain insights about the entanglement Hamiltonian in this complicated regime, 
in Appendix\;\ref{app-massive:1block} we report the results of some numerical analyses for the case of a single block
whose outcomes might provide some handle to explore the continuum limit.

\section{Free chiral current model}
\label{sec-chiral-current}

In this section, we explore the entanglement Hamiltonian of the union of two disjoint intervals $A=A_1 \cup A_2$
 in the line for the free chiral current $j (x)$ given by the derivative of the massless scalar field,
 whose commutation relation is $ \big[\, j(x)\, , j(y) \,\big]=\textrm{i}\,\delta'(x-y)$,
when the entire system is in its ground state. 
This model has been explored e.g. in  \cite{Sonnenschein:1988ug, Arias:2018tmw, Berenstein:2023tru}
and this quantity has been studied in \cite{Arias:2018tmw}, both in the continuum and in the lattice. 
%
%
In Appendix \ref{app:chiral}
the formulas for the entanglement entropies and the entanglement Hamiltonian 
in the lattice model of the chiral current \cite{Arias:2018tmw} are reviewed.
In Appendix\;\ref{app:mutual-info} we discuss the R\'enyi mutual information
for both the massless scalar and the chiral current.


The Hamiltonian of the lattice model on the circle made by $L$ sites reads
\begin{equation}
\label{eq:chiral_lattice_H}
    \widehat{H}  = \frac{1}{2} \sum_{i \,= \,1}^L  \hat{b}^2_i  \,. 
\end{equation}
The operators $\hat{b}_i$ satisfy the commutation relations
corresponding to the discretization of the commutation relation for the chiral current $j(x)$ in the continuum,
namely
\begin{equation}
\label{eq:chiral_lattice_commutator}
    \big[\, \hat{b}_i \, ,\hat{b}_j \, \big] = \,  \textrm{i} \,Y_{i,j} 
    \;\;\;\;\qquad\;\;\;
    Y_{i,j}  \equiv \, \delta_{j,i+1}-\delta_{j,i-1}   \,. 
\end{equation}
Since periodic boundary conditions  are imposed, 
the matrix $Y$ is non-invertible (indeed, its spectrum contains two vanishing eigenvalues).

The generic element of the correlation matrix is $B_{i,j}\equiv \langle \hat{b}_i \,\hat{b}_j\rangle$  
for finite $L$ when the system is in its ground state
has been found in \cite{Arias:2018tmw} and its limit $L \to +\infty$ reads
\begin{equation}
\label{B-mat-element}
B_{i,j}=
\left\{ \begin{array}{ll}
\displaystyle 
-\frac{1+(-1)^{i-j}}{\pi\left( |j-i|^2-1\right)} \hspace{1cm} & |j-i| \neq 1 
\\
\rule{0pt}{.8cm}
\displaystyle \hspace{.2cm}
\frac{\textrm{i}}{2} \, Y_{i,j}  \hspace{1cm} & |j-i|=1 
\end{array}
\right.
\end{equation}
whose dependence on $|j-i|$ originates from the translational invariance of the model.
We remark the continuum limit of the lattice model described by (\ref{eq:chiral_lattice_H})
contains two free chiral currents \cite{Arias:2018tmw,Berenstein:2023tru}, as discussed in Appendix\;\ref{app:chiral}.
This leads us to introduce factor of $2$ in a proper way in the process of comparing
the lattice results against the ones in the continuum.

%

Considering  the bipartition of the line characterised by the subsystem $A$ made by $L_A$ sites
(which are not necessarily contiguous),
the crucial quantities related to the bipartite entanglement 
are  obtained by restricting the correlation matrix (\ref{B-mat-element}) to $A$.
In particular, in this way one obtains 
the $L_A \times L_A$ matrices $Y_A$ and $B_A$ 
from (\ref{eq:chiral_lattice_commutator})  and (\ref{B-mat-element}) respectively,
that satisfy $\textrm{Im}(B_A) = Y_A /2$.
The matrix $Y_A$ is invertible only for even values of $L_A$.
Indeed, since $Y_A$ is a real, antisymmetric and tridiagonal Toeplitz matrix, 
its eigenvalues are given by $ \lambda_j=-2\,\textrm{i}\cos [\pi j / (L_A+1)] $,
which implies that $Y_A$ is not invertible when $L_A$ is odd; 
hence only even values for $L_A$ are considered in our analysis.

The technical details underlying the numerical evaluation 
of the entanglement entropies and of the matrix characterising the entanglement Hamiltonian 
from the matrices $Y_A$ and $B_A$
are reviewed in Appendix\;\ref{app:chiral}, following \cite{Arias:2018tmw}.
Some results about the R\'enyi mutual information of this model are discussed in Appendix\;\ref{app:mutual-info}.

In particular, by introducing  the following $L_A \times L_A$ imaginary matrix
\begin{equation}
\label{eq:Rmatrix}
    R_A \equiv -\textrm{i}\,Y_A^{-1} B_A -  \frac{1}{2}\, \mathbf{1} 
    =  -\textrm{i}\,Y_A^{-1} \, \textrm{Re} (B_A)
\end{equation}
where  $\mathbf{1}$ is  the $L_A \times L_A$ identity matrix,
the entanglement Hamiltonian can be written as
\begin{equation}
\label{eq:H_A_in_terms_of_BA}
        \widehat{K}_A
        \equiv   
        \hat{ \boldsymbol{b}}^{\textrm{t}} M \, \hat{\boldsymbol{b}}
        \;\;\;\qquad\;\;\;
        M \equiv -  \,\frac{\textrm{i}}{2}\; h\big(R_A^2\big) \, R_A \,Y_A^{-1} 
\end{equation}
in terms of $\hat{\boldsymbol{b}}^{\textrm{t}} =\big(\, \hat{b}_1,\dots, \hat{b}_{L_A} \big)$ 
and of the function $h(y)$  defined in \eqref{eq:hfunction}. 

\begin{figure}[t!]
\vspace{-.5cm}
\includegraphics[width=1.05\textwidth]{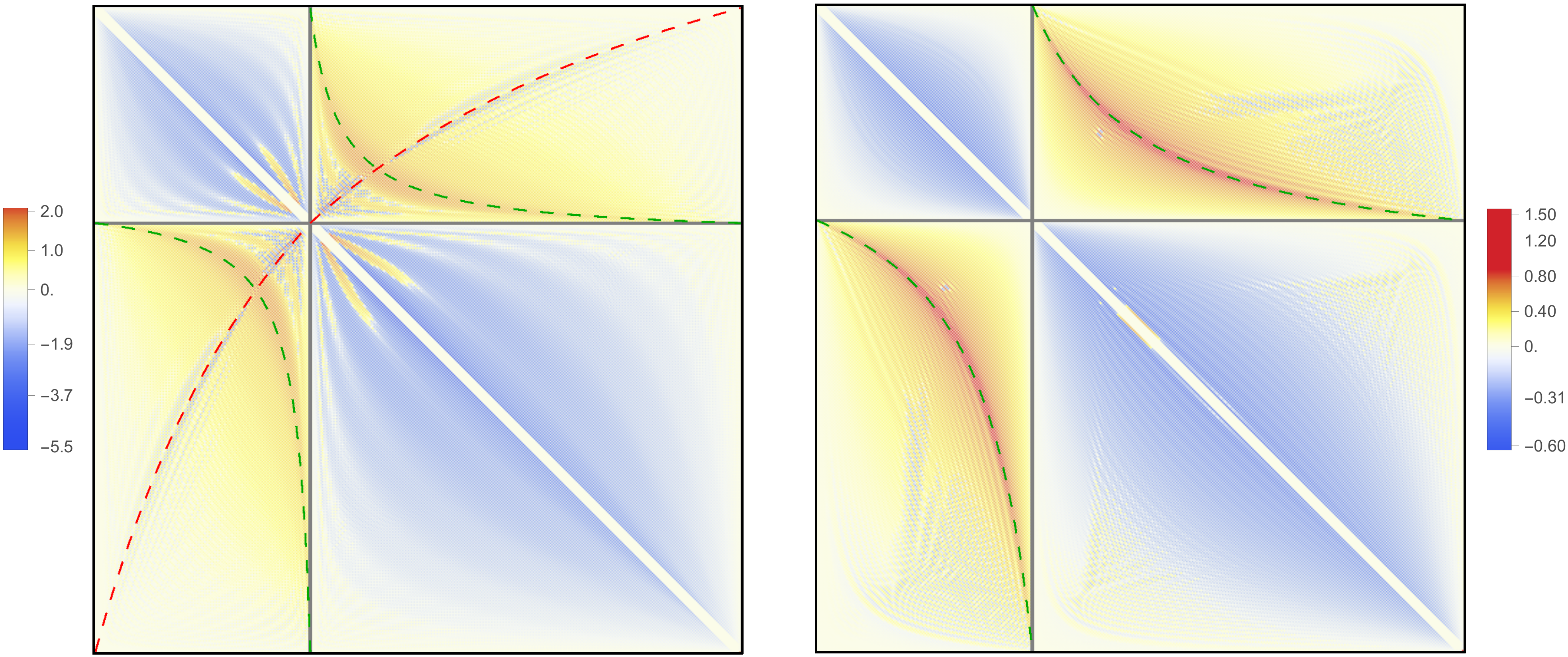}
\\
\rule{0pt}{7.7cm}
\hspace{-.19cm}
\includegraphics[width=1.046\textwidth]{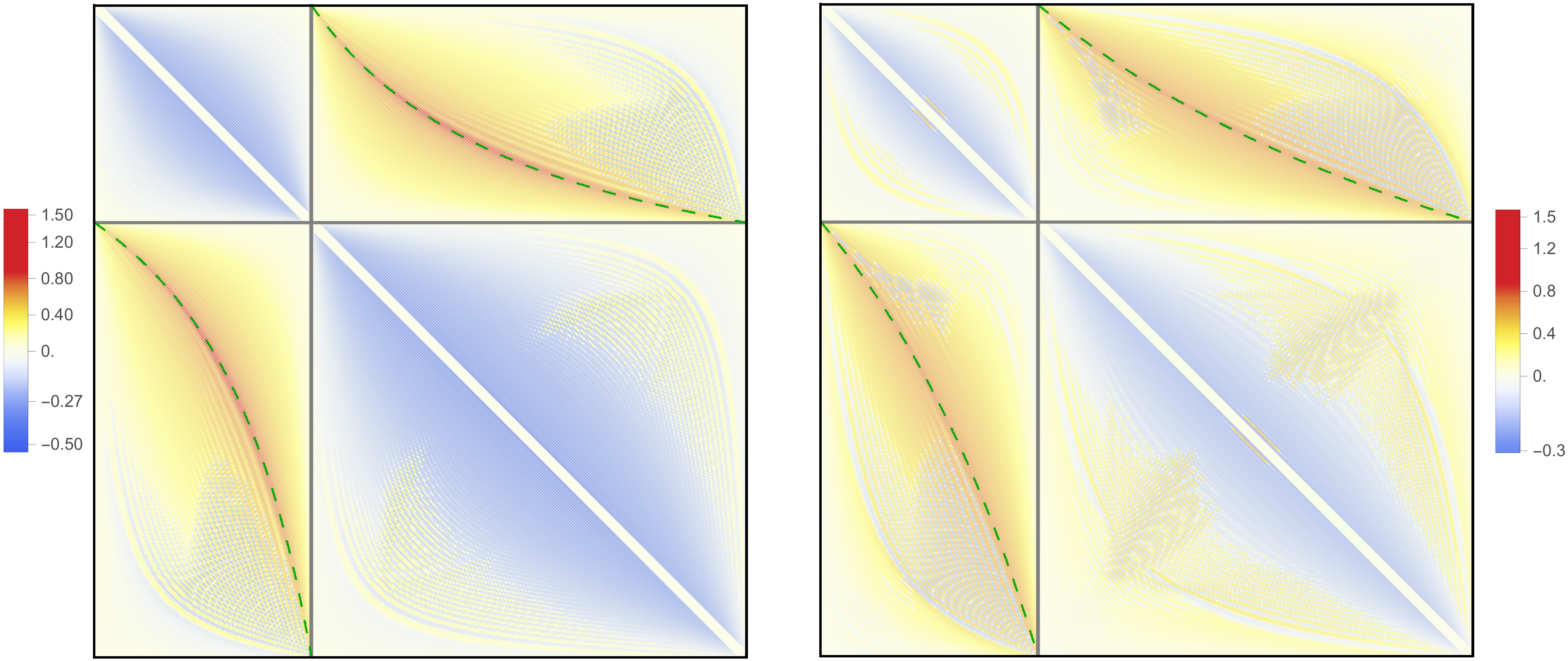}
\vspace{-.3cm}
\caption{Matrix $M$ in (\ref{eq:H_A_in_terms_of_BA}) for $L_2 = 2L_1$ with $L_1=150$ 
and separation distance given by $\delta =1/15$ (top left),
$\delta =4/15$ (top right),  $\delta =8/15$ (bottom left) and $\delta =3/2$ (bottom right).
}
\label{density-plots-M-chiral-current}
\end{figure}

Since we consider the case where the subsystem $A=A_1 \cup A_2$ is made by two disjoint blocks; 
the real and symmetric $L_A \times L_A$ matrix $M$ in (\ref{eq:H_A_in_terms_of_BA})
is naturally decomposed as follows
\begin{equation}
\label{eq:MatrixM}
M=\left(
\begin{array}{c|c}
M^{\textrm{\tiny  $(1,\!1)$} } & \, M^{\textrm{\tiny  $(1,\!2)$} } \\
\hline M^{\textrm{\tiny  $(2,\!1)$} } & \,M^{\textrm{\tiny  $(2,\!2)$} }
\end{array}
\right)
\end{equation}
where the diagonal blocks $M^{\textrm{\tiny  $(1,\!1)$} }$ and $M^{\textrm{\tiny  $(2,\!2)$} }$ 
correspond to $A_1$ and $A_2$ respectively, 
hence  they are $L_1 \times L_1$ and $L_2 \times L_2$ symmetric matrices respectively.

\begin{figure}[t!]
	\vspace{-.4cm}
	\hspace{-.5cm}	
	\includegraphics[width=1.05\textwidth]{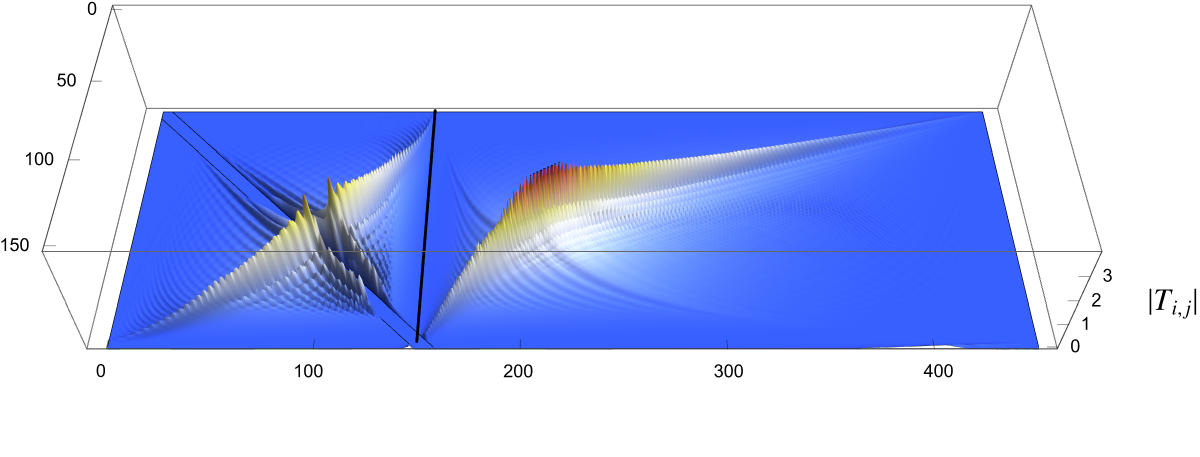}
	\vspace{-.8cm}
	\\
	\includegraphics[width=1.05\textwidth]{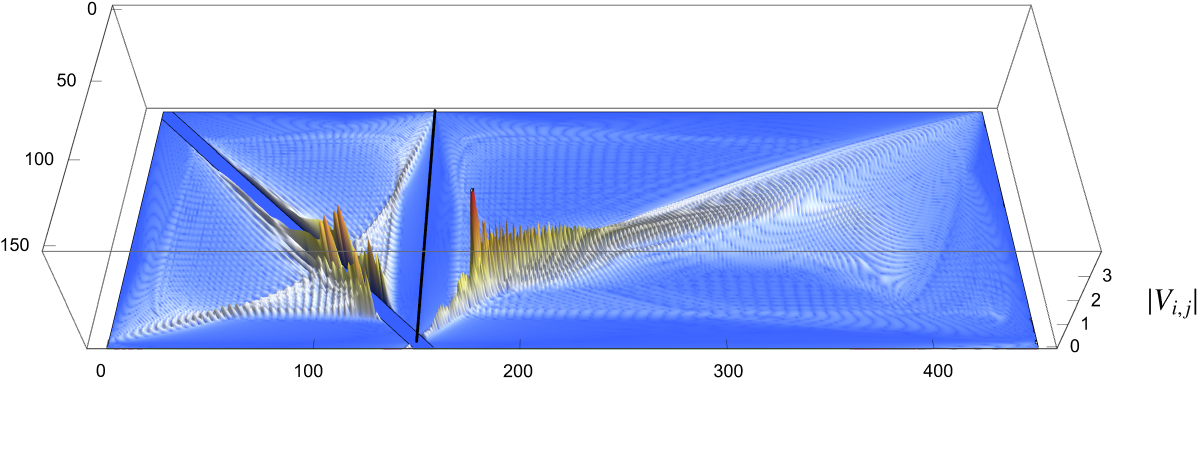}
	\vspace{.6cm}
	\\
	\includegraphics[width=1.05\textwidth]{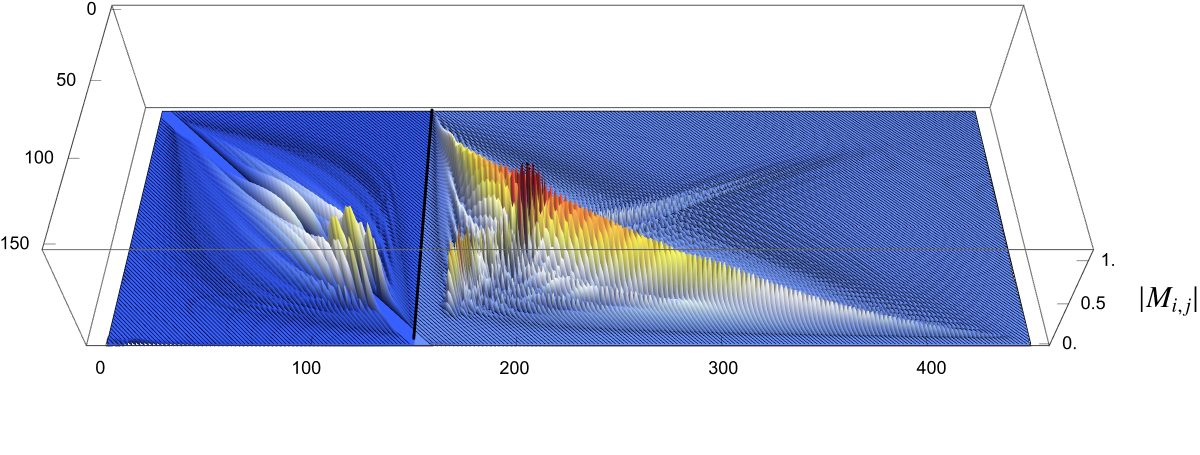}
	\vspace{-.9cm}
	\caption{
	Matrix elements $|T_{i,j}|$ (top), $|V_{i,j}|$ (middle) and $|M_{i,j}|$ (bottom) 
	with $i \in A_1 \cup A_2$ and $j \in A_1$ 	(see \eqref{eq:MatrixTandVdecomposition} and \eqref{eq:MatrixM}),
	for $L_2 = 2 L_1$ with $L_1=150$ 
	and separation distance given by $\delta = 2/15$.
	The data in the top and middle panels are obtained for $\omega = 10^{-500}$. 
	While the front corresponding to $x_{_\chi}$  (see \eqref{x-chi-def}) 
	is visible in the diagonal block of $T$ and $V$ (top and middle panel),
	it does not occur in the diagonal block of $M$ (bottom panel).
	}
	\label{3dplots-MatrixTVMabs}
\end{figure}


In Fig.\,\ref{density-plots-M-chiral-current} 
we show the matrix $M$ in (\ref{eq:H_A_in_terms_of_BA}) for $L_2= 2 L_1$ with $L_1=120$ and various separation distances,
identifying the blocks occurring in the  decomposition (\ref{eq:MatrixM}) through  grey segments. 
The main diagonal and the other two diagonals next to it in both directions (hence five diagonals in total),
which contain the elements with large amplitudes,
have been removed in order to make visible  the remaining non vanishing elements. 
The green and red dashed lines correspond to $x_c$ and $x_{_\Gamma}$  in \eqref{x-conj-def} and \eqref{x-gamma-def} respectively,
like in Fig.\,\ref{fig:EH-density}.

Comparing the matrix $M$ in Fig.\,\ref{density-plots-M-chiral-current} for the chiral current 
with the matrices $T$ and $V$  in Fig.\,\ref{fig:EH-density} for the massless scalar,
the most relevant feature to highlight is the absence of the front given by 
$x_{_\chi}$  (see \eqref{x-chi-def}) in the diagonal blocks of $M$,
which occurs in the diagonal blocks of $T$ and $V$ instead
(see the dashed black curves in Fig.\,\ref{fig:EH-density}).
A comparison among the matrices $T$, $V$ and $M$ has been performed also 
in Fig.\,\ref{3dplots-MatrixTVMabs} 
by considering the absolute value of their elements 
and showing a three dimensional representation of their upper blocks 
(see \eqref{eq:MatrixTandVdecomposition} and \eqref{eq:MatrixM})
for  $L_2 = 2 L_1$, $L_1=150$  and $\delta = 2/15$
(the largest diagonals have been removed,
as done in Fig.\,\ref{fig:EH-density} and in Fig.\,\ref{density-plots-M-chiral-current}).
Also from Fig.\,\ref{3dplots-MatrixTVMabs} one realises that 
the front for $x_{_\chi}$  (see \eqref{x-chi-def}) 
does not occur in the diagonal block in the bottom panel,
while it is clearly visible in the other two diagonal blocks,
as already remarked above.


We remark that the diagonal blocks $M^{(1,1)} $ and $M^{(2,2)} $ (see (\ref{eq:MatrixM})) in Fig.\,\ref{density-plots-M-chiral-current}
display the same qualitative structure observed from the analytic expression, as shown in Fig.\,5 of \cite{Arias:2018tmw}.

\begin{figure}[t!]
	\vspace{-.5cm}
	\hspace{-.45cm}
	\includegraphics[width=1.07\textwidth]{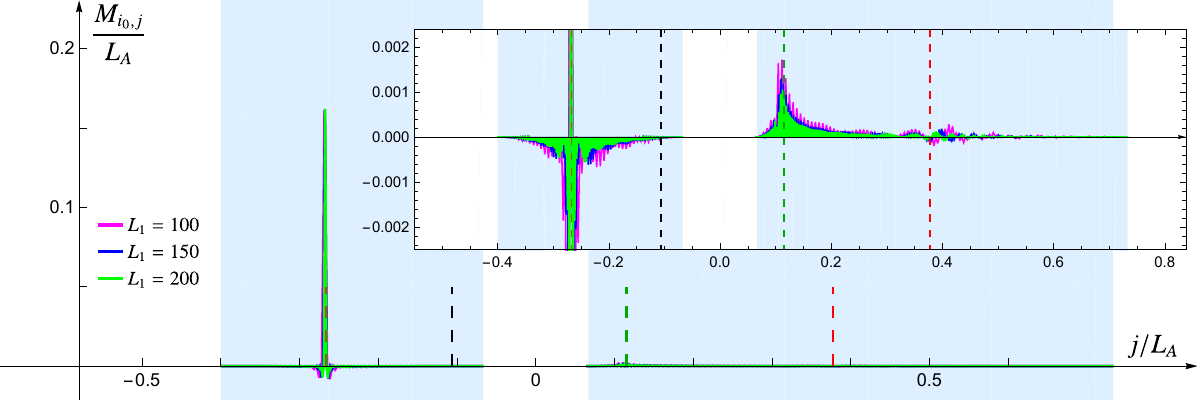}
	\vspace{.2cm}
	\caption{ 
	Matrix elements of  $M$ along the row labelled by $i=i_0$.
	Here $L_2=2L_1$, $\delta=2/15$ and $i_0=\tfrac{2}{5} L_1$.
	 The green, red, black and brown vertical dashed segments denote the intersections of the row with 
	 \eqref{x-conj-def}, \eqref{x-gamma-def}, \eqref{x-chi-def} and the main diagonal respectively
	 (see also the dashed lines in Fig.\,\ref{fig:EH-density} and Fig.\,\ref{density-plots-M-chiral-current}).
	}
\label{fig:2int-chiral-row}
\end{figure}


In Fig.\,\ref{fig:2int-chiral-row} we display some numerical data for the matrix elements $M_{i_0,j}/L_A$
along the line labelled by $i=i_0$,
for $L_2 = 2L_1$, $\delta = 2/15$, $i_0=\tfrac{2}{5} L_1$ and various values of $L_1$.
The dashed vertical lines denote the intersections of the row with $x_c$, $x_{_\Gamma}$, $x_{_\chi}$ 
(see \eqref{x-conj-def}, \eqref{x-gamma-def} and \eqref{x-chi-def} respectively)
and the main diagonal, like in  Fig.\,\ref{fig:all_elements}.
From the inset of Fig.\,\ref{fig:2int-chiral-row}, we observe that the amplitude of the matrix elements
increase around the intersection with the main diagonal, $x_c$ and  $x_{_\Gamma}$,
but a significant change in the amplitude does not occur around $x_{_\chi}$,
like in the fermionic case discussed in Appendix\;\ref{app-fermions}
(see Fig.\,\ref{fig:2int-fermion-row}).
It is worth comparing Fig.\,\ref{fig:2int-chiral-row} with Fig.\,\ref{fig:all_elements},
in order to appreciate the qualitative differences in these profiles.

\begin{figure}[t!]
	\vspace{-.5cm}
	\hspace{-.45cm}
	\includegraphics[width=1.\textwidth]{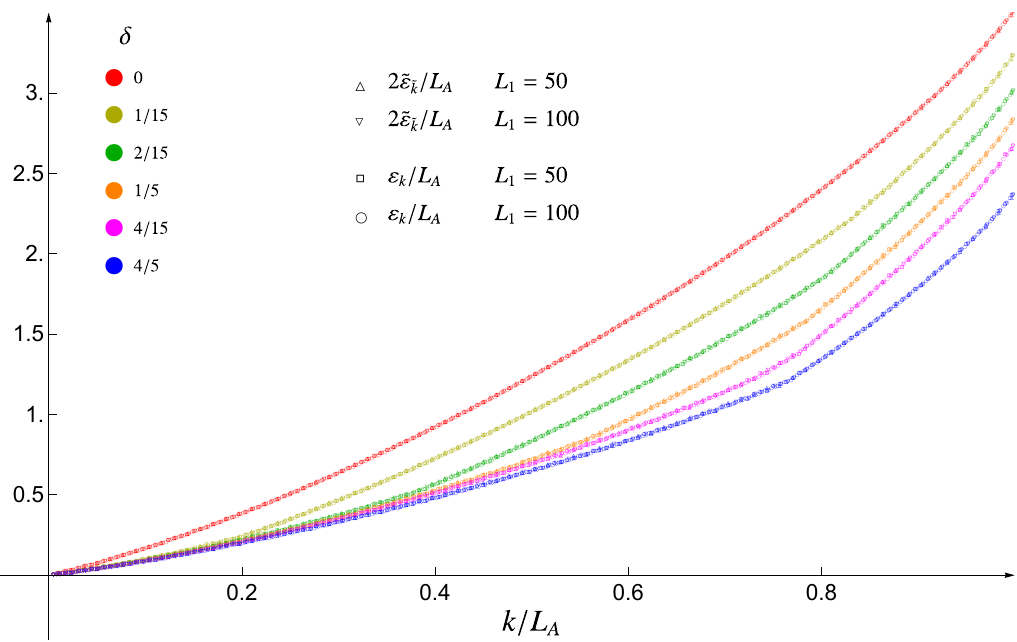}
	\vspace{.2cm}
	\caption{Single-particle entanglement spectrum $\tilde{\varepsilon}_{\tilde{k}}$ for the chiral current 
		and single-particle entanglement spectrum $\varepsilon_k$ for the massless ($\omega = 10^{-50}$), when $A$ is made by two blocks $L_2 = 2L_1$
		(see also the bottom panel of Fig.\,\ref{fig:EntSpec_massless_lratio1}, where the data points for $\varepsilon_k$ in this setup are also displayed). 
		The relation $\tilde{k} =  k/2 $ has been used in the comparison of these two single-particle entanglement spectra.
		}
\label{fig:ES-comparison}
\end{figure}

It is instructive to consider the single-particle entanglement spectrum 
of the entanglement Hamiltonian for the chiral current that we are exploring 
and comparing the result with the corresponding one
for the harmonic chain in the massless regime (see Fig.\,\ref{fig:EntSpec_massless_lratio1}).
This comparison has been performed in Fig.\,\ref{fig:ES-comparison}, for configurations having $L_2 = 2L_1$.
Taking into account the doubling in the proper way (see the caption of the figure), 
a nice match of the data points is observed between these two quantities. 
Since the two underlying models are not equivalent  \cite{Arias:2018tmw, Berenstein:2023tru},
it would be insightful to understand this numerical agreement analytically. 

Another instructive comparison between the massless harmonic chain 
and the lattice model for the chiral current is based 
on the mutual information and its generalisation given by the R\'enyi mutual information, 
which involves the R\'enyi entropies,
whose continuum limit has been explored in \cite{Calabrese:2009ez, Arias:2018tmw},
finding analytic results. 
We remark that, in the continuum limit, these two bosonic lattice models are different.
Indeed, for instance, the massless scalar satisfies the Haag duality,
while the chiral current does not possess this property \cite{Arias:2018tmw}.
The comparison between the R\'enyi mutual information in these models
is discussed in Appendix\;\ref{app:mutual-info}, 
where we also provide some numerical checks of the analytic expressions through lattice computations
(see Fig.\,\ref{fig:CH_mutualinformation-scalar} and Fig.\,\ref{fig:CH_mutualinformation-current}). 
%

\begin{figure}[t!]
\vspace{-.2cm}
	\begin{minipage}{0.53\textwidth}
		\hspace{-.92cm}
		\includegraphics[width=1.0\linewidth]{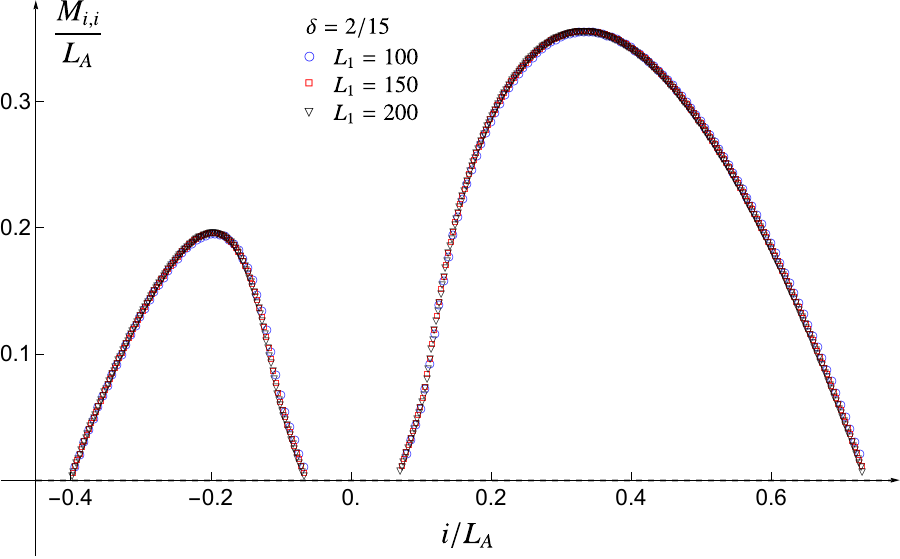}
	\end{minipage}
	\hspace{-0.3cm}
	\begin{minipage}{0.53\textwidth}
		\vspace{0.14cm}
		\includegraphics[width=1.0\linewidth]{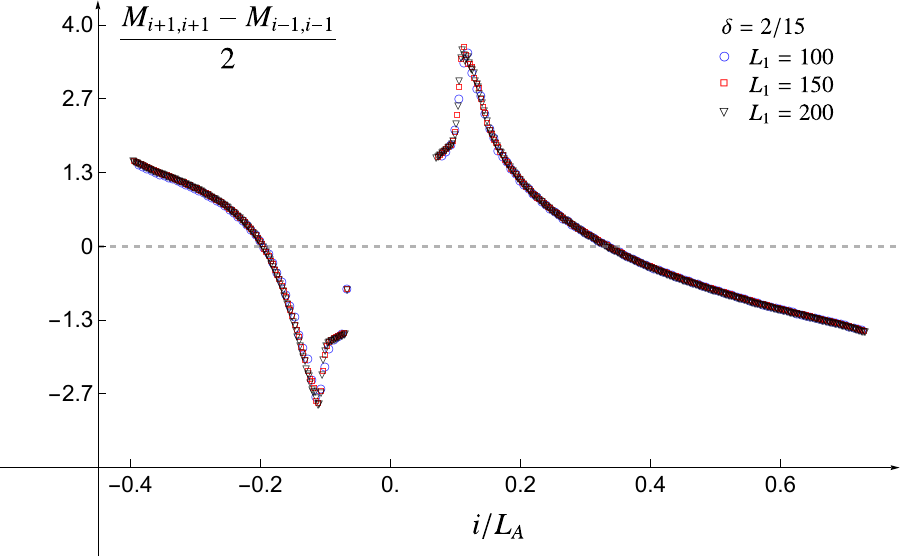}
	\end{minipage}
\vspace{.7cm}
	\\
	\begin{minipage}{0.53\textwidth}
		\hspace{-1.04cm}
		\includegraphics[width=1.0\textwidth]{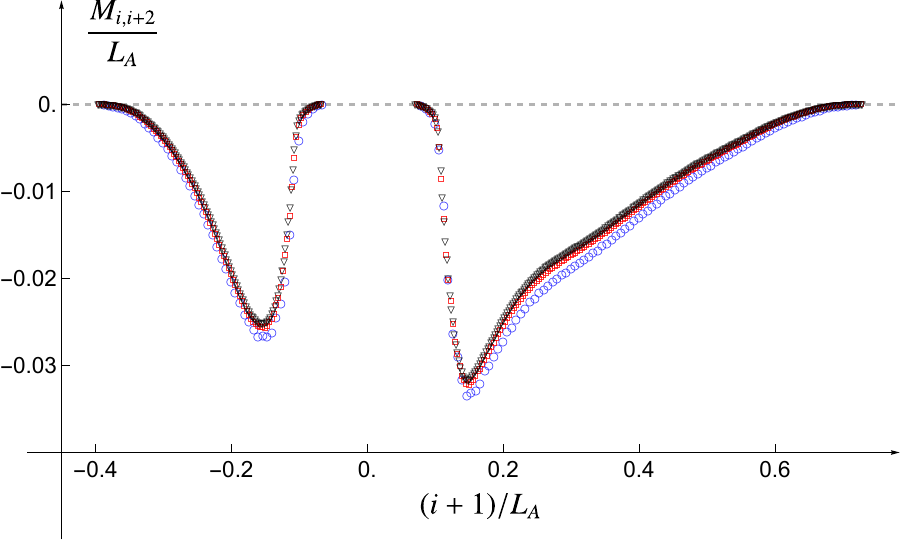}
	\end{minipage}
	\hspace{.9cm}
	\begin{minipage}{0.53\textwidth}
		\hspace{-1.28cm}
		\vspace{-0.05cm}
		\includegraphics[width=1.0\linewidth]{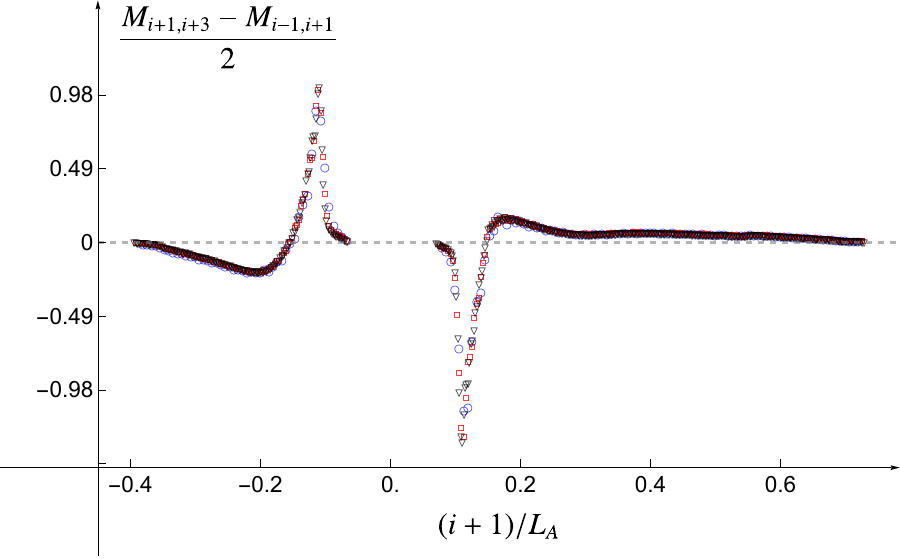}
	\end{minipage}
\vspace{.2cm}
	\caption{Diagonals $M_{i,i}$ (top left) and $M_{i,i+2}$ (bottom left) and their discrete first derivative in the corresponding right panels,
		for $L_2 = 2L_1$. 
	}
	\label{MainDiagonals-M-lratio2}
\end{figure}

In Fig.\,\ref{MainDiagonals-M-lratio2} we display some numerical data for the main diagonal $M_{i,i}$
and the first non vanishing diagonal $M_{i,i+2}$ of the matrix $M$ defined in (\ref{eq:H_A_in_terms_of_BA}).
Indeed, all the odd diagonals of $M$ vanish, namely $M_{i,i+2k+1}=0$ for $k \geqslant 0$.
This feature agrees with the fact that this model exhibits a parity symmetry 
(see $\hat{b}_i\rightarrow (-1)^i \, \hat{b}_i$ in Appendix\,\ref{app:chiral}) \cite{Berenstein:2023tru},
which also implies that $B_{i,i+2k+1} = 0$ in \eqref{B-mat-element}.
%
Notice that, instead,  
all the diagonals of the matrices $T$ and $V$ explored in Sec.\,\ref{sec-2int-massless}
are typically non vanishing. 
The collapses of the data points in the left panels of Fig.\,\ref{MainDiagonals-M-lratio2}
indicate that $M_{i,i}/L_A$ and  $M_{i,i+2}/L_A$
provide well defined continuous functions as $L_A \to +\infty$.
However, the corresponding right panels show that 
these continuous functions are not smooth because their second derivative is discontinuous.
This lack of smoothness has also been observed for the matrices $T$ and $V$ (see the right panels of Fig.\,\ref{MainDiagonals-lratio2}).

\begin{figure}[t!]
	\vspace{-.3cm}
	\hspace{-.54cm}
     \includegraphics[scale=.8]{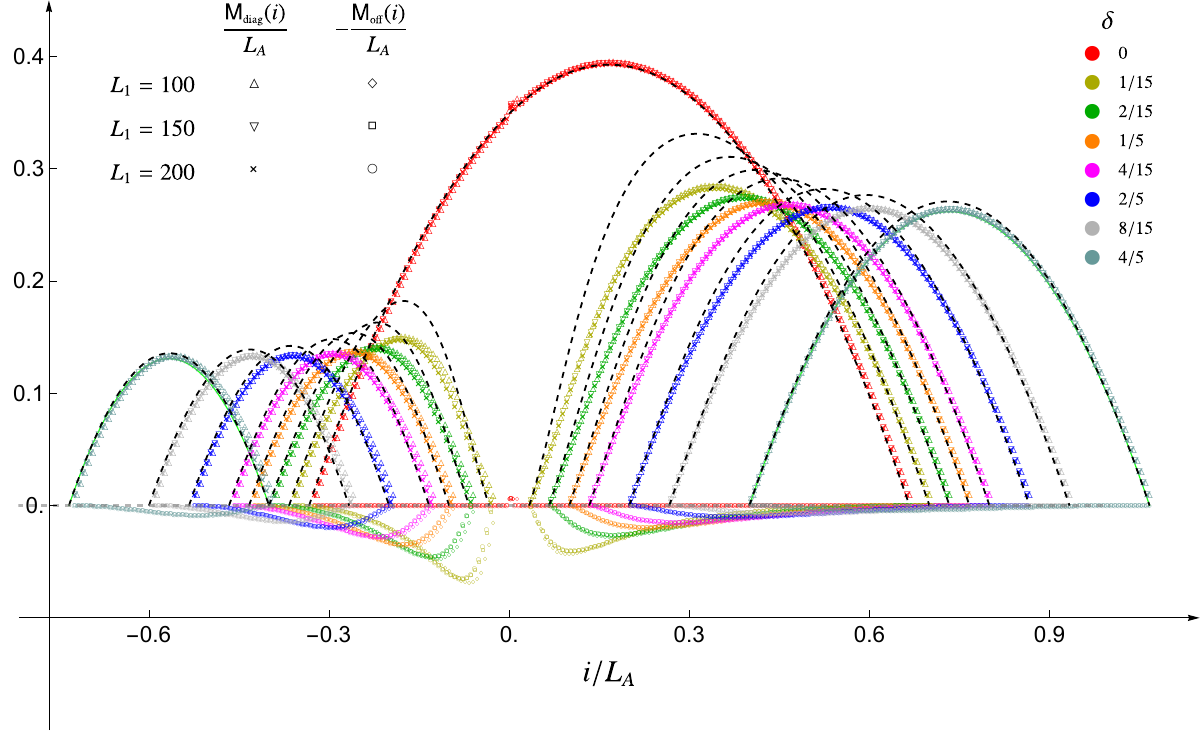}
	\\
	\rule{0pt}{11.2cm}
	\hspace{-.45cm}
        \includegraphics[scale=.8]{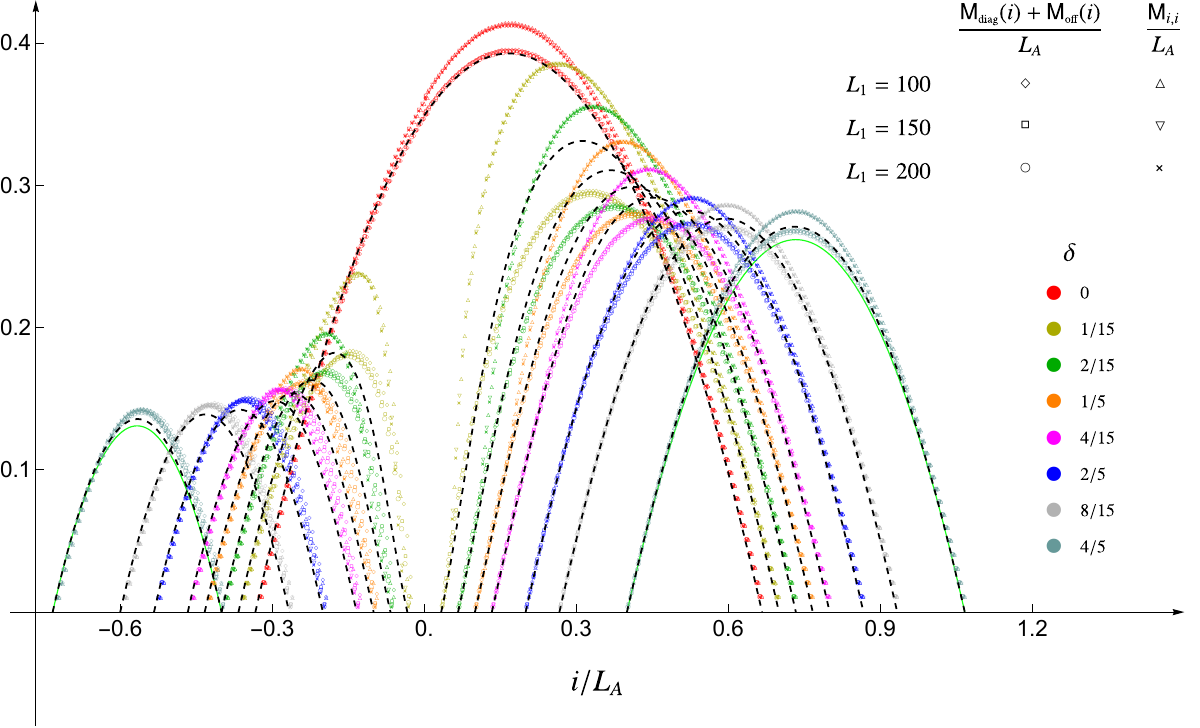}
	\vspace{.0cm}
	\caption{Combinations of elements in $M$ introduced in \eqref{eq:MatrixMdiag} and \eqref{eq:MatrixMoff}, for $L_2 = 2L_1$. 
	}
	\label{fig:CH_Block_Row_Sum}
\end{figure}


\begin{figure}[t!]
	\vspace{-.3cm}
	\hspace{-1cm}
    \includegraphics[scale=0.85]{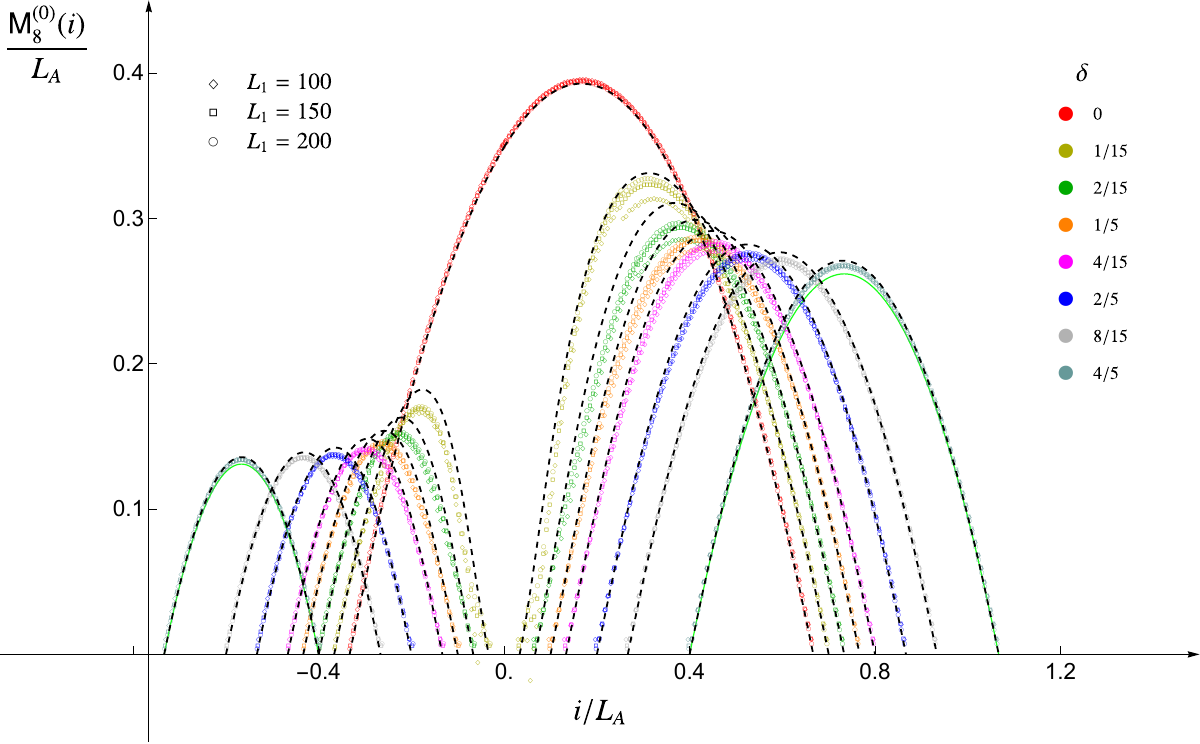}
    	\vspace{-.4cm}
    \caption{
    The combinations of diagonals   in \eqref{M-summation-0}  
	for an optimal $k_\textrm{\tiny max}  \ll L_A$ when $L_2 = 2L_1$.}
    \label{fig:CH_DiagonalSummation}
\end{figure}


It is worth adapting to the chiral current model 
the analysis performed in Sec.\,\ref{sec-2int-massless} for the massless harmonic chain.
This leads us to consider the sum of all the elements of a block of $M$ along a row.
In particular, in the diagonal and off-diagonal blocks of $M$ (see (\ref{eq:MatrixM})),
let us  introduce respectively
\bea
\label{eq:MatrixMdiag}
    \mathsf{M}_{\textrm{\tiny diag}}(i)
    &=&
    \sum_{j\in{A_n}} 
    M^{\textrm{\tiny  $(n,\! n)$} }_{i,j}
    \hspace{1cm}
    i \in A_n
    \\
    \rule{0pt}{.6cm}
    \label{eq:MatrixMoff}
    \mathsf{M}_{\textrm{\tiny off} }(i)
    &=&
    \sum_{j\in A_{n}}  M^{\textrm{\tiny  $(m,\!n)$} }_{i,j}
    \hspace{1cm}
    i \in A_m
        \hspace{1cm}
            m \neq n  \,. 
\eea

Some numerical results for these quantities are reported in Fig.\,\ref{fig:CH_Block_Row_Sum}
and the perfect collapses of the data points for increasing values of $L_1$ indicate that
the ratios $\mathsf{M}_{\textrm{\tiny diag}}(i)/L_A$ and $ \mathsf{M}_{\textrm{\tiny off} }(i)/L_A$
provide well defined functions in the thermodynamical limit $L_A \to +\infty$, for a given value of $L_2/L_1$.
The results of the corresponding analyses for the matrices $T$ and $V$
have been reported in Fig.\,\ref{fig:DiagOffDiag_massless_lratio2_Tmat} and \ref{fig:DiagOffDiag_massless_lratio2_Vmat}.
The dashed black curves correspond to $\beta_{\textrm{\tiny loc}}(x) / 4$, from \eqref{beta-loc-2int},
and the solid green curves to its limit as the  large separation distance diverges. 
%


By comparing the numerical results for the row-wise summations (\ref{eq:MatrixMdiag})-(\ref{eq:MatrixMoff}) displayed in Fig.\,\ref{fig:CH_Block_Row_Sum}
with the corresponding ones for the row-wise summations of the matrix $T$ in  (\ref{row-wise-sums-diag}), reported in Fig.\,\ref{fig:DiagOffDiag_massless_lratio2_Tmat},
a remarkable qualitative similarity is observed.


It is worth also adapting the procedure discussed in Sec.\,\ref{sec-2int-massless} for the massless harmonic chain
to the chiral current model that we are exploring. 
Thus, in analogy with \eqref{T-V-summation-0}, let us introduce
\be
\label{M-summation-0}
\mathsf{M}_{k_\textrm{\tiny max}}^{(0)}(i)
\equiv
\left\{\begin{array}{ll}
\displaystyle	
M_{i,i} + 2 \sum_{k=1}^{k_\textrm{\tiny max}} M_{i,i+k} 
\hspace{1cm}& \tilde{i} \in A_<
	\\
	\rule{0pt}{.9cm}
\displaystyle		
M_{i,i} + 2 \sum_{k=1}^{k_\textrm{\tiny max}} M_{i-k,i} 
\hspace{1cm}& \tilde{i} \in A_>
\end{array}
\right.
\ee 
where $A_<$ and $A_>$ are defined in \eqref{A-sets-matrix-indices-def};
hence only the elements in the diagonal blocks of $M$ occur. 
In Fig.\,\ref{fig:CH_DiagonalSummation} some numerical results  for (\ref{M-summation-0}) are shown
and they are compared against the same dashed black and solid green curves occurring in Fig.\,\ref{fig:CH_Block_Row_Sum}.
The value of $k_\textrm{\tiny max} \ll L_A$ has been chosen in the optimal way, in order to have stable curves. 
Comparing Fig.\,\ref{fig:CH_DiagonalSummation} 
with the top panel of Fig.\,\ref{fig:EHmassless_lratio2} for the matrix $T$,
a remarkable similarity is observed.

The discrepancy occurring in both Fig.\,\ref{fig:CH_Block_Row_Sum} and Fig.\,\ref{fig:CH_DiagonalSummation} 
between the curves defined by the data points and the dashed black curves
could be understood by taking into account properly the non-local terms in the entanglement Hamiltonian
found in \cite{Arias:2018tmw}.
%

\newpage
\section{Conclusions}
\label{sec-conclusions}

%
In this paper we have reported some numerical analyses for
the entanglement Hamiltonian of the union of two disjoint blocks $A = A_1 \cup A_2$
in the infinite harmonic chain in its ground state (Sec.\,\ref{sec-correlators-EH}). 
Our main results are obtained in the two limiting regimes of 
large mass (Sec.\,\ref{sec-2int-massive-equal} and Sec.\,\ref{sec-2int-generic-EH})
and vanishing mass (Sec.\,\ref{sec-2int-massless} and Sec.\,\ref{sec-chiral-current}).
Some observations have been made also in the crossover regime 
where the mass parameter takes finite and non vanishing values
(Sec.\,\ref{sec-crossover} and Appendix\;\ref{app-massive:1block}).
%
Since this entanglement Hamiltonian has the quadratic form \eqref{KA-T-V-matrices},
it is characterised by the $L_A \times L_A$ real and symmetric matrices $T$ and $V$,
where $L_A$ is the size of the subsystem,
namely $L_A = L_1 +L_2$ for two disjoint blocks,
where $L_j$ is the number of sites in the block $A_j$.


In the large mass regime (Sec.\,\ref{sec-2int-massive-equal} and Sec.\,\ref{sec-2int-generic-EH}), 
we have extended the analysis of the entanglement Hamiltonian of the single block 
reported in \cite{Eisler:2020lyn} 
to the case of two disjoint blocks.
Also for this configuration the matrices $T$ and $V$ become tridiagonal in this regime, 
hence they are fully described by the profiles of the diagonals whose elements are
$T_{i,i}$, $V_{i,i}$ and $V_{i,i+1}$.
We observe that the analytic expressions for these profiles are described by 
(\ref{T-diag-2int})-(\ref{V-diag-2int}) when  $L_1 = L_2$ 
and by  (\ref{T-diag-2int-different})-(\ref{V-diag-2int-different}) when $L_1 < L_2$,
in terms of piecewise linear functions
whose slopes are written explicitly in terms of the analytic result 
for the entanglement Hamiltonian of the half chain found in \cite{Peschel:1999xeo} 
(see Sec.\,\ref{sec-single-interval-massive}).
These piecewise linear functions can be discontinuous and their 
disconnected terms are represented pictorially
in Fig.\,\ref{fig:2int-equal-aux-functions-shapes} for  $L_1 = L_2$ 
and in Fig.\,\ref{fig:2int-generic-aux-functions-shapes} for $L_1 < L_2$.
A remarkable agreement with the numerical data points is observed, 
as shown in Fig.\,\ref{fig:2int-equal-massive-diag} when $L_1 = L_2$ 
and in Fig.\,\ref{fig:2int-massive-EH-diag-chi2-chi4} and Fig.\,\ref{fig:2int-massive-EH-diag-chi3} when $L_1 < L_2$.
The main feature to highlight about the large mass regime 
is the occurrence of sharp transitions in the profiles of the diagonals 
as the dimensionless parameter $\delta\equiv D/L_A$ changes,
being  $D$ the number of sites separating the two blocks.
Furthermore, 
we have observed two qualitatively different behaviours
depending on the value of the dimensionless ratio $\rho \equiv L_1 / L_A$
characterising the relative size of the two blocks, 
introduced in (\ref{rho-chi-def}),
whose critical value corresponds to $L_2 = 3L_1$ (see (\ref{critical-distances-L12}))
(see Fig.\,\ref{fig:2int-skeleton} for a pictorial representation).
We remark that the mutual information does not display these transitions;
indeed, the area law holds in this regime. 
Instead, sharp transitions corresponding to the same critical values observed for the entanglement Hamiltonian
occur in the single-particle entanglement spectrum.
This quantity is well described by continuous piecewise linear functions 
given in (\ref{es-2int-equal}) and (\ref{es-2int-different}), for  $L_1 = L_2$  and $L_1 < L_2$ respectively.
Also for the entanglement spectra a nice agreement with the numerical data points has been found,
as shown in Fig.\,\ref{fig:2int-equal-massive-spectrum} for $L_1 = L_2$ 
and in Fig.\,\ref{fig:2int-equal-massive-spectrum-chi2-chi4} and Fig.\,\ref{fig:2int-equal-massive-spectrum-chi3} for $L_1 < L_2$.
A heuristic picture explaining the critical values of $\delta$ observed in our numerical analyses 
has been described in Appendix\;\ref{app-disk-picture}.


In the regime of vanishing mass, 
both the harmonic chain in the massless limit  \cite{Casini:2009sr}
and the free chiral current \cite{Arias:2018tmw} have been explored,
in Sec.\,\ref{sec-2int-massless} and Sec.\,\ref{sec-chiral-current} respectively. 
While in the harmonic chain 
the entanglement Hamiltonians is characterised by the pair of $L_A \times L_A$ matrices $T$ and $V$ discussed above,
in the free chiral current model it is fully described 
by the $L_A \times L_A$ matrix $M$ given in  (\ref{eq:H_A_in_terms_of_BA}).
All these matrices display inhomogeneities and long-range couplings
(see  Fig.\,\ref{fig:EH-density}, Fig.\,\ref{density-plots-M-chiral-current} and Fig.\,\ref{3dplots-MatrixTVMabs}).
Besides the largest contributions to the entanglement Hamiltonian given by the elements around the main diagonal of these matrices, 
the subleading contributions are observed along specific fronts 
described by the analytic expressions in  \eqref{x-conj-def}, \eqref{x-gamma-def} and \eqref{x-chi-def}
(see the dashed curves in Fig.\,\ref{fig:EH-density} and Fig.\,\ref{density-plots-M-chiral-current}
and also the vertical dashed lines in Fig.\,\ref{fig:all_elements} and Fig.\,\ref{fig:2int-chiral-row}).
We remark that the fronts corresponding to \eqref{x-chi-def} in the diagonal blocks 
are not observed for the chiral current
and this provides an interesting qualitative difference between the two models 
that would be worth explaining in future studies.

 The thermodynamic limit of the main contributions to the entanglement Hamiltonian,
 that are given by the main diagonal and the nearby diagonals of the matrices mentioned above,  
 have been explored 
(see Fig.\,\ref{MainDiagonals-lratio2} for the harmonic chain  and Fig.\,\ref{MainDiagonals-M-lratio2} for the chiral current),
 as done in \cite{Eisler:2017cqi, DiGiulio:2019cxv} for the single block in infinite chains, 
 finding continuous curves with discontinuous second derivative. 
 The latter feature has not been observed in the case of the single block. 
 It would be insightful to obtain analytic expressions for these non-smooth curves,
 as done in \cite{Eisler:2017cqi} for the single block in an infinite fermionic hopping chain.

%
 
 Following the numerical analyses reported in \cite{Arias:2016nip, Eisler:2022rnp} to recover 
 the weight functions in the entanglement Hamiltonians of two disjoint blocks on the line in the continuum limit, 
 we have considered the summations of the matrix elements along a given row 
 (see (\ref{row-wise-sums-diag})-(\ref{row-wise-sums-off-diag}) and \eqref{eq:MatrixMdiag}-\eqref{eq:MatrixMoff})
 and also some specific combinations of diagonals 
 (see (\ref{T-V-summation-0})-(\ref{V-summation-2}) and (\ref{M-summation-0})).
 Remarkable collapses of the numerical data points are observed for these row-wise summations,  as shown in 
 Fig.\,\ref{fig:DiagOffDiag_massless_lratio2_Tmat} and Fig.\,\ref{fig:DiagOffDiag_massless_lratio2_Vmat}  for the harmonic chain 
 and in  Fig.\,\ref{fig:CH_Block_Row_Sum} for the chiral current. 
 Instead, for the combinations of diagonals the collapses are less clean,
 (see  Fig.\,\ref{fig:EHmassless_lratio2} and Fig.\,\ref{fig:EHmassless_V0term_lratio2}  for the harmonic chain 
 and Fig.\,\ref{fig:CH_DiagonalSummation} for the chiral current).
 These results should be compared against analytic expressions of the weight functions 
 occurring in the entanglement Hamiltonian of two disjoint intervals in the continuum limit,
 which are available in the literature for the chiral current \cite{Arias:2018tmw}  
 but still unknown for the massless scalar. 
 The entanglement Hamiltonian of two disjoint intervals is non-local in both these models
 and it worth exploring a possible relation between these non-local operators. 
While for the chiral current a fully non-local expression has been proposed \cite{Arias:2018tmw},
 for the massless scalar  we cannot established whether this quadratic operator is fully non-local as well 
 or just bilocal, like for the massless Dirac field \cite{Casini:2009vk}.

 
 The single-particle entanglement spectra for two disjoint blocks in these two infinite bosonic chains have also been investigated, 
 finding the numerical results reported in Fig.\,\ref{fig:EntSpec_massless_lratio1} for the harmonic chain
 and in Fig.\,\ref{fig:ES-comparison} for the chiral current. 
Furthermore, in Fig.\,\ref{fig:ES-comparison} it has been also shown that a simple relation occurs between these two single-particle entanglement spectra.
 Explaining such numerical agreement would gain  relevant insights 
 in the comprehension of the relation between the harmonic chain and the chiral current model. 

 
Analytic results for the continuum limit are available for the mutual R\'enyi information,
both in the harmonic chain in the massless regime \cite{Calabrese:2009ez} 
and in the chiral current model \cite{Arias:2018tmw}.
Further numerical checks of these analytic expressions have been reported in Appendix\;\ref{app:mutual-info}
(see Fig.\,\ref{fig:CH_mutualinformation-scalar} and Fig.\,\ref{fig:CH_mutualinformation-current}).


In the regime where the mass parameter is neither very large nor vanishing,
some numerical results about the entanglement Hamiltonians in the harmonic chain have been reported,
both for a single block (Appendix\;\ref{app-massive:1block}) and for two disjoint blocks (Sec.\,\ref{sec-crossover}).
In the latter case,  in comparison with the large mass regime, 
we observe that the three-diagonals approximation does not hold (see Fig.\,\ref{fig:EH-density-massive})
and also the sharp transitions do not occur anymore (see Fig.\,\ref{fig:2int-equal-massive-diag-crossover}).
For a single block,  
the limiting procedure employed in \cite{DiGiulio:2019cxv} for the massless regime 
has been considered (see Fig.\,\ref{Crossover-oldsum-kmaxnonscaling})
and also alternative procedures have been proposed which display better data collapses
(see Fig.\,\ref{Crossover-oldsum} and Fig.\,\ref{Crossover-newsum}).
Although these results provide some interesting insights for the continuum limit, 
they do not provide a prediction for the entanglement Hamiltonian 
of a single interval for the massive scalar field. 
This operator is fully non-local and finding its explicit expression is still an important open problem.

The results presented in this manuscript suggest some interesting questions.
For instance, in the large mass regime,
since the mutual information does not capture the sharp transitions 
observed in the entanglement Hamiltonian and in the single-particle entanglement spectrum, 
it would be interesting to find some simpler entanglement quantifier that can detect this feature of the large mass regime. 
We find it worth also trying to extend the analysis recently described in \cite{Baranov:2024vru} 
for the single block to the case of two disjoint blocks,
in order to see whether the piecewise linear profiles that we have obtained 
occur also through this approach. 

Our analyses can be extended in various directions. 
First, the quantities and the various regimes that we have discussed for the harmonic chain 
should be explored also in some fermionic chains (see e.g. \cite{ParisenToldin:2018uzz} for an interacting case).
Furthermore, one could consider the possibility to explore 
entanglement Hamiltonians for an arbitrary number of disjoint blocks \cite{Casini:2009vk},
in systems with boundaries \cite{Mintchev:2020uom, Eisler:2022rnp, Rottoli:2022plr, Estienne:2023ekf}
or with point-like defects \cite{Mintchev:2020jhc},
in inhomogeneous systems \cite{Tonni:2017jom, Bonsignori:2024gky, Bernard:2024pqu}
and in time-dependent scenarios, like e.g. the ones provided by quantum quenches
\cite{Cardy:2016fqc, Wen:2018svb, DiGiulio:2019lpb, Rottoli:2022ego, Rottoli:2024ylt}.
We find it worth investigating also 
the entanglement spectra associated to these entanglement Hamiltonians \cite{Cardy:2016fqc,Lauchli:2013jga,Surace:2019mft}
and extending all these questions to models in higher dimensions \cite{Javerzat:2021hxt, Huerta:2022tpq, Huerta:2023dqt}
or to other non-relativistic quantum systems 
\cite{Mintchev:2022xqh,Mintchev:2022yuo,Eisler:2023yys,Eisler:2024okk}.

\vskip 20pt 
\centerline{\bf Acknowledgments} 
\vskip 5pt

We are grateful to Giuseppe Di Giulio for collaboration at the initial stage of this project.
It is our pleasure to thank Jérôme Dubail, Mihail Mintchev, Christoph Minz, Diego Pontello
and especially Viktor Eisler and Ingo Peschel for useful discussions or correspondence. 


\vskip 20pt 

\appendix

\section{A heuristic picture for the transitions}
\label{app-disk-picture}

In the regime of large $\omega$, 
the three-diagonals approximation for the matrices $T$ and $V$ in (\ref{KA-T-V-matrices}) 
holds, even when the subsystem $A$ is the union of disjoint blocks,  
as discussed in Sec.\,\ref{sec-2int-massive-equal} and Sec.\,\ref{sec-2int-generic-EH},
for $L_1 = L_2$ and $L_1 \leqslant  L_2$ respectively. 
In this appendix we describe a simple heuristic geometric picture
that provides the three critical values (\ref{critical-distances-values}) 
(see also (\ref{delta-critical-values-rho}), in terms of (\ref{delta-chi-ratios}) and (\ref{rho-chi-def}))
observed for the separation distance $D$ between the two blocks.

\begin{figure}[t!]
\vspace{-.5cm}
\hspace{0.cm}
\includegraphics[width=1\textwidth]{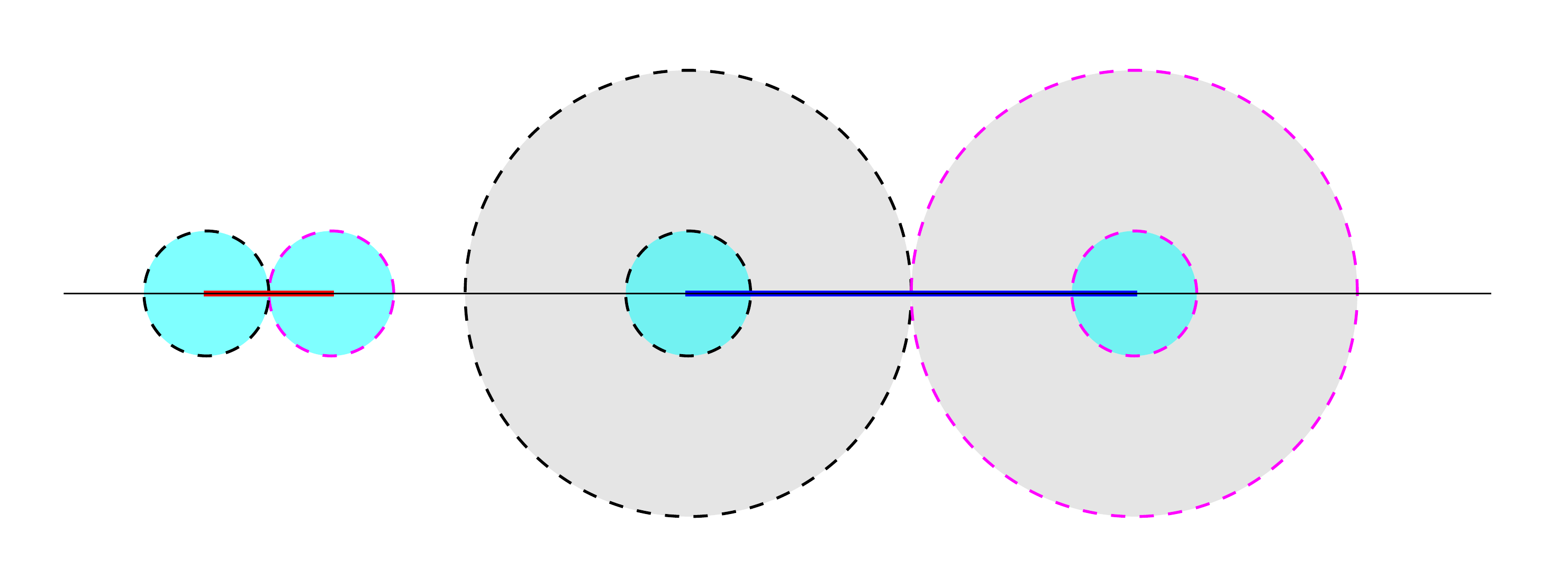}
\vspace{-.7cm}
\caption{Configuration of disks 
in the heuristic picture providing  the critical distances (\ref{critical-distances-values}),
described in Appendix\;\ref{app-disk-picture}.
}
\label{fig:2int-disk-picture}
\end{figure}

In this heuristic picture, the two endpoints of each interval play different roles,
as also suggested by the twist fields method \cite{Calabrese:2004eu,Cardy:2007mb},
where the entanglement entropies of the bipartition of the line
associated to $N$ disjoint interval is computed 
as the $2N$-point function of twist fields
and the endpoints of each interval support different (although related) kinds of fields
(for the massless regime, see e.g. also  \cite{Calabrese:2009ez,Calabrese:2010he,Coser:2013qda}). 
In the following, for the sake of simplicity we focus on the case where the subsystem $A= A_1 \cup A_2$
is the union of two disjoint intervals $A_1$ and $A_2$ (i.e. $N=2$) 
whose lengths are $L_1$ and $L_2$ respectively, 
where $L_1 \leqslant L_2$ without loss of generality
(see the red and blue segments in Fig.\,\ref{fig:2int-disk-picture}). 

For each endpoint of the interval $A_j =\big[a_j , b_j\big]$, with $j \in \big\{1, 2\big\}$,
consider a one-dimensional domain of influence of length $L_j = b_j - a_j$ centered in the endpoint itself;
hence two of such domains of influence occur for the interval $A_j$, centered in $a_j$ and $b_j$.
Since these two domains are of different types,
depending on whether they are centered either in the first or the second endpoint of the interval,
let us denote them by $\mathcal{D}_j^{+}(a_j)$ and $\mathcal{D}_j^{-}(b_j)$ respectively,
where the argument indicates the center of the corresponding domain of influence. 
While for $L_1 = L_2$, 
only the four domains of influence $\mathcal{D}_1^{+}(a_1)$, $\mathcal{D}_1^{-}(b_1)$, $\mathcal{D}_2^{+}(a_2)$ and $\mathcal{D}_2^{-}(b_2)$ occur, 
when $L_1 < L_2$, besides these four domains of influence, 
it is natural to introduce also $\mathcal{D}_1^{+}(a_2)$ and  $\mathcal{D}_1^{-}(b_2)$, with length $L_1$, but centered in $a_2$ and $b_2$ respectively. 
The six different domains $\mathcal{D}_j^{\pm}(y)$ above mentioned for $L_1 < L_2$ 
can be obtained by drawing for each of them
a circle centered in the corresponding endpoint $y$ with radius $L_j/2$. 
In Fig.\,\ref{fig:2int-disk-picture} we show these six circles in a typical configuration, 
where the black and magenta dashed lines
correspond to $\mathcal{D}^{+}$ and $\mathcal{D}^{-}$ type of domain respectively. 
Thus, in this heuristic geometric picture, the largest interval detects the occurrence of the smallest one
because its endpoints support the centers of two domains of influence of the same type but with different radii.

\begin{figure}[t!]
\vspace{.5cm}
\hspace{-1.5cm}
\includegraphics[width=1.2\textwidth]{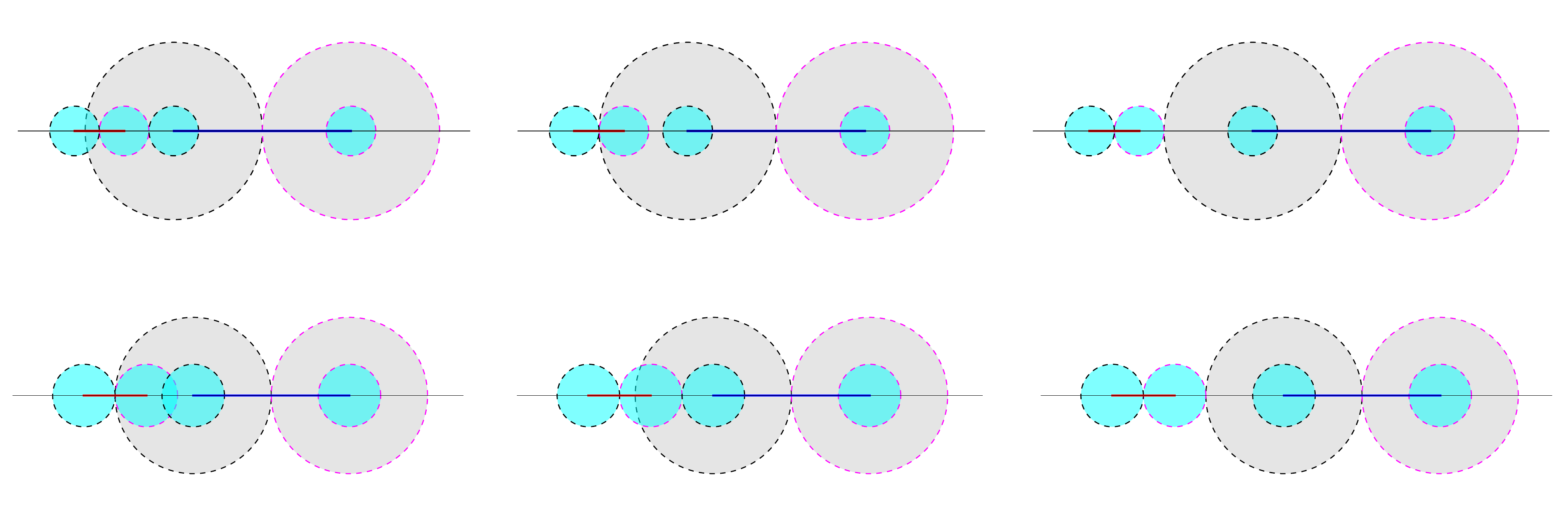}
\vspace{-.5cm}
\caption{Configurations corresponding to the critical distances (\ref{critical-distances-values})
and their disks in the heuristic picture discussed in Appendix\;\ref{app-disk-picture},
for $\rho < \rho_{\textrm{c}} $ (top panels) and  $\rho > \rho_{\textrm{c}} $ (bottom panels).
On each row, the distance decreases from the rightmost to the leftmost panel.
}
\label{fig:2int-disk-picture-critical}
\end{figure}

When $D > (L_1 + L_2)/2$, the above mentioned disks do not intersect,
but their intersections become non trivial for $D \leqslant (L_1 + L_2)/2$.
It is not difficult to realise that the critical values (\ref{critical-distances-values})
correspond to the values of $D$
where the number of intersections between the circles of different types changes. 

This simple geometric criterion is illustrated in Fig.\,\ref{fig:2int-disk-picture-critical},
where  $\rho < \rho_{\textrm{c}} $ and $\rho > \rho_{\textrm{c}} $
(namely $L_1 < (L_2 - L_1)/2$ and $L_1 > (L_2 - L_1)/2$ respectively)
in the top and bottom panels respectively, 
being $\rho_{\textrm{c}} $ the critical value for $\rho$ defined in (\ref{critical-distances-L12}).
In each line of Fig.\,\ref{fig:2int-disk-picture-critical},
the three panels display the three configurations of disks corresponding to
the critical separation distances in (\ref{critical-distances-values}), 
ordered in an increasing way, going from left to right. 
%

\section{Large $\omega$ regime: An equivalent description of the diagonals}
\label{app-A1-fixed}

In this appendix we provide the definition of $F_2( \delta, \rho \, ; x ) $ to employ 
in (\ref{T-diag-2int-different}) and (\ref{V-diag-2int-different})
for the parameterisation of the endpoints of $A = A_1 \cup A_2$ given in (\ref{endpoints-A1-fixed-A2-move}).
%
The result is obtained by adapting the procedure described in Sec.\,\ref{sec-EH-2int-different},
where the parameterisation (\ref{endpoints-A-symm-move}) is chosen
and therefore (\ref{tilde-F2-def-different-2int-below-rhoc}) and (\ref{tilde-F2-def-different-2int-above-rhoc})
must be employed.


Setting the origin $O$ in the second endpoint of $A_1$,
in order to fix the parameters in the auxiliary functions (\ref{s-value-data})-(\ref{tilde-p-q-lambda-def}),
here we consider the rhombus $\mathcal{R}_0$ having vertices in $O$, $P_0$ and $P_\pm$, where
\be
\label{P-points-rhombi-A1-fixed}
P_0 \equiv (0 \,,  \rho )
\;\;\;\;\;\qquad \;\;\;\;
P_{-} \equiv ( - \rho/2 \,,  \rho/2 )
\;\;\;\;\;\qquad \;\;\;\;
P_{+} \equiv ( 3\rho/2 \,,  \rho/2 )
\ee
hence the edges of $\mathcal{R}_0$ have slopes $\pm 1$ and $\pm 1/3$.
The other rhombus $\widetilde{\mathcal{R}}$ to consider has vertices in $P_0$, $P_{+}$, $Q_<$ and $Q_>$, where
\be
\label{Q-points-rhombi-A1-fixed}
Q_< \equiv (  1-2\rho \, , \, 1/2 )
\;\;\;\;\;\qquad \;\;\;\;
Q_> \equiv \bigg( \frac{2-\rho}{2}  \, , \, \frac{1-\rho}{2} \bigg)
\ee
implying that its edges have slopes $\pm 1/2$ and $\pm 1/3$ for any value of $\rho$.
In the limiting case of equal blocks, i.e. when $\rho=1/2$, we have that $Q_<  = P_0$ and $Q_>  = P_+$;
hence only $\mathcal{R}_0$ occurs also for this parameterisation of the endpoints of $A$.


The parameters in the auxiliary functions (\ref{s-value-data})-(\ref{tilde-p-q-lambda-def})
occurring in the definition of $F_2( \delta, \rho \, ; x ) $ for the four phases I, II, III and IV
are determined from the rhombi $\mathcal{R}_0$ and $\widetilde{\mathcal{R}}$
introduced above, 
according to the same principle adopted in Sec.\,\ref{sec-EH-2int-different} (see Fig.\,\ref{fig:2int-skeleton}).
This analysis allows to construct the function $F_2(\delta,\rho; x)$
corresponding to the parameterisation (\ref{endpoints-A1-fixed-A2-move}) for $A$.

When $\rho  \in \big(0 ,\rho_{\textrm{c}} \big]$, we find 
\be
\label{tilde-F2-def-different-2int-below-rhoc-app}
F_2( \delta, \rho \, ; x ) 
\,\equiv\,
\left\{
\begin{array}{ll}
\Delta_1( x) + \Delta_2(x)
 & 
\delta \in \big[\,\delta_{\textrm{c}}^{\,\textrm{\tiny I/II}} , +\infty\,\big)
\\
\rule{0pt}{.6cm}
\Delta_1( x) +  \bar{\Delta}_2( 2a_2-\rho/2 ; x)
 & 
\delta \in \big[\,\delta_{\textrm{c}}^{\,\textrm{\tiny II/III}} , \delta_{\textrm{c}}^{\,\textrm{\tiny I/II}} \,\big]
\\
\rule{0pt}{.6cm}
\Delta_1( x) + \tilde{\Delta}_2( 2a_2-\rho/2 ,  2a_2 ; x)
 & 
\delta \in \big[\,\delta_{\textrm{c}}^{\,\textrm{\tiny III/IV}} , \delta_{\textrm{c}}^{\,\textrm{\tiny II/III}} \,\big]
\\
\rule{0pt}{.6cm}
\lambda_1(-a_2/2 ;x) + \tilde{\lambda}_2( 3a_2/2 ,\rho-a_2 , 2a_2 ; x )
 \hspace{.8cm}
 & 
\delta \in \big[\,0\,, \delta_{\textrm{c}}^{\,\textrm{\tiny III/IV}} \,\big]
\end{array}
\right.
\ee
while for $\rho  \in \big[\rho_{\textrm{c}}\,, 1/2 \big]$ 
the function $F_2(\delta,\rho; x)$ is
\be
\label{tilde-F2-def-different-2int-above-rhoc-app}
F_2( \delta, \rho \, ; x ) 
\,\equiv\,
\left\{
\begin{array}{ll}
\Delta_1( x) + \Delta_2(x)
 & 
\delta \in \big[\,\delta_{\textrm{c}}^{\,\textrm{\tiny I/II}} , +\infty\,\big)
\\
\rule{0pt}{.6cm}
\Delta_1( x) +  \bar{\Delta}_2( 2a_2-\rho/2  ; x)
 & 
\delta \in \big[\,\delta_{\textrm{c}}^{\,\textrm{\tiny II/III}} , \delta_{\textrm{c}}^{\,\textrm{\tiny I/II}} \,\big]
\\
\rule{0pt}{.6cm}
\lambda_1( -a_2/2 ;x) + \bar{\lambda}_2(3a_2/2 ,\rho-a_2  ; x )
 & 
\delta \in \big[\,\delta_{\textrm{c}}^{\,\textrm{\tiny III/IV}} , \delta_{\textrm{c}}^{\,\textrm{\tiny II/III}} \,\big]
\\
\rule{0pt}{.6cm}
\lambda_1( -a_2/2 ;x) + \tilde{\lambda}_2( 3a_2/2 ,\rho-a_2 , 2a_2 ; x )
 \hspace{.8cm}
 & 
\delta \in \big[\,0\,, \delta_{\textrm{c}}^{\,\textrm{\tiny III/IV}} \,\big]  \,. 
\end{array}
\right.
\ee
This function satisfies the consistency conditions (\ref{adj-cond-diverse2int}) and (\ref{distant-cond-diverse2int})
in the limiting regimes of vanishing and large separation distance, as already mentioned in the main text
just below these requirements.

\section{Entanglement Hamiltonian in the infinite fermionic chain}
\label{app-fermions}

In order to make a comparison with the bosonic lattice models discussed in the main text, 
in this Appendix we consider the entanglement Hamiltonian of two disjoint blocks in the infinite fermionic hopping chain
and review the formulas underlying the numerical analysis of its continuum limit \cite{Eisler:2022rnp},
which provides the entanglement Hamiltonian of two disjoint intervals 
for the massless Dirac field on the line and in its ground state, found in \cite{Casini:2009vk}.


The free massless Dirac field is a prototypical example of two-dimensional CFT with central charge $c=1$.
This fermionic field is a doublet made by two chiral complex fields,
in terms of  are the light-cone coordinates $u_\pm \equiv x \pm t$. 
In the case where this model is defined on the line and in its ground state, 
the entanglement Hamiltonian of the subsystem $A$ given by the union of two disjoint intervals
(i.e. $A \equiv A_1 \cup A_2$, where $A_j \equiv [a_j, b_j]$ with $j \in \{1,2\}$)
is the sum of a local term and a bilocal term
 \cite{Casini:2009vk}
\be
\label{mod-ham-2int-KA}
K_A  = K_{A, \textrm{\tiny loc}} + K_{A, \textrm{\tiny biloc}}  \,. 
\ee
The local term is
\be
\label{K_A-2int-terms-local}
K_{A, \textrm{\tiny loc}}
=
\int_A \beta_{\textrm{\tiny loc}}(x) \, T_{tt}(x)\, \rd x 
\ee
where the weight function $\beta_{\textrm{\tiny loc}}(x) $  has been introduced in (\ref{beta-loc-2int}) 
and $T_{tt}(x)$ is the energy density  of the massless Dirac field, namely
\be
\label{T00-lambda-def}
T_{tt}(x)  
\equiv 
\,\frac{\textrm{i}}{2}
\left[
\Big (\! 
:\!  \psi^\ast_\textrm{\tiny R}\, (\partial_x \psi_\textrm{\tiny R} ) \! : 
-
:\! (\partial_x \psi^\ast_\textrm{\tiny R})\, \psi_\textrm{\tiny R} \! :
\! \Big) (x)
-
\Big (\! 
:\!  \psi^\ast_\textrm{\tiny L}\, (\partial_x \psi_\textrm{\tiny L}) \! : 
-
:\! (\partial_x \psi^\ast_\textrm{\tiny L})\, \psi_\textrm{\tiny L} \! :
\! \Big) (x)
\right]
\ee
where $\psi_\textrm{\tiny R}$ and $\psi_\textrm{\tiny L}$ 
are the right and left chiral components of the massless Dirac field.
The bilocal term is
\be
\label{beta-bi-loc-2int}
K_{A, \textrm{\tiny biloc}}
=
\int_A 
\beta_{\textrm{\tiny biloc}}(x) \, T_{\textrm{\tiny biloc}}(x, x_{\textrm{\tiny c}} ) \, \rd x 
\ee
whose weight function can be written through (\ref{beta-loc-2int}) and the conjugate point (\ref{x-conj-def}) as follows
\be
\label{velocity_fund-2int}
\beta_{\textrm{\tiny biloc}}(x) =
\frac{\beta_{\textrm{\tiny loc}} (x_{\textrm{c}} )}{ x - x_{\textrm{c}} } 
\ee
and the bilocal quadratic operator $T_{\textrm{\tiny biloc}}(x, y) $ is defined as 
\be
\label{T-bilocal-fermions-app}
T_{\textrm{\tiny biloc}}(x, y)  \equiv 
\frac{\textrm{i}}{2}
\left[\,
:\!\! \Big( \psi^\ast_\textrm{\tiny R} (x) \,  \psi_\textrm{\tiny R} (y) - \psi^\ast_\textrm{\tiny R} (y) \,  \psi_\textrm{\tiny R} (x)  \Big) \!\!: 
-
:\!\! \Big( \psi^\ast_\textrm{\tiny L} (x) \,  \psi_\textrm{\tiny L} (y) - \psi^\ast_\textrm{\tiny L} (y) \,  \psi_\textrm{\tiny L} (x)  \Big) \!\!: 
\right]  . 
\ee


\begin{figure}[t!]
\vspace{-.5cm}
\hspace{-0.2cm}
\includegraphics[width=1.058\textwidth]{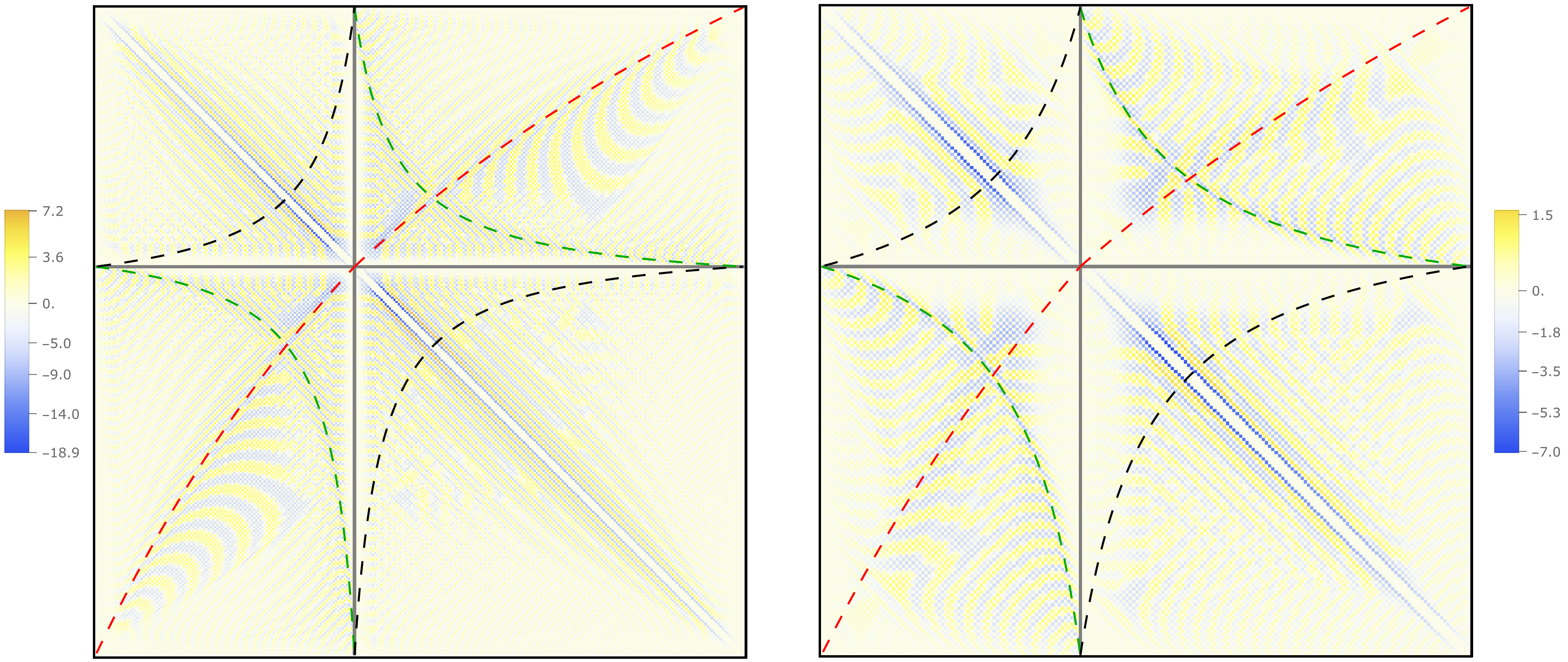}
\vspace{-.2cm}
\caption{ Matrix $H$ for $L_2 = 1.5L_1$ with $L_1=120$ and $\delta = 1/10$ (left) or $\delta = 1/4$ (right).
}
\label{fig:2int-fermion-density}
\end{figure}

\begin{figure}[t!]
	\vspace{0.5cm}
	\hspace{-.45cm}
	\includegraphics[width=1.07\textwidth]{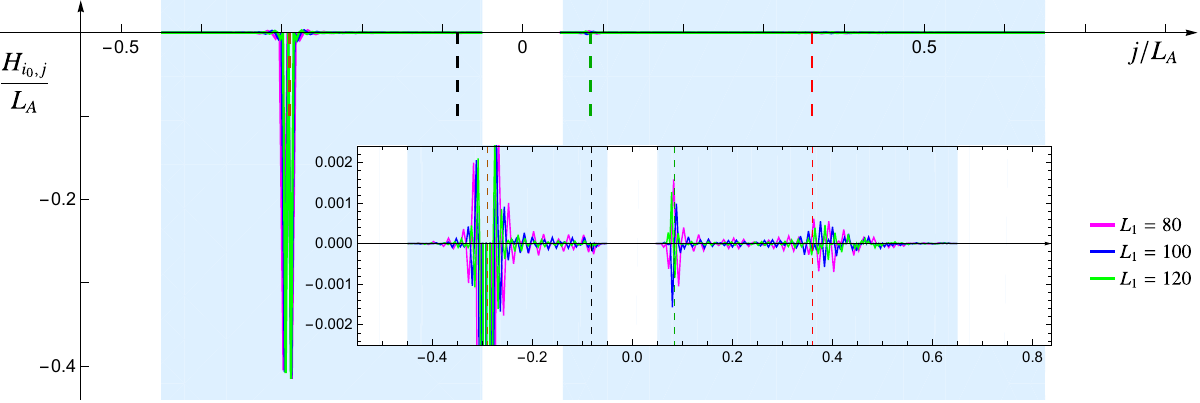}
	\vspace{-.2cm}
	\caption{Matrix elements of  $H$ along a row corresponding to the site labelled by $i=i_0$.
	 Here $L_2=1.5L_1$ and $i_0= 0.4  L_1$, and $\delta=1/10$.
	 The green, red, black and brown vertical dashed segments highlight  the intersections of the row with the curves corresponding to 
	 \eqref{x-conj-def}, \eqref{x-gamma-def}, \eqref{x-chi-def} and the main diagonal respectively
	 (see the dashed lines in Fig.\,\ref{fig:2int-fermion-density}). 
	 The inset zooms in to highlight the oscillatory behaviour and the relative amplitudes around these intersections.
	}
\label{fig:2int-fermion-row}
\end{figure}

The operator (\ref{mod-ham-2int-KA}) can be obtained from a lattice model  through a continuum limit.
\\
Consider the infinite fermionic hopping chain characterised by the following  Hamiltonian
\begin{equation}
\label{hamiltonian-free-fermion}
    \widehat{H}=- \frac{1}{2} 
    \sum_{i \,\in\, \mathbb{Z}} \big(c_i^{\dagger} c_{i+1}+c_{i+1}^{\dagger} c_i\big)
    +
    \mu \sum_{i \,\in\, \mathbb{Z}}  c_i^{\dagger} c_i
\end{equation}
in terms of the fermionic creation and annihilation operators 
and of the chemical potential $\mu=\cos q_{\textrm{\tiny F}}$,
whose ground state is a Fermi sea with occupied momenta $q \in [ -q_{\textrm{\tiny F}} , q_{\textrm{\tiny F}}]$.
The generic element of the correlation matrix is
\begin{equation}
  C_{i,j} \equiv \langle c_i^{\dagger} c_j\rangle =\frac{\sin \!\big[q_{\textrm{\tiny F}}(i-j)\big] }{ \pi(i-j) }
\end{equation}
and its restriction to the subsystem $A$ defines the reduced correlation matrix $C_A$, 
which is the crucial quantity to explore in order to study the bipartite entanglement in this setup.

We are interested in the free fermionic chain (\ref{hamiltonian-free-fermion}) in its ground state
when the bipartition of the infinite chain is given by the union of two disjoint blocks $A= A_1 \cup A_2$, 
made by $L_1$ and  $L_2 $ contiguous sites.
The entanglement Hamiltonian of $A$ reads \cite{Peschel:2002yqj, Eisler:2009vye}
\begin{equation}
    \widehat{K}_A=\sum_{i,j \in A} H_{i,j} \,c_i^{\dagger} c_j  
\end{equation}
where the $L_A \times L_A$ matrix $H$ can be written as 
\begin{equation}
\label{H-mat-decomposition}
H_{i, j}=\sum_{k=1}^{L_A} \phi_k(i) \,\varepsilon_k \,\phi_k(j)
\;\;\;\qquad \;\;\;
\varepsilon_k=\log\! \left( \frac{1-\zeta_k}{\zeta_k} \right)
\end{equation}
where $\varepsilon_k$ are the single-particle entanglement energies,
while $\zeta_k$  and $\phi_k(i)$
are  the eigenvalues $\zeta_k$ and the eigenvectors of $C_A$, 
which can be diagonalised numerically.

In Fig.\,\ref{fig:2int-fermion-density} we show the matrix $H_{i,j}$ in (\ref{H-mat-decomposition})
for $L_2=1.5L_1$ and $q_{\textrm{\tiny F}} = \pi/2$.
The green, red and black dashed curves, 
corresponding to \eqref{x-conj-def}, \eqref{x-gamma-def} \eqref{x-chi-def} respectively, 
are shown to facilitate the comparison with Fig.\,\ref{fig:EH-density}, Fig.\,\ref{density-plots-M-chiral-current} and Fig.\,\ref{3dplots-MatrixTVMabs}
for the bosonic models explored in this manuscript. 
While the green and the red dashed hyperbolae can be observed in the profile of the matrix elements of $H$,
as already highlighted in \cite{Eisler:2022rnp},
the black dashed hyperbola does not occur, in contrast with the case of the massless harmonic chain
(see Fig.\,\ref{fig:EH-density}).
This observation is further supported by Fig.\,\ref{fig:2int-fermion-row}, 
where the elements of a single row $H_{i_0, j}$ are shown and 
the magnitude does not increase around the vertical dashed black segment,
in contrast with the case of the massless harmonic chain
(see Fig.\,\ref{fig:all_elements}).

\begin{figure}[t!]
	\hspace{-1.cm}
	\begin{minipage}{0.5\textwidth}
		\centering
		\includegraphics[width=1.0\textwidth]{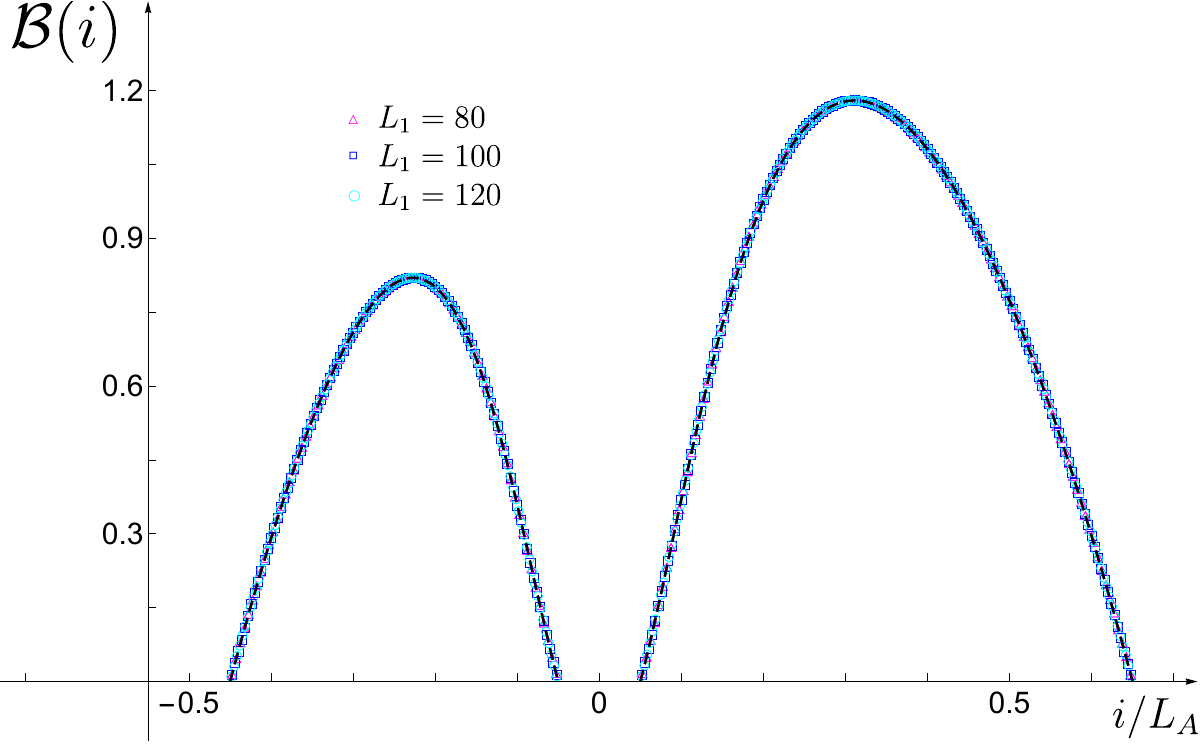}
	\end{minipage}
	\hspace{1cm}
	\begin{minipage}{0.5\textwidth}
		\centering
		\includegraphics[width=1.0\linewidth]{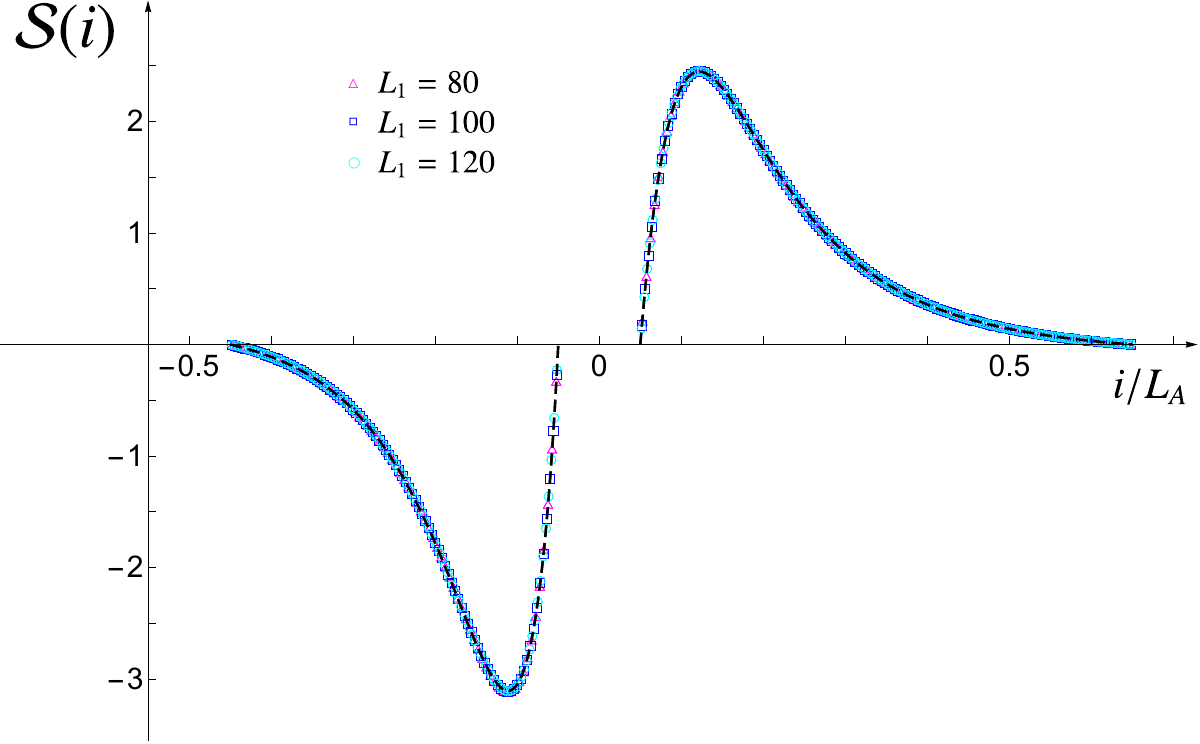}
	\end{minipage}
	\vspace{.2cm}
	\caption{The combinations (\ref{eq:fermion-EH-diag-lattice}) (left) and (\ref{eq:fermion-EH-offdiag-lattice}) (right)
	for $L_2=1.5 L_1$ and $\delta=1/10$, for various values of $L_A$. 
	The dashed curves are given by the weight functions (\ref{beta-loc-2int}) and (\ref{velocity_fund-2int}) respectively.
	}	
	\label{fig:2int-fermion-beta}
\end{figure}

The continuum limit procedure requires us to introduce the continuum spatial coordinate $x = i\, s$ in terms of the infinitesimal lattice spacing $s$
and to perform the following replacement
\begin{equation}
\label{eq:fermionic_continuum_limit}
    c_i \rightarrow \sqrt{s}\left(\mathrm{e}^{\mathrm{i} q_{\mathrm{F}} x} \psi_\textrm{\tiny R}(x) + \mathrm{e}^{-\mathrm{i} q_{\mathrm{F}} x} \psi_\textrm{\tiny L}(x)\right)
\end{equation}
where $\psi_\textrm{\tiny R}$ and $\psi_\textrm{\tiny L}$ denote the right-moving and the left-moving chiral fields respectively.

Considering the half-filling case, where $ q_{\mathrm{F}}  s= \pi/2$,
in \cite{Eisler:2022rnp} it has been shown that proper combinations of the matrix elements of $H$ 
provide the weight functions (\ref{beta-loc-2int}) and (\ref{velocity_fund-2int}) in the continuum limit. 
In particular, for the diagonal blocks, one considers the following combination of the matrix elements on the $i$-th row
\be
\label{eq:fermion-EH-diag-lattice}
 \mathcal{B}(i) \equiv \,
  - \!\sum_{j\in A_k} (-1)^{(j-i)/2}(j-i) \,H_{i,j}
  \;\;\;\;\qquad\;\;\;
  i \in A_k
  \qquad
  k \in \{1,2\}
\ee
which is slightly different from the combination defined in Eq.\,(31) of \cite{Eisler:2022rnp}, 
where a summation over the diagonals occurs. 
In the off-diagonal blocks, the combination of the matrix elements on the $i$-th row introduced in \cite{Eisler:2022rnp} reads
\be
\label{eq:fermion-EH-offdiag-lattice}
    \mathcal{S}(i)\equiv \!
    \sum_{j\in A_r } \! (-1)^{(j-i-1)/2} \,H_{i,j}
      \;\;\;\;\qquad\;\;\;
  i \in A_k
    \qquad
    r\neq k
  \qquad
  r,k \in \{1,2\}  \,. 
\ee

In Fig.\,\ref{fig:2int-fermion-beta}, we show that 
(\ref{eq:fermion-EH-diag-lattice}) and (\ref{eq:fermion-EH-offdiag-lattice}) 
nicely reproduce the weight functions 
(\ref{beta-loc-2int}) and (\ref{velocity_fund-2int}) respectively in the scaling limit,
as already found in \cite{Eisler:2022rnp}.

\section{Massless regime: Further details}
\label{app-massless}


In this appendix we report further numerical results supporting  the observations
made in  Sec.\,\ref{sec-2int-massless}
for the entanglement Hamiltonian of two disjoint blocks in the massless regime.

\begin{center}
\begin{figure}[t!]
	\vspace{-.4cm}
	\hspace{-1.cm}
	\begin{minipage}{0.5\textwidth}
		\includegraphics[width=1.02\textwidth]{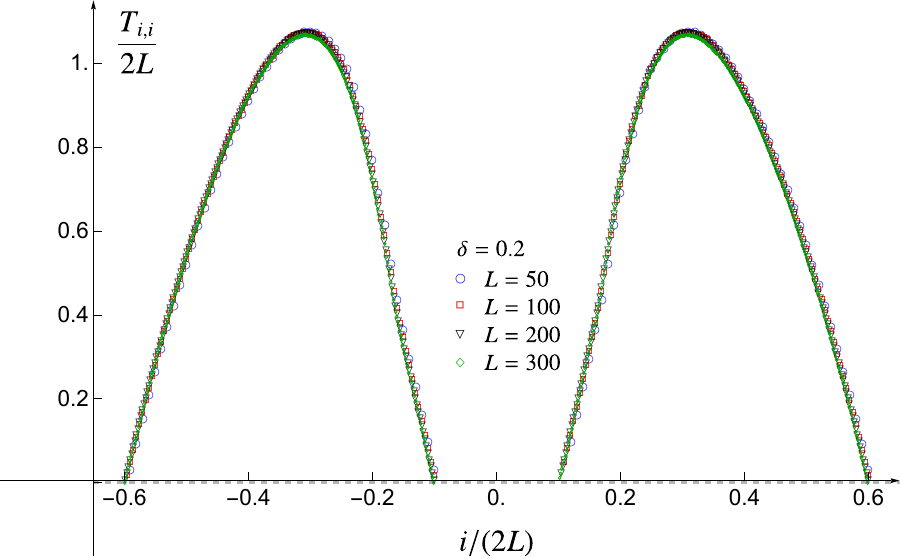}
	\end{minipage}
	\hspace{1cm}
	\begin{minipage}{0.5\textwidth}
		\includegraphics[width=1.02\textwidth]{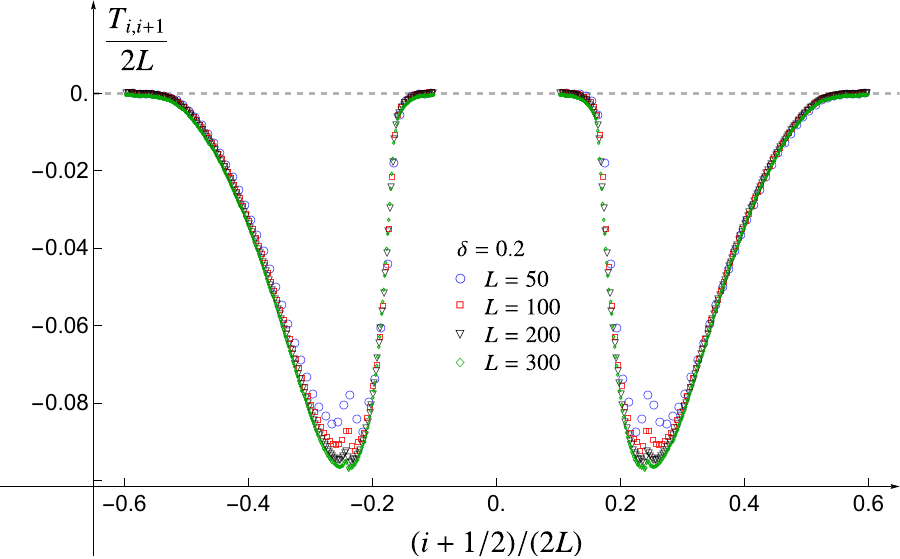}
	\end{minipage}
	\vspace{.5cm}
	\\
	\vspace{.5cm}
	\hspace{-1.08cm}
	\begin{minipage}{0.5\textwidth}
		\includegraphics[width=1.02\textwidth]{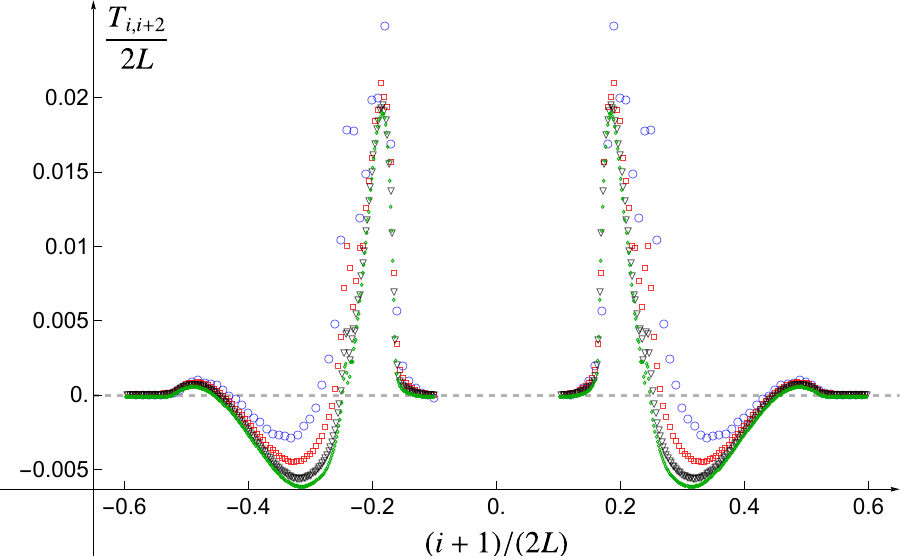}
	\end{minipage}
	\hspace{1cm}
	\begin{minipage}{0.5\textwidth}
		\includegraphics[width=1.02\textwidth]{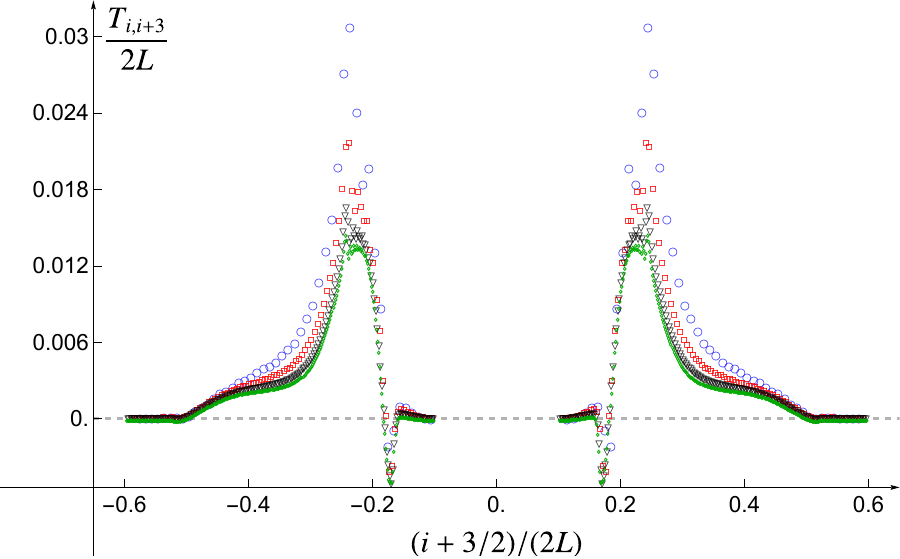}
	\end{minipage}
	\\
	\vspace{-.0cm}
	\hspace{-1.08cm}
	\begin{minipage}{0.5\textwidth}
		\includegraphics[width=1.02\textwidth]{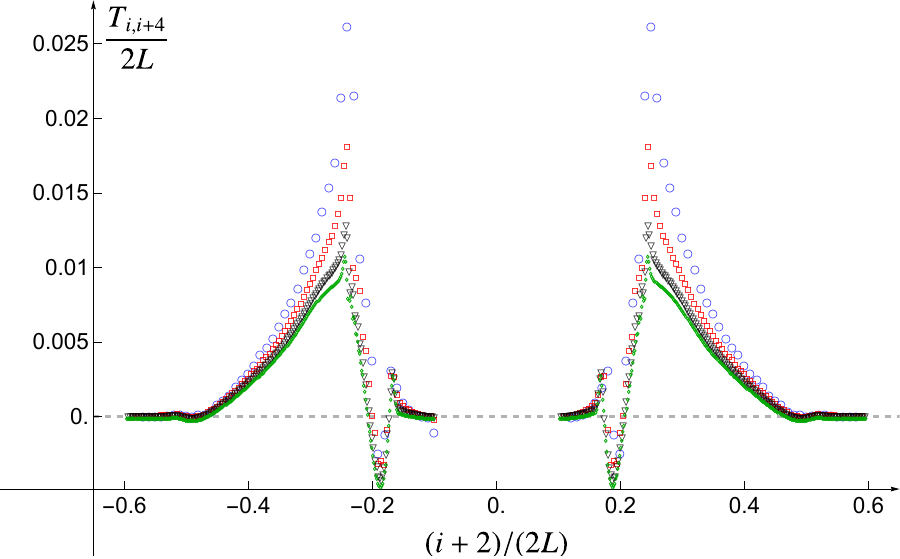}
	\end{minipage}
	\hspace{1cm}
	\begin{minipage}{0.5\textwidth}
		\includegraphics[width=1.02\textwidth]{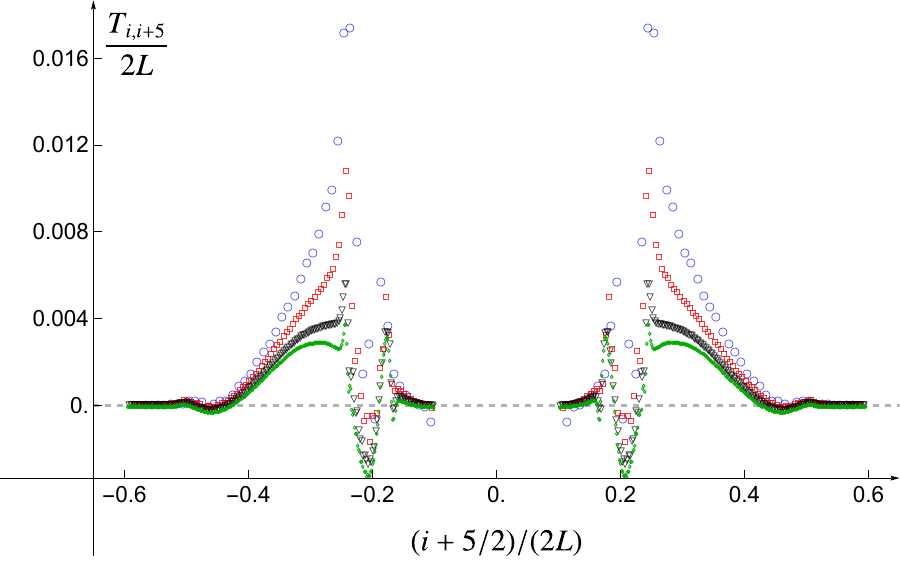}
	\end{minipage}
	\vspace{.5cm}
	\\
	\vspace{.5cm}
	\hspace{-1.08cm}
	\begin{minipage}{0.5\textwidth}
		\includegraphics[width=1.02\textwidth]{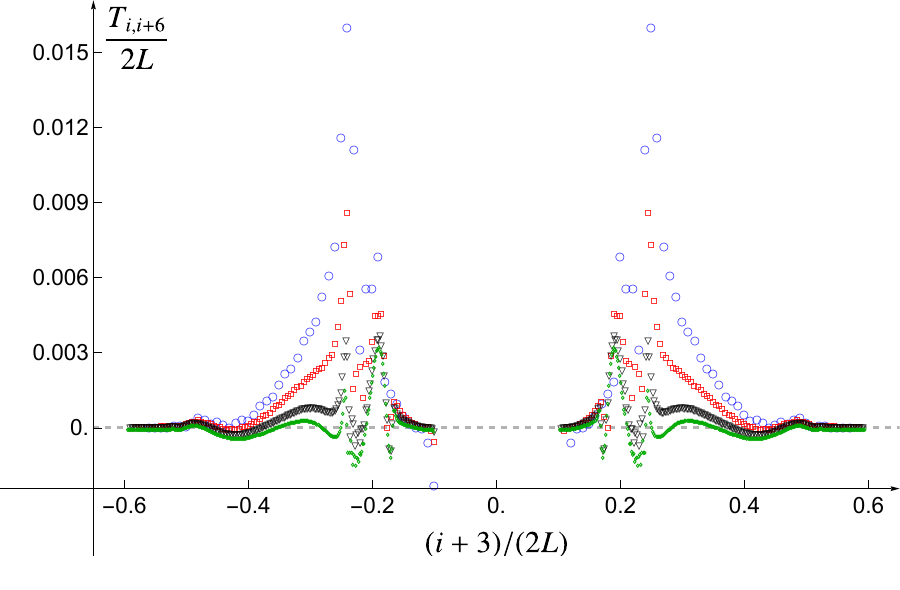}
	\end{minipage}
	\hspace{1cm}
	\begin{minipage}{0.5\textwidth}
		\includegraphics[width=1.02\textwidth]{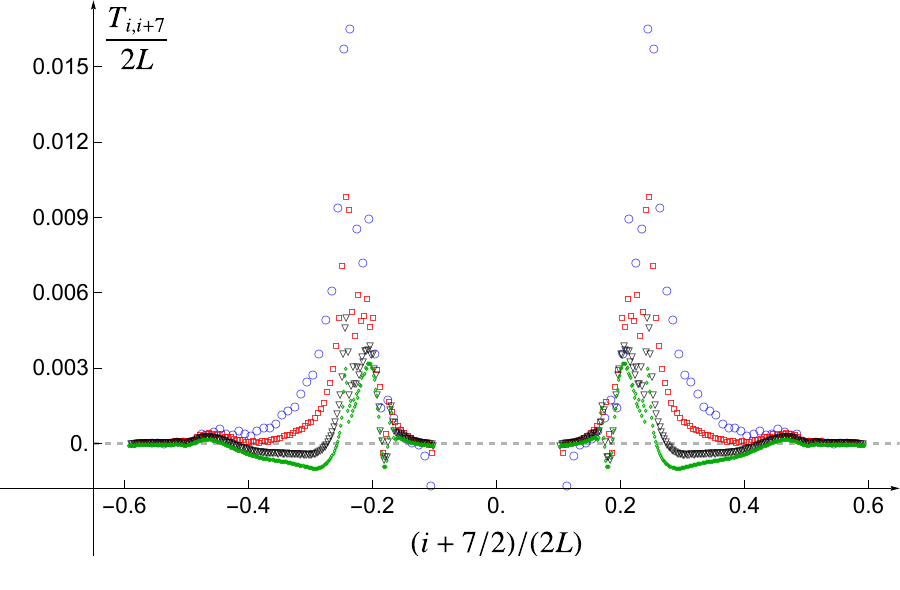}
	\end{minipage}
	\vspace{-.3cm}
	\caption{Diagonals of the matrix $T$ for $L_1 = L_2$ and $\omega L_A=10^{-50}$.}
	\label{Tdiagonals}
\end{figure}
\end{center}
\begin{figure}[t!]
	\vspace{-.4cm}
	\hspace{-1.cm}
	\begin{minipage}{0.5\textwidth}
		\includegraphics[width=1.02\textwidth]{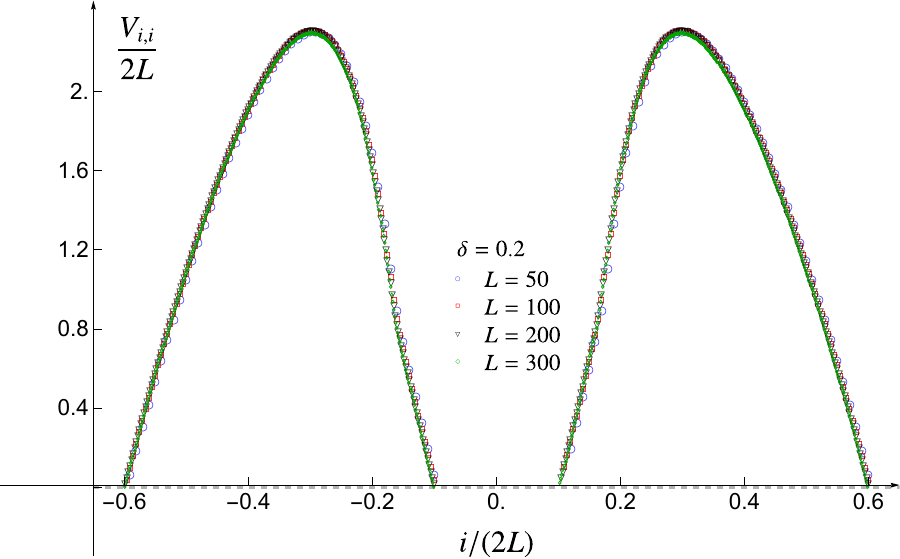}
	\end{minipage}
	\hspace{1cm}
	\begin{minipage}{0.5\textwidth}
		\includegraphics[width=1.02\textwidth]{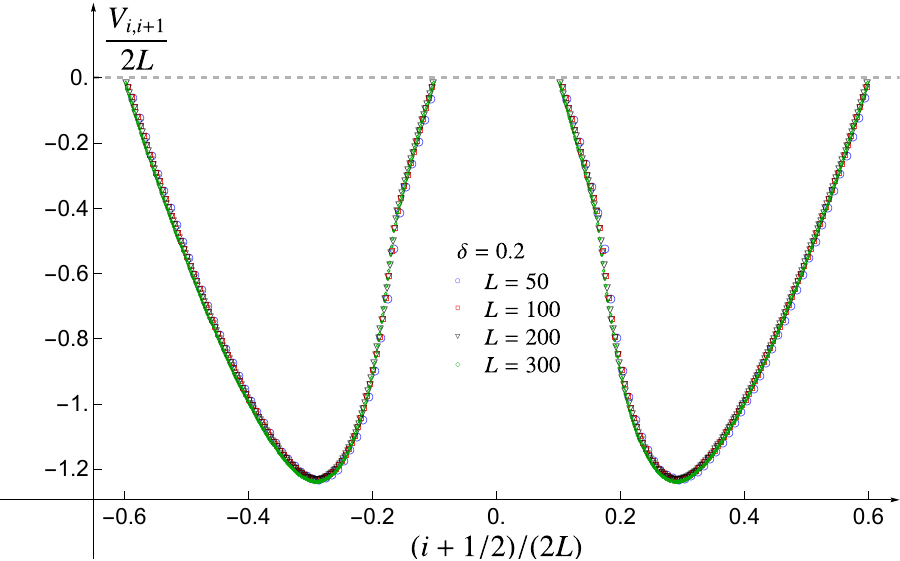}
	\end{minipage}
	\vspace{.5cm}
	\\
	\vspace{.5cm}
	\hspace{-1.08cm}
	\begin{minipage}{0.5\textwidth}
		\includegraphics[width=1.02\textwidth]{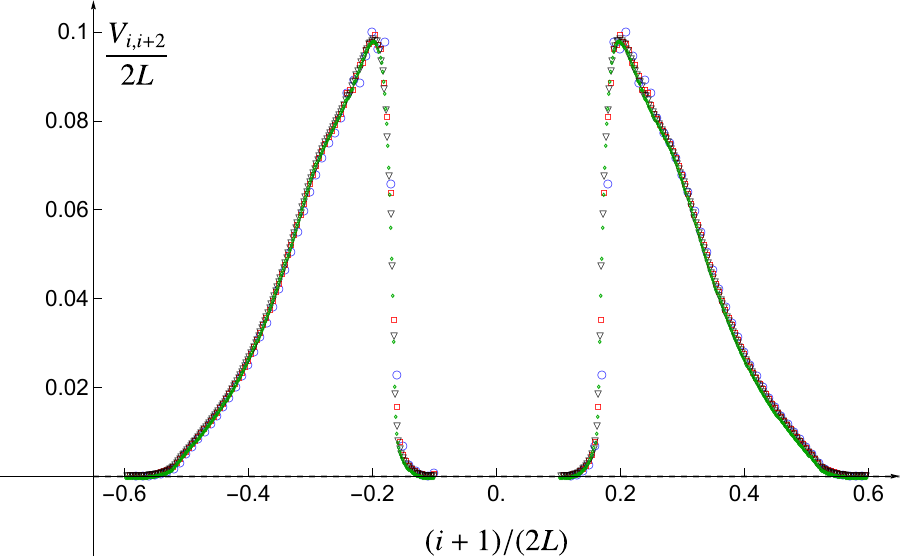}
	\end{minipage}
	\hspace{1cm}
	\begin{minipage}{0.5\textwidth}
		\includegraphics[width=1.02\textwidth]{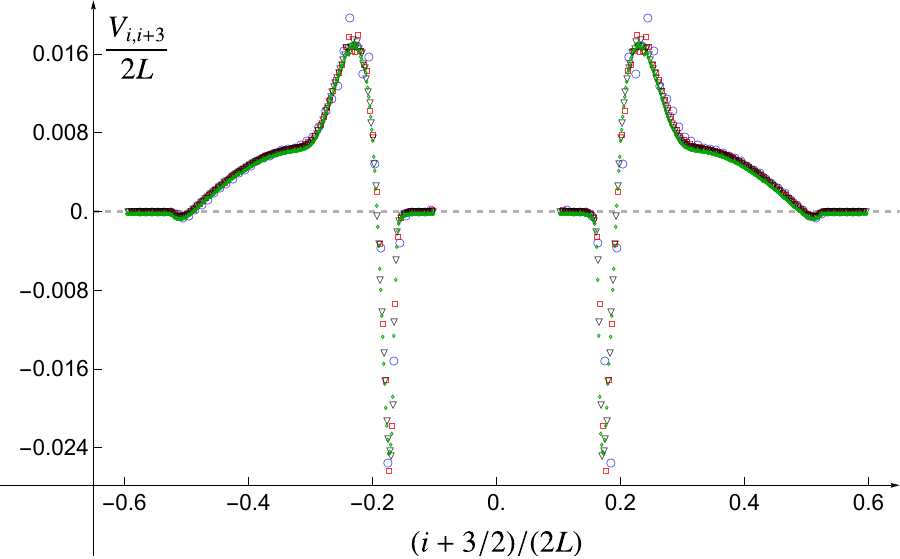}
	\end{minipage}
	\\
	\vspace{-.0cm}
	\hspace{-1.08cm}
	\begin{minipage}{0.5\textwidth}
		\includegraphics[width=1.02\textwidth]{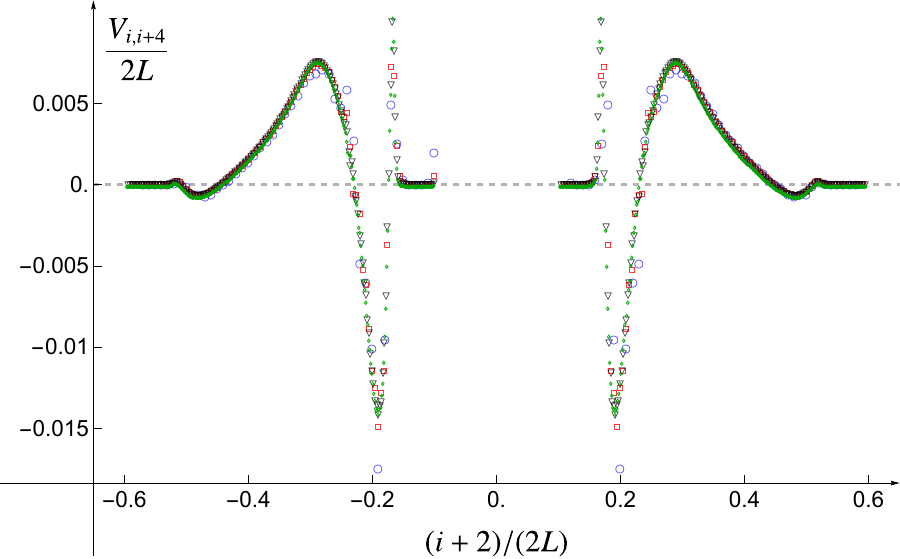}
	\end{minipage}
	\hspace{1cm}
	\begin{minipage}{0.5\textwidth}
		\includegraphics[width=1.02\textwidth]{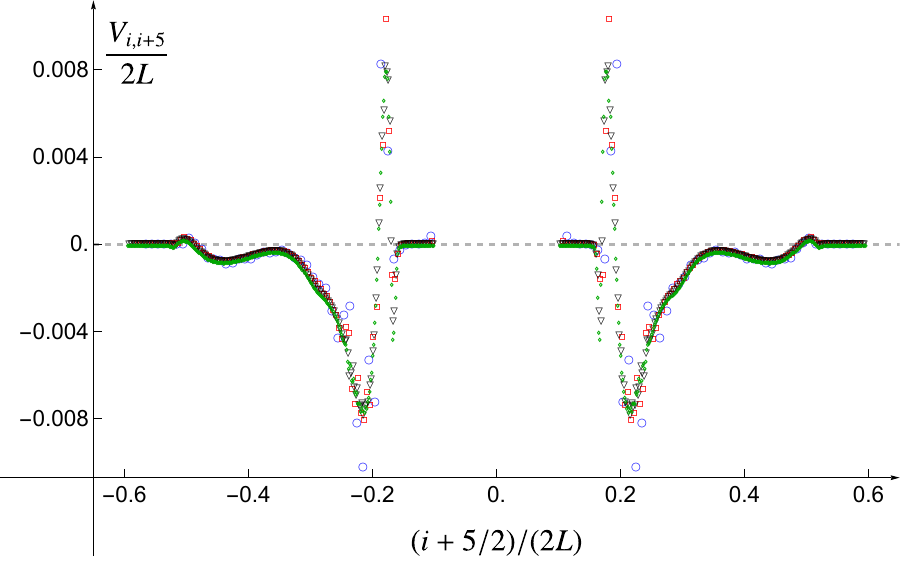}
	\end{minipage}
	\vspace{.5cm}
	\\
	\vspace{.5cm}
	\hspace{-1.08cm}
	\begin{minipage}{0.5\textwidth}
		\includegraphics[width=1.02\textwidth]{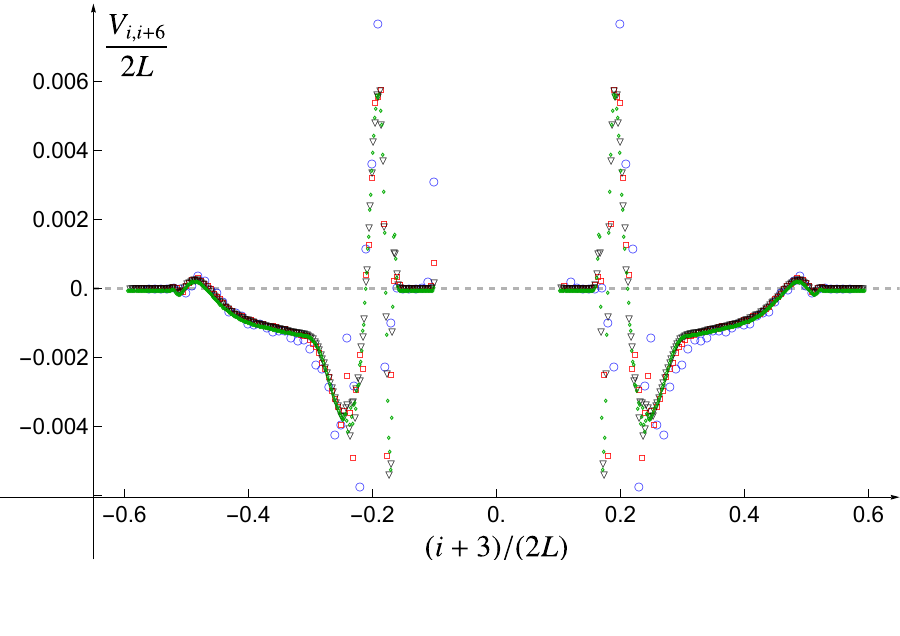}
	\end{minipage}
	\hspace{1cm}
	\begin{minipage}{0.5\textwidth}
		\includegraphics[width=1.02\textwidth]{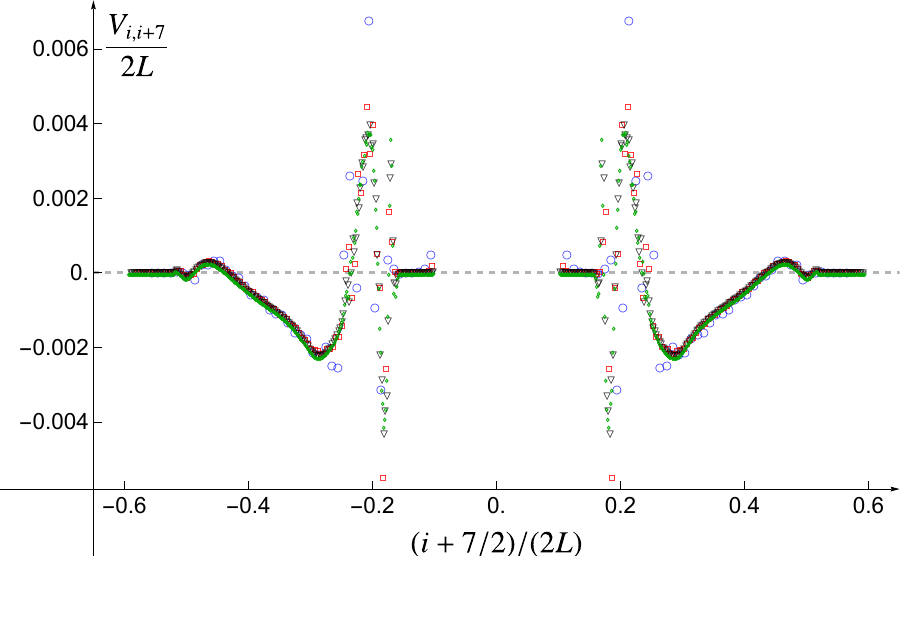}
	\end{minipage}
	\vspace{-.6cm}
	\caption{Diagonals of the matrix $V$ for $L_1 = L_2$ and $\omega L_A=10^{-50}$.}
	\label{Vdiagonals}
\end{figure}

Considering the simplest configuration of equal intervals with $L_1 = L_2 \equiv L$ in the massless regime with $\omega L_A=10^{-50}$
and choosing $\delta=0.2$,
in Fig.\,\ref{Tdiagonals} and Fig.\,\ref{Vdiagonals} we report some numerical data 
showing that the functions $\tau_k(x_k)$ and $\nu_k(x_k)$,
introduced (\ref{diagonal-collapse-assumption})  and which involve only the elements in the diagonal blocks of $T$ and $V$,
are well defined. 
In particular, eight diagonals have been explored, i.e. $0 \leqslant k \leqslant 7$ in (\ref{diagonal-collapse-assumption}).
An example with  $L_2 = 2L_1$ has been considered in Fig.\,\ref{MainDiagonals-lratio2},
only for the diagonals providing the largest contributions. 
The numerical results displayed in Fig.\,\ref{Tdiagonals} and Fig.\,\ref{Vdiagonals} 
suggest that the functions $\tau_k(x_k)$ and $\nu_k(x_k)$ are continuous but not smooth.
This lack of smoothness has been also checked by exploring  
the discrete first derivative of $\tau_k(x_k)$ and $\nu_k(x_k)$
from the numerical data points, as done the right panels of Fig.\,\ref{MainDiagonals-lratio2},
but we do not find it worth reporting these plots here.
%

\begin{figure}[t!]
	\vspace{-.5cm}
	\hspace{0.cm}
	\includegraphics[width=1\textwidth]{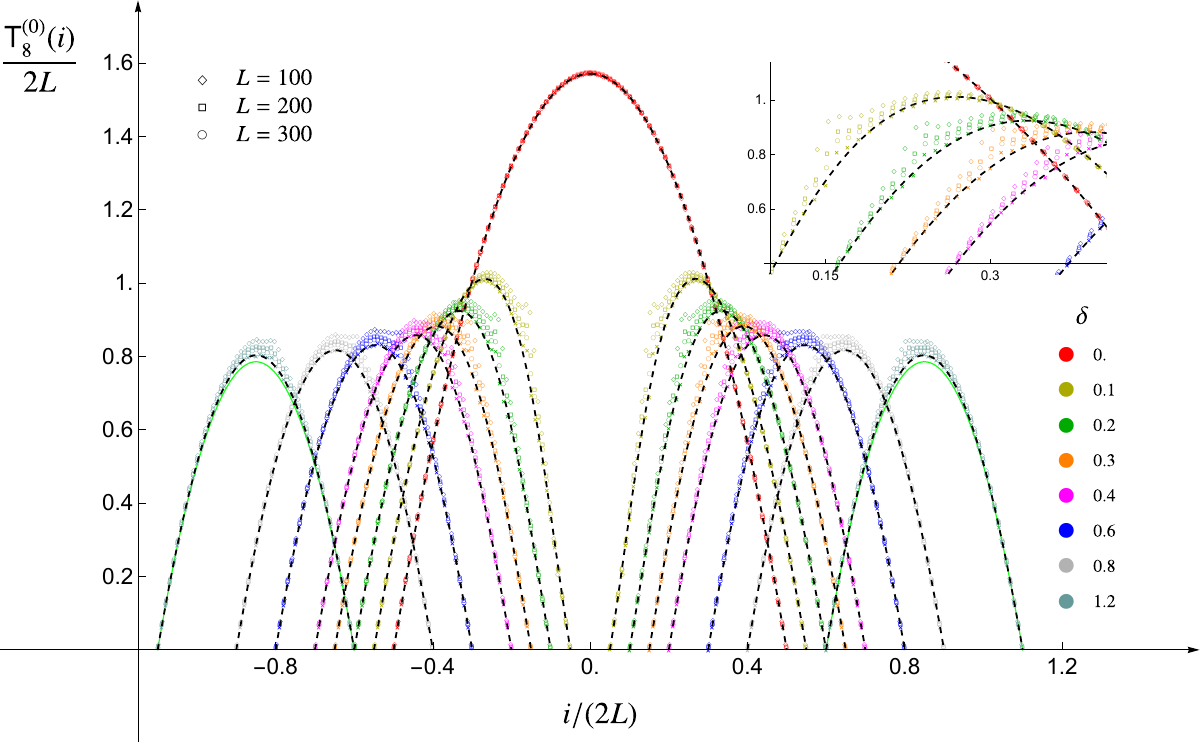}
	\\
	\rule{0pt}{10.2cm}
	\hspace{-.339cm}
	\includegraphics[width=1\textwidth]{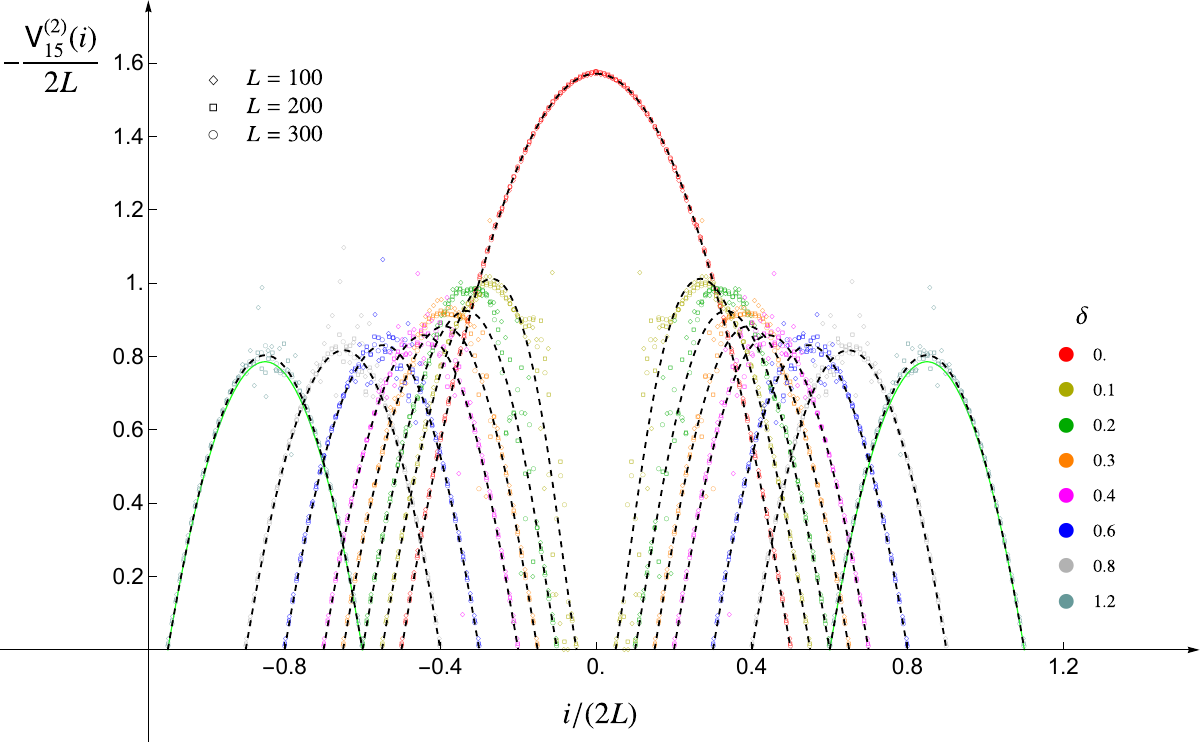}
	\vspace{-.2cm}
	\caption{The combinations of diagonals defined 
	in the first expression of (\ref{T-V-summation-0}) (top) and in (\ref{V-summation-2}) (bottom) 
	for $k_\textrm{\tiny max}  \ll L_A$ (see also (\ref{T0-V0-large-L})), when $L_2 = L_1$
	(see also Fig.\,\ref{fig:EHmassless_lratio2}).
	The dashed black curve corresponds to (\ref{beta-loc-2int})
	and the solid green curve to its limit $d \to +\infty$.
	}
	\label{fig:EHmassless_lratio1}
\end{figure}
\begin{figure}[t!]
	\vspace{-.5cm}
	\hspace{0.cm}
	\includegraphics[width=1\textwidth]{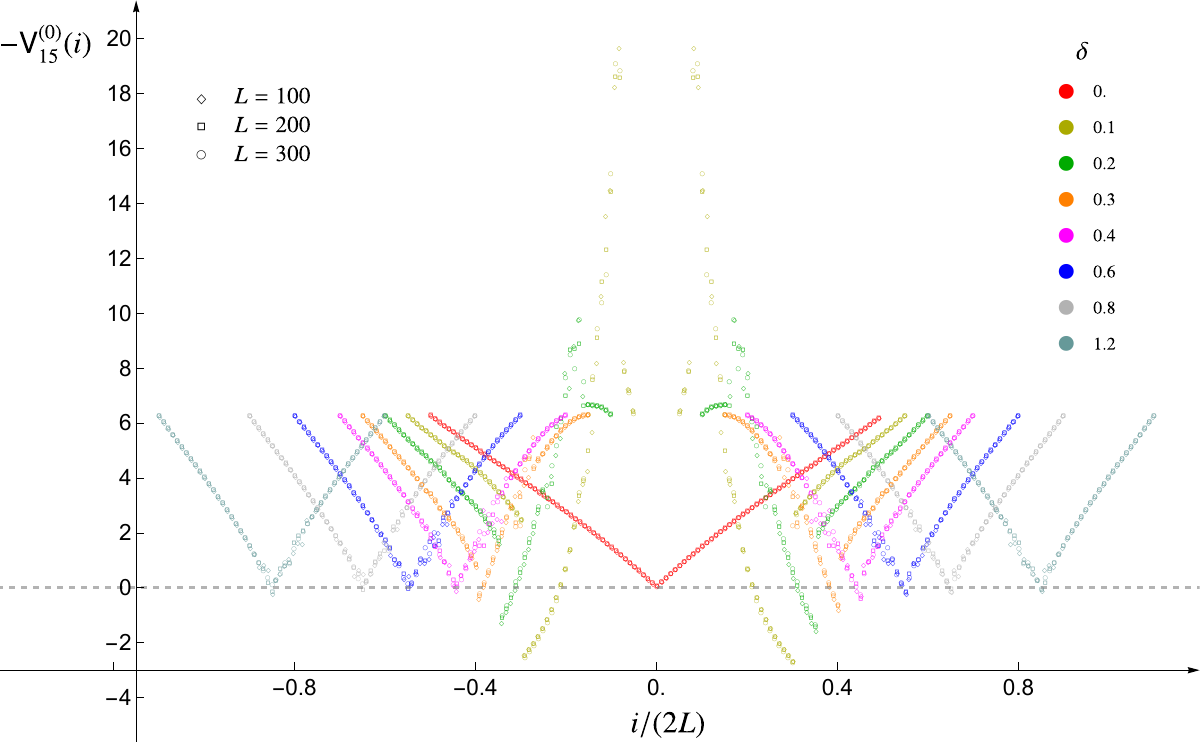}
	\vspace{-.2cm}
	\caption{The combination of diagonals defined in the second expression of (\ref{T-V-summation-0}) 
	for $k_\textrm{\tiny max}  \ll L_A$ (see also (\ref{T0-V0-large-L})), when $L_2 = L_1$ (see also Fig.\,\ref{fig:EHmassless_V0term_lratio2}).
	}
	\label{fig:EHmassless_V0term_lratio1}
\end{figure}

Considering configurations of equal intervals
and in the massless regime with $\omega L_A=10^{-50}$,
in Fig.\,\ref{fig:EHmassless_lratio1} and Fig.\,\ref{fig:EHmassless_V0term_lratio1}
we report some numerical results to test the functions 
introduced in the l.h.s.'s of (\ref{T0-V0-large-L-large-k}),
as done in Fig.\,\ref{fig:EHmassless_lratio2} and Fig.\,\ref{fig:EHmassless_V0term_lratio2}
for $L_2 = 2L_1$;
hence we refer the reader to the main text for the discussion of the quantities involved in this analysis. 
Comparing the results displayed in these four figures, 
one observes that the convergence of the data points is slightly better when 
the subsystem $A$ is made by equal intervals.

\begin{figure}[t!]
	\vspace{-.5cm}
	\hspace{0.cm}
	\includegraphics[width=1\textwidth]{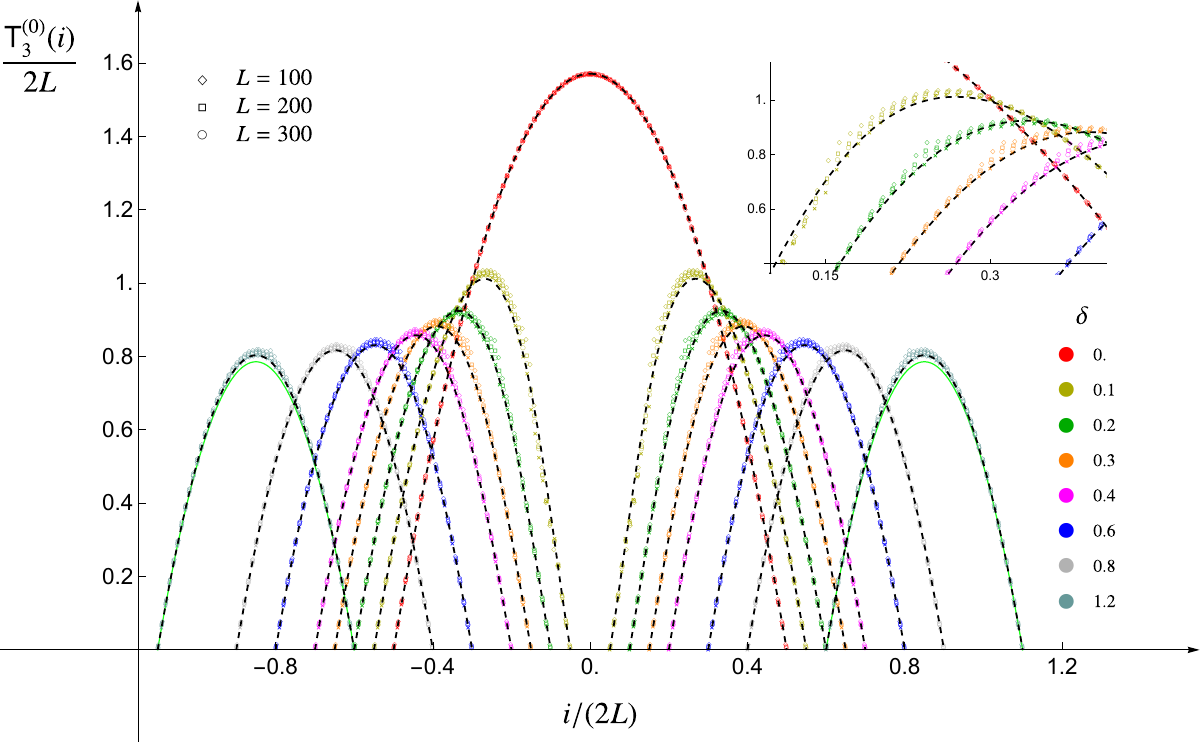}
	\\
	\rule{0pt}{10.2cm}
	\hspace{-.3cm}
	\includegraphics[width=1\textwidth]{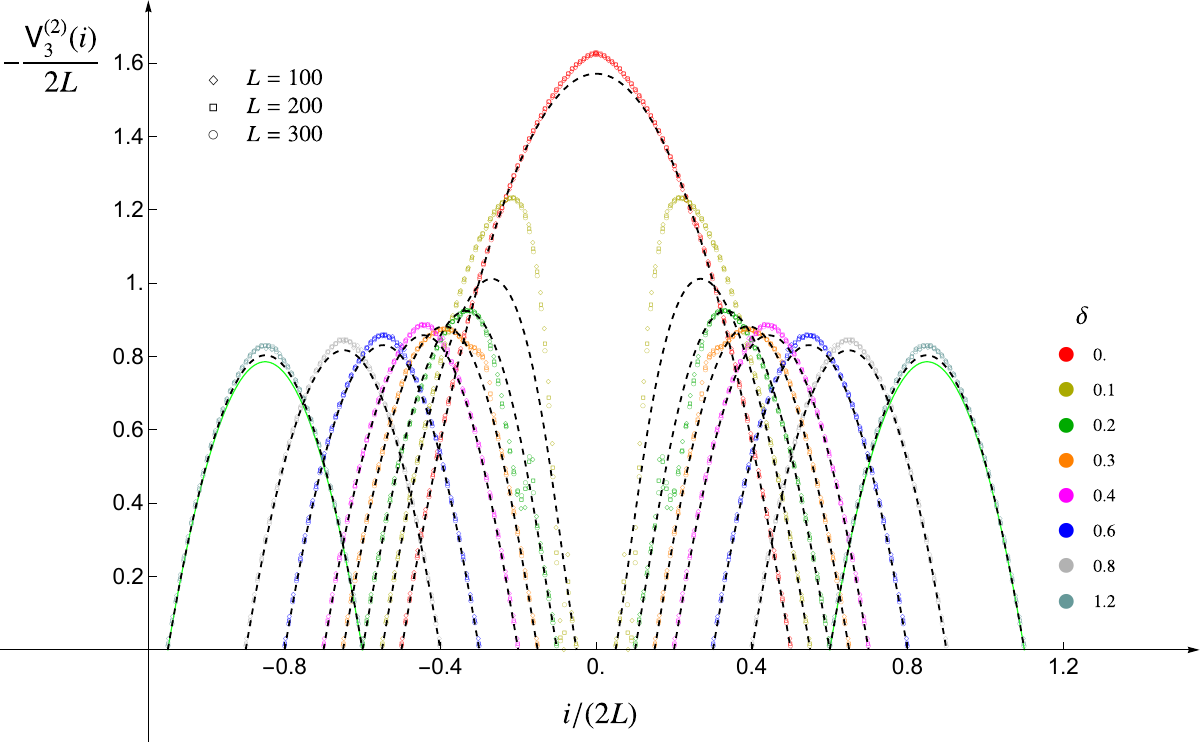}
	\vspace{-.2cm}
	\caption{The combinations of diagonals displayed in Fig.\ref{fig:EHmassless_lratio1}, for $k_\textrm{\tiny max} = 3$.
	}
	\label{fig:EHmassless_lratio1_kmax3}
\end{figure}
\begin{figure}[t!]
	\vspace{-.5cm}
	\hspace{0.cm}
	\includegraphics[width=1\textwidth]{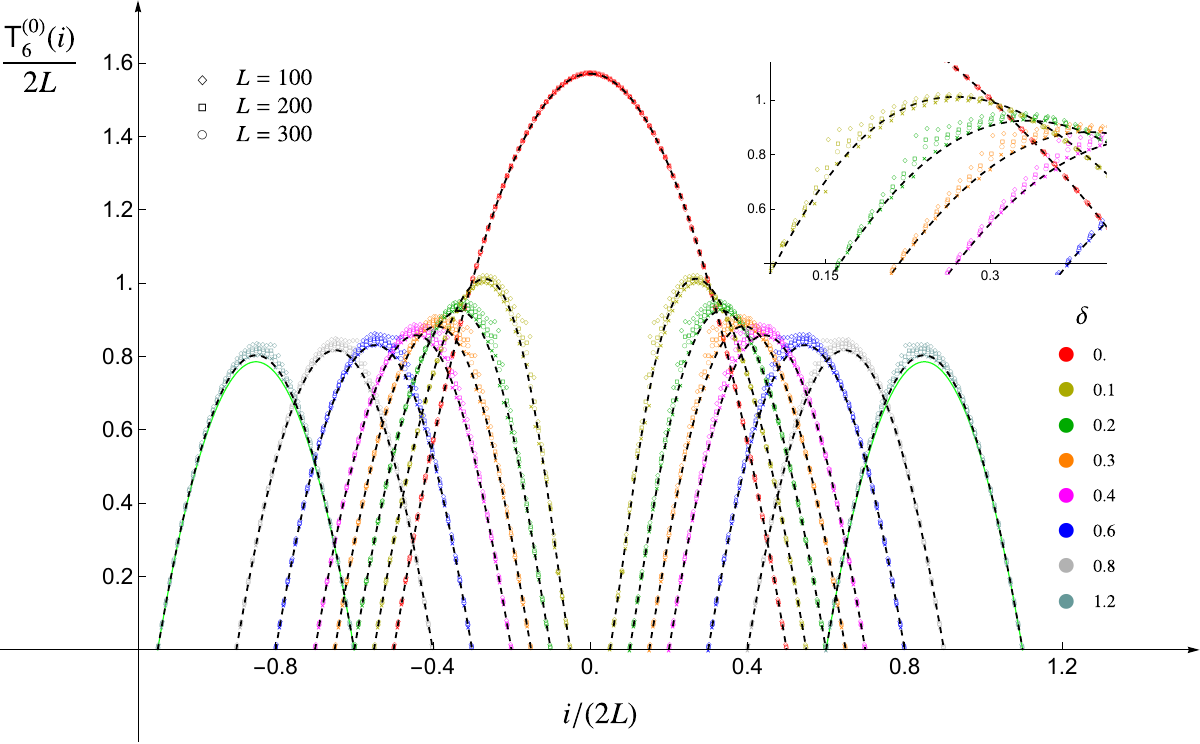}
	\\
	\rule{0pt}{10.2cm}
	\hspace{-.33cm}
	\includegraphics[width=1\textwidth]{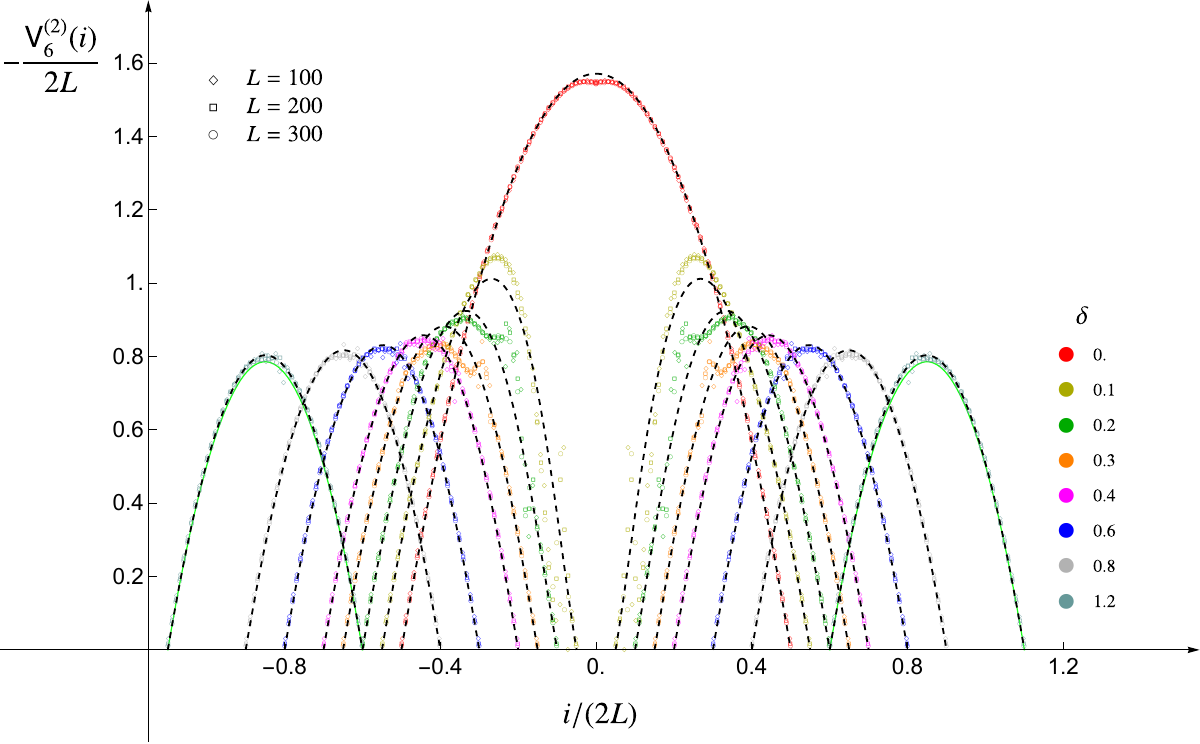}
	\vspace{-.2cm}
	\caption{The combinations of diagonals displayed in Fig.\ref{fig:EHmassless_lratio1} 
	and Fig.\,\ref{fig:EHmassless_lratio1_kmax3} for $k_\textrm{\tiny max} = 6$.
	}
	\label{fig:EHmassless_lratio1_kmax6}
\end{figure}

An important aspect of the limiting procedure discussed in Sec.\,\ref{sec-2int-massless}
to explore the local term,
based on the combinations of diagonals introduced in (\ref{T-V-summation-0})-(\ref{T0-V0-large-L}),
is the consistency condition $k_\textrm{\tiny max}  \ll L_A$.
In the simple case of equal intervals, 
the role of $k_\textrm{\tiny max} $ can be investigated by 
comparing the numerical data points reported
in Fig.\,\ref{fig:EHmassless_lratio1}, Fig.\,\ref{fig:EHmassless_lratio1_kmax3} and Fig.\,\ref{fig:EHmassless_lratio1_kmax6}.
In Fig.\,\ref{fig:EHmassless_lratio1} the values of $k_\textrm{\tiny max} $
correspond to the ones where the best convergence has been observed. 
The most significant improvement provided by the optimal choice of $k_\textrm{\tiny max} $ occurs
in $ \mathsf{V}_{k_\textrm{\tiny max}}^{(2)}(i) / (2L) $ (see the bottom panels of these three figures).
Indeed, notice that already for adjacent intervals (i.e. for $\delta=0$)
when $k_\textrm{\tiny max} = 3$ and $k_\textrm{\tiny max}=6$ this quantity does not provide 
the expected parabola (see also the analysis about $k_\textrm{\tiny max} $ 
performed in \cite{DiGiulio:2019cxv} for the single block).
The occurrence of a non-local term coming from the diagonal blocks of $T$ and $V$
might require a non trivial modification of the limiting procedure to extract the local term. 
%

%

\clearpage

As for the off-diagonal blocks in the matrices $T$ and $V$,
the numerical results in Fig.\,\ref{fig:EH-density} and Fig.\,\ref{fig:all_elements}
(see also the top and middle panel of Fig.\,\ref{3dplots-MatrixTVMabs})
indicate that a large contribution comes from the elements close to the front 
corresponding to $x_{_\Gamma}(x)$ in \eqref{x-gamma-def}, with $x\in A$.
We find it worth exploring the elements of these matrices along this front. 

\begin{figure}[t!]
	\vspace{-.7cm}
	\hspace{-.67cm}
	\includegraphics[width=1.04\textwidth]{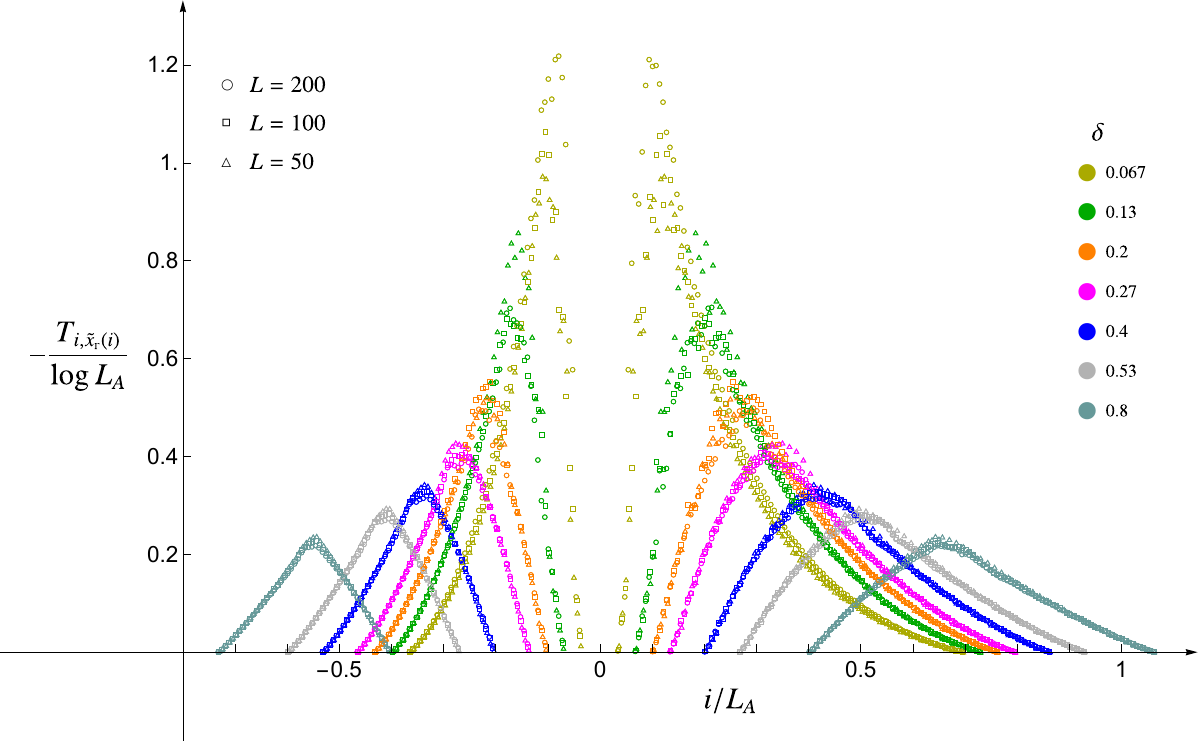}
	\\
	\rule{0pt}{10.7cm}
	\hspace{-.76cm}
	\includegraphics[width=1.04\textwidth]{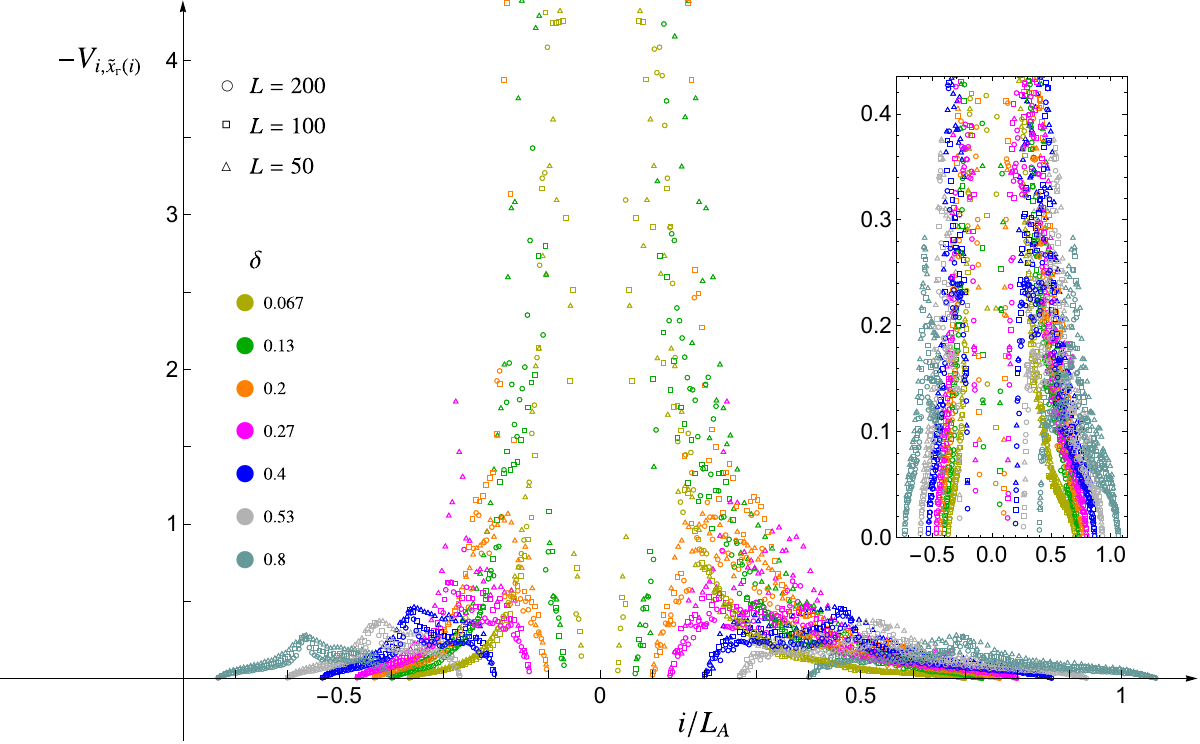}
	\vspace{-.2cm}
	\caption{	
	Elements in the off-diagonal blocks of $T$ (top) and $V$ (bottom) 
	around the front identified by \eqref{x-gamma-def} as discussed in Appendix\;\ref{app-massless},
	for $L_2=2L_1$ and various separations.
	}
	\label{fig:T_hyperbola_massless_lratio2}
\end{figure}

Since for $L_1 \neq L_2$ the function $x_{_\Gamma}(i)$ with $i \in A$ does not provide integer numbers,
a way to define this front on the lattice must be introduced. 
Along the $i$-th row of either $T$ or $V$, 
we consider the element labelled by $\lfloor x_{_\Gamma}(i) \rfloor$ 
and the four elements around it, namely the five elements corresponding to 
$\lfloor x_{_\Gamma}(i) \rfloor + k $ with $k \in \{-2, -1, 0, 1, 2\}$.
The minimum among these five values provides $\tilde{x}_{_\Gamma}(i)$,
that is employed in Fig.\,\ref{fig:T_hyperbola_massless_lratio2}.
When $L_1 = L_2$, we have that  $\tilde{x}_{_\Gamma}(i) = 2L+d-i$,  
which corresponds to the element of the antidiagonal. 

\begin{figure}[t!]
	\vspace{-.7cm}
	\hspace{-.61cm}
	\includegraphics[width=1.04\textwidth]{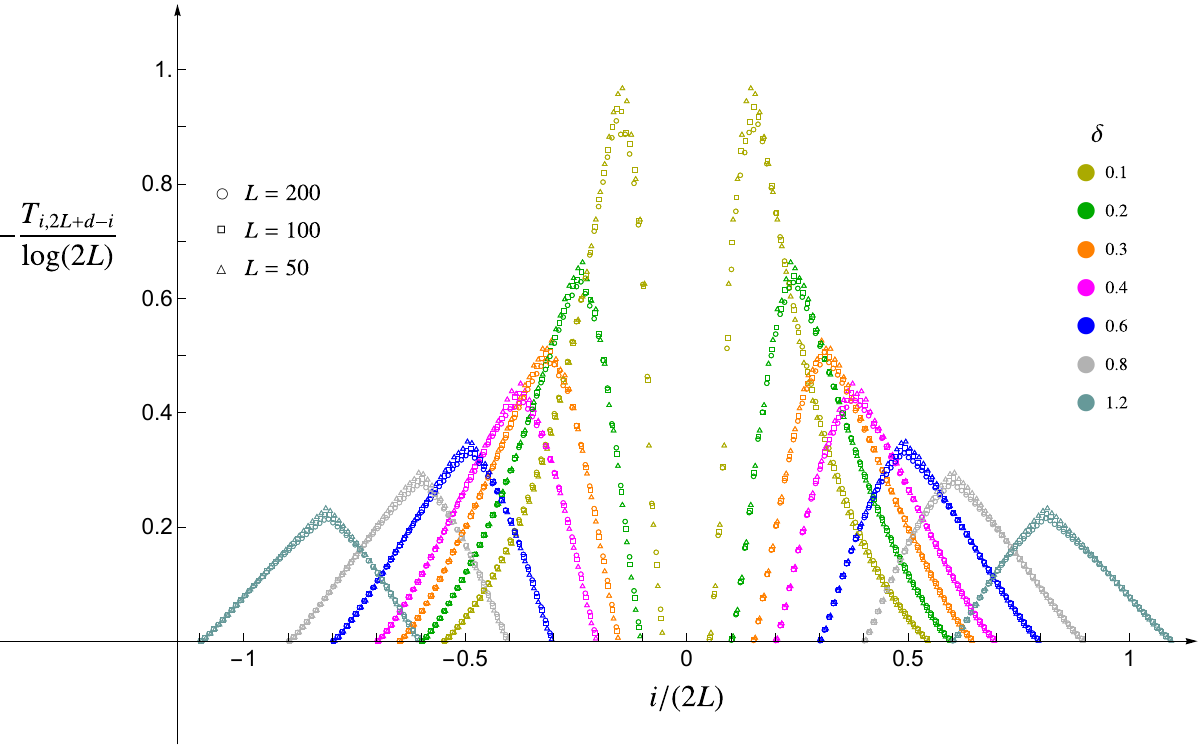}
	\\
	\rule{0pt}{10.7cm}
	\hspace{-.7cm}
	\includegraphics[width=1.04\textwidth]{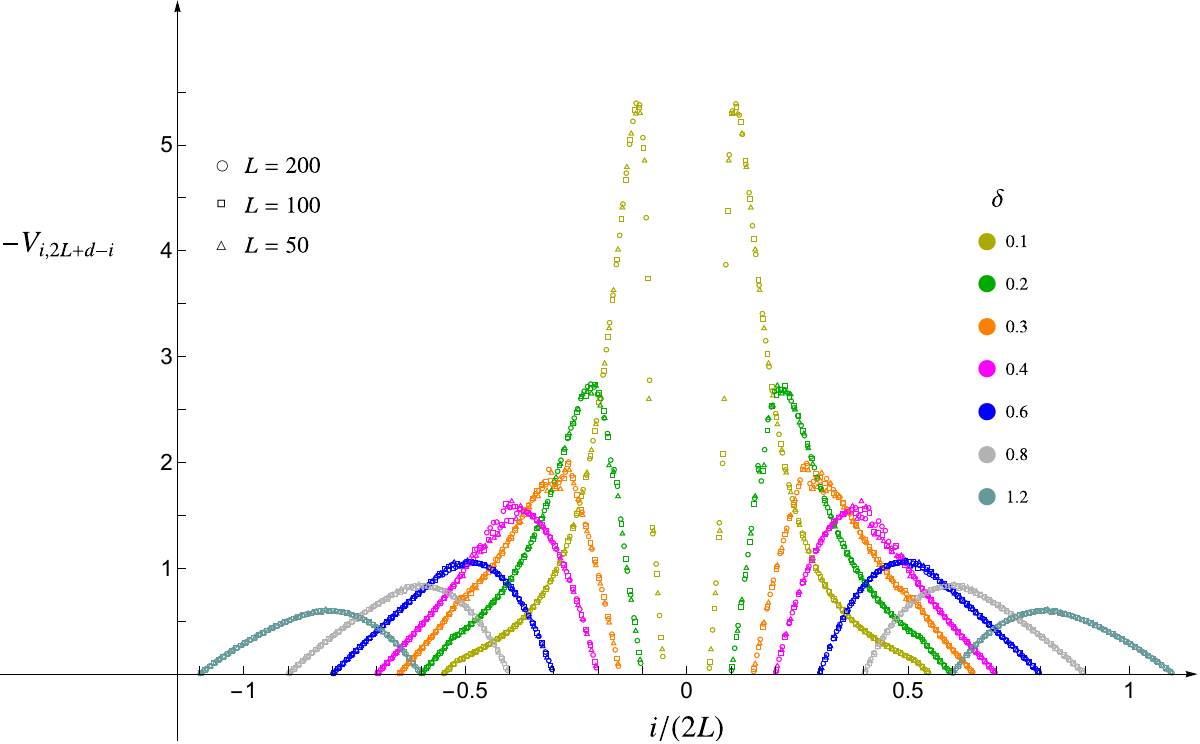}
	\vspace{-.2cm}
	\caption{	
	Antidiagonals of $T$ (top) and $V$ (bottom) for $L_1=L_2$ and various separations. 
	}
	\label{fig:AntiDiag_massless_lratio1}
\end{figure}

When $L_1=L_2$ the expression \eqref{x-gamma-def} provides  the antidiagonal of the matrices $T$ and $V$.
In this case $\tilde{x}_{_\Gamma}(i)=2L+d-i$
and the corresponding numerical results are shown in Fig.\,\ref{fig:AntiDiag_massless_lratio1}. 
The collapses of these data points suggest that the leading scaling behaviour with respect to the total size $2L$ of the subsystem is given by 
$T_{i, 2L+d-i}\sim \log(2L) $ and $V_{i,2L+d-i}\sim \mathcal{O}(1)$. 
In \cite{Arias:2018tmw}, a logarithmic scaling along the antidiagonal has also been observed for the chiral current (see Sec.\,\ref{sec-chiral-current}).
The convergence of the  numerical results obtained for unequal blocks is worse with respect to the case of equal blocks. 
In Fig.\,\ref{fig:T_hyperbola_massless_lratio2} we report some data obtained for $L_2=2L_1$
and, as for the matrix $T$, it seems that $T_{i,x_{_\Gamma}(i)}\sim \log L_A$
(although the data collapses are not as good as in the equal blocks case),
while  the data obtained from the matrix $V$ do not display any clear convergence.

\clearpage 

\clearpage

\clearpage

\section{Massive regime: Single block}
\label{app-massive:1block}

In order to gain some insights about the massive regime, 
in this appendix we investigate the entanglement Hamiltonian (\ref{KA-T-V-matrices})
of a single block $A$ made by $L$ contiguous sites in the infinite harmonic chain 
when the mass parameter takes finite and non vanishing values. 
Some observations about this regime have been also made in \cite{Eisler:2020lyn}.


Following \cite{DiGiulio:2019cxv, Javerzat:2021hxt}, 
let us consider the combinations of diagonals given by 
$\mathsf{T}_{k_\textrm{\tiny max}}^{(0)}$, $\mathsf{V}_{k_\textrm{\tiny max}}^{(0)}$ and $\mathsf{V}_{k_\textrm{\tiny max}}^{(2)}$, 
introduced in \eqref{T-V-summation-0}-\eqref{V-summation-2},
where, in order to take into account the simpler case of the single block that we are considering, 
(\ref{A-sets-matrix-indices-def}) is replaced by
\be
\label{A-sets-matrix-indices-def-1int}
A_< \equiv \big[1\,, L/2  \big] 
\;\;\;\;\qquad\;\;\;\;
\\
\rule{0pt}{.6cm}
A_>  \equiv  \big[ L /2 + 1\, , L \big]   \,. 
\ee
In our analysis even values of $L$ are chosen; hence $A_<$ and $A_>$ are well defined.
Notice that the definition (\ref{A-sets-matrix-indices-def-1int}) 
leads to a maximum allowed value for $k_\textrm{\tiny max}$ given by  $L/2$.

\begin{figure}[t!]
	\hspace{-.9cm}
	\begin{minipage}{0.5\textwidth}
		\centering
		\includegraphics[width=1.0\textwidth]{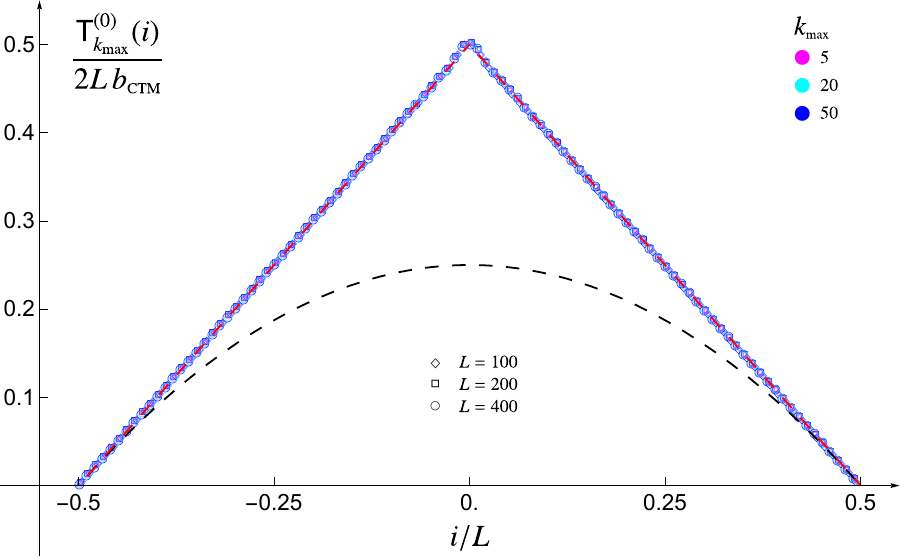}
	\end{minipage}
	\hspace{1cm}
	\begin{minipage}{0.5\textwidth}
		\centering
		\includegraphics[width=1.0\textwidth]{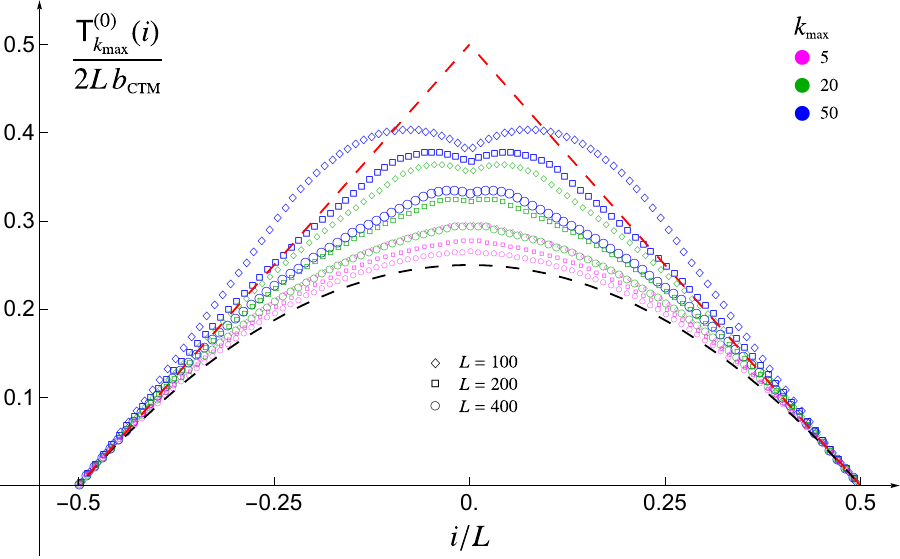}
	\end{minipage}
	\vspace{1.0cm}
	\\
	\vspace{1.0cm}
	\hspace{-1.cm}
	\begin{minipage}{0.5\textwidth}
		\centering
		\includegraphics[width=1.0\linewidth]{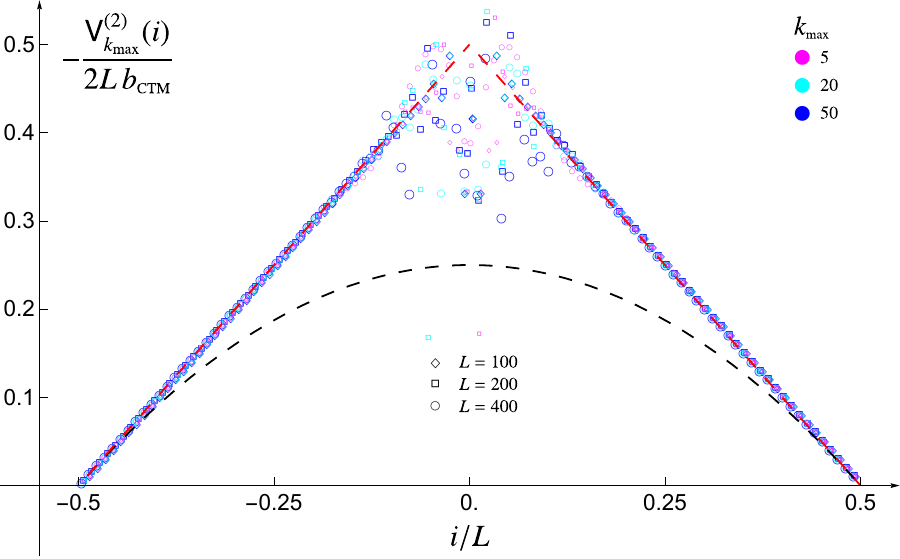}
	\end{minipage}
	\hspace{1cm}
	\begin{minipage}{0.5\textwidth}
		\centering
		\includegraphics[width=1.0\linewidth]{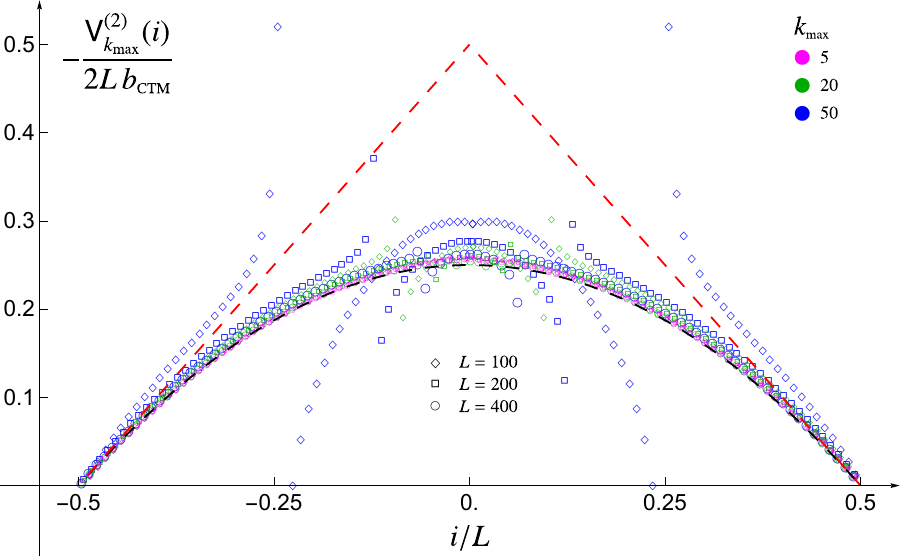}
	\end{minipage}
	\\
		\vspace{1.0cm}
	\hspace{-1.cm}
	\begin{minipage}{0.5\textwidth}
		\centering
		\includegraphics[width=1.0\textwidth]{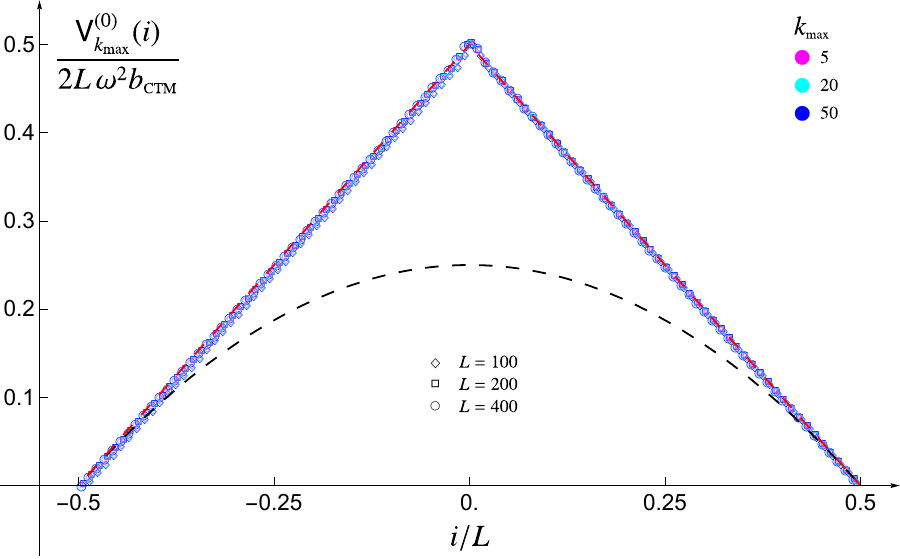}
	\end{minipage}
	\hspace{1cm}
	\begin{minipage}{0.5\textwidth}
		\centering
		\includegraphics[width=1.0\linewidth]{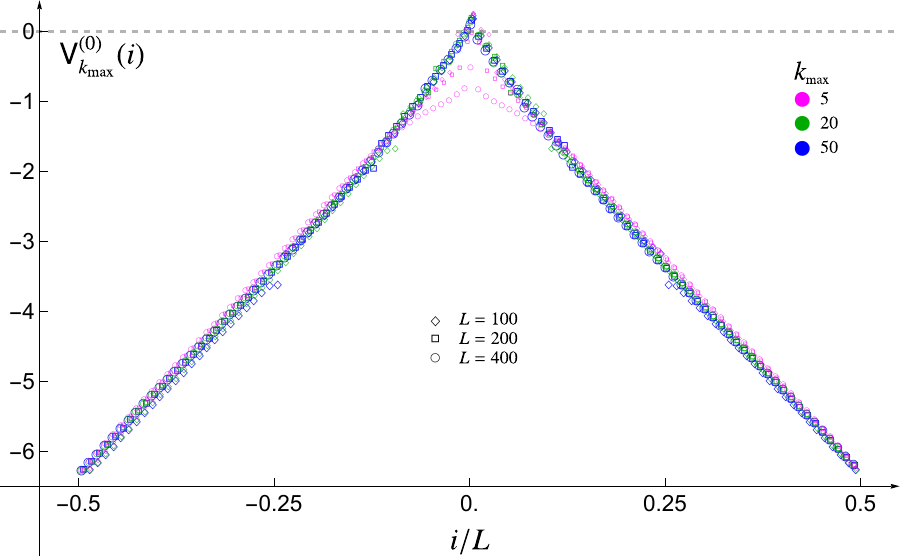}
	\end{minipage}
	\vspace{-.5cm}
	\caption{Combinations of diagonals given by \eqref{T-V-summation-0}-\eqref{V-summation-2} and (\ref{A-sets-matrix-indices-def-1int})
	for a fixed value of $k_\textrm{\tiny max} \ll L$,
	when either $\omega L=500$ (left) or $\omega L=1$ (right).
	}
	\label{Crossover-oldsum-kmaxnonscaling}
\end{figure}

In Fig.\,\ref{Crossover-oldsum-kmaxnonscaling} 
we report some numerical results for these combinations of diagonals 
when $k_{\textrm{\tiny max}} \ll L $ is kept fixed  while $L$ increases. 
In particular, we have that $\omega L= 500$ and $\omega L= 1$ in the left and right panels respectively. 
The red and black dashed curves in Fig.\,\ref{Crossover-oldsum-kmaxnonscaling} 
correspond to the triangular function \cite{Eisler:2020lyn} 
and to the parabolic function \cite{Hislop:1981uh} respectively,
that occur in the large mass (see Sec.\,\ref{sec-single-interval-massive}) 
and in the massless regime respectively. 
In particular, the data points in the left panels are nicely described 
by the function employed in \eqref{T-diag-1int}-\eqref{F1-from-Delta}.
The quantity considered in the bottom left panel
comes from the proper combination of both the expressions in \eqref{V-diag-1int}
and only $\omega^2$ occurs in the coefficient characterising its slope. 
Moreover, we remark that the three-diagonals approximation cannot be applied for Fig.\,\ref{Crossover-oldsum-kmaxnonscaling}
(see also Fig.\,10 of \cite{Eisler:2020lyn}).
Taking the limit $L \to +\infty$ with $\omega L = 1$ provides a probe of the massive regime of the quantum field theory in the continuum. 
In the top and middle right panels of Fig.\,\ref{Crossover-oldsum-kmaxnonscaling},
a collapse of the numerical data points in this regime as $L \to +\infty$ is not observed. 
Instead, in the bottom right panel of the same figure, 
the data points of $\mathsf{V}_{k_\textrm{\tiny max}}^{(0)}(i)$ nicely provide a triangular shape function,
but we remark that this quantity scales in a different way 
with respect to the corresponding panel on the left, where the large mass regime is considered. 
%
It would be interesting to explore further this dramatic change in the scaling for this quantity.

\begin{figure}[t!]
	\hspace{-.9cm}
	\begin{minipage}{0.5\textwidth}
		\centering
		\includegraphics[width=1.0\textwidth]{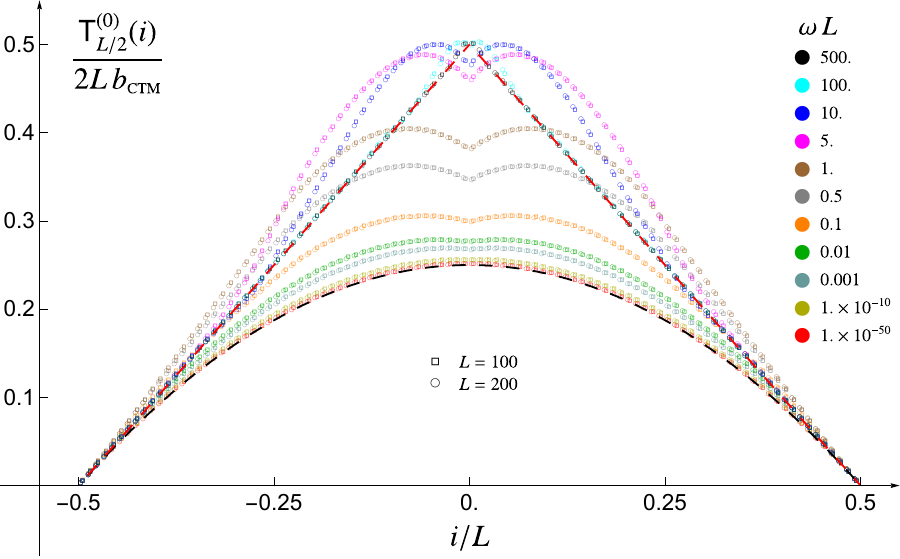}
	\end{minipage}
	\hspace{1cm}
	\begin{minipage}{0.5\textwidth}
		\centering
		\includegraphics[width=1.0\linewidth]{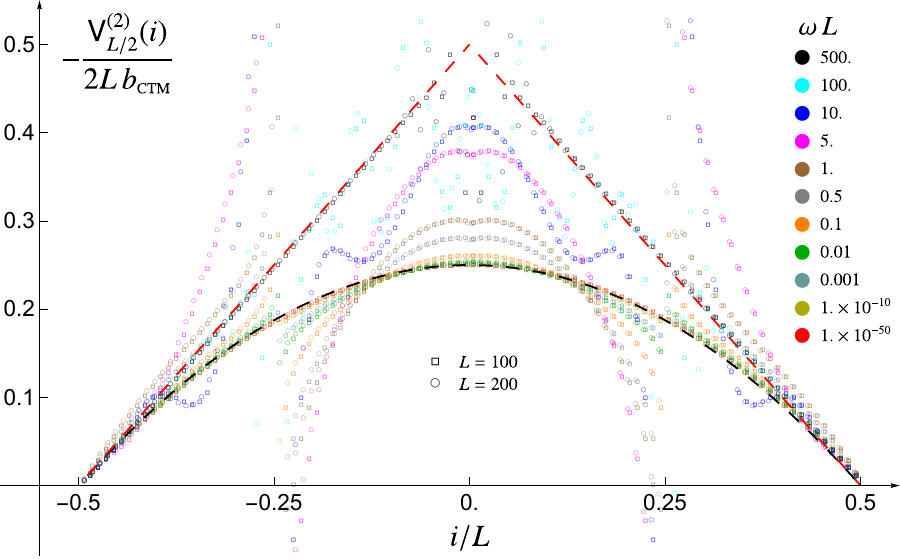}
	\end{minipage}
	\vspace{1.0cm}
	\\
	\vspace{1.0cm}
	\hspace{-1.cm}
	\begin{minipage}{0.5\textwidth}
		\centering
		\includegraphics[width=1.0\textwidth]{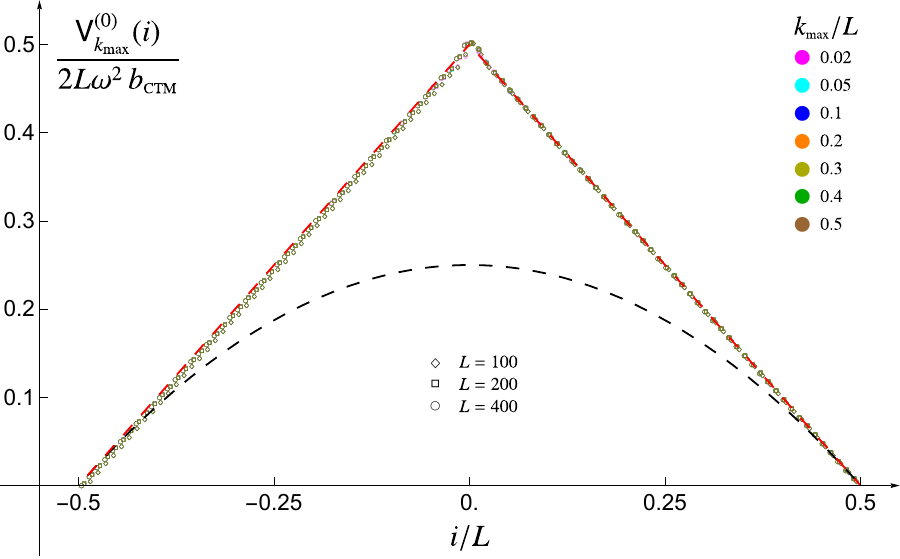}
	\end{minipage}
	\hspace{1cm}
	\begin{minipage}{0.5\textwidth}
		\centering
		\includegraphics[width=1.0\linewidth]{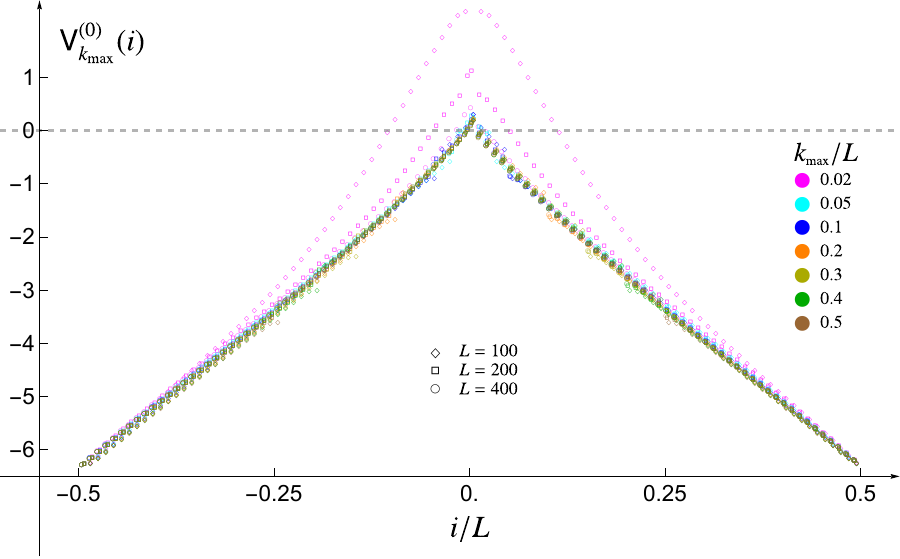}
	\end{minipage}
	\\
	\vspace{1.0cm}
	\hspace{-1.cm}
	\begin{minipage}{0.5\textwidth}
	\centering
	\includegraphics[width=1.0\textwidth]{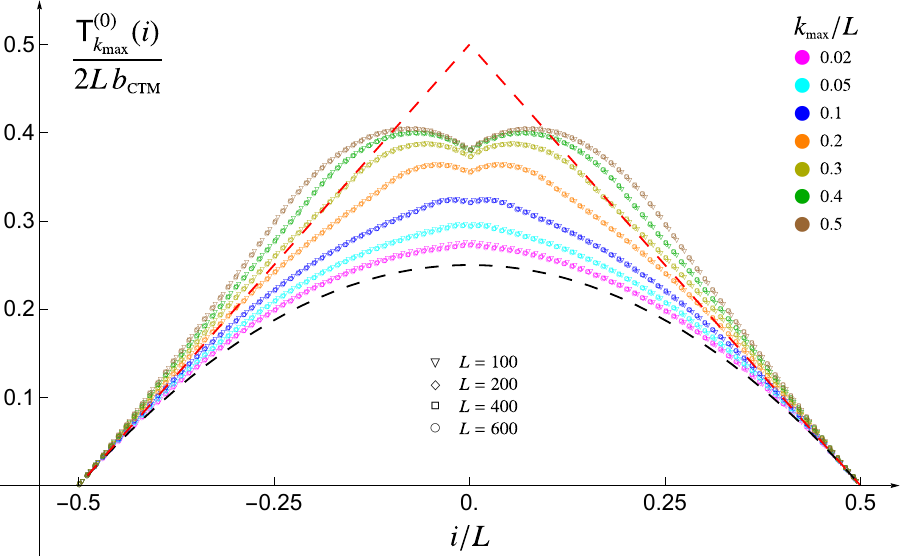}
	\end{minipage}
	\hspace{1cm}
	\begin{minipage}{0.5\textwidth}
	\centering
	\includegraphics[width=1.0\linewidth]{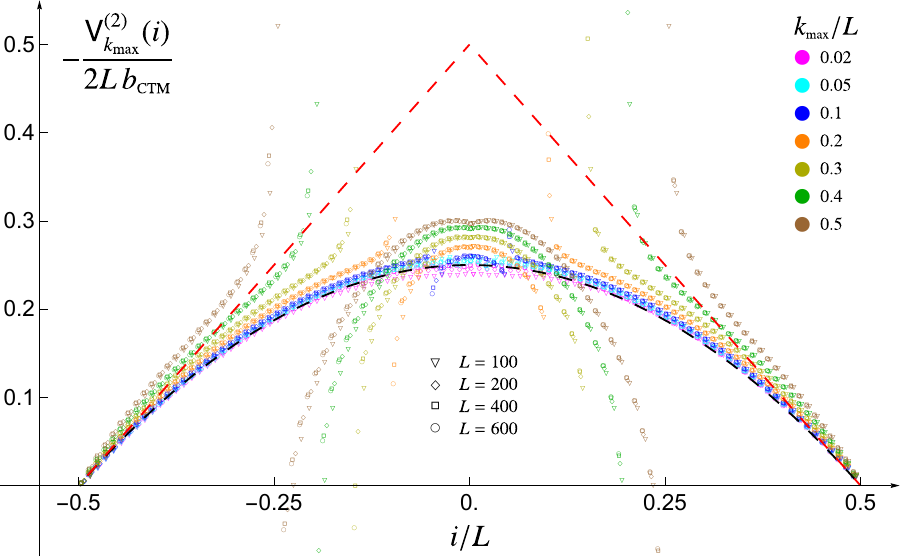}
	\end{minipage}
	\vspace{-.5cm}
	\caption{Combinations of diagonals in \eqref{T-V-summation-0}-\eqref{V-summation-2} and (\ref{A-sets-matrix-indices-def-1int})
	for fixed  $k_\textrm{\tiny max} / L$ and $\omega L$.
	}
	\label{Crossover-oldsum}
\end{figure}

\begin{figure}[t!]
	\hspace{-.9cm}
	\begin{minipage}{0.5\textwidth}
		\centering
		\includegraphics[width=1.0\textwidth]{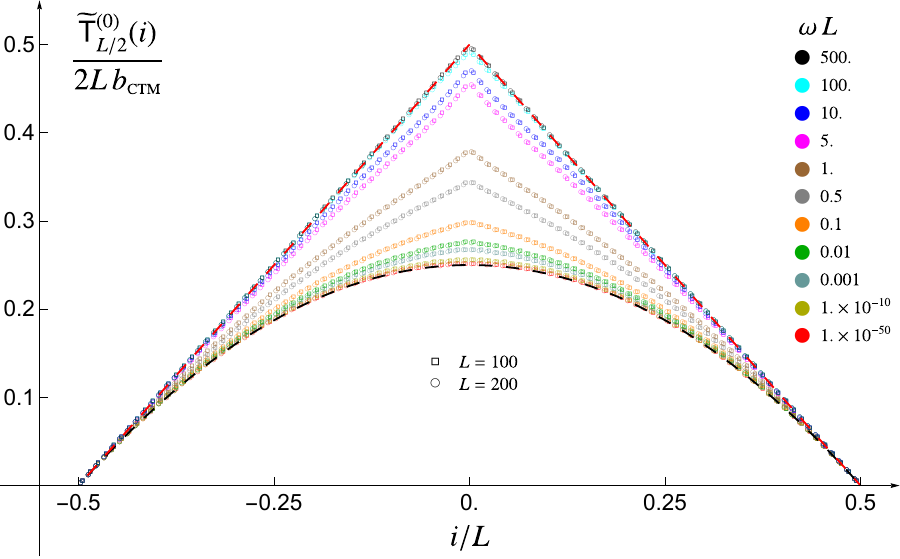}
	\end{minipage}
	\hspace{1cm}
	\begin{minipage}{0.5\textwidth}
		\centering
		\includegraphics[width=1.0\linewidth]{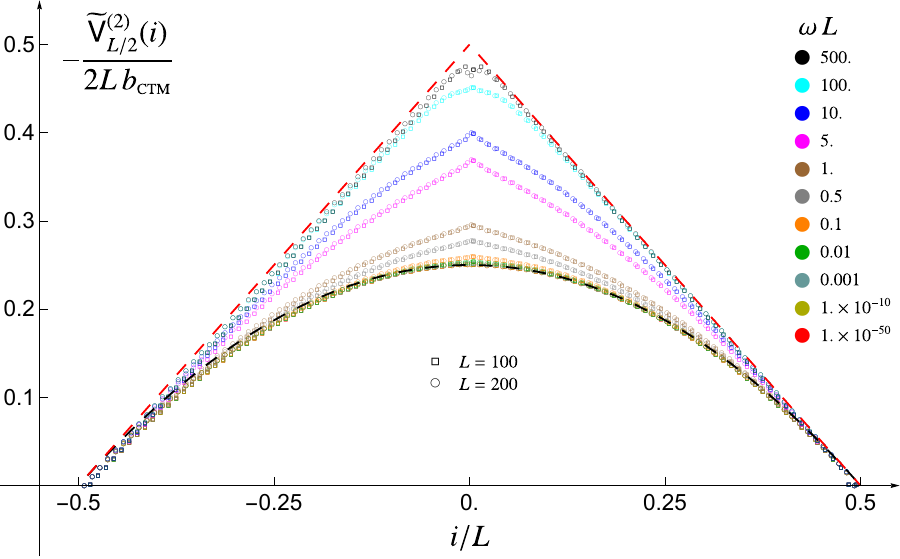}
	\end{minipage}
	\vspace{1.0cm}
	\\
	\vspace{1.0cm}
	\hspace{-1.cm}
	\begin{minipage}{0.5\textwidth}
		\centering
		\includegraphics[width=1.0\textwidth]{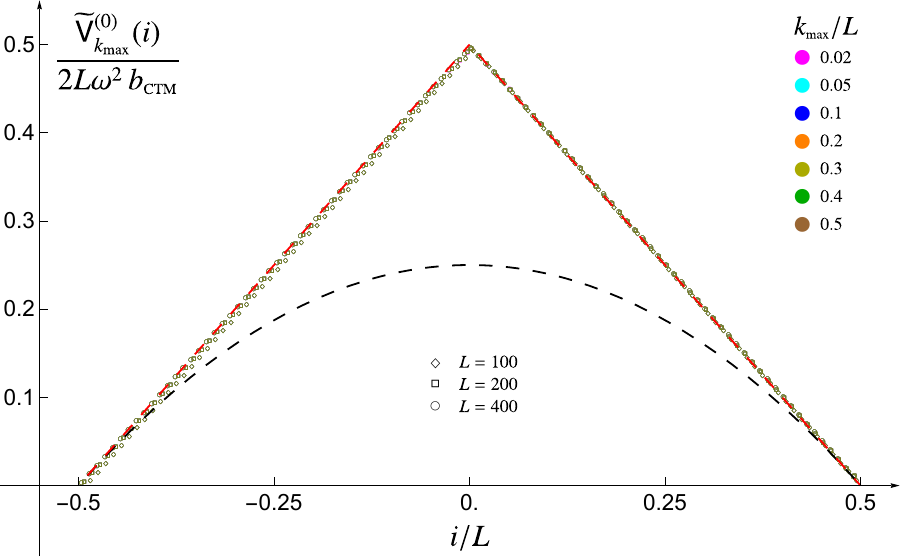}
	\end{minipage}
	\hspace{1cm}
	\begin{minipage}{0.5\textwidth}
		\centering
		\includegraphics[width=1.0\linewidth]{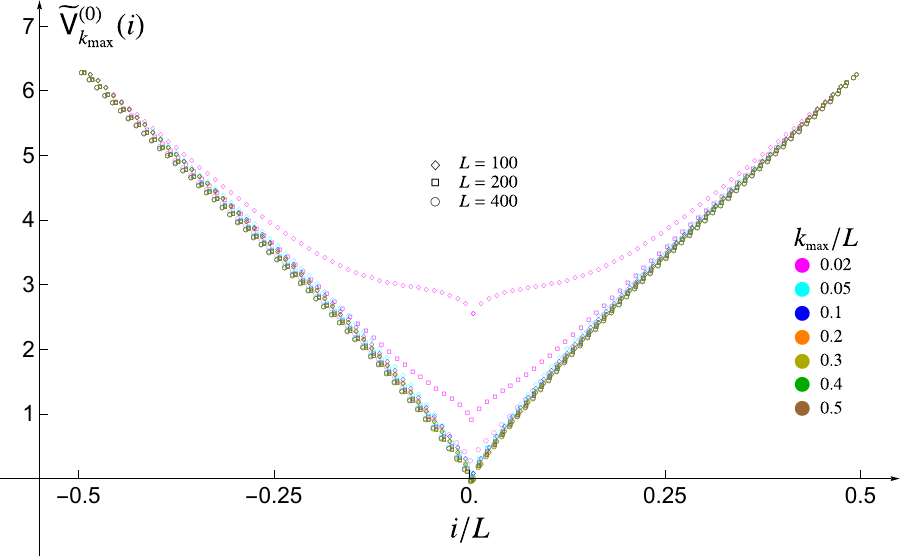}
	\end{minipage}
	\\
	\vspace{1.0cm}
	\hspace{-1.cm}
	\begin{minipage}{0.5\textwidth}
		\centering
		\includegraphics[width=1.0\textwidth]{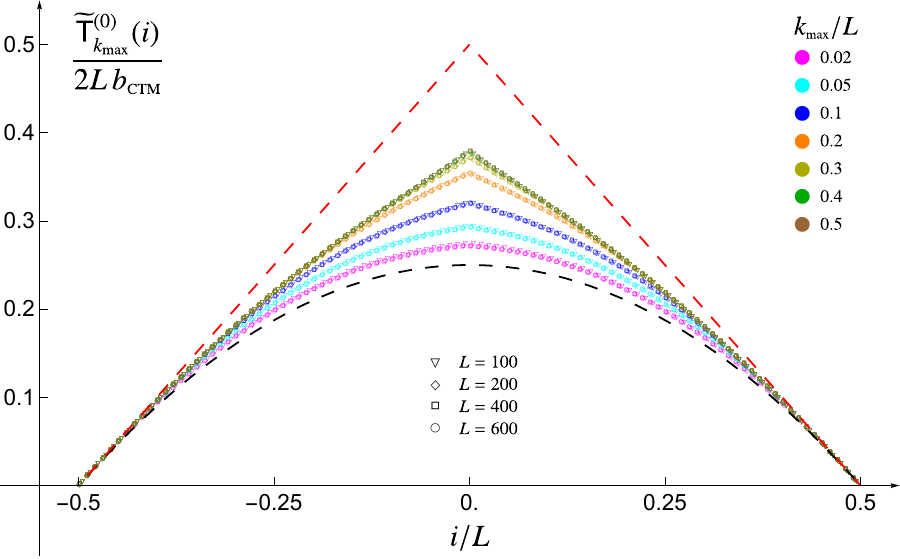}
	\end{minipage}
	\hspace{1cm}
	\begin{minipage}{0.5\textwidth}
		\centering
		\includegraphics[width=1.0\linewidth]{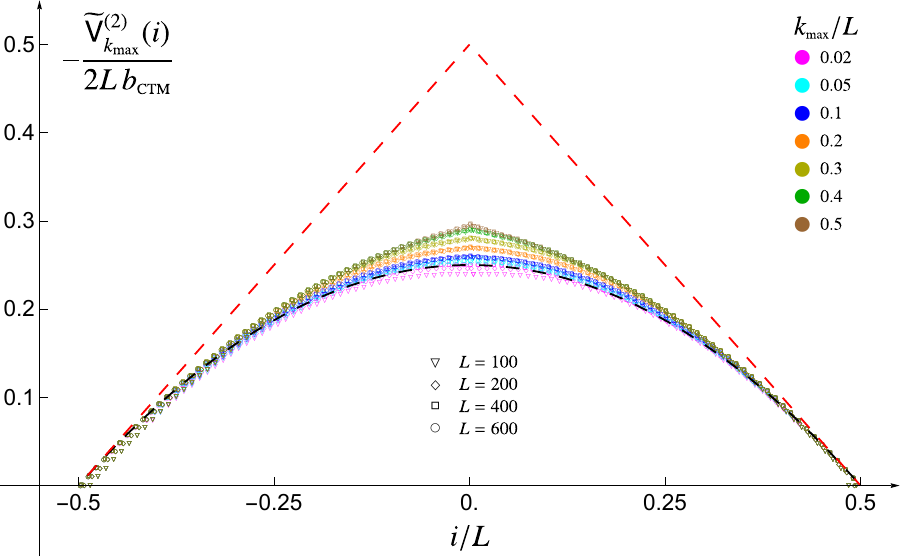}
	\end{minipage}
	\vspace{-.5cm}
	\caption{Combinations  \eqref{T-V-summation-0-tilde}-\eqref{V-summation-2-tilde} and (\ref{A-sets-matrix-indices-def-1int})
	for $k_\textrm{\tiny max} / L$ and $\omega L$ kept fixed   (see also Fig.\,\ref{Crossover-oldsum}). 	
	}
	\label{Crossover-newsum}
\end{figure}


The lack of data points collapses for increasing values of $L$ when $k_\textrm{\tiny max} $ is kept fixed,
observed in Fig.\,\ref{Crossover-oldsum-kmaxnonscaling}, 
leads  us to consider a different scaling limit where 
the ratio $k_\textrm{\tiny max} / L$ is kept fixed as $L$ increases. 
In Fig.\,\ref{Crossover-oldsum} we report some numerical results for the combinations of diagonals
$\mathsf{T}_{k_\textrm{\tiny max}}^{(0)}$, $\mathsf{V}_{k_\textrm{\tiny max}}^{(0)}$ and $\mathsf{V}_{k_\textrm{\tiny max}}^{(2)}$ 
in this scaling limit. 
In the top panels $k_\textrm{\tiny max}/L = 1/2$ and different values of $\omega L$ are considered.
Instead, in the remaining panels $\omega L$ is kept fixed while $k_\textrm{\tiny max}/L$ takes different values
($\omega L = 1$ in the middle right panel and in both the bottom panels, and $\omega L = 500$ in the  middle left panel). 
%
%
Besides the fact that the expected results are recovered in the limiting regimes of $\omega L \to 0^+$ and $\omega L \gg 1$,
the main feature to highlight in Fig.\,\ref{Crossover-oldsum} is given by the nice collapses of the data points for increasing values of $L$.
These collapses suggest that this scaling limit can provide useful insights 
to explore the continuum limit of the entanglement Hamiltonian in the massive regime.
In the middle panels, 
where either $\omega L = 500$ (left panel) or  $\omega L = 1$ (right panel),
we show that this quantity scales differently as one goes from a finite value of $\omega L$ to $\omega L \gg 1$
(see also the bottom panels of  Fig.\,\ref{Crossover-oldsum-kmaxnonscaling}).
It would be worth exploring further this change in the scaling also in the setting where the ratio $k_\textrm{\tiny max} / L$ is kept fixed
while $L \to +\infty$.
In the bottom panels of Fig.\,\ref{Crossover-oldsum}
the combinations of diagonals $\mathsf{T}_{k_\textrm{\tiny max}}^{(0)}$ and $\mathsf{V}_{k_\textrm{\tiny max}}^{(2)}$ are considered, 
we obtain well defined curves for different values of $k_\textrm{\tiny max} / L$ for both these quantities, 
but the ones corresponding to $\mathsf{V}_{k_\textrm{\tiny max}}^{(2)}$ display two singularities.


The data collapses, the singular behaviour of the curves 
and the  fact that the triangular function is not an upper bound for them,
observed in Fig.\,\ref{Crossover-oldsum},
suggest to consider the same scaling limit for different combinations of diagonals. 
A possibility is to modify (\ref{T-V-summation-0}) and (\ref{V-summation-2}) by introducing respectively 
\be
\label{T-V-summation-0-tilde}
\widetilde{\mathsf{T}}_{k_\textrm{\tiny max}}^{(0)}(i)
\equiv
\left\{\begin{array}{l}
\displaystyle	T_{i,i} + 2 \sum_{k=1}^{k_\textrm{\tiny max}} T_{i,i-k} 
	\\
	\rule{0pt}{.9cm}
\displaystyle		T_{i,i} + 2 \sum_{k=1}^{k_\textrm{\tiny max}} T_{i+k,i} 
\end{array}
\right.
\hspace{1.cm}
\widetilde{\mathsf{V}}_{k_\textrm{\tiny max}}^{(0)}(i)
\equiv
\left\{\begin{array}{ll}
\displaystyle	V_{i,i} + 2 \sum_{k=1}^{k_\textrm{\tiny max}} V_{i,i-k}  
\hspace{.8cm}& i \in A_<
	\\
	\rule{0pt}{.9cm}
\displaystyle		V_{i,i} + 2 \sum_{k=1}^{k_\textrm{\tiny max}} V_{i+k,i}  
& i \in A_>
\end{array}
\right.
\ee
and 
\be
\label{V-summation-2-tilde}
\widetilde{\mathsf{V}}_{k_\textrm{\tiny max}}^{(2)}(i)
\equiv
\left\{\begin{array}{ll}
\displaystyle	\sum_{k=1}^{k_\textrm{\tiny max}} k^2\,V_{i,i-k} 
\hspace{.8cm}& i \in A_<
	\\
	\rule{0pt}{.9cm}
\displaystyle		\sum_{k=1}^{k_\textrm{\tiny max}} k^2 \, V_{i+k,i} 
& i \in A_>
\end{array}
\right.
\ee
where (\ref{A-sets-matrix-indices-def-1int}) is employed to define the range of the index $i$, 
imposing that the sums stop when $i-k\leqslant 0$ or $i+k>L$.


In Fig.\,\ref{Crossover-newsum}, we report numerical results for the combinations of diagonals 
$\widetilde{\mathsf{T}}_{k_\textrm{\tiny max}}^{(0)}$, $\widetilde{\mathsf{V}}_{k_\textrm{\tiny max}}^{(0)}$ and $\widetilde{\mathsf{V}}_{k_\textrm{\tiny max}}^{(2)}$, 
introduced in (\ref{T-V-summation-0-tilde})-(\ref{V-summation-2-tilde}),
in the same setups, choices of the parameters and scheme for the panels 
considered in Fig.\,\ref{Crossover-oldsum}.
Also for these combinations, data collapses are observed and the expected curves are recovered in the limiting regimes of 
$\omega L \to 0^+$ and $\omega L \gg 1$.
Comparing Fig.\,\ref{Crossover-newsum} and Fig.\,\ref{Crossover-oldsum},
we remark that the triangular and parabolic functions,
characterising the $\omega L \to 0^+$ and $\omega L \gg 1$ regimes respectively, 
provide respectively  a lower bound and an upper bound
for the curves obtained from (\ref{T-V-summation-0-tilde})-(\ref{V-summation-2-tilde}) 
when $\omega L $ takes finite and non vanishing values. 
When $\omega L = 1$, the curve in the middle right panel of Fig.\,\ref{Crossover-newsum}
corresponds to the one obtained 
in the middle right panel of Fig.\,\ref{Crossover-oldsum} or in the bottom right panel of Fig.\,\ref{Crossover-oldsum-kmaxnonscaling}
with the sign flipped.

\section{Details on the free chiral current model}
\label{app:chiral}

In this Appendix 
we discuss some features of the free chiral current model \cite{Berenstein:2023tru}
and, following \cite{Arias:2018tmw}, we review  some technical details 
about the numerical evaluation of the entanglement Hamiltonian matrix $H$,
whose results are discussed in Sec.\,\ref{sec-chiral-current}.



Consider the lattice Hamiltonian $H=\frac{1}{2} \sum_{i=1}^L b_i$ on a circle \cite{Berenstein:2023tru}.
At classical level, the variables $b_i$ satisfy the unusual Poisson brackets given by 
$\big\{b_i,b_j \big\}=\delta_{i,j+1}-\delta_{i,j-1}$
and the equation of motion reads $ \dot{b}_i= \big\{b_i,H \big\}=b_{i-1}-b_{i+1}$.
%
This equation admits plane wave solutions $b_i= \exp [\textrm{i}(ki-Et)]$ 
with dispersion relation $E(k)=2\sin(k)$, 
whose periodicity allows to restrict the domain of the quasi-momentum to $k\in[0,2\pi]$.
Since $E(0)=E(\pi)=0$, two values of $k$ correspond to the same extremal value of $E(k)$. 
Moreover, the group velocities  $\partial_kE|_{k=0}=-\partial_kE|_{k=\pi}=2$ around these values have opposite sign;
hence the model contains  both left-moving and right-moving excitations, corresponding to the modes around $k=0$ and $k=\pi$ respectively. 
This feature can be clarified by defining the variables $\tilde{b}_i=(-1)^i \, b_i$, 
which satisfy $\big\{\tilde{b}_i,\tilde{b}_j \big\}=- \big\{b_i,b_j \big\}=\delta_{i,j-1}-\delta_{i,j+1}$
and give $\tilde{b}_i=\exp [\textrm{i}( (k+\pi)i-Et)]$ for the plane wave solution.
Each left-moving excitation having  $k\in [0,\pi/2]\cup[3\pi/2,2\pi]$ encoded in $b_i$
can be associated to a right-moving solution encoded in $\tilde{b}_i$. 
The above mentioned Hamiltonian can be written  in terms of both types of excitations as 
$H=\frac{1}{4}\sum_{i=1}^L \! \big(b^2_i+\tilde{b}_i^2\big)$.

The quantisation procedure, 
which transforms the classical variables into operators $b_i\rightarrow \hat{b}_i$
and the Poisson brackets into  commutators,
preserves these features of the excitation spectrum. 
Considering also the operators $\hat{\tilde{b}}_i$ coming from $\tilde{b}_i$, 
we have that $[\hat{b}_i,\hat{\tilde{b}}_j]=0$.
%
In the continuum limit, the relevant values of $k$ are the minima of $|E(k)|$, namely $k=0$ and $k=\pi$.
The resulting model is the sum of two chiral quantum field theories 
corresponding to the left and right moving excitations mentioned above. 


In order to explore the entanglement Hamiltonian in the free chiral current model, following \cite{Arias:2018tmw},
we consider  the $ L_A \times L_A$ matrices $Y_A$ and $B_A$ introduced in Sec.\,\ref{sec-chiral-current}, with $L_A$ even.
As for $Y_A$, since $L_A$ is even
the $ L_A \times L_A$ orthogonal matrix $O$  and the matrix $E$ with the same size can be constructed such that
\begin{equation}
\label{eq:CAtoJ}
   E\,O \,Y_A \, O^{\textrm{t}} E \,=
   \bigg(
   \begin{array}{cc}
\boldsymbol{0} \; & \boldsymbol{1}  \\
-\boldsymbol{1}  \; & \boldsymbol{0} 
\end{array}
\bigg)
\equiv J
\;\;\;\;\qquad\;\;\;
E = 
   \bigg(
   \begin{array}{cc}
X & \boldsymbol{0}  \\
 \boldsymbol{0} & X
\end{array}
\, \bigg)
\end{equation}
where $X$ is a diagonal matrix with strictly positive real numbers on its main diagonal.
%
The matrices $O$ and $E$ allow us to introduce the following canonical operators 
\begin{equation}
  \hat{r}_i = \widetilde{O}_{i,j} \, \hat{b}_j  
  \;\;\;\;\;\qquad\;\;\;
  \widetilde{O} \equiv E\,O
    \;\;\;\;\;\qquad\;\;\;
    \big[ \hat{r}_i, \hat{r}_j \big]=\textrm{i} J_{i,j}
\end{equation}
where  $\widetilde{O}$ is neither an orthogonal nor a symplectic matrix. 
The corresponding correlators $\langle  \hat{r}_i \, \hat{r}_j   \rangle$ 
provide the $L_A \times L_A$ real and symmetric matrix $\tilde{\gamma}_A  $ as follows (notice that $E^{\textrm{t}} = - E$)
\begin{equation}
\label{eq:ReFA_to_GammaA}
    \langle  \hat{r}_i \, \hat{r}_j   \rangle=\, \widetilde{O}  \, B_A \, \widetilde{O}^{\textrm{t}} 
    \;\;\qquad\;\;
       (\tilde{\gamma}_A)_{i,j} \equiv \textrm{Re} \big[ \langle \hat{r}_i \,\hat{r}_j \rangle \big] 
    \;\;\qquad\;\;
    \tilde{\gamma}_A    
    =
    \,\widetilde{O}  \, \big[ \textrm{Re}\, B_A \big] \, \widetilde{O} ^{\textrm{t}} 
    \equiv 
       \bigg(
    \begin{array}{cc}
\widetilde{Q} & S \\
S^{\textrm{t}} & \widetilde{P}
\end{array}
   \,\bigg)  \,. 
\end{equation}
The matrix $\tilde{\gamma}_A $ is positive definite for all the cases that we have considered, 
although this feature has been checked only numerically. 
Assuming that $\tilde{\gamma}_A $ is positive definite, its Williamson decomposition reads
\begin{equation}
    \tilde{\gamma}_A=W^{\textrm{t}} \big(\widetilde{D}\oplus\widetilde{D} \big)W
\end{equation} 
where $W$ is a  $L_A \times L_A$ real symplectic matrix  and $\widetilde{D}$
is the $(L_A/2) \times (L_A/2)$ diagonal matrix containing the symplectic spectrum 
$\{ \tilde{\sigma}_1,\dots,\tilde{\sigma}_{L_A/2} \}$ of $\tilde{\gamma}_A$ on its diagonal,
satisfying $\tilde{\sigma}_k > 0$,
that can be obtained  from the spectrum of $(\textrm{i}J\tilde{\gamma}_A)^2$ in a standard way \cite{Weedbrook:2012cjy}.
This symplectic spectrum provides the R\'enyi entropies and the entanglement entropy 
through (\ref{renyi-from-ss}) and (\ref{EE-from-ss}) respectively.

The matrix $R_A$ introduced  in \eqref{eq:Rmatrix} is related to $\tilde{\gamma}_A$.
Indeed, by inverting \eqref{eq:CAtoJ},  one first observes that $Y_A^{-1}=-\,\widetilde{O}^{\textrm{t}} J^{\textrm{t}} \,\widetilde{O} $.
Then, combining this expression with \eqref{eq:Rmatrix}, we get 
\begin{equation}
\label{eq:VCinverse}
    R_A \, Y_A^{-1}
    =
    \textrm{i} \, \widetilde{O} ^{\textrm{t}} \, J^{\textrm{t}}\, \tilde{\gamma}_A J  \, \widetilde{O} 
\end{equation}
and 
\begin{equation}
\label{eq:Vsquared}
    R_A^2
    =
    -Y_A^{-1} ( \textrm{Re} B_A) \, Y_A^{-1} (\textrm{Re} B_A)
    = 
    \widetilde{O}^{\textrm{t}} \, (\textrm{i}J\tilde{\gamma}_A)^2 \, \widetilde{O} ^{-\textrm{t}}  \,. 
\end{equation}

The entanglement Hamiltonian can then be written 
in terms of either $\hat{r}_i $ or $\hat{b}_i $ as follows
\begin{equation}
\label{eq:entanglementhamiltonian}
  \widehat{K}_A
  \,=\,
  \frac{1}{2} \, \hat{\boldsymbol{r}}^{\mathrm{t}} \,N_A \, \hat{\boldsymbol{r}} 
  \,=\,
   \hat{ \boldsymbol{b}}^{\textrm{t}} M \, \hat{\boldsymbol{b}}
\end{equation}
where $N_A$ reads
\cite{Casini:2009sr} (see also \cite{DiGiulio:2019lpb})
\begin{equation}
\label{eq:HAmatrix}
 N_A  \,=\, h\big(\sqrt{ (\mathrm{i} J \tilde{\gamma}_A )^2}\,\big) \, J^{\mathrm{t}} \, \tilde{\gamma}_A \,J 
\end{equation}
in terms of $h(y)$ defined in \eqref{eq:hfunction};
while the matrix $M$ has been introduced  in \eqref{eq:H_A_in_terms_of_BA}
and it can be obtained by using \eqref{eq:VCinverse} and \eqref{eq:Vsquared} in \eqref{eq:HAmatrix}.


The entanglement Hamiltonian \eqref{eq:entanglementhamiltonian} can be written as follows
%
\begin{equation}
    \widehat{K}_A= \left(\begin{array}{c}
            \hat{ \boldsymbol{b}}^{\textrm{t}} \\
            \hat{ \tilde{\boldsymbol{b}}}^{\textrm{t}}
    \end{array}\right)
    \left(\begin{array}{cc}
      M/2  &  0\\
      0   & \widetilde{M}/2
    \end{array}\right) 
    \left(\begin{array}{c}
            \hat{ \boldsymbol{b}} \\
            \hat{ \tilde{\boldsymbol{b}}}
    \end{array}\right)
\end{equation}
where $\widetilde{M}_{i,j}=(-1)^{i+j} M_{i,j} $ and $\hat{\tilde{{b}}}_i=(-1)^i\,\hat{b}_i$.
Notice that $M_{i,i+2k+1}=0$ because of the parity symmetry \cite{Berenstein:2023tru}; hence $\widetilde{M}=M$.

In the continuum limit, that requires the UV cutoff $a$, 
we have $\hat{b}_i \to a j(x) $ and $\hat{\tilde{{b}}}_i \to a \bar{j}(x)$ for the operators
and $M_{ij} / 2 \to H(x,y)$ for the matrix, where $i a \to x$ and $j a \to y$.
The chiral fields coming from the left and right moving excitations are decoupled also in the entanglement Hamiltonian, 
which is therefore given by the sum of the corresponding operators associated to these two different types of excitations.

\section{R\'enyi mutual information}
\label{app:mutual-info}

The R\'enyi mutual information 
for the bipartition of the line given by $A = A_1 \cup A_2$
is a UV finite quantity defined as the following combination of  R\'enyi entropies
\be
\label{renyi-mutual-def}
     I^{(n)}_{A_1,A_2} = S^{(n)}_{A_1}+S^{(n)}_{A_1}-S^{(n)}_{A_1\cup A_2}
\ee
where $n \geqslant 0$ is an integer and its analytic continuation $n \to 1$ gives the mutual information $I_{A_1,A_2}$.
In this Appendix we discuss the R\'enyi mutual information (\ref{renyi-mutual-def}) 
for the (uncompactified) massless scalar field \cite{Calabrese:2009ez}
and the chiral current \cite{Arias:2018tmw},
whose underlying lattice models have been described 
in Sec.\,\ref{sec-correlators-EH} and Sec.\,\ref{sec-chiral-current}  respectively. 
In these CFT models, 
the R\'enyi mutual information (\ref{renyi-mutual-def}) 
is a function of the harmonic ratio of the four endpoints of  $A$, namely
\begin{equation}
\label{eta-ratio-def}
    \eta\equiv \frac{(v_1-u_1)(v_2-u_2)}{(u_2-u_1)(v_2-v_1)}
\end{equation}
where $A_j = (u_j , v_j)$ for $j \in \{1,2\}$ \cite{Furukawa:2008uk, Calabrese:2009ez}.

%
The R\'enyi mutual information (\ref{renyi-mutual-def}) for the massless scalar reads \cite{Calabrese:2009ez}
 \begin{equation}
 \label{eq:full_renyi_mutual_info}
     I^{(n) }_{A_1,A_2; \textrm{\tiny (ms)} }
     =
     -\frac{n+1}{6\,n}\log (1-\eta)+\frac{ D_n(\eta)+ D_n(1-\eta) }{2(1-n)} 
 \end{equation}
where 
\begin{equation}
\label{Dn-def}
    D_n(\eta) \equiv \sum_{k=1}^{n-1} \log \! \big[ { }_2 F_1( k/n ,1- k/n ; 1 ; \eta) \big]  \,. 
\end{equation}
Taking the analytic continuation $n\rightarrow 1$ of \eqref{eq:full_renyi_mutual_info},
one obtains the mutual information 
\begin{equation}
\label{eq:full_mutual_information_n1}
    I_{A_1,A_2; \textrm{\tiny (ms)} }
    =
    -\frac{1}{3}\log (1-\eta)+\frac{D'_1(\eta)+D'_1(1-\eta) }{2}
\end{equation}
where 
\begin{equation}
\label{eq:D1prime_eta}
   D_1^{\prime}(\eta) 
   \equiv 
   \lim_{n\rightarrow1} \frac{D_n(\eta)}{1-n}
   =\,
   -\int_{-\textrm{i}  \infty}^{\textrm{i}  \infty} \frac{\pi z}{\textrm{i}\, [ \sin (\pi z) ]^2} \;
    \log \!\big[ { }_2 F_1(z,1-z ; 1 ; \eta) \big]\,  \rd z  \,. 
\end{equation}
Discarding the logarithmic term in the r.h.s.'s of (\ref{eq:full_renyi_mutual_info}) 
and (\ref{eq:full_mutual_information_n1}),
the resulting expression is invariant under exchange $\eta \leftrightarrow 1-\eta$,
indicating that this model satisfies the Haag duality \cite{Calabrese:2009ez}.

\begin{figure}[t!]
	\vspace{-.5cm}
	\hspace{-1cm}
	\includegraphics[width=1.07\textwidth]{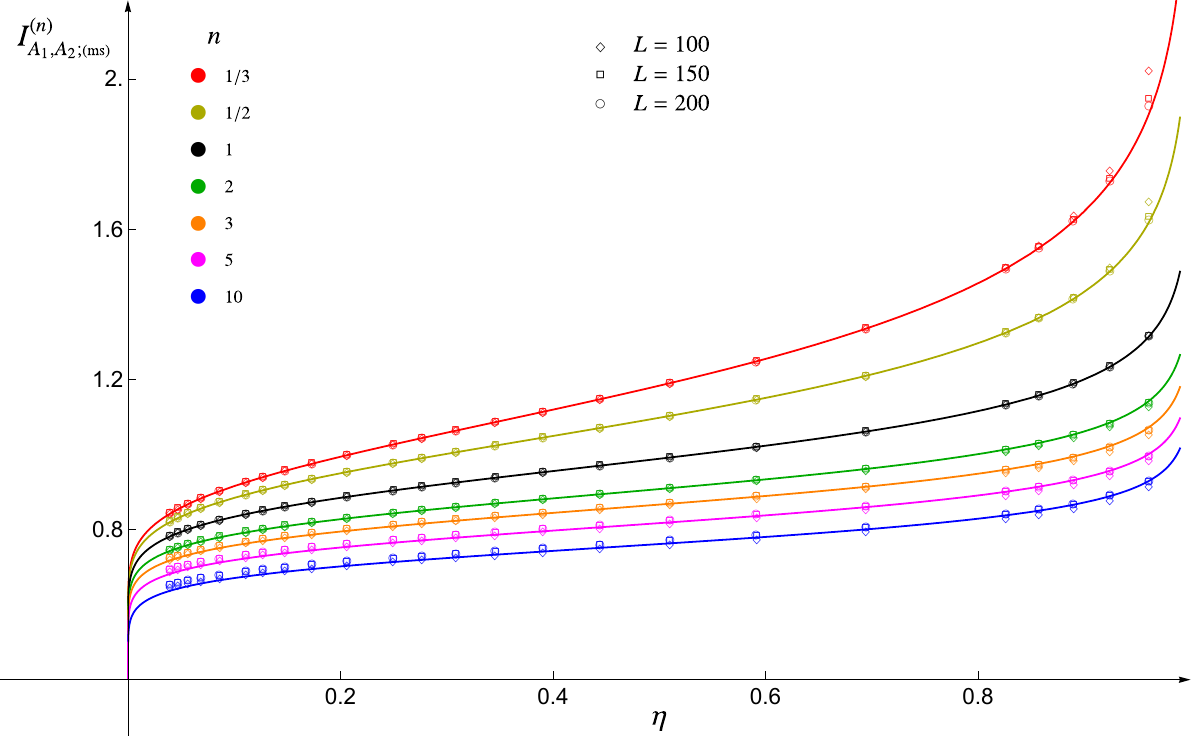}
	\vspace{-.2cm}
	\caption{R\'enyi mutual information of equal blocks of length $L$ for the massless scalar. Here $\omega L_A=10^{-50}$.}
	\label{fig:CH_mutualinformation-scalar}
\end{figure}

In Fig.\,\ref{fig:CH_mutualinformation-scalar} we show some numerical data 
for the R\'enyi mutual information (\ref{renyi-mutual-def}) of equal blocks of size $L$
for the massless harmonic chain,
evaluated by using (\ref{renyi-from-ss}), with $\omega L_A=10^{-50}$.
The agreement between the data points and the solid curves obtained from 
 \eqref{eq:full_renyi_mutual_info} and  \eqref{eq:full_mutual_information_n1}, 
 for $n\neq 1$ and $n=1$ respectively, 
provides an important check for the analysis of \cite{Calabrese:2009ez}.

The R\'enyi mutual information (\ref{renyi-mutual-def}) 
in the chiral current model (see Sec.\,\ref{sec-chiral-current}) reads \cite{Arias:2018tmw} 
\begin{equation}
\label{eq:chiral_renyi_mutual_info}
     I^{(n) }_{A_1,A_2; \textrm{\tiny (cc)} }
    =
    -\frac{n+1}{12\,n}\log (1-\eta)+U_n(\eta)
\end{equation}
where 
\begin{equation}
\label{eq:Un_eta}
U_n(\eta) \equiv  
\frac{\textrm{i} \, n }{2(n-1)} 
\int_0^{+\infty} 
\big[ \operatorname{coth}(\pi s n)-\operatorname{coth}(\pi s) \, \big]
\,\log \! \left(\frac{{ }_2 F_1(1+i s,-i s ; 1 ; \eta)}{{ }_2 F_1(1-i s, i s ; 1 ; \eta)}\right)
\rd s  \,. 
\end{equation}
The analytic continuation $n\rightarrow 1$ of \eqref{eq:chiral_renyi_mutual_info} 
provides the mutual information for this model
\begin{equation}
\label{eq:chiral_mutual_info_n1}
     I_{A_1,A_2; \textrm{\tiny (cc)} }
    =
    -\frac{1}{6}\log(1-\eta) +U(\eta)
\end{equation}
where 
 \begin{equation}
 \label{eq:U_eta}
  U(\eta)\equiv 
  \lim_{n\rightarrow 1} U_n(\eta)
  =
  -\frac{\textrm{i}\, \pi}{2} \int_0^{+\infty} \! \frac{s}{[ \sinh(\pi s)]^2} \;
  \log \!\left(\frac{{ }_2 F_1(1+i s,-i s ; 1 ; \eta)}{{ }_2 F_1(1-i s, i s ; 1 ; \eta)}\right)
  \rd s  \,. 
\end{equation}

\begin{figure}[t!]
	\vspace{-.5cm}
	\hspace{-1cm}
    \includegraphics[width=1.07\textwidth]{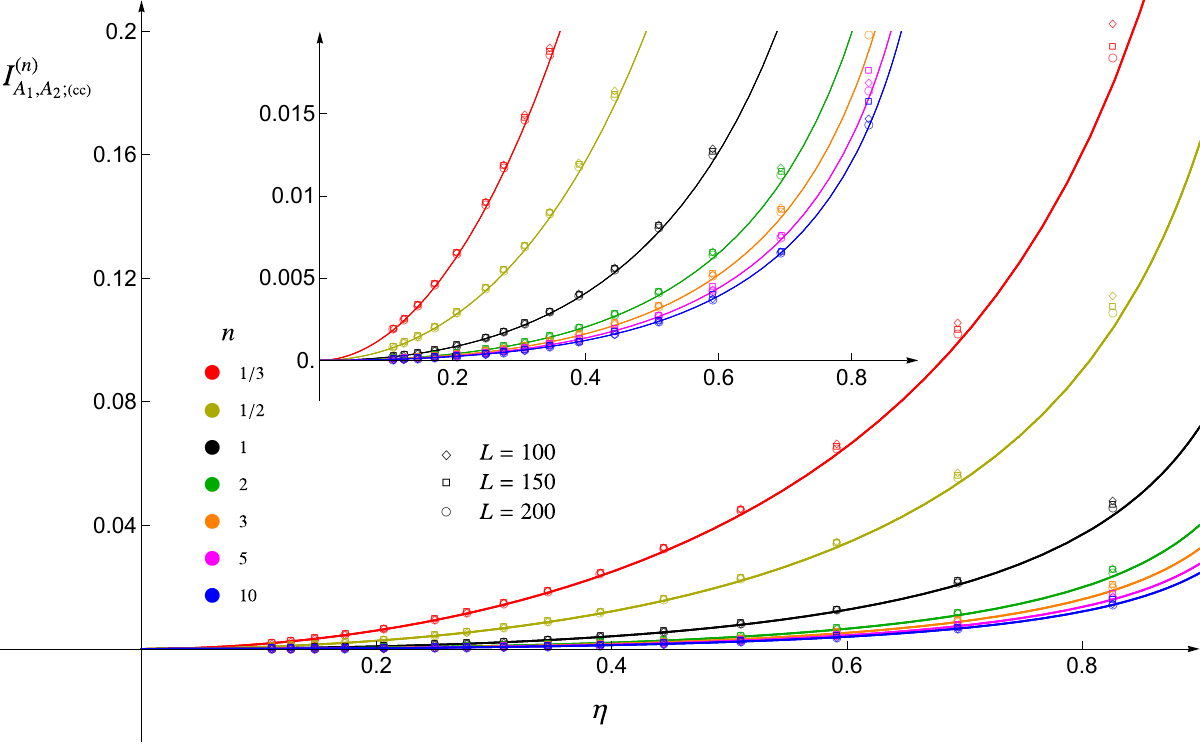}
    	\vspace{-.2cm}
    \caption{R\'enyi mutual information of equal blocks of length $L$ for the chiral current. 
    }
    \label{fig:CH_mutualinformation-current}
\end{figure}

In Fig.\,\ref{fig:CH_mutualinformation-current} 
we consider the Rényi mutual information for various values of $n$, including the $n \to 1$ case corresponding to the mutual information, 
by comparing the numerical data points obtained from the lattice model 
(by using \eqref{renyi-mutual-def}, \eqref{renyi-from-ss} and (\ref{EE-from-ss}), as discussed in Appendix\;\ref{app:chiral})
against the analytic result in the continuum \cite{Arias:2018tmw}
given by \eqref{eq:chiral_renyi_mutual_info} and \eqref{eq:chiral_mutual_info_n1}.
The lattice results have been divided by a factor of $2$, 
in order to take into account the fact that the continuum limit 
of the lattice model \eqref{eq:chiral_lattice_H}
is described by two chiral currents (see Appendix\;\ref{app:chiral}).
A good agreement between the lattice data points as $L$ increases and  
the corresponding analytic expressions in the continuum is observed.


It is straightforward to realise that (\ref{Dn-def}) and (\ref{eq:Un_eta}) are related as follows
\begin{equation}
\label{Un-Dn-relation}
    U_n(\eta)=\frac{D_n(\eta)}{2(1-n)}
\end{equation}
that can be checked numerically.
The special case of (\ref{Un-Dn-relation}) corresponding to $n \to 1$, namely $U(\eta)=D'_1(\eta)/2$,
can be easily shown by performing the change of variable $s = \textrm{i}z$.
This relation, combined with (\ref{eq:full_mutual_information_n1}) and (\ref{eq:chiral_mutual_info_n1}),
leads to the following relation between the mutual information of these two models
\begin{equation}
\label{eq:diff_information}
I_{A_1,A_2; \textrm{\tiny (ms)} } - I_{A_1,A_2; \textrm{\tiny (cc)} }
= - \frac{1}{6}\log(1-\eta)+U(1-\eta)  \,. 
\end{equation}
Considering the r.h.s.'s of (\ref{eq:chiral_renyi_mutual_info}) and (\ref{eq:chiral_mutual_info_n1}),
since the functions $U_n(1-\eta)$ and $U(1-\eta)$ are not invariant under the exchange $\eta \leftrightarrow 1-\eta$,
the chiral current model does not satisfy Haag duality, as discussed in  \cite{Arias:2018tmw}.


\bibliographystyle{nb}

\bibliography{refsEH2int}

\end{document}